\documentclass[twocolumn,preprintnumbers,pre,floatfix,superscriptaddress,amsmath,amssymb,notitlepage]{revtex4-1}

\usepackage{subfigure}
\usepackage{amsmath}
\usepackage{color}
\usepackage{amssymb}
\usepackage{setspace}
\usepackage[pdftex]{graphicx}% Include figure files
\usepackage{dcolumn}% Align table columns on decimal point
\usepackage{bm}% bold mathe
\usepackage{times}
\usepackage{amssymb,amsfonts,amsmath}
\usepackage{cite}

\newcommand{\lsim}{\lower.7ex\hbox{$\;\stackrel{\textstyle<}{\sim}\;$}}

\begin{document}
%
%\author{Dane Taylor}
%\email{dane.r.taylor@gmail.com}
%\affiliation{Statistical and Applied Mathematical Sciences Institute, Research Triangle Park, NC, 27709, USA}
%\affiliation{Carolina Center for Interdisciplinary Applied Mathematics, Department of Mathematics, University of North Carolina at Chapel Hill, NC, 27599, USA}
%\author{Florian Klimm}
%\affiliation{Potsdam Institute for Climate Impact Research, 14473 Potsdam, Germany}
%\affiliation{Department of Physics, Humboldt-Universit\"at zu Berlin, 12489 Berlin, Germany}
%\affiliation{Mathematical Institute, University of Oxford, OX2 6GG, UK}
%\author{Heather A. Harrington}
%\affiliation{Mathematical Institute, University of Oxford, OX2 6GG, UK}
%\author{Miroslav Kram\'ar}
%\affiliation{Department of Mathematics, Rutgers, The State University of New Jersey, Piscataway, NJ,  08854, USA}
%\author{Konstantin Mischaikow}
%\affiliation{Department of Mathematics, Rutgers, The State University of New Jersey, Piscataway, NJ,  08854, USA}
%\affiliation{BioMaPS Institute, Rutgers, The State University of New Jersey, Piscataway, NJ,  08854, USA}
%\author{Mason A. Porter}
%\affiliation{Mathematical Institute, University of Oxford, OX2 6GG, UK}
%\affiliation{CABDyN Complexity Centre, University of Oxford, Oxford OX1 1HP, UK}
%\author{Peter J. Mucha}
%\affiliation{Carolina Center for Interdisciplinary Applied Mathematics, Department of Mathematics, University of North Carolina at Chapel Hill, NC, 27599, USA}

\title{Topological data analysis of contagion maps for examining spreading processes on networks}

\author{Dane Taylor$^{1,2}$, Florian Klimm$^{3,4,5}$, Heather A. Harrington$^{5}$, Miroslav Kram\'ar$^{6}$, \\
Konstantin Mischaikow$^{6,7}$, Mason A. Porter$^{5,8}$, and Peter J. Mucha$^{2}$}

\maketitle

{
\noindent$^1${Statistical and Applied Mathematical Sciences Institute, Research Triangle Park, NC, 27709, USA}\\
$^2${Carolina Center for Interdisciplinary Applied Mathematics, Department of Mathematics, University of North Carolina at Chapel Hill, NC, 27599, USA}\\
$^3${Potsdam Institute for Climate Impact Research, 14473 Potsdam, Germany}\\
$^4${Department of Physics, Humboldt-Universit\"at zu Berlin, 12489 Berlin, Germany}\\
$^5${Mathematical Institute, University of Oxford, OX2 6GG, UK}\\
$^6${Department of Mathematics, Rutgers, The State University of New Jersey, Piscataway, NJ,  08854, USA}\\
$^7${BioMaPS Institute, Rutgers, The State University of New Jersey, Piscataway, NJ,  08854, USA}\\
$^8${CABDyN Complexity Centre, University of Oxford, Oxford OX1 1HP, UK}\\
%$^9${Department of Applied Physical Sciences, University of North Carolina at Chapel Hill, NC 27599, USA}\\
}

\begin{abstract}

Social and biological contagions are influenced by the spatial embeddedness of networks. Historically, many epidemics spread as a wave across part of the Earth's surface; however, in modern contagions long-range edges---for example, due to airline transportation or communication media---allow clusters of a contagion to appear in distant locations. Here we study the spread of contagions on networks through a methodology grounded in topological data analysis and nonlinear dimension reduction. We construct ``contagion maps'' that use multiple contagions on a network to map the nodes as a point cloud. By analyzing the topology, geometry, and dimensionality of manifold structure in such point clouds, we reveal insights to aid in the modeling, forecast, and control of spreading processes. Our approach highlights contagion maps also as a viable tool for inferring low-dimensional structure in networks.

\end{abstract}

%\title{Topological data analysis of contagion maps for examining spreading processes on networks} 

%\maketitle

%~\clearpage

%%%
\clearpage

Considerable research during the past few decades has aimed to understand spreading dynamics on networks \cite{Barrat2008,Easley2010,Mollison1977,porter2014,review2014}---a widespread phenomenon that occurs in diverse settings that range from biological epidemics \cite{Brockmann2013,Colizza2006,Hufnagel2004} to collective social processes such as social movements \cite{Hedstrom1994} and innovation diffusion \cite{Rogers2003}. To study spreading, it is useful to contrast two classes of networks: ``geometric networks,'' in which nodes lie in a metric space and are connected by short-range ``geometric edges'' that are constrained by the nodes' locations (e.g., lattices that describe discretized partial differential equations \cite{Chong2008}), and networks that are not geometric, in the sense that their edges are not constrained or defined by distances between nodes. Although the embedding of nodes in a metric space is ubiquitous for spatial networks on Earth's surface \cite{Barthelemy2011}, recent studies have explored the mapping of nodes in a network to locations in a (potentially) latent and (typically) low-dimensional metric space for an extensive variety of applications.  Such applications {include inferring} missing and spurious edges in networks \cite{Liben2007,Clauset2008,Guimera2009,Boguna2012}; 
efficiently routing information across the internet \cite{Boguna2010,Serrano2008}; 
identifying node-specific attributes that are responsible for edge formation in social networks \cite{Hoff2002}; 
and nonlinear dimension reduction of proximity networks inferred from point-cloud data (e.g., images, videos, and time series) for data-storage and signal-processing applications \cite{Singer2011,Singer2013,Tenenbaum2000,Belkin2003,Coifman2005,Gerber2007,Sorzano2014,Lafon2006}.  

\begin{figure}[h]
\centering
\includegraphics[width=.65\linewidth]{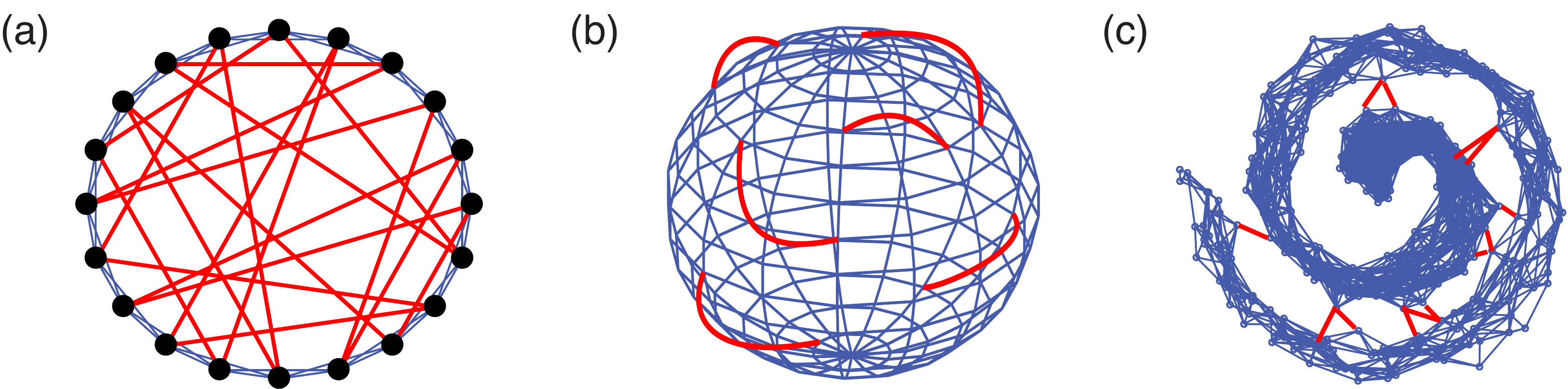} 
\vskip -5pt
\caption{ 
{\emph{Examples of noisy geometric networks.}} Nodes are embedded in three manifolds: (a)~a ring (1D) embedded as a circle in $\mathbb{R}^2$; (b)~a spherical surface (2D) in $\mathbb{R}^3$; and (c)~a bounded plane (2D) embedded (nonlinearly) in $\mathbb{R}^3$ in a configuration known as the ``swiss roll'' \cite{Tenenbaum2000}. {Given 
a network with ``geometric edges'' (blue), in panels (a)--(b), we add ``non-geometric edges''} (red) uniformly at random. {In panel (c), by contrast,} we add noise to the nodes' locations in the ambient space and place edges between nodes that are nearby in {that} space. In this {scenario}, we interpret edges between nodes that are nearby with respect to the ambient space but not the manifold as the non-geometric edges.
}
\label{fig:noisy_geometric}
\vskip -5pt
\end{figure}

\begin{figure}[t!]
\centering
\includegraphics[width=.7\linewidth]{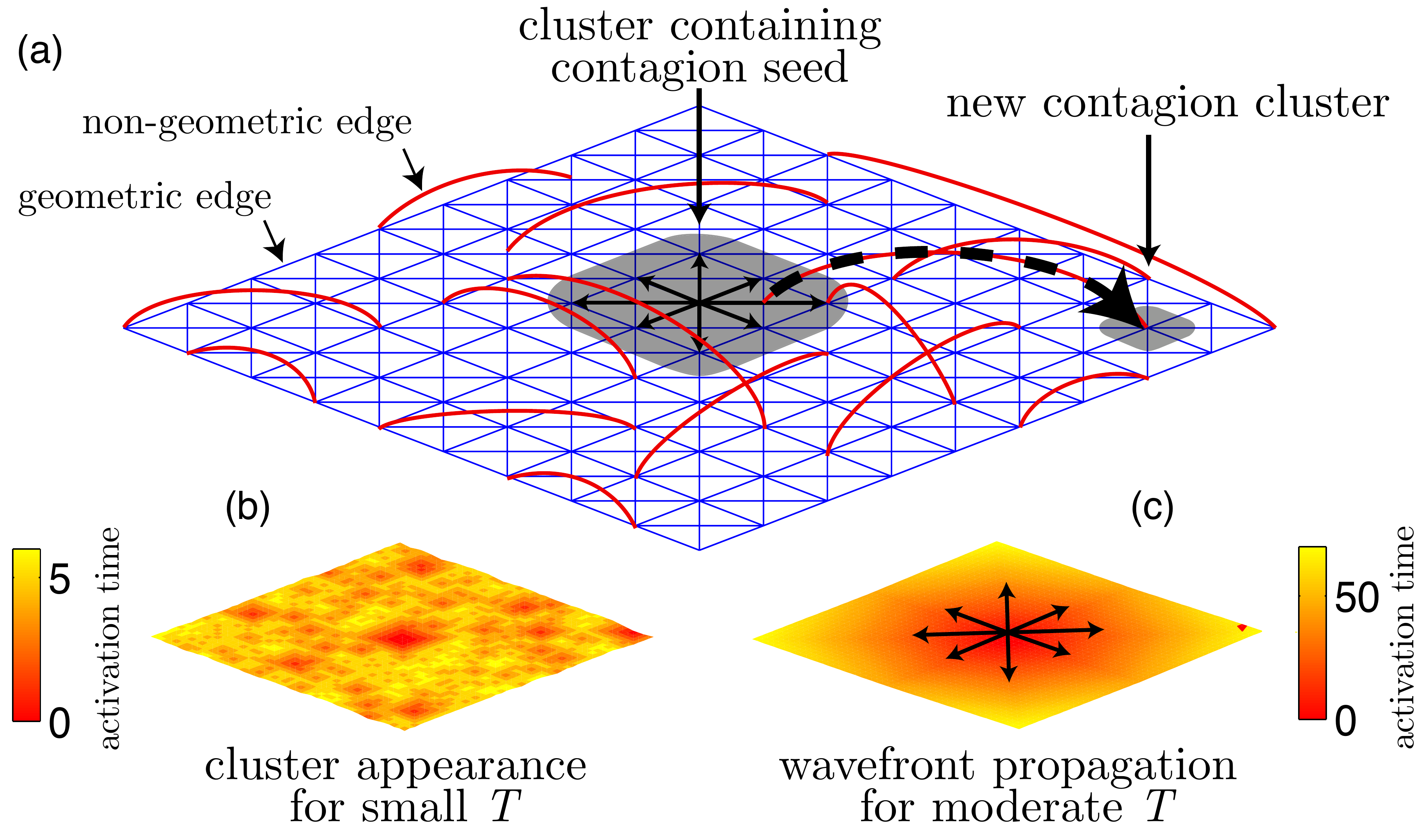}
\vskip -5pt
\caption{\emph{Wavefront propagation and the appearance of new clusters.}
(a) Contagions on a noisy geometric network containing geometric edges along a manifold (in this case, a two-dimensional {lattice, which we indicate with the blue edges}) and non-geometric edges {(red edges), which} introduce shortcuts in the network. We study two phenomena in the evolution of contagion clusters (shaded areas): {``wavefront propagation''} (WFP) describes the outward expansion of a contagion cluster's boundary, and the {``appearance of a new contagion cluster''} (ANC) occurs when a contagion spreads exclusively along non-geometric edges (dashed arrow). 
{(b,c)} We {examine} WFP and ANC {for the Watts threshold model (WTM) \cite{Watts2002} for complex contagions by studying} node activation times (i.e., the times at which nodes adopt the contagion), {which depend on the WTM thresholds $\{T_i\}$, which we {take to} be identical for every node (i.e., $T_i=T$ for all $i$).} (b) For small $T$, frequent ANC leads to rapid dissemination of a contagion. (c) For moderate $T$, little to no ANC occurs and WFP leads to slow dissemination. For large $T$, there is no spreading. For a given network, activation times across multiple realizations of a contagion (with varying initial conditions) map the nodes to a point cloud via what we call a {``WTM map''.} 
}
\label{fig:2Dcontagion}
\end{figure}

When dynamics such as contagions occur on a geometrically-embedded network, it is fundamental to question the extent to which the dynamics follows such underlying low-dimensional structure. This question is particularly important and difficult for geometric networks that are supplemented with long-range ``non-geometric edges,'' which directly connect nodes that are distant from each other with respect to an underlying metric space. Long-range edges arise in numerous applications, either by chance (e.g., subways that connect distant parts of cities) \cite{Barthelemy2011} {or} as a result of merging distinct layers in multilayer networks \cite{Kivela2014}. {In some scenarios, they can also be construed as a source of ``noise'' in an otherwise geometric network (e.g.,} when edges arise due to the presence of noise for {inferred proximity networks} \cite{Gerber2007,Sorzano2014}). They also play important roles in small-world network models \cite{Porter2012} such as Watts-Strogatz \cite{Watts1998}, Newman-Watts \cite{Newman2000}, and Kleinberg \cite{Kleinberg2000} networks. Because we are interested in the geometric embeddedness of such networks, we use the term ``noisy geometric networks'' for networks that include non-geometric edges as supplements to geometric edges. {(See Figs.~\ref{fig:noisy_geometric}--\ref{fig:2Dcontagion} for examples).}

The presence of long-range edges can significantly alter how processes spread \cite{mona1999,Watts1998,Newman2000,Kleinberg2000}. For example, it is traditional to characterize contagions in a geometric setting using {``wavefront propagation'' (WFP)} \cite{Mollison1977}, which agrees with the qualitative properties of historical epidemics such as the Black Death \cite{Noble1974}. By contrast, Refs.~\cite{Brockmann2013,Hufnagel2004,Colizza2006} (and numerous other sources \cite{Barrat2008}) have highlighted that modern biological epidemics tend to be dominated by long-range transportation networks, such as airline networks or railway networks, rather than by geographic proximity. Spreading across long-range edges can result in {the ``appearance of new clusters'' (ANC) of a contagion that are} spatially-distant, which is an important phenomenon in the dynamics of {recent} global epidemics \cite{Marvel2013}.  Indeed, it has been reported that prominent strains of influenza (e.g., H1N1/09) exhibited a pattern of ``skip-and-resurgence'' (in which some countries avoided outbreaks in some years) during recent worldwide outbreaks \cite{He2014}. In addition, long-range edges can also have significant effects on social contagions \cite{Centola2010,Centola2007,Centola2007b,Ghasemiesfeh2013}. Given the (either implicit or explicit) geometric embeddedness for so many of the networks on which ideas and diseases spread \cite{Barthelemy2011,Barrat2008}, an improved understanding of contagions on noisy geometric networks is important for numerous applications, which range from the identification of influential spreaders of information \cite{Kempe2003} to control of biological epidemics \cite{Pastor2002,Rhodes1997}.

{WFP and ANC can be very different in social versus biological contagions. One important difference arises from phenomena such as social reinforcement \cite{Centola2010,Centola2007,Centola2007b,Ghasemiesfeh2013}, which occurs only for social contagions.} {In Fig.~\ref{fig:2Dcontagion}, we illustrate the prominent effect of social reinforcement for the Watts threshold model (WTM) \cite{Watts2002} of social contagions. The WTM is} {a} generalization of bootstrap percolation \cite{Aizenman1988} { and is} based on the idea that each node $i$ has some threshold $T_i\ge0$ \cite{Granovetter1978} for adopting a social contagion (i.e., for becoming infected). The threshold dynamics gives rise to the characterization of the WTM as a so-called ``complex'' contagion because the dynamics at each node $i$ depend on the states of all neighboring nodes, and it might be necessary for multiple neighbors to be infected before node $i$ adopts a contagion. Importantly, for some threshold values, WFP can dominate ANC even in the presence of many ``noisy'' edges---a phenomenon that has widespread applications (see Sec.~\ref{sec:discuss} and our Supplementary Discussion).

In the present paper, we study bifurcations in {WFP and ANC dynamics by examining data} that are generated by several contagions on a given noisy geometric network. Our methodology is grounded in the field of computational topology \cite{Kaczynski2004,edels2010}, and we note that there has been rapidly intensifying interest (see, e.g., \cite{Petri2012,Petri2013,Kahle2013,Bobrowski2014}) in using tools from computational topology to study {structural features} in networks and for machine learning \cite{Machine}. In taking this perspective, we introduce a map from the network nodes to points in a metric space based on contagion dynamics. By analogy to diffusion maps \cite{Coifman2005} and similar ideas in nonlinear dimension reduction {and manifold learning} \cite{Singer2011,Singer2013,Tenenbaum2000,Belkin2003,Coifman2005,Gerber2007,Sorzano2014,Lafon2006}, we use the term {``contagion maps''} for these maps. 
We investigate the topology, geometry, and dimensionality properties of these maps, and we find for the contagion regime that predominantly exhibits WFP versus ANC that these properties correspond to the manifold that underlies the noisy geometric network. We examine both synthetic and empirical networks, including a transit system in London {(see Sec.~\ref{sec:london}, Supplementary Note 1, and Supplementary Figs.~1--5).} 
Given that the manifold structure in a contagion map can reflect the underlying manifold structure of a noisy geometric network, contagion maps also help for the identification {of} such underlying structure. This has numerous applications, including for the denoising of networks {(see Supplementary Note 2 and Supplementary Figs. 6--7)}.

%%%%%%%%%%%%%%%%%%%%%%%%%%%%%%%%%%%%%%%%%%%%%%%%%%%
\section{Results}
%%%%%%%%%%%%%%%%%%%%%%%%%%%%%%%%%%%%%%%%%%%%%%%%%%%

%%%%%%%%%%%%%%%%%%%%%%%%%%%%%%%%%%%%%%%%%%%%%%%%%%%
\subsection{Noisy Geometric Networks}\label{sec:NGN}
%%%%%%%%%%%%%%%%%%%%%%%%%%%%%%%%%%%%%%%%%%%%%%%%%%%

Noisy geometric networks are a class of networks that arise from geometric networks \cite{Barthelemy2011} but also include non-geometric, ``noisy'' edges. Consider a set $\mathcal{V}$ of network nodes that have intrinsic locations $\{\bold w^{(i)}\}_{i\in\mathcal{V}}$ in a metric space. We restrict our attention to nodes that lie on a manifold $\mathcal{M}$ that is embedded in an ambient space $\mathcal{A}$ (i.e., $\bold w^{(i)}\in\mathcal{M}\subset \mathcal{A}$). We use the term ``node-to-node distance'' to refer to the distance between nodes in this embedding space $\mathcal{A}$, which we equip with the Euclidean norm $\|\cdot\|_2$ (although one can also use other metrics \cite{Boguna2010}). To create a noisy geometric network, we place the nodes in the underlying manifold and add two families of edges: (1) a set $\mathcal{E}^{\rm{(G)}}$ of {\it geometric edges}, such that $(i,j)\in\mathcal{E}^{\rm{(G)}}$ {when} nodes $i$ and $j$ are sufficiently close to one another {(i.e., the length of shortest path along the manifold $\mathcal{M}$ that connects the two nodes is less than some distance threshold);} and (2) a set $\mathcal{E}^{\rm{(NG)}}$ of {\it non-geometric edges}, which we place using some random process between pairs of nodes $(i,j)$, where $i\not=j$ and $(i,j)\not\in \mathcal{E}^{\rm{(G)}}$. 
In Figs.~\ref{fig:noisy_geometric}(a,b), we show examples of constructing noisy geometric networks by adding non-geometric edges uniformly at random. In Fig.~\ref{fig:noisy_geometric}(c), we show a construction that is motivated by nonlinear dimension reduction of point-cloud data \cite{Tenenbaum2000,Belkin2003,Coifman2005,Gerber2007,Sorzano2014}.

As an illustrative example, consider the noisy ring lattice in Fig.~\ref{fig:noisy_geometric}(a), which is similar to the Newman-Watts variant of the Watts-Strogatz small-world model \cite{Watts1998,Newman2000}. Specifically, we consider $N$ nodes that are uniformly spaced along the unit circle in $\mathbb{R}^2$. We then add geometric edges so that every node $i$ is connected to its $d^{\rm{(G)}}$ nearest-neighbor nodes. (Note that there are no self-edges.) We then add $d^{\rm{(NG)}}$ non-geometric edges to each node and connect the ends of {these} edges (i.e., the stubs) uniformly at random while avoiding self-edges and multi-edges. The resulting network is a $(d^{\rm{(G)}}+d^{\rm{(NG)}})$-regular network that contains $Nd^{\rm{(G)}}/2$ geometric edges and $Nd^{\rm{(NG)}}/2$ non-geometric edges. We can thus specify {this class of random networks} using three parameters: $N$, $d^{\rm{(G)}}$, and $d^{\rm{(NG)}}$. {It is also useful to define the ratio {$\alpha=d^{\rm{(NG)}}/d^{\rm{(G)}}$} of non-geometric to geometric edges.} Our construction assumes that $N$ and $d^{\rm{(G)}}$ are even. In Fig.~\ref{fig:noisy_geometric}(a), we depict a noisy ring network with $N=20$ and $(d^{\rm{(G)}},d^{\rm{(NG)}})=(4,2)$. In {Supplementary Note~3} {(see also Supplementary Figs.~8--9)}, we {study models} of noisy geometric networks on a ring manifold that incorporate heterogeneity in the nodes' degrees and/or locations.

%%%%%%%%%%%%%%%%%%%%%%%%%%%%%%%%%%%%%%%%%%%%%%%%%%%
\subsection{Watts Threshold Model (WTM)}\label{sec:WTM}
%%%%%%%%%%%%%%%%%%%%%%%%%%%%%%%%%%%%%%%%%%%%%%%%%%%

We analyze a well-known dynamical system for social contagions: the Watts threshold model (WTM) for complex contagions \cite{Watts2002}. In addition to allowing analytical tractability, we have two other motivations for using the WTM. First, WTM contagions yield {``filtrations''} of a network and thereby allow us to develop a methodology grounded in computational topology \cite{Kaczynski2004,edels2010,Petri2012,Petri2013,Kahle2013,Bobrowski2014}. 
Second, the WTM is a simple-but-insightful model for social influence that has the virtue of explicitly considering social reinforcement  \cite{Centola2010,Centola2007,Centola2007b}. 

We define a WTM contagion as follows. Given an unweighted network (which we represent using an adjacency matrix $A$) with a set $\mathcal{V}$ of nodes and a set $\mathcal{E}$ of edges, we let $\eta_i(t)$ denote the state of node $i\in\mathcal{V}$ at time $t$, where $\eta_i(t)=1$ indicates adoption (i.e., infection) and $\eta_i(t)=0$ indicates non-adoption. We initialize a contagion at time $t=0$ by choosing a set of nodes $\mathcal{S}\subset\mathcal{V}$ and setting $\eta_i(0)=1 $ for $i\in \mathcal{S}$ and $\eta_i(0)=0$ for all other nodes. We refer to $\mathcal{S}$ as the ``contagion seed.'' We consider synchronous updating in discrete time \cite{porter2014}, so a node $i$ that has not already adopted the contagion at time $t$ [i.e., $\eta_i(t)=0$] will adopt it during the next time step [i.e., $\eta_i(t+1)=1$] if and only if $f_i> T_i$, {where $T_i$ is a node-specific adoption threshold and} $f_i=d_i^{-1}\sum_j A_{ij}\eta_j(t)$ denotes the fraction of neighbors that are infected and $d_i=\sum_jA_{ij}$ is the degree of node $i$. {(Note that this is a slight modification from the original WTM \cite{Watts2002}, which uses the adoption criterion $f_i\geq T_i$.)} We repeat this process until the system reaches an equilibrium point at some time $t^*< N$ (i.e., no further adoptions occur). For each node $i$, we let $x^{(i)}$ denote the node's ``activation time,'' which is the time $t$ at which the node adopts the contagion. {Given $\{T_i\}$ and the contagion seed, a WTM contagion on a network is a deterministic process. Additionally,} a node's adoption of the contagion is irreversible (i.e., there is no unadoption in this model), so the dynamics are monotonic in the sense that the subset $\mathcal{I}(t)\subseteq\mathcal{V}$ of infected nodes at time $t$ is non-decreasing with time [i.e., $\mathcal{I}(t)\subseteq \mathcal{I}(t+1)$]. One can thus use the contagion to construct a ``filtration'' of the network nodes $\mathcal{V}$. (See Refs.~\cite{Kaczynski2004,edels2010} and our discussion in {Supplementary Note~4}.)

%%%%%%%%%%%%%%%%%%%%%%%%%%%%%%%%%%%%%%%%%%%%%%%%%%%
\subsection{Contagion Maps}\label{WTMmap}
%%%%%%%%%%%%%%%%%%%%%%%%%%%%%%%%%%%%%%%%%%%%%%%%%%%

We study contagion maps based on WTM contagions, and we refer to these maps as ``WTM maps.'' A WTM map is a nonlinear map of nodes in a network to a point cloud in a metric space based on the activation times from several realizations of a WTM contagion. Given $J$ realizations of a WTM contagion on a network with different initial conditions, the associated WTM map is a function from $\mathcal V$ to $\mathbb{R}^J$ that records the activation time $x^{(i)}_j$ of the $i$th node in the $j$th realization.  More precisely, we define a ``regular'' WTM map as $\mathcal{V}\mapsto \{\bold x^{(i)}\}_{i\in\mathcal{V}}\in \mathbb{R}^J$, where $\bold x^{(i)} = [x^{(i)}_1,x^{(i)}_2,\dots,x^{(i)}_J]^T$.
In practice, we enumerate the contagions $j=1,2,\dots,J\le N$, and we initialize the $j$th contagion at a contagion seed $\mathcal{S}^{(j)}$ such that $\{j\}\subseteq\mathcal{S}^{(j)}$ for each $j$. 
(Note that one can select any $J$ nodes as seeds by relabeling the nodes.) In addition to the regular WTM map $\mathcal{V}\mapsto \{\bold x^{(i)}\}_{i\in\mathcal{V}}$, we also define ``reflected'' and ``symmetric'' versions of the WTM map for the subset of nodes $\mathcal{J}= \{1,2,\dots,J\}\subseteq \mathcal{V}$.
Letting $y_j^{(i)}=x_i^{(j)}$ and $z_j^{(i)}=x_j^{(i)} +x_i^{(j)}$, we define the reflected WTM map $\mathcal{J}\mapsto \{\bold y^{(i)} \}_{i\in\mathcal{J}} \in\mathbb{R}^N$ and symmetric WTM map $\mathcal{J}\mapsto \{\bold z^{(i)} \}_{i\in\mathcal{J}}\in\mathbb{R}^J$. For a given network and thresholds $\{T_i\}$, the regular, reflected, and symmetric WTM maps are deterministic.

\begin{figure}[t]
\centering
\includegraphics[width=.6\linewidth]{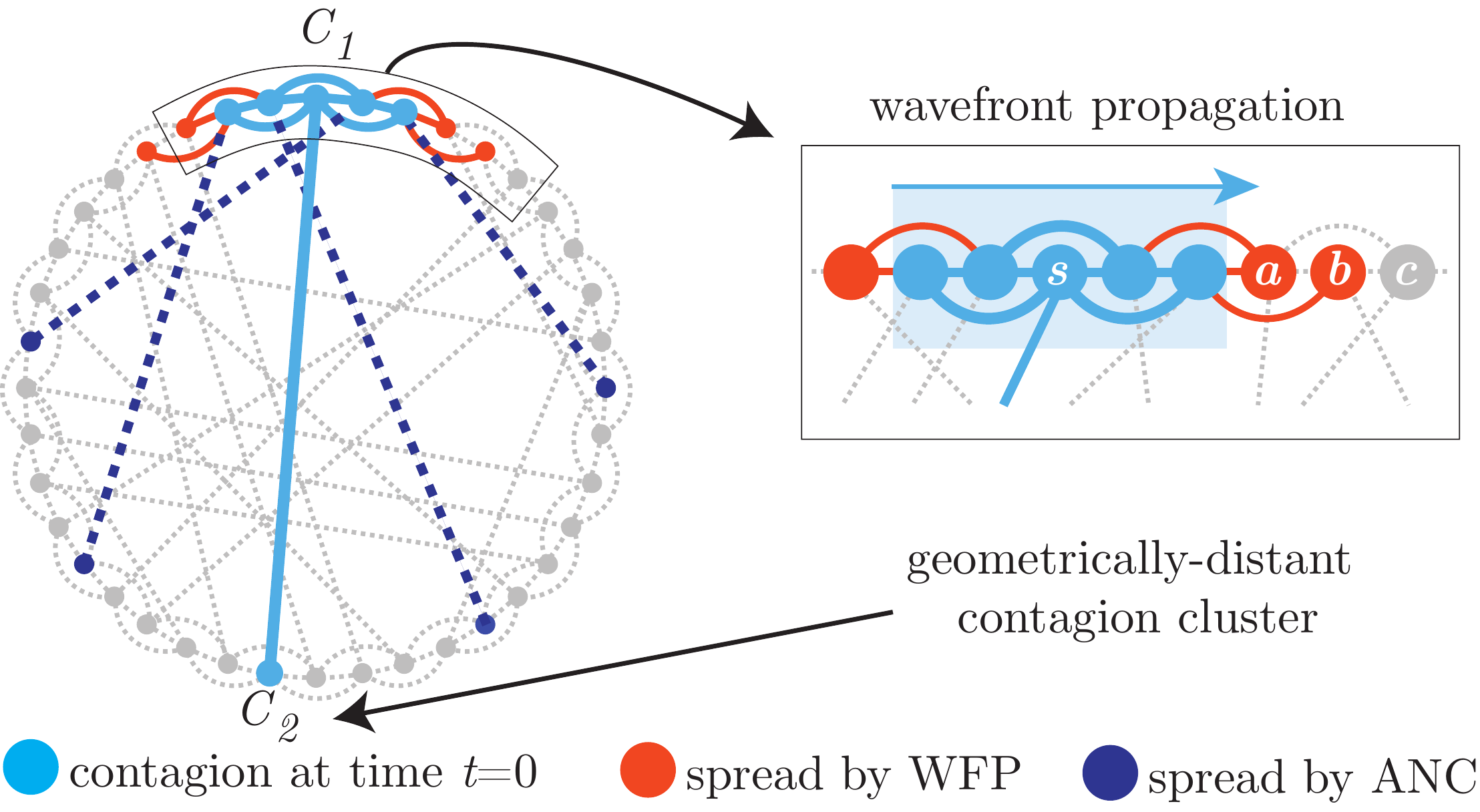}
\caption{ {\emph{Contagion initialized with cluster seeding.}}
A WTM contagion on a noisy ring lattice {in which each node has $d^{\rm{(G)}}=4$ geometric edges and $d^{\rm{(NG)}}=1$ non-geometric edge.} We initialize the contagion at time $t=0$ by setting node $s$ and its network neighbors as infected ({indicated by the light blue nodes and edges}). This results in two contagion clusters: $C_1$ and $C_2$. At time $t = 1$, {depending on the WTM thresholds $\{T_i\}$,} additional nodes can adopt the contagion either via WFP {and/or} via ANC. {As indicated by the orange nodes and edges, nodes that are in the ``boundary'' of $C_1$ can adopt the contagion via WFP traveling around the underlying ring lattice. {We illustrate this idea further in the magnifying box, where nodes $a$ and $b$ in the boundary of $C_1$ can potentially become infected} in the first time step. Alternatively, nodes that share only a non-geometric edge with {a contagion seed can potentially} become infected via ANC (as indicated by the dark blue nodes and dashed edges).} 
}
\label{fig:ring_1}
\end{figure}

The choice of contagion seeds $\{\mathcal{S}^{(j)}\}$ plays a crucial role in determining the dynamics of WTM contagions and a WTM map. In practice, we use $J=N$ realizations of a WTM contagion for an $N$-node network, for which we initialize the $j$th realization with a contagion seed $\mathcal{S}^{(j)}=\{j\}\cup\{k|A_{jk}\not=0\}$ that includes node $j$ and its network neighbors. We use the term ``cluster seeding'' to describe this type of initial condition, {which we illustrate in Fig.~\ref{fig:ring_1}.} By contrast, we use the term ``node seeding'' to refer to the initialization of a contagion at a single node: $\mathcal{S}^{(j)}=\{j\}$. Additionally, note that setting $J=N$ yields $\mathcal{J}=\mathcal{V}$, and then the complete set of nodes is mapped by all versions of a WTM map. {In Supplementary Note~5 {(see also Supplementary Fig.~10 and Supplementary Table 1)} we show that the typical computational complexity for constructing a WTM map is $\mathcal{O}(NM)$, where $M$ is the number of edges. {See Sec.~\ref{sec:code} for code, which we have made available, that constructs WTM maps.}}

We now motivate our choice for contagion initialization. The {requirement that $\{j\}\subseteq\mathcal{S}^{(j)}$} is convenient because it allows us to think of the activation time $x^{(i)}_j$ as a notion of distance from node $j$ to node $i$ (i.e., it describes the time that is required for a contagion to travel from node $j$ to $i${)}. This choice is akin to the diffusion distance \cite{Coifman2005} and commute-time distance \cite{Doyle1984} derived from diffusion dynamics (although the latter is known to have shortcomings for certain classes of networks \cite{Radl2009}). To illustrate this point, suppose that contagion seeds are individual nodes (i.e., $\mathcal{S}^{(j)}=\{j\}$ for $j\in\mathcal{V}$), and suppose that we construct the WTM map $\mathcal{V}\mapsto\{\tilde{x}^{(i)}\}_{i\in\mathcal{V}}$ for {$T_i=T=0$ for each node $i\in\mathcal{V}$}. In this case, the activation time $\tilde{x}^{(i)}_j=\tilde{x}^{(j)}_i$ exactly recovers the length of the shortest path between nodes $i$ and $j$, and this in turn defines a metric on the discrete space $\mathcal{V}$. In fact, the $N\times N$ matrix $\tilde X=[\tilde{\bold x}^{(1)},\dots,\tilde{\bold x}^{(N)}]$ is a dissimilarity matrix, which is central to many algorithms for dimension reduction \cite{Tenenbaum2000,Belkin2003,Coifman2005,Gerber2007,Sorzano2014} (including Isomap \cite{Tenenbaum2000}, which implements the mapping of nodes based on shortest paths). Letting $T_i>0$ and still assuming that each $\tilde x_j^{(i)}$ is finite, we show in {Supplementary Note~4} that the symmetric WTM map induces a metric on $\mathcal{V}$. More generally, we show that a set of ``filtrations'' induces a metric under certain conditions. Consequently, we find that one can also use topological data analysis of networks to study the {embedding} geometry of networks.

Although node seeding has wonderful mathematical properties, cluster seeding is very useful in practice because it can allow a contagion to infect a larger fraction of the nodes in a network. When $T_i>0$ for each $i\in\mathcal{V}$, it is common for WTM contagions to reach equilibria that do not saturate the network with a contagion. This implies that $x_j^{(i)}=\infty$ for some $i,j\in\mathcal{V}$. Activation times of infinity pose a problem, because WTM maps are well-defined only for activation times $x^{(i)}_j$ {that are finite} (see Sec.~\ref{sec:practical}). 
{Contagions initialized with clusters of a contagion are more likely to spread than those that are initialized at a single node \cite{Gleeson2007}, so cluster seeding increases the range of threshold choices that yield activation times that are finite. 
 Although WTM maps that we construct using cluster seeding no longer automatically induce a metric on the node set $\mathcal{V}$, one can still construe $x_j^{(i)}$ as a distance from node $j$ to $i$ if the contagion seeds are sufficiently small, $|\mathcal{S}^{(j)}| \ll |\mathcal{V}|$.}

%~\clearpage

%%%%%%%%%%%%%%%%%%%%%%%%%%%%%%%%%%%%%%%%%%%%%%%%%%%
\subsection{WTM Contagions on Noisy Ring Lattices}\label{sec:bif_results}
%%%%%%%%%%%%%%%%%%%%%%%%%%%%%%%%%%%%%%%%%%%%%%%%%%%

{To guide our experiments on using WTM maps to study {WFP and ANC} on noisy ring lattices, we conduct a bifurcation analysis for WTM contagions that are initialized with cluster seeding. We present our analysis in detail in Sec.~\ref{sec:bif} and Supplementary Note 6, and we summarize our results here. 

Our primary results are two sequences of critical values for the WTM threshold $T$ that depend on the non-geometric degree $d^{\textrm{(NG)}}$ and geometric degree $d^{\textrm{(G)}}$. These critical values determine the presence versus absence of WFP and ANC as well as their rates. 
The qualitative features of ANC behavior are determined by the thresholds
\begin{equation}
	T_k^{\mathrm{(ANC)}} \triangleq \frac{d^{\rm{(NG)}}-k}{d^{\rm{(G)}}+d^{\rm{(NG)}}}\,, \quad k=0,1,\dots,d^{\rm{(NG)}}\,.
	\label{eq:ANC_crits}
\end{equation}
Whenever $T \in \left[T_{k+1}^{\mathrm{(ANC)}},T_k^{\mathrm{(ANC)}}\right)$, a node requires at least $(d^{\textrm{(NG)}}-k)$ neighbors from non-geometric edges to be infected before it adopts the contagion. This subsequently determines the rate at which new clusters of contagion appear. For $T\ge T_0^{\mathrm{(ANC)}}$, there is no ANC. 
The qualitative features of WFP are determined by the thresholds
\begin{equation}
	T_k^{\mathrm{(WFP)}} \triangleq \frac{d^{\rm{(G)}}/2-k}{{d^{\rm{(G)}}+d^{\rm{(NG)}}}}\,, \quad k=0,1,\dots,\frac{d^{\rm{(G)}}}{2}\label{eq:WFP_crits},
\end{equation}
where a wavefront propagates at a speed of {$k+1$} nodes per time step for $T \in \left[T_{k+1}^{\mathrm{(WFP)}},T_k^{\mathrm{(WFP)}}\right)$. For $T\ge T_0^{\mathrm{(WFP)}}$, there is no WFP. 
} 

\begin{figure}[t]
\centering
\includegraphics[width=.65\linewidth]{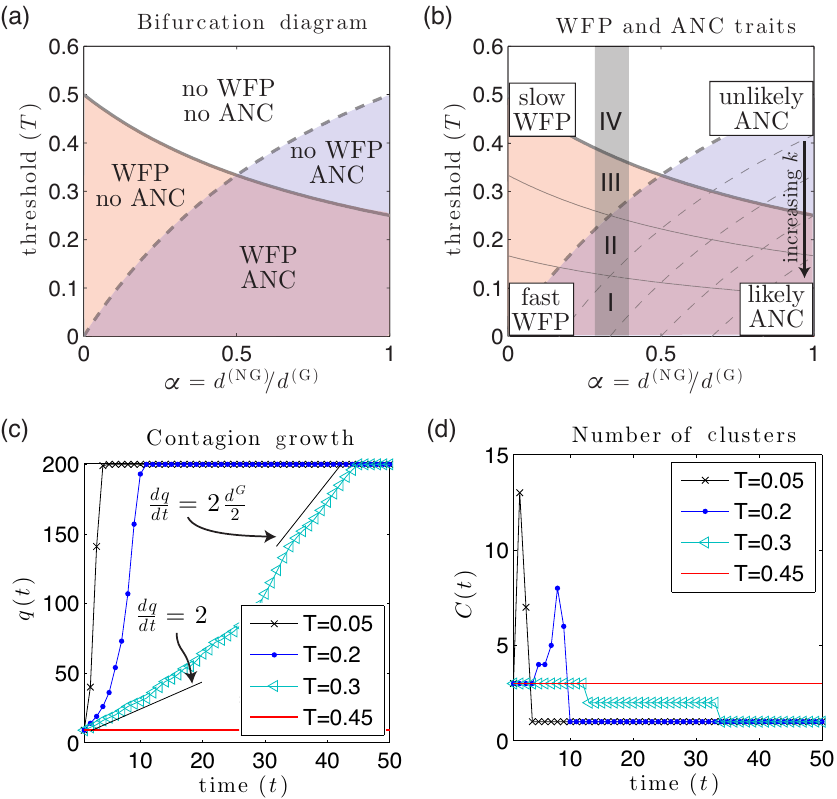}
\caption{ {\emph{Bifurcation analysis for WTM contagions on a noisy ring lattice.}}
(a) We plot the critical thresholds for {$k=0$ given by Eq.~\eqref{eq:ANC_crits} (dashed curve) and Eq.~\eqref{eq:WFP_crits} (solid curve) versus the ratio $\alpha = d^{\rm{(NG)}}/d^{\rm{(G)}}$ of non-geometric to geometric edges.} These curves divide the parameter space into four qualitatively different contagion regimes, which we characterize by the presence versus absence of WFP and ANC.
(b)~Equations~\eqref{eq:ANC_crits} and \eqref{eq:WFP_crits} for other values of $k$ further describe WFP and ANC, {and we show them} for $d^{\rm{(G)}}=6$. Note that the curves become lower with increasing $k$. Fixing $(d^{\rm{(G)}},d^{\rm{(NG)}})=(6,2)$, which yields $\alpha=1/3$, we find four contagion regimes (which we label using the symbols I--IV), where increasing $T$ corresponds to slower WFP and less frequent ANC. 
(c)~For $N=200$ and $T\in\{0.05,0.2,0.3,0.45\}$, we plot the contagion size $q(t)$ versus time $t$ for one realization of a WTM contagion with cluster seeding [i.e., $q(0)=1+d^{\rm{(G)}}+d^{\rm{(NG)}}=9$]. We observe, as expected, that the growth rate decreases with $T$. In particular, for regime III (e.g., $T=0.3$), the contagion spreads strictly via WFP, which initially spreads at a rate of $1$ node per time step (both clockwise and counterclockwise along the ring) but eventually accelerates to $d^{\rm{(G)}}/2$ nodes per time step. 
{As we show using the labeled black lines, we predict and observe linear growth for $q(t)$ when the contagion spreads by WFP and no ANC and either $q(t)\approx 1 $ or $q(t)\approx N$. (See Sec.~\ref{sec:bif} and Supplementary Note 6.)}
(d)~We plot the number of contagion clusters $C(t)$ versus $t$. As expected, $C(t)$ only increases above its initial value of $C(0)=1+d^{\rm{(NG)}}=3$ for regimes I and II (for which $T<T_0^{\mathrm{(ANC)}})$. There is no spreading in regime IV. 
}
\label{fig:bifurcation_ring_lattice}
\end{figure}

In Fig.~\ref{fig:bifurcation_ring_lattice}(a), we show a bifurcation diagram that summarizes the WTM dynamics for various values of the contagion threshold $T$ and ratio $\alpha = d^{\rm{(NG)}}/d^{\rm{(G)}}$ of non-geometric edges to geometric edges. The solid and dashed curves, respectively, describe Eq.~\eqref{eq:ANC_crits} and Eq.~\eqref{eq:WFP_crits} for $k=0$.  That is, $T^{\mathrm{(WFP)}}_0={1}/({2+2\alpha})$ and $T^{\mathrm{(ANC)}}_0={\alpha}/({\alpha+1})$, which intersect at $(\alpha,T)=(1/2,1/3)$ and yield four regimes of contagion dynamics that we characterize by the presence versus absence of WFP and ANC. In Fig.~\ref{fig:bifurcation_ring_lattice}(b), we plot Eqs.~\eqref{eq:ANC_crits} and \eqref{eq:WFP_crits} with other $k$ values for $d^{\rm{(G)}}=6$, where we note that lower curves correspond to larger $k$. Observe that increasing $T$ for fixed $\alpha$ leads to slower WFP and less frequent ANC. In particular, for $(d^{\rm{(G)}},d^{\rm{(NG)}})=(6,2)$ (which implies that $\alpha=1/3$), we find four qualitatively different regimes of WFP and ANC traits (see the regions that we label I--IV). 

\clearpage

In Figs.~\ref{fig:bifurcation_ring_lattice}(c,d), we illustrate dynamics from these regimes by choosing $T\in\{0.05,0.2,0.3,0.45\}$ and plotting the size $q(t)$ of the contagion [see Fig.~\ref{fig:bifurcation_ring_lattice}(c)] and the number of contagion clusters $C(t)$ [see Fig.~\ref{fig:bifurcation_ring_lattice}(d)] versus time $t$. Note that the number $C(t)$ of contagion clusters is equal to the number of connected components in the subgraph of the original network that only includes infected nodes and geometric edges. The values of $q(t)$ and $C(t)$ that we determine numerically (for $N = 200$) agree with our analysis. For $T=0.05$, the WTM contagion saturates the network [i.e., $q(t)\to N$] very rapidly due in part to the appearance of many contagion clusters early in the contagion process. For $T=0.2$, the contagion saturates the network relatively rapidly due to the appearance of some new contagion clusters. For $T=0.3$, the contagion saturates the network slowly, as no new contagion clusters appear and the contagion spreads only via WFP. For $T=0.45$, the contagion does not saturate the network, as neither WFP nor ANC occur.

%~\clearpage

%%%%%%%%%%%%%%%%%%%%%%%%%%%%%%%%%%%%%%%%%%%%%%%%%%%
\subsection{{Analyzing WTM Maps {for} Noisy Ring Lattices}}\label{sec:num}
%%%%%%%%%%%%%%%%%%%%%%%%%%%%%%%%%%%%%%%%%%%%%%%%%%%

In this section, we analyze symmetric WTM maps $\mathcal{V}\mapsto\{\bold z^{(i)}\}$ for noisy ring lattices in several ways: geometrically, topologically, and in terms of dimensionality. Our point-cloud analytics identify parameter regimes in which characteristics of a network's underlying manifold also appear in the WTM maps. {This makes it possible to do manifold learning} and {to} assess the extent to which a contagion exhibits WFP {(along a network's underlying manifold) versus ANC. 

In Fig.~\ref{fig:pointcloud}, we study WTM maps for a noisy ring lattice with $N=200$ and $(d^{\rm{(G)}},d^{\rm{(NG)}})=(6,2)$. We give each node $i$ an intrinsic location $\bold w^{(i)}=[\cos(2\pi i/N),\sin(2\pi i/N)]^T$ on the unit circle $\mathcal{M}=\{(a,b)|a^2+b^2=1\}\subset\mathbb{R}^2$. In Fig.~\ref{fig:pointcloud}(a), we illustrate the point clouds $\{\bold z^{(i)}\}_{i\in\mathcal{V}}\in\mathbb{R}^N$ that result from WTM maps with thresholds of $T\in\{0.05,0.2,0.3,0.45\}$, which correspond to the four regimes of contagion dynamics that are predicted by Eqs.\ \eqref{eq:ANC_crits} and \eqref{eq:WFP_crits} {for $\alpha=1/3$}. [See labels I--IV in Fig.~\ref{fig:bifurcation_ring_lattice}(b).] To visualize the $N$-dimensional point clouds $\{\bold z^{(i)}\}_{i\in\mathcal{V}}$, we use principal component analysis (PCA) to project onto $\mathbb{R}^2$ \cite{Tenenbaum2000,Sorzano2014,Cox2010}. The color of each node at location $\bold w^{(i)}$ and point $\bold z^{(i)}$ reflects the activation time for node $i$ during one realization of the WTM contagion that we use to generate the WTM map. In particular, dark blue nodes (points) indicate the contagion seed under cluster seeding. Gray nodes (points) never adopt the contagion and thus have activation times that are infinite. For practical purposes, we set these activation times to be $2N$ rather than $\infty$. (See {Sec.~\ref{sec:practical}} for additional discussion.) Regime III is the regime for which the point cloud $\{\bold z^{(i)}\}$ appears to best resemble (up to rotation) the nodes' intrinsic locations $\{\bold w^{(i)}\}$. This is expected, as this regime corresponds to WFP and no ANC. (In other words, the contagion follows the network's underlying manifold $\mathcal{M}$.)

\begin{figure}[h!]
\centering
\includegraphics[width=\linewidth]{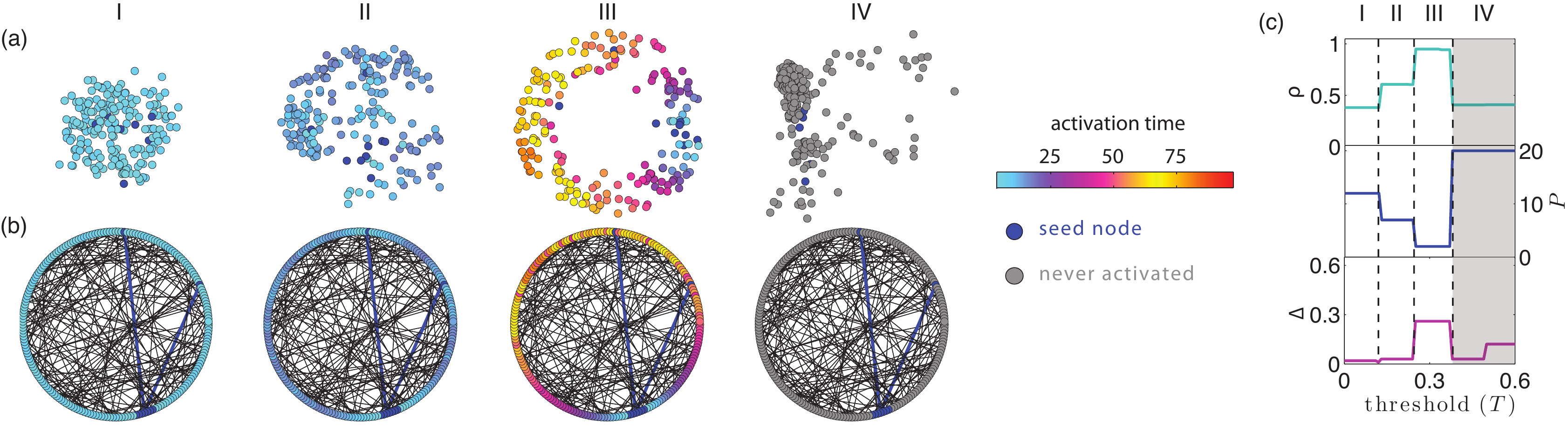}
\caption{ {\emph{Contagion maps applied to noisy ring lattices.}}
Symmetric WTM maps were applied to a noisy ring lattice with $N=200$ {and $(d^{\rm{(G)}},d^{\rm{(NG)}})=(6,2)$. }
(a) We show point clouds $\{\bold z^{(i)}\}\in\mathbb{R}^N$ for WTM maps with $T\in\{0.05,0.2,0.3,0.45\}$, which correspond, respectively, to regimes I--IV in Fig.~\ref{fig:bifurcation_ring_lattice}(b). For visualization purposes, we show two-dimensional projections of the $N$-dimensional point clouds after applying principal component analysis (PCA) \cite{Sorzano2014,Cox2010}. 
(b)~{We show} one realization of the contagion that we used to construct the WTM maps in panel (a). The color of each point in panel (a)---and corresponding node in panel (b)---indicates the node's activation time for one realization. Nodes in the contagion seed are dark blue, and nodes that never adopt the contagion are gray. 
(c)~{As {we discuss} in the text, we} analyze point clouds that result from WTM maps with respect to three criteria: 
geometry through a Pearson correlation coefficient $\rho$;
dimensionality through the embedding dimension $P$; and 
topology through the difference $\Delta$ of lifetimes. (See the main text as well as the Methods section.) The vertical dashed lines in panel (c) indicate the predicted bifurcations in contagion dynamics from Eqs.~\eqref{eq:ANC_crits} and \eqref{eq:WFP_crits} [see Fig.~\ref{fig:bifurcation_ring_lattice}(b)]. Note that there are activation times that are infinite for $T\ge T_0^{\rm{(WFP)}}= 3/8$ [shaded region in (c)]. As expected for regime III, {$\rho\approx1$, $P\approx2$, and large $\Delta$ indicate that} the geometry, dimensionality and topology of the point cloud recover those of {a} ring manifold $\mathcal{M}$.
{See Secs.~\ref{sec:methods2}, \ref{sec:methods3}, and \ref{sec:methods4} as well as Supplementary Note 7 for discussions of these approaches for analyzing point clouds.}
}
\label{fig:pointcloud}
\end{figure}

In Fig.~\ref{fig:pointcloud}(c), we summarize the characteristics of WTM maps for different thresholds $T\in[0,0.6]$. For each threshold, we analyze manifold structure in a point cloud by studying geometry {through} a Pearson correlation coefficient $\rho$; dimensionality through an approximate embedding dimension $P$; and topology through $\Delta$, which denotes the difference in lifetimes for the two most persistent $1$-cycles in a Vietoris-Rips filtration \cite{Kaczynski2004,edels2010}. Large values of $\Delta$ indicate the presence of a single dominant $1$-cycle (i.e., a ring) in a point cloud. 
See Secs.~\ref{sec:methods2}, \ref{sec:methods3}, and \ref{sec:methods4} as well as Supplementary Note~7 {and Supplementary Figs.~11--13} for additional discussion of our analysis of point clouds. %{For example, further details on the Vietoris-Rips filtration are provided in Supplementary Figs.~11-13.}

As expected by our analysis, for regime III (which exhibits WFP but no ANC), we identify characteristics of the manifold $\mathcal{M}$ in the point clouds that result from WTM maps. Namely, for regime III, the point cloud has similar geometry (i.e., indicated by large $\rho$), embedding dimension (i.e., indicated by $P=2$), and topology (i.e., indicated by large $\Delta$) as the network's underlying manifold $\mathcal{M}${(i.e., a ring)}. 

In Fig.~\ref{fig:correlations_1}, we analyze WTM maps applied to noisy ring lattices for various values of $\alpha=d^{\rm{(NG)}}/d^{\rm{(G)}}$. Specifically, we show values for $\rho$, $P$, and $\Delta$ for $N=200$, $d^{\rm{(G)}}=20$, various $T$, and various $d^{\rm{(NG)}}$. We show using the solid and dashed curves, respectively, that the transitions between the qualitatively different regions of these properties closely resemble the bifurcation structure from Eqs.~\eqref{eq:ANC_crits} and \eqref{eq:WFP_crits} with $k=0$. In particular, there is WFP but no ANC, we are able to consistently identify the geometry, embedding dimension, and topology of the underlying manifold of the noisy ring lattice using the WTM map. When there is both WFP and ANC, the extent to which a contagion adheres to the network's underlying manifold depends on $\alpha$ and $T$, and we can quantify this extent using the point-cloud measures $\rho$, $P$, and $\Delta$. We illustrate our observations further in Fig.~\ref{fig:correlations_1}(d) by fixing $\alpha=1/3$ and plotting $\rho$, $P$, and $\Delta$ as a function of the threshold $T$. We show results for $(d^{\rm{(G)}},d^{\rm{(NG)}})=(6,2)$ (blue dashed curves) and $(d^{\rm{(G)}},d^{\rm{(NG)}})=(24,8)$ (red solid curves). Observe that the latter curve is smoother than the former one.  The latter curve yields values of $\rho$, $P$, and $\Delta$ that better reflect the underlying ring manifold $\mathcal{M}$. By contrast, increasing the number $N$ of nodes increases the contrast {(i.e., as observed through $\rho$, $P$, and $\Delta$)} between the region that predominantly exhibits WFP and the other regions.

To give some perspective on the performance of WTM maps for identifying a noisy geometric network's underlying manifold even in the presence of many non-geometric edges, we use the arrows in Fig.~\ref{fig:correlations_1}(d) to indicate the values of $\rho$, $P$, and $\Delta$ for a mapping of nodes based on shortest paths, which one can construe as a variant of the dimension-reduction algorithm Isomap \cite{Tenenbaum2000} (which we apply to an unweighted network rather than to a point cloud). Specifically, we map $\mathcal{V}\mapsto\{{\tilde{\bold x}}^{(i)}\}$ with $T=0$ (as we discussed in the section ``WTM Maps''). 

In {Supplementary~Note~8}, we describe additional {numerical} results that compare a WTM map to Isomap \cite{Tenenbaum2000} and a Laplacian eigenmap \cite{Belkin2003} for generalizations of the noisy ring lattice by (1) allowing the node locations to be a random sampling of points on the unit circle and (2) allowing heterogeneity in their geometric and non-geometric degrees. {We define these other network structures in Supplementary Note~2. Our} results {(see Supplementary Figs.~14--21)} reveal large parameter regimes
% [e.g., $T\approx0.3$ in Fig.~\ref{fig:correlations_1}(d)] 
in which the {ring manifold that underlies the noisy ring lattice} is much more apparent (i.e., as indicated by large $\rho$, small $P$, and large $\Delta$) for maps based on WTM contagions versus those based on shortest-path or diffusion dynamics {(i.e., as in the Laplacian eigenmap)}. We stress that any applications of dimension reduction (e.g., manifold learning) in networks should use an approach that is appropriate for the question of interest. This is why we use contagions in this paper instead of other types of spreading dynamics.

\begin{figure}[t]
\centering
\includegraphics[width=\linewidth]{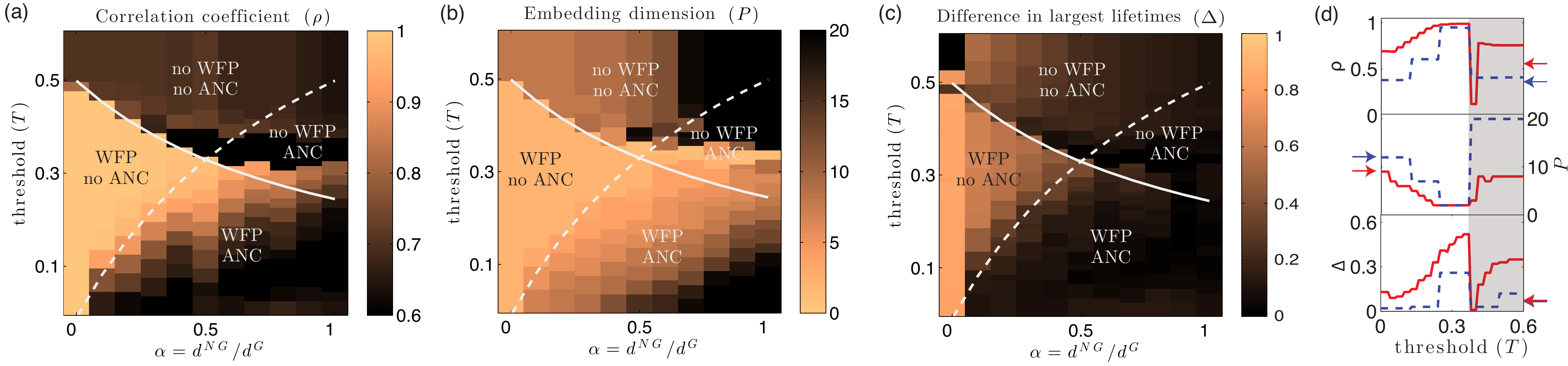}
\caption{ {\emph{Analyzing manifold structure in contagion maps.}}
We analyze the point clouds of WTM maps for various thresholds $T$ for noisy ring lattices with $N=200$ and various ratios $\alpha=d^{\rm{(NG)}}/d^{\rm{(G)}}$. (As an example, we show results for $d^{\rm{(G)}}=20$ and various values of $d^{\rm{(NG)}}$.) For each point cloud, we study {(a)~geometry through $\rho$, (b) dimensionality through $P$, and (c) topology through $\Delta$} (see the text and the Methods section). The transitions between qualitatively different structures in the WTM maps (i.e., as seen through $\rho$, $P$, and $\Delta$) closely resemble the bifurcation structure from Eqs.~\eqref{eq:ANC_crits} and \eqref{eq:WFP_crits}, which we show for $k=0$ using solid and dashed curves, respectively. In panel (d), we fix $\alpha=1/3$ and plot $\rho$, $P$, and $\Delta$ as a function of threshold $T$. We show results for $(d^{\rm{(G)}},d^{\rm{(NG)}})=(6,2)$ (blue dashed curves) and $(24,8)$ (red solid curves). Note that there are activation times that are infinite for $T\ge T_0^{\rm{(WFP)}}= 3/8$ [shaded region in (d)]. The arrows indicate the $\rho$, $P$, and $\Delta$ values that we obtain for the embedding of nodes based on shortest paths, which {(as we discuss in the text)} one can construe as a variant of the dimension-reduction algorithm Isomap \cite{Tenenbaum2000}. 
}
\label{fig:correlations_1}
\end{figure}

\begin{figure}[t]
\includegraphics[width=\linewidth]{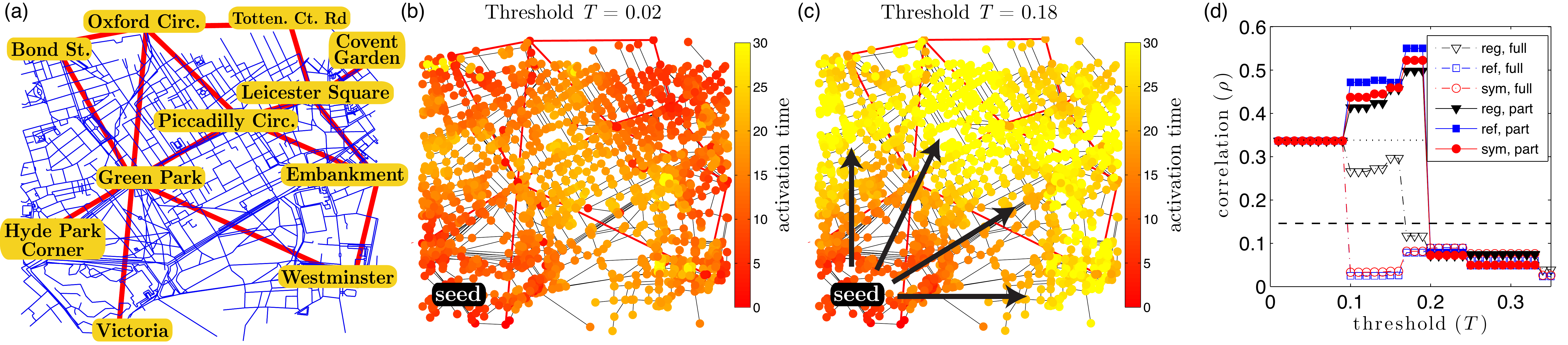}
\vskip -10pt
\caption{  {\emph{Complex contagions on a London transit system.}}
(a) London transit network with $N=2217$ nodes (i.e., intersections), $2854$ roads \cite{Lee2012} (which we interpret as geometric edges), and $15$ metropolitan lines \cite{Rombach2014} (which we interpret as non-geometric edges). 
(b) Node activation times for a WTM contagion initialized with cluster seeding illustrate for small $T$ that contagions quickly spread by skipping across the metro lines; this leads to ANC. 
(c)~In contrast, for moderate $T$, the contagion spreads via slow WFP. 
(d) Although not all contagions exhibit such extreme sensitivity to $T$ (see {Supplementary Note 5}), the dependence of ANC and WFP on $T$ is captured by the geometry of WTM maps if one appropriately handles the activation times that are infinite (i.e., nodes that never adopt the contagion). See the discussion in the text. {The curves with symbols indicate the values of $\rho$ for WTM maps (curves with symbols), and the horizontal dotted and dashed lines, respectively, indicate $\rho$ for the mapping of nodes based on shortest-path distances (i.e., as in the Isomap algorithm \cite{Tenenbaum2000}) and a 2D Laplacian eigenmap \cite{Belkin2003}.}
}
\label{fig:london}
\end{figure}

%%%%%%%%%%%%%%%%%%%%%%%%%%%%%%%%%%%%%%%%%%%%%%%%%%%
\subsection{Contagions on a London Transit Network}\label{sec:london}
%%%%%%%%%%%%%%%%%%%%%%%%%%%%%%%%%%%%%%%%%%%%%%%%%%%

In addition to synthetic networks, we study WTM maps for a London transit network [see Fig.~\ref{fig:london}(a)]. Nodes in the network represent intersections of known latitude and longitude (their coordinates are $\{\bold w^{(i)}\}$), geometric edges represent roads (from data used in Ref.~\cite{Lee2012}), and non-geometric edges represent metropolitan lines (from data used in Ref.~\cite{Rombach2014}). {We have made the network publicly available (see Sec.~{\ref{sec:code})}.} We present our results in detail in {Supplementary Note 1}, and we summarize them here. 

Our central finding is that the qualitative dynamical regimes that we observe for synthetic noisy geometric networks also occur in the London transit network. 
More specifically, we observe both WFP and ANC. Additionally, as we illustrate in Fig.~\ref{fig:london}(b)--(c), these phenomena can be very sensitive to the WTM threshold $T$. We study WFP and ANC by examining the geometry of WTM maps. However, we do not study their one-dimensional homology, as computations of homology (which remains a very active area of research \cite{Bauer2014,Vejdemo2014}) have a much higher computational cost than our calculations of geometry and dimensionality.

In Fig.~\ref{fig:london}(d), we plot the Pearson correlation coefficient $\rho$ that compares the distance between mapped nodes to their actual distance from each other (according to latitude and longitude) for various values of $T$. We show results for the regular, reflected, and symmetric versions of a WTM map {(curves with symbols); and the horizontal dotted and dashed lines, respectively, give $\rho$ for the mapping of nodes based on {shortest-path distances} (i.e., as in the Isomap algorithm \cite{Tenenbaum2000}) and a 2D Laplacian eigenmap \cite{Belkin2003}}. For each type of WTM map, we handle the activation times that are infinite (see Sec.~\ref{sec:practical}) using two methods. In the method that we label ``full,'' we keep the entire matrix that encodes activation times, and we set the activation times that are infinite to be $2N$. (Recall that we used this approach when studying WTM maps for synthetic networks.) In the method that we label ``part,'' we neglect contagions that do not saturate a network, so we use only a portion of the values in the matrix that encode activation times. In Fig.~\ref{fig:london}(d), we see that these choices give contrasting results. For the ``full'' option, activation times of infinity (which arise when $T\gtrapprox0.1$) distort the WTM map and lead to a drop in $\rho$. In contrast, the ``part'' method neglects activation times of infinity, and we find that there is a range of $T$ values for which there is a pronounced increase in $\rho$. Such improved agreement between the geometry of WTM contagions and the transit network's inherent latitudinal and longitudinal embedding on Earth's surface is characteristic of an increase in WFP versus ANC.  Interestingly, we find that the small node degrees (e.g., $\langle d_i\rangle \approx 2.59$) and the significant heterogeneity (e.g., with respect to node locations, node degrees, and the length of roads) in the London transit network causes WFP and ANC to be extremely sensitive to the value of $T$ for only a few of the contagion seeds $S^{(j)}$ (see {Supplementary Note~1} {and Supplementary Fig.~5}). Nevertheless, as we have demonstrated, such minority cases still have a significant effect on WTM maps.

Our numerical experiments for the London transit network highlight additional complexities that can arise for networks that are constructed from empirical data, and they offer complementary insights to our investigation of synthetic networks. In particular, the synthetic networks that we examine either are homogeneous or are only slightly heterogeneous, so the WFP and ANC behavior tends to be similar for contagions that are initialized in different parts of a network. This is not the case for the London transit network, which has significant heterogeneity and very small node degrees (which seems to exacerbate the effect of heterogeneity). Infections that start in some parts of the network have rather different properties than those than start in others, and one also needs to consider multiple strategies for how to handle activation times of infinity. There are also other interesting phenomena that our approach can examine for heterogeneous networks.  For example, in {Supplementary Note~1}, we study the geometry of WTM contagions for individual nodes (rather than averaging our results over an entire network) in what amounts to an ``egocentric'' analysis of geometry. We find that the local geometry of WTM maps (and hence of contagions) at a given node relates strongly to its proximity to a metro line.

%%%%%%%%%%%%%%%%%%%%%%%%%%%%%%%%%%%%%%%%%%%%%%%%%%%
\section{Discussion}\label{sec:discuss}
%%%%%%%%%%%%%%%%%%%%%%%%%%%%%%%%%%%%%%%%%%%%%%%%%%%

Many empirical networks include a combination of geometric edges between nearby nodes and non-geometric, long-range edges \cite{Barthelemy2011}.  Such situations can arise when nodes are restricted by their locations in a physical space (such as in a city) or in terms of latent underlying spaces \cite{Boguna2012,Boguna2010,Serrano2008,Hoff2002,Singer2011,Singer2013,Tenenbaum2000,Belkin2003,Coifman2005,Gerber2007,Sorzano2014}. When considering a spreading process on a noisy geometric network, it is important to understand the extent to which a contagion follows such underlying structure. To address this question, we conducted a detailed investigation using the Watts threshold model (WTM) of complex contagions (with uniform threshold $T$) on noisy geometric networks. The spreading dynamics exhibit both wavefront propagation (WFP) that follows the underlying manifold structure of a network as well as the appearance of new clusters (ANC) of contagions in distant locations. To investigate the extent to which a WTM contagion adheres to a network's underlying manifold, we introduced the notion of WTM maps (and contagion maps more generally) and showed when a contagion predominantly spreads via WFP that WTM maps recover the topology, geometry, and dimensionality of a network's underlying manifold even in the presence of many non-geometric (i.e., ``noisy'') edges. 

Our methodology of constructing and analyzing contagion maps has important implications not only for the analysis, modeling, and control of contagions, but also for other dynamics that can be used to construct filtrations of networks. Moreover, by studying manifold structure in contagion maps, we have shown that such maps can also be used to identify and study manifold structure in networks. We have compared WTM maps to Laplacian eigenmaps \cite{Belkin2003} and Isomaps \cite{Tenenbaum2000} (see {Supplementary Note~8} for additional discussion) and found that WTM maps---which are based on a nonlinear and nonconservative dynamical process---yield results that contrast with those from the other methods. This is {sensible}, as nonconservative and conservative dynamics (e.g., diffusion) are known to give different results for which nodes are central \cite{ghosh2012} and what network structures constitute bottlenecks to the dynamics \cite{lerman2012pre}. 

In {the} Supplementary Discussion, we further consider the implications of our work on three important fields of research: 
(i)~studying contagions and other dynamics from the perspective of high-dimensional data analysis (i.e., computational topology and nonlinear dimension reduction), 
(ii) identifying low-dimensional {(e.g., manifold)} structure in networks,
(iii) identifying low-dimensional {(e.g., manifold)} structure in point-cloud data.

%%%%%%%%%%%%%%%%%%%%%%%%%%%%%%%%%%%%%%%%%%%%%%%%%%%
\section{Methods}\label{sec:methods}
%%%%%%%%%%%%%%%%%%%%%%%%%%%%%%%%%%%%%%%%%%%%%%%%%%%

%%%%%%%%%%%%%%%%%%%%%%%%%%%%%%%%%%%%%%%%%%%%%%%%%%%
\subsection{{Data and Code Availability}}\label{sec:code}
%%%%%%%%%%%%%%%%%%%%%%%%%%%%%%%%%%%%%%%%%%%%%%%%%%%

The London transit network that we study in Sec.~\ref{sec:london} and the code that we use to construct WTM maps are available as Supplementary Files 1 and 2.

%%%%%%%%%%%%%%%%%%%%%%%%%%%%%%%%%%%%%%%%%%%%%%%%%%%
\subsection{Bifurcation Analysis}\label{sec:bif}
%%%%%%%%%%%%%%%%%%%%%%%%%%%%%%%%%%%%%%%%%%%%%%%%%%%

To guide our study of WTM maps, {we set $T_i=T$ for each node $i\in\mathcal{V}$}, and we perform a bifurcation analysis of WTM contagions on noisy ring lattices. In particular, we investigate the dependence of ANC and WFP on the contagion threshold $T$ and on the network parameters $d^{\rm{(G)}}$, $d^{\rm{(NG)}}$, and $N$. {In Fig.~\ref{fig:ring_1}, we illustrate ANC and WFP for this class of networks with $d^{\rm{(G)}}=4$, $d^{\rm{(NG)}}=1$, and $N=40$ by considering a WTM contagion at time $t=0$.} The light blue nodes are in the contagion seed $S=s\cup \{k|A_{sk}\not=0\}$, which is centered at node $s\in\mathcal{V}$. Because node $s$ is incident to both geometric and non-geometric edges, the contagion is initialized with $1+d^{\rm{(NG)}}=2$ contagion clusters. We denote these clusters by $C_1$ and $C_2$. Cluster $C_1$ is more likely to grow via WFP than $C_2$. The orange nodes in Fig.~\ref{fig:ring_1} are what we call contagion cluster $C_1$'s ``boundary''---the set of nodes that have yet to adopt the contagion but which are exposed to it via a geometric edge that is incident to an infected node in $C_1$. As we show in the magnification on the right, nodes in the boundary can adopt the contagion via WFP. Nodes that are not infected and not on the boundary can become infected via ANC. (See the dark blue nodes and dashed edges.)

If node $i$ adopts a contagion via ANC, then by definition it is not in the boundary of a contagion cluster, so its neighbors due to geometric edges have yet to adopt the contagion. Consequently, node $i$ potentially has $0,1,\dots,d^{\textrm{(NG)}}$ neighbors that are infected, and its fraction of infected neighbors is restricted to $f_i\in\{0,\frac{1}{d^{\textrm{(G)}}+d^{\textrm{(NG)}}},\frac{2}{d^{\textrm{(G)}}+d^{\textrm{(NG)}}},\dots,\frac{d^{\textrm{(NG)}}}{d^{\textrm{(G)}}+d^{\textrm{(NG)}}}\}$. This observation yields the critical thresholds
\begin{equation}
	T_k^{\mathrm{(ANC)}} \triangleq \frac{d^{\rm{(NG)}}-k}{d^{\rm{(G)}}+d^{\rm{(NG)}}}\,, \quad k=0,1,\dots,d^{\rm{(NG)}}\,.
	\nonumber%\label{eq:ANC_crits}
\end{equation}
The contagion dynamics changes abruptly at the critical values of $T$, so {the qualitative dynamics of ANC for} any $T \in \left[T_{k+1}^{\mathrm{(ANC)}},T_k^{\mathrm{(ANC)}}\right)$ are similar to each other, but there are abrupt changes at the endpoints of the interval. 
In particular, whenever $T \in \left[T_{k+1}^{\mathrm{(ANC)}},T_k^{\mathrm{(ANC)}}\right)$, a node requires at least $(d^{\textrm{(NG)}}-k)$ neighbors due to non-geometric edges to be infected before it adopts the contagion. In {Supplementary Note~6}, we study the probability that a node has exactly {$(d^{\textrm{(NG)}}-k)$} infected non-geometric neighbors at time $t$. For large networks, this probability is approximately $\left(d^{\rm{(NG)}}\atop {d^{\rm{(NG)}}-k}\right)[q(t)/N]^{d^{\rm{(NG)}}-k}[1-q(t)/N]^{k}$, where $q(t)$ denotes the number of nodes that have adopted the contagion at or before time $t$. Note that the probability is an expectation over the ensemble of noisy ring lattices, because it uses the fact that non-geometric edges are generated uniformly at random in our model. Therefore, it does not matter which of the $q(t)$ nodes happen to be infected.

Turning to WFP, we now study contagion transmissions exclusively across geometric edges. That is, given a node $i$ in a contagion cluster's boundary, we assume that the node's neighbors due to non-geometric edges are not infected. Naturally, this assumption does not always hold, but it is insightful to first examine this ideal case and then consider more general situations as perturbations of such a baseline analysis of WFP.

To facilitate our discussion, we will use the example contagion illustrated in Fig.~\ref{fig:ring_1}. In particular, we consider WFP in the clockwise direction for cluster $C_1$. Nodes $a$, $b$, and $c$ are exposed, respectively, to 2, 1, and 0 nodes that have adopted the contagion, so their fractions of neighbors that are infected are $f_a=2/5$, $f_b=1/5$, and $f_c=0/5$. Note that we assume that the non-geometric edges for nodes $a$, $b$, and $c$ are incident to nodes that are not infected (i.e., which have not adopted the contagion). Because $f_i > T$ for node $i$ to adopt the contagion, one of three situations can occur at time $t = 1$: (1) if $0\le T<1/5$, then nodes $a$ and $b$ adopt the contagion; (2) if $1/5\le T<2/5$, then node $a$ adopts the contagion; and (3) if $2/5\le T$, then the contagion cluster {$C_1$} does not increase in size via WFP. Node $c$ cannot adopt the contagion via WFP at time $t = 1$ for any $T\ge0$. We find that WFP is governed by the critical thresholds
\begin{equation}
	T_k^{\mathrm{(WFP)}} \triangleq \frac{d^{\rm{(G)}}/2-k}{{d^{\rm{(G)}}+d^{\rm{(NG)}}}}\,, \quad k=0,1,\dots,\frac{d^{\rm{(G)}}}{2}\,,
	\nonumber%\label{eq:WFP_crits},
\end{equation}
where a wavefront propagates at a speed of {$k+1$} nodes per time step for $T \in \left[T_{k+1}^{\mathrm{(WFP)}},T_k^{\mathrm{(WFP)}}\right)$. For $T\ge T_0^{\mathrm{(WFP)}}$, there is no WFP.

We now include additional discussion of} the assumptions in our analysis of WFP. Specifically, when considering whether or not node $i$ in a contagion cluster's boundary will become infected, we assumed that its non-geometric edges are not incident to an infected node. Obviously, this assumption is valid for $d^{\textrm{(NG)}}=0$. However, as we discuss in Supplementary Note 6, the expected probability (over an ensemble of noisy geometric networks with non-geometric edges generated uniformly at random) that a node's non-geometric edge is incident to an infected node is $q(t)/(N-1)$. Similarly, the probability that a node has $d^{\textrm{(NG)}}$ non-geometric neighbors and that none of them are infected is approximately $[1-q(t)/N]^{d^{\rm{(NG)}}}$, which is therefore the probability that our assumption is valid. In particular, whenever $q(t)\ll N$, which necessarily requires $N\gg 1$ and describes the scenario of an early stage of a contagion on a large network, the probability that our assumption is valid is approximately equal to 1. Therefore, Eq.~\eqref{eq:WFP_crits} accurately describes the speed of WFP in this scenario with high probability. (Note that we also assume that $d^{\textrm{(NG)}}\ll N$, so there cannot be too many non-geometric edges.)
 
Equation~\eqref{eq:WFP_crits}, which one can construe as a ``local'' result, is also very useful for predicting the ``global'' behavior of WFP. To see this, we make the following two observations: (1) If a contagion cannot spread when $q(t)\ll N$, then it will not reach a state in which $q(t) = \mathcal{O}(N)$; and (2) if $q(t)$ does spread for $q(t) \ll N$, then it will also spread when $q(t) = \mathcal{O}(N)$ because an increase in $q(t)$ will help promote further spreading. Specifically, the presence of a node in the boundary with infected non-geometric neighbors can accelerate WFP by allowing the node to adopt the contagion with fewer infected geometric neighbors than Eq.~\eqref{eq:WFP_crits} would predict. In fact, when the contagion size is large [i.e., when $q(t)\approx N$], we find that the WFP speed accelerates up to $d^{\rm{(G)}}/2$ nodes per time step (i.e., all nodes in the boundary on one side of the contagion cluster become infected upon each time step). Similar accelerated WFP has also been observed for other applications including species dispersion \cite{Hallatscheka2014}. See {Supplementary Note 6} for further discussion. 

In {Supplementary Note 3}, we use a perturbative approach to generalize our bifurcation analysis to a family of synthetic noisy geometric networks with slight heterogeneities. In our generalizations, we examine the WFP and ANC behavior of WTM contagions at each node. When the nodes are identical (i.e., as in the synthetic ring lattice), the contagion behavior is uniform across a network; this leads to the bifurcation diagram in Fig.~\ref{fig:bifurcation_ring_lattice}. When there is heterogeneity, the contagion behavior at each node varies across a network. However, if the amount of heterogeneity is small, then one can construct a perturbed bifurcation diagram in which the boundaries between contagion regimes are thickened. That is, as one varies $T$ or $\alpha$, the transition from one regime (e.g., WFP and no ANC) to another (e.g., WFP and ANC) still occurs, but it does not occur simultaneously for each node.

%%%%%%%%%%%%%%%%%%%%%%%%%%%%%%%%%%%%%%%%%%%%%%%%%%%
\subsection{{Activation Times of Infinity in WTM Maps}}\label{sec:practical}
%%%%%%%%%%%%%%%%%%%%%%%%%%%%%%%%%%%%%%%%%%%%%%%%%%%
When studying WTM maps, one needs a strategy for dealing with activation times that are infinite (which in some cases might be useful for identifying outliers and in other cases might be problematic). After constructing a map such as $\mathcal{V}\mapsto \{\bold x^{(i)}\}\in\mathbb{R}^J$, the distance between points $\bold x^{(i)}$ and $\bold x^{(j)}$ for $i,j\in\mathcal{V}$ can be infinite or even undefined, which complicates any subsequent analyses of the point cloud $\{\bold x^{(i)}\}$. 
Such an issue can also arise for distances that are derived from shortest paths or the commute time for diffusion, so algorithms for mapping networks often assume that a network consists of a single connected component \cite{Tenenbaum2000,Belkin2003}. Distances that are infinite are not an issue for diffusion maps \cite{Coifman2005}, because the nodes are mapped to a bounded metric space whose diameter is equal to twice the maximum of the heat kernel.

For complex contagions, activation times {that are infinite} arise not only due to disconnected networks, but also for networks that are ``disconnected'' with respect to the contagion dynamics. In the present work, we use two methods for handling activation times that are infinite: we either set these activation times to be large but finite (i.e., to be $2N\ll \infty$), or we neglect the contagions that lead to activation times that are infinite by restricting the map to a subset of contagions (i.e., $j\in \mathcal{J}'\subset \mathcal{J}$, where $\mathcal{J}'=  \{ j\in \mathcal{J} | \bold x_j^{(i)} <\infty ~\forall \, i\}$). 
We note in passing (though we do not explore the strategy in the present manuscript) that there exist maps such as $d\mapsto d/ (d+1) \in [0,1]$ that map an unbounded metric space to a topologically-equivalent metric space that is bounded. This ought to be useful for some situations.

%%%%%%%%%%%%%%%%%%%%%%%%%%%%%%%%%%%%%%%%%%%%%%%%%%%
\subsection{Geometry of WTM Maps}\label{sec:methods2}
%%%%%%%%%%%%%%%%%%%%%%%%%%%%%%%%%%%%%%%%%%%%%%%%%%%

To quantify the similarity of the geometry of a WTM map to that of the nodes on the underlying manifold of a noisy geometric network, we calculate the Pearson correlation coefficient $\rho$ to relate node-to-node distances for the WTM map. In Fig.~\ref{fig:pointcloud}, we compare the geometry of $\{\bold z^{(i)}\}$ [see panel (a)] to that of the nodes' locations $\{\bold w^{(i)}\}\in\mathcal{M}$ [see panel (b)] by computing a Pearson correlation coefficient $\rho$ to compare the node-to-node distances for the two point clouds (i.e., $\|\bold z^{(i)}-\bold z^{(j)}\|_2$ and $\|\bold w^{(i)}-\bold w^{(j)}\|_2$ for $(i,j)\in\mathcal{V}\times\mathcal{V}$). We conduct our comparison with respect to the dimension of the ambient spaces in which the points lie (i.e., $\mathbb{R}^N$ for $\{\bold z^{(i)}\}$ and $\mathbb{R}^2$ for $\{\bold w^{(i)}\}$). See {Supplementary Note~7} for further discussion.

%%%%%%%%%%%%%%%%%%%%%%%%%%%%%%%%%%%%%%%%%%%%%%%%%%%
\subsection{Dimensionality of WTM Maps}\label{sec:methods3}
%%%%%%%%%%%%%%%%%%%%%%%%%%%%%%%%%%%%%%%%%%%%%%%%%%%

We study the dimensionality by examining the residual variance \cite{Tenenbaum2000,Cox2010} of the point cloud $\{\bold z^{(i)}\}$ and computing the smallest dimension such that we lose less than 5\% of the variance when projecting to a lower dimension using PCA \cite{Sorzano2014,Tenenbaum2000,Cox2010}. We refer to this dimension as the ``embedding dimension'' $P$. 
Specifically, we estimate the embedding dimension $P$ of a WTM map by studying $p$-dimensional projections of the WTM map obtained via PCA for different values of 
$p \in \{1,2,\dots\}$. For each projection, we calculate the residual variance $R_p=1-(\rho^{(p)})^2$ \cite{Tenenbaum2000,Cox2010}, where $\rho^{(p)}$ denotes the Pearson correlation coefficient that relates the geometric similarity between the $p$-dimensional projection and the unprojected WTM map (see Sec.~\ref{sec:methods2}). We define the embedding dimension $P$ as the smallest dimension $p$ such that $R_p<0.05$. See {Supplementary Note~7} for further discussion.

%%%%%%%%%%%%%%%%%%%%%%%%%%%%%%%%%%%%%%%%%%%%%%%%%%%
\subsection{Topology of WTM Maps}\label{sec:methods4}
%%%%%%%%%%%%%%%%%%%%%%%%%%%%%%%%%%%%%%%%%%%%%%%%%%%

We study the topology of a WTM map by examining the persistence diagram of a Vietoris-Rips filtration that is generated by the point cloud $\{\bold z^{(i)}\}$ \cite{Kaczynski2004,edels2010}. 
For our experiments involving a noisy ring lattice, we are interested primarily in assessing the presence versus absence of a ring topology in a WTM map.
We thus study the persistent homology of a WTM map by examining a Vietoris-Rips filtration using the software package {\sc Perseus} \cite{Mischaikow2013}. We calculate persistent 1D features (i.e., 1-cycles) for the point cloud and record the difference $\Delta=l_1-l_2$ between the two largest lifetimes of such 1D features. We normalize all lifetimes by the diameter of the point cloud so that $\Delta,l_1,l_2 \in[0,1]$.
(Note that sometimes it can be preferable to use the ``bottleneck distance'' between persistence diagrams \cite{Kramar} rather than $\Delta$.) See {Supplementary Note~7} for further discussion.

\section{Acknowledgements}
DT and PJM were partially supported by the Eunice Kennedy Shriver National Institute of Child Health \& Human Development of the National Institutes of Health under Award Number R01HD075712.  
DT was also funded by the National Science Foundation under Grant DMS-1127914 to the Statistical and Applied Mathematical Sciences Institute (SAMSI). 
 DT also acknowledges an Institute of Mathematics and its Applications (IMA) travel grant to attend the workshop Topology and Geometry of Networks and Discrete Metric Spaces. 
MAP was supported by the European Commission FET-Proactive project PLEXMATH (Grant No. 317614) and also acknowledges a grant (EP/J001759/1) from the EPSRC. FK's stay in Oxford was supported in part by the latter grant. 
HAH gratefully acknowledges funding from EPSRC Fellowship EP/K041096/1, King Abdullah University of Science and Technology (KAUST) KUK-C1-013-04, a SAMSI Low-Dimensional Structure in High-Dimensional Data workshop travel grant, and an AMS Simons travel grant.
KM and MK were partially supported by NSF grants NSF-DMS-0915019, 1125174,
1248071, and contracts from AFOSR and DARPA.
We thank Yannis Kevrekidis and Barbara Mahler for discussions and numerous helpful comments on a version of this manuscript.
We thank James Gleeson, Ezra Miller, Sayan Mukherjee, and Hal Schenck for helpful discussions. 
The content is solely the responsibility of the authors and does not necessarily represent the official views of any of the funding agencies.

\section{Author Contributions}
All authors designed the research and wrote the text. DT and FK developed the analytical results and conducted the numerical experiments.

\section{Financial Interest Statement}
The authors declare no competing financial interests.

\end{document}

\title{Supplementary Information: Topological data analysis of contagion maps for examining spreading processes on networks}

\author{Dane Taylor$^{1,2}$, Florian Klimm$^{3,4,5}$, Heather A. Harrington$^{5}$, Miroslav Kram\'ar$^{6}$, \\
Konstantin Mischaikow$^{6,7}$, Mason A. Porter$^{5,8}$, and Peter J. Mucha$^{2}$}

\maketitle

\noindent$^1${Statistical and Applied Mathematical Sciences Institute, Research Triangle Park, NC, 27709, USA}\\
$^2${Carolina Center for Interdisciplinary Applied Mathematics, Department of Mathematics, University of North Carolina at Chapel Hill, NC, 27599, USA}\\
$^3${Potsdam Institute for Climate Impact Research, 14473 Potsdam, Germany}\\
$^4${Department of Physics, Humboldt-Universit\"at zu Berlin, 12489 Berlin, Germany}\\
$^5${Mathematical Institute, University of Oxford, OX2 6GG, UK}\\
$^6${Department of Mathematics, Rutgers, The State University of New Jersey, Piscataway, NJ,  08854, USA}\\
$^7${BioMaPS Institute, Rutgers, The State University of New Jersey, Piscataway, NJ,  08854, USA}\\
$^8${CABDyN Complexity Centre, University of Oxford, Oxford OX1 1HP, UK}

\tableofcontents
\newpage

%%%%%%%%%%%%%%%%%%%%%%%%%%%%%%%%%%%%%%%%
%%%%%%%%%%%%%%%%%%%%%%%%%%%%%%%%%%%%%%%%
\section{Supplementary Note 1: Complex Contagions on a London Transit Network}\label{sec:London}
%%%%%%%%%%%%%%%%%%%%%%%%%%%%%%%%%%%%%%%%
%%%%%%%%%%%%%%%%%%%%%%%%%%%%%%%%%%%%%%%%

The primary goal of our work has been to develop the notion of a WTM map and to demonstrate the utility of using such maps for examining WTM contagions on noisy geometric networks.  Specifically, we conducted a detailed examination that contrasts wavefront propagation (WFP) along geometric edges versus the appearance of new contagion clusters (ANC) due to the presence of non-geometric, ``noisy'' edges. We have focused on synthetic networks---and, in particular, on noisy geometric networks on a ring manifold---and we conducted a bifurcation analysis to guide our study. However, one can use WTM maps on far more general types of networks such as noisy geometric networks that are constructed from empirical data.  (More generally, one can also use contagion dynamics that one constructs from other types of spreading processes.) This allows two important applications to real systems: (1) one can study the extent to which a contagion on a network exhibits spatial phenomena such as WFP versus non-spatial phenomena such as ANC; and (2) one can infer (potentially) unknown low-dimensional structure in a network. In this section, we highlight these ideas for an empirical network that describes transit infrastructure in a part of London.

\begin{figure*}[b!]
\centering
\includegraphics[width=.9\linewidth]{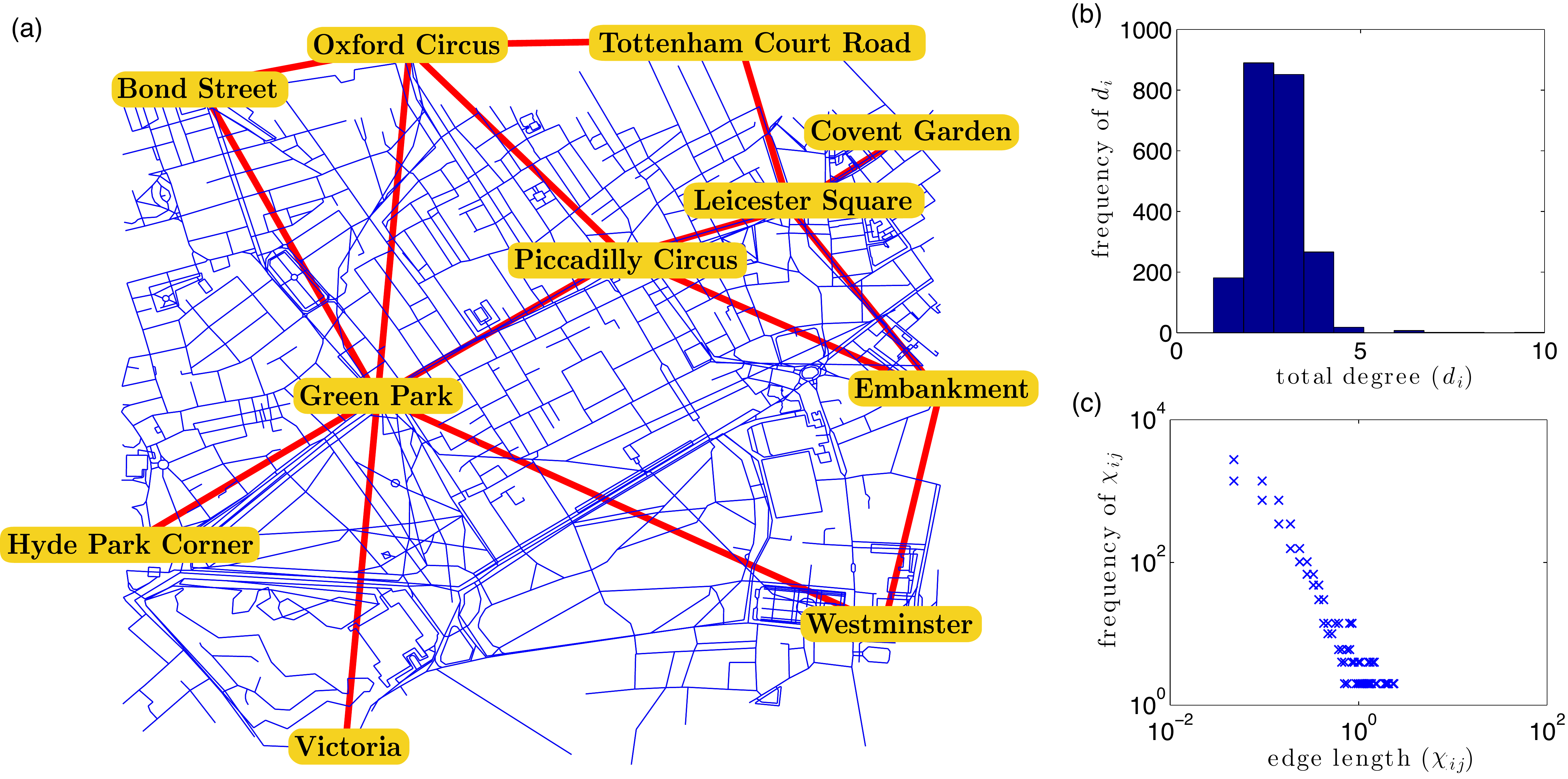}
\caption{
London transit network with $N=2217$ nodes, $2854$ geometric edges, and $15$ non-geometric edges {(which we have made publicly available, as we discussed in Sec.~III~A} of the main manuscript).
(a)~The geometric edges (blue), which we take from Ref.~\cite{Lee2012}, are roads between intersections; and the non-geometric edges (red), which we take from Ref.~\cite{Rombach2014}, give connections between metro stations. Some nodes ($i \in \mathcal{P}\subset\mathcal{V}$, where $|\mathcal{P}|=11$) correspond to both intersections and metro stations, whereas other nodes ($i \in \mathcal{V}\setminus\mathcal{P}$) correspond only to intersections. Each node $i \in \mathcal{V}$ has an intrinsic location $\{\bold w^{(i)}\}$ based on its latitude and longitude. 
(b)~Histogram of the frequencies of the nodes' total degrees $\{d_i\}$ (i.e., $d_i=d_i^{\rm{(G)}}+ d_i^{\rm{(NG)}}$), where the mean is $\langle d_i\rangle \approx 2.59$. 
(c)~Histogram of the frequencies of the edge lengths $\{\chi_{ij}\}$, where $\chi_{ij}=m(i,j)$ is the Euclidean distance between locations $\bold w^{(i)}$ and $\bold w^{(j)}$ for each edge $(i,j)\in\mathcal{E}$ [see Eq.~\eqref{eq:distance1}]. {See \ref{sec:London} for further discussion.}
}
\label{fig:london}
\end{figure*}

~\clearpage

%%%%%%%%%%%%%%%%%%%%%%%%%%%%%%%%%%%%%%%%
~\\{\bf \Large {Description of the} London Transit Network}\\
%%%%%%%%%%%%%%%%%%%%%%%%%%%%%%%%%%%%%%%%

\begin{figure*}[b!]
\centering
\includegraphics[width=.7\linewidth]{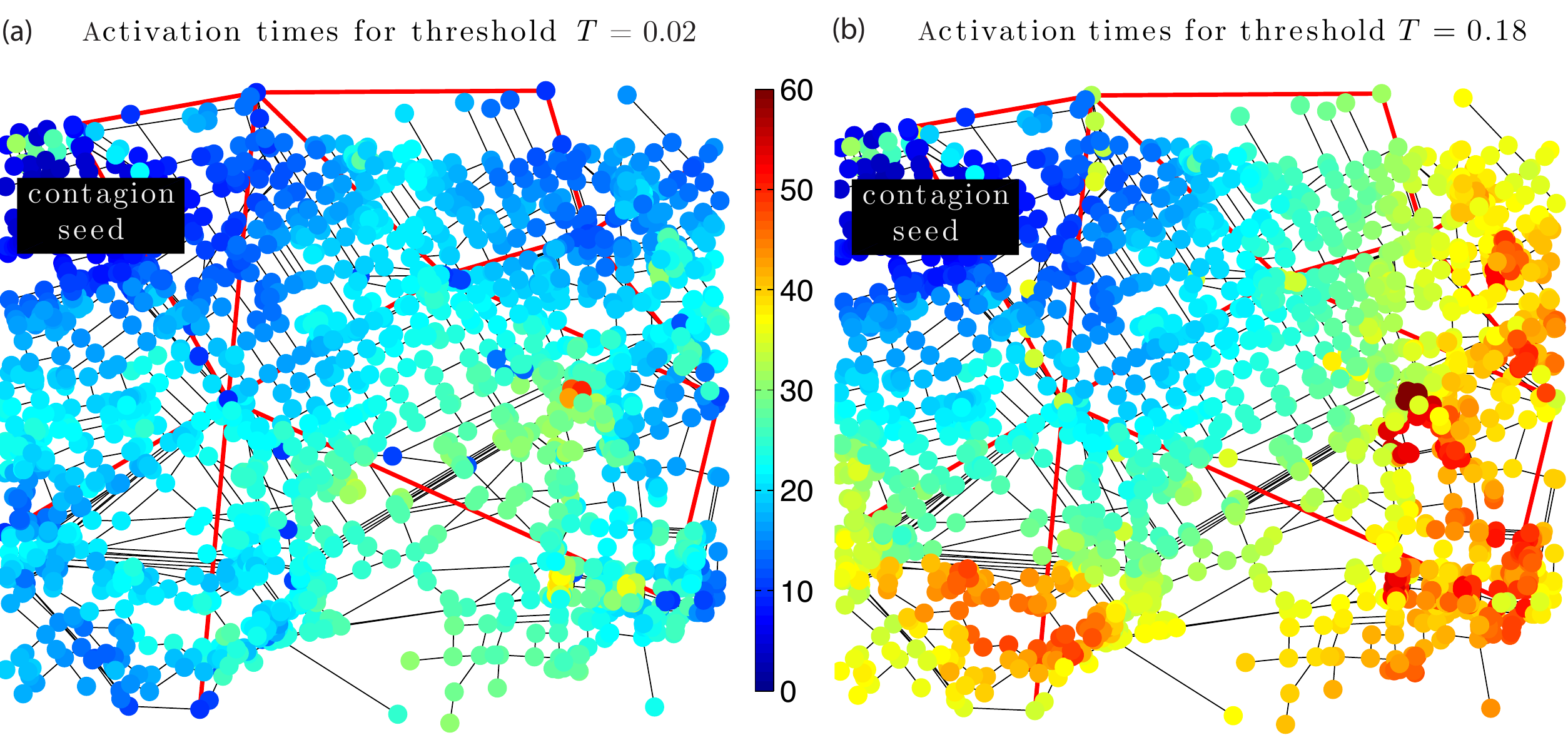}
\caption{
Activation times $\{\bold x_j^{(i)}\}$ for nodes $i\in\mathcal{V}$ for a WTM contagion on the London transit network, which we initiate with cluster seeding centered at a node $j$ near the Bond Street Station. (a) For small thresholds, such as $T=0.02$, nodes near metro stations have small activation times, so the contagion does not follow the geometric edges (i.e., the roads). (b) For moderate threshold values, such as $T=0.18$, the activation times have a large positive correlation with the Euclidean distances between the intrinsic node locations $\{\bold w^{(i)}\}$ (given by latitude and longitude). Therefore, the WFP and ANC phenomena of WTM contagions with this initialization depend significantly on the value of $T$. Although this is not ``typical'' of all WTM contagions on this network, such situations have a significant effect on the resulting WTM maps. See \ref{sec:London} for further discussion.
}
\label{fig:london_0}
\end{figure*}

{As we illustrate in Supplementary} Fig.~\ref{fig:london}(a), we study WTM contagions on a London transit network that includes both roads (which we interpret as short-range, geometric edges) and metro lines (which we interpret as long-range, non-geometric edges). The nodes $\mathcal{V}=\{1,\dots,N\}$ (where $N=2217$) in the network correspond to intersections, and we obtain the edges from Refs.~\cite{Lee2012} (road data) and \cite{Rombach2014} (metro data). {We construct the merged network {(which we have posted, as we discussed in Sec.~III~A of the main text),} by utilizing} the latitudinal and the longitudinal coordinates to place the locations of metro stations at the nearest intersection of roads. Thus, the nodes $\mathcal{V}$ consist of two sets: (1) nodes $\mathcal{P}\subset\mathcal{V}$ that correspond to both metro stations and intersections and thus have both geometric and non-geometric edges; and (2) nodes $\mathcal{V}\setminus\mathcal{P}$ that correspond to intersections and have only geometric edges (i.e., roads). Additionally, because the network of metro lines in Ref.~\cite{Rombach2014} covers a much larger spatial area than the road network in Ref.~\cite{Lee2012}, we include only metro stations in the convex hull of the road network. (There are $|\mathcal{P}|=11$ such stations.) In {Supplementary} Fig.~\ref{fig:london}(b), we show histograms of the frequencies of the nodes' total degrees $\{d_i\}$, where $d_i=d_i^{\rm{(G)}}+ d_i^{\rm{(NG)}}$ and the mean is $\langle d_i\rangle  \approx 2.59$. In {Supplementary} Fig.~\ref{fig:london}(c), we show histograms of the frequencies of the edge lengths $\{\chi_{ij}\}$, where $\chi_{ij}=m(i,j)$ is the Euclidean distance between locations $\bold w^{(i)}$ and $\bold w^{(j)}$ [see Eq.~\eqref{eq:distance1}] for each edge $(i,j)\in\mathcal{E}$. In practice, we give node $i$ an intrinsic location of $\bold w^{(i)}=[w_1^{(i)},w_2^{(i)}]^T$, where $w_1^{(i)}$ and $w_2^{(i)}$ denote, respectively, the intersection's latitudinal and longitudinal coordinate. We normalize each set of coordinates to have unit variance [see {Supplementary} Fig.~\ref{fig:london}(a)]. In general, such a projection from a patch on the surface of a sphere (e.g., the Earth's surface) to a 2D plane might not be justified. However, the effect of this projection to a plane is negligible in this case due to the very small size of the patch. 

Before analyzing WTM contagions and WTM maps for the London transit network, let's consider the following experiment. In {Supplementary} Fig.~\ref{fig:london_0}, we illustrate that the extent to which a WTM contagion adheres to the network's underlying manifold--the Earth's surface---can be very sensitive to a variety of factors, including the contagion seed and the WTM threshold $T$. We plot the London transit network and color each node $i\in\mathcal{V}$ according to its activation time $x_j^{(i)}$ for a single contagion that we initialize with cluster seeding centered at a node $j$, which we take to be near the Bond Street Station. 
%No, Mr. Bond. I expect you to die.
In panels (a) and (b), we show $\{x_j^{(i)}\}$ for nodes $i\in\mathcal{V}$ with thresholds of $T=0.02$ and $T=0.18$, respectively. Note for $T=0.02$ that the contagion spreads via both roads and metro lines, so the contagion includes ANC. By contrast, for $T=0.18$, the contagion does not spread across the metro lines; rather, it spreads via WFP along the roads. As we shall see, this extreme sensitivity to the threshold $T$ for the behavior of WTM contagions is not typical for all contagion seeds. Nevertheless, we find that such rare cases can have a large impact on the network's WTM maps. 

%%%%%%%%%%%%%%%%%%%%%%%%%%%%%%%%%%%%%%%%
~\\{\bf {\Large Numerical Results for the Geometry of WTM Maps}}\\
%%%%%%%%%%%%%%%%%%%%%%%%%%%%%%%%%%%%%%%%

In this section, we study the geometry of WTM contagions on the London transit network that we studied in {Sec.~I~F} of the main text by examining the geometry of WTM maps. As before, we study geometry through the Pearson correlation coefficient $\rho$ given by Eq.~\eqref{eq:RHO}. We do not study the dimensionality and topology because of the large computational time that it would entail.

To guide our investigation, we first study the equilibrium sizes of contagions (i.e., the number of infected nodes after the contagion stops spreading \cite{Gleeson2007}). Our motivation is as follows. Recall that for WTM maps to be well-defined, all activation times $\{x_j^{(i)}\}$ must be finite. In our numerical experiments for synthetic networks, we therefore focused on this situation (e.g., see the main text and \ref{sec:numerics}), and we chose to handle activation times that were infinite by setting them to be $2N$. Even with the restriction to finite activation times, we found a rich set of diverse qualitative dynamics. However, for the London transit network, the most interesting WTM maps occur for threshold values $T$ that involve activation times of infinity. For this example, we must account for activation times of infinity more carefully to be able to study WTM contagions with WTM maps in such situations. 

We thus begin by studying the equilibrium sizes of WTM contagions that we initialize with cluster seeding centered at each node $i\in\mathcal{V}$. Specifically, for a given threshold $T$, we study the size ${C}_i^{(\rm{target)}}$ of each node $i$'s ``target node set'' (which we define as the set of nodes $\{j\}$ such that $x_i^{(j)}$ is finite) and the size ${C}_i^{(\rm{source)}}$ of its ``source node set'' (which we define as the set of nodes $\{j\}$ such that $x_j^{(i)}$ is finite). In other words, node $j$ is in the target node set for node $i$ if a contagion that is initialized at node $i$ eventually spreads to node $j$, and node $j$ is in the source node set for node $i$ if a contagion that is handle at node $j$ eventually spreads to node $i$.

\begin{figure*}[t!]
\centering
\includegraphics[width=.9\linewidth]{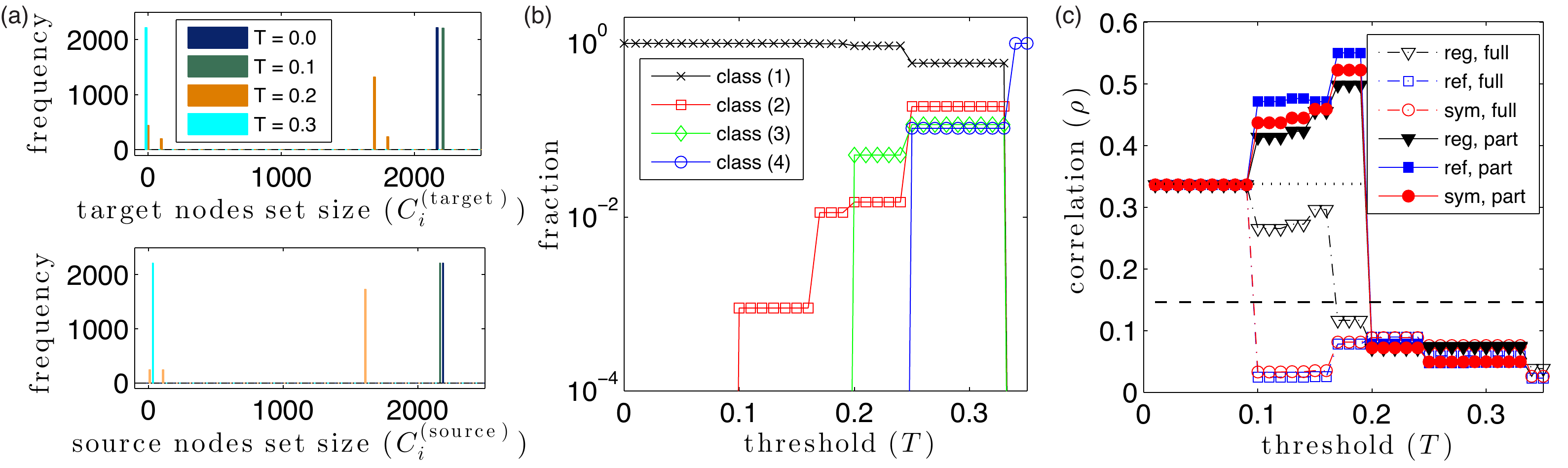}
\caption{
Equilibrium contagion sizes and geometry of WTM maps on the London transit network. 
(a)~Histogram of the frequency of sizes ${C}_i^{(\rm{target)}}$ for the target node sets and sizes ${C}_i^{(\rm{source)}}$ for the source node sets. Nodes tend to have either very large or very sizes of the target and source node sets, so we assign nodes into four classes: (1) large ${C}_i^{(\rm{target)}}$ and large ${C}_i^{(\rm{source)}}$, (2) small ${C}_i^{(\rm{target)}}$ and large ${C}_i^{(\rm{source)}}$, (3) large ${C}_i^{(\rm{target)}}$ and small ${C}_i^{(\rm{source)}}$, and (4) small ${C}_i^{(\rm{target)}}$ and small ${C}_i^{(\rm{source)}}$. In practice {(as we discuss in {\ref{sec:London}}),} we compare these values to $N/2$ to assign nodes to classes.
(b) Fraction of nodes in classes (1)--(4). All nodes shift from class (1) to class (4) as $T$ increases; however, for the approximate range $T\in(0.1,0.25)$, nodes are only in classes (1)--(3). 
(c) Pearson correlation coefficient $\rho$ for the WTM map (solid curves), Isomap (horizontal dotted line), and a 2D Laplacian eigenmap (horizontal dashed line). For the WTM map, we show results for the regular (``reg''), reflected (``ref''), and symmetric (``sym'') versions of the WTM map. For each version, we handle the activation times of infinity in two ways: we either (1) set these activation times to be $2N$ and consider the complete matrix of activation times (``full'') as we proceeded with our studies of synthetic networks; or (2) we neglect these values and examine only the remaining submatrix of activation times (``part'') after removing appropriate rows and columns. For the values of $T$ for which nodes are exclusively in classes (1) and (2) [i.e., for $T$ in the approximate range $(0.1,0.2)$], we find that $\rho$ increases for the WTM maps when we neglect the activation times of infinity. For the WTM maps in which we set the activation times of infinity to $2N$, the values of $\rho$ for $T \gtrapprox 0.1$ are considerably smaller than those for $T \lessapprox 0.1$.  This is especially prominent in the symmetric and reflected WTM maps. See \ref{sec:London} for further discussion.
}
\label{fig:london_007}
\end{figure*}

In Supplementary Fig.~\ref{fig:london_007}(a), we show histograms of the frequencies of (top panel) ${C}_i^{(\rm{target)}}$ and (bottom panel) ${C}_i^{(\rm{source)}}$ for the network nodes for WTM contagions with threshold values of $T\in\{0,0.1,0.2,0.3\}$ that we initialize with cluster seeding.  As expected, the WTM contagions infect almost all (or all) of the nodes when $T$ is small, whereas they spread to just a small number of nodes (or even $0$ nodes) when $T$ is sufficiently large. For example, observe for most nodes that ${C}_i^{(\rm{target)}}$ and ${C}_i^{(\rm{source)}}$ are approximately $N=2217$ for $T\le0.2$, whereas they are very small for most nodes for $T=0.3$. Additionally, the target and source node sets seem to exhibit dichotomous behavior in our experiments: they are often either very large (i.e., equal to or only a bit smaller than $N$) or very small (i.e., approximately $1$). We observe this feature both for the values of $T$ that we depict as well as for other values of $T$. (We examined $T \in \{0.01,0.02,\dots,0.5\}$.) Motivated by this observation, we assign the nodes to four classes for a given $T$: (1) nodes $i$ with large $C_i^{(\rm{target})}$ and large $C_i^{(\rm{source})}$ that can initiate large contagions and also adopt most contagions; (2) nodes $i$ with small $C_i^{(\rm{target})}$ and large $C_i^{(\rm{source})}$ that do not initialize large contagions but adopt almost all contagions; (3) nodes $i$ with large $C_i^{(\rm{target})}$ and small $C_i^{(\rm{source})}$ that initialize large contagions but almost never adopt contagions; and (4) nodes $i$ with small $C_i^{(\rm{target})}$ and small $C_i^{(\rm{source})}$ that neither initialize large contagions nor adopt many contagions.  In this classification, we arbitrarily take $N/2$ to be the boundary between ``large'' and ``small'' for both types of division. 

In Supplementary Fig.~\ref{fig:london_007}(b), we examine the fraction of nodes in each class as a function of $T$. For sufficiently small $T$ (e.g., $T<0.1$), almost all network nodes are in class (1) and almost all WTM contagions saturate the entire network. However, for large $T$ (e.g., $T>0.35$), all nodes are in class (4) because no contagions spread if the threshold is sufficiently large. The transitions between the different classes are interesting. Specifically, observe for the approximate range $T\in(0.1,0.2)$ that a small fraction of nodes moves from class (1) to class (2). Moreover, for the approximate range $T\in(0.2,0.25)$, class (2) and class (3) each contain only a small fraction of the nodes. Class (4) remains empty until $T\ge 0.25$, and it {then} grows as we increase $T$ until all nodes are in class (4) for $T\gtrapprox0.35$.

In Supplementary Fig.~\ref{fig:london_007}(c), we plot the Pearson correlation coefficient $\rho$ from Eq.~\eqref{eq:RHO} to compare the geometry of the nodes' original locations $\{\bold w^{(i)}\}$ to point clouds that result from WTM maps. We show results for the regular, reflected, and symmetric versions of the WTM map. (See {Sec.~I~C} of the main text.) We consider two different methods for handling the activation times of infinity [which necessarily arise whenever nodes are in classes (2)--(4)]. We either set the activation times to $2N$ and investigate the complete matrix of activation times, or we consider only finite activation times by using only the associated submatrix of activation times (after removing appropriate rows and columns that contain activation times of infinity). To illustrate our analysis, consider the latter case for the map $\mathcal{V}\mapsto\{\bold x^{(i)}\}$. We project each point $\bold x^{(i)}\in\mathbb{R}^N$ onto $\mathbb{R}^J$ with $J\le N$ by ignoring the dimensions that correspond to WTM contagions that are initialized with cluster seeding at nodes in class (2). This corresponds to considering the point cloud $\{\hat{\bold x}^{(i)}\}$, where $\hat{\bold x}^{(i)}=\Omega\bold x^{(i)}$ and the $J\times N$ projection matrix $\Omega$ has entries $\Omega_{jk_j}=1$, where $j \in \{1,\dots,J\}$, the set $\{k_1,k_2,\dots,k_J\}$ indicates the nodes that are not in class (2), and all other entries $\Omega_{jk}$ are equal to $0$.  For the reflected WTM map, we consider the map $\{i\}\mapsto\{{\bold y}^{(i)}\}$ only for nodes $i$ that are not in class (2). Finally, for the symmetric WTM map, we consider the map $\{i\}\mapsto\{\hat{\bold z}^{(i)}\}$, where $\hat{\bold z}^{(i)}=\Omega\bold z^{(i)}$, and we only map nodes $i\in\mathcal{V}$ that are not in class (2). 

{Returning our attention to {Supplementary} Fig.~\ref{fig:london_007}(c), note} for the reflected and symmetric WTM maps that we calculate the Pearson correlation coefficients $\rho$ only for the mapped points. As expected, $\rho$ for the WTM maps depends significantly on $T$, and one can observe that shifts in $\rho$ are well-aligned with changes in $C_i^{(\rm{target})}$ and $C_i^{(\rm{source})}$. The approximate range of thresholds $T\in(0.1,0.2)$ is particularly interesting, as we observe that values of $\rho$ for WTM maps increase when we neglect the activation times that are infinite. These larger $\rho$ values, in turn, indicate an improved agreement between the nodes' original locations and the geometry of the point clouds that result from the WTM maps. By contrast, for WTM maps in which we handle activation times of infinity by setting them to $2N$, we find that the values of $\rho$ are smaller for $T\gtrapprox 0.1$ than they are for $T\lessapprox 0.1$. {That is, when we handle the activation times of infinity in this way, we find that the WTM map becomes significantly distorted away from the known spatial embedding on Earth's surface.}

\begin{figure*}[t!]
\centering
\includegraphics[width=.9\linewidth]{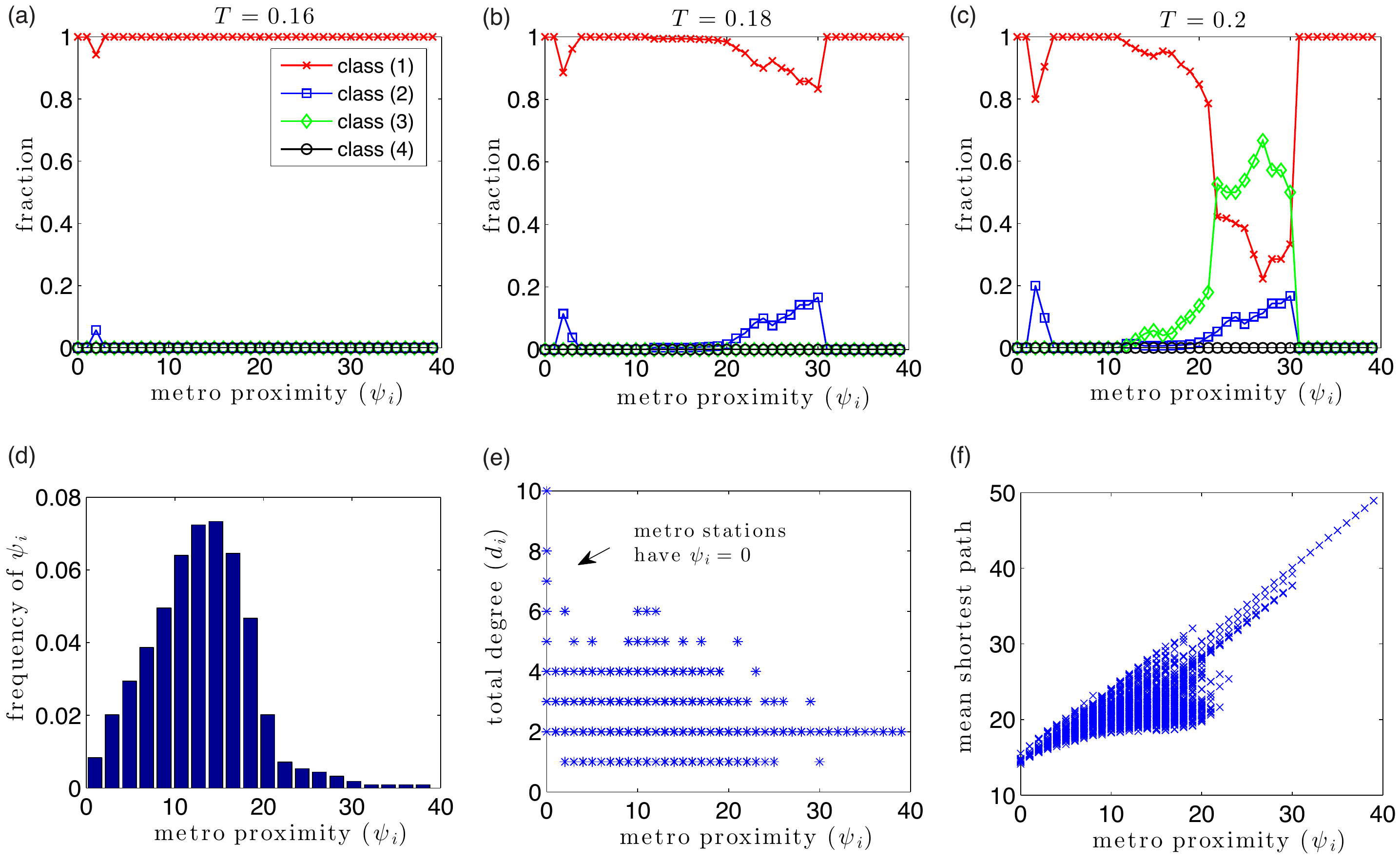}
\caption{
(Top row) For a given value of the metro proximity $\psi_i$ (i.e., the length of the shortest path from a node $i$ to a metro station), we show the fraction of nodes that we assign to the four classes (1)--(4) based on the sizes of their target and source node sets (see the text). (a) For $T=0.16$, almost all nodes are in class (1), though a few nodes are in class (2).  The latter are located $\psi_i=2$ edges from a metro station. (b) For $T=0.18$, most nodes are in class (1), but some nodes are in class (2). These are either 2--3 edges from a metro station, or they are ``isolated'' nodes that are distant from the other nodes (including, by definition, metro stations). (c) For $T=0.2$, nodes are in classes (1)--(3). As before, nodes in class (2) are either 2--3 edges from a metro station or are isolated. The nodes in class (3) are relatively isolated. 
(Bottom row) Properties of the metro proximities $\{\psi_i\}$.
 (d) Frequency of nodes with a given metro proximity $\psi_i$. (e) Scatter plot of the nodes' total degrees $\{d_i\}$ versus their metro proximities $\{\psi_i\}$. (f) For each node $i$, we plot the mean length of the shortest paths from that node to the remaining nodes versus its metro proximity $\psi_i$.  Note that isolated nodes, which are by definition distant from metro stations, are also distant from other nodes. See \ref{sec:London} for further discussion.
}
\label{fig:london_008}
\end{figure*}

We now attempt to gain some insight into which nodes we assign to classes (1)--(4). In {Supplementary} Fig.~\ref{fig:london_008}, we investigate the importance of the nodes' {\it metro proximities} $\{\psi_i\}$, where $\psi_i$ denotes the length of a shortest path on the London transit network from node $i$ to a metro station (i.e., $\psi_i=0$ for nodes that are metro stations, $\psi_i=1$ for their neighbors, and so on). We consider nodes that are at least 20 edges from any metro station to be ``isolated.'' In the top row, for a given value of the metro proximity $\psi_i$, we plot the fraction of nodes at that proximity in each of the four classes. Panels (a), (b), and (c), respectively, give results for threshold values of $T=0.16$, $T=0.18$, and $T=0.2$, which are characteristic of the range of $T$ in which we observe large values of $\rho$ for the WTM maps [see {Supplementary} Fig.~\ref{fig:london_007}(c)]. Note for $T=0.16$ that almost all nodes are in class (1), but several are in class (2). {Interestingly, all nodes in class (2) are} located $\psi_i=2$ edges from metro stations. {It follows that a} WTM contagion tends not to spread very far when we initialize it with cluster seeding centered at such nodes. For $T=0.18$, we again find that some nodes are in class (2), whereas the majority of nodes are in class (1). However, the nodes in class (2) are either 2--3 edges from a metro station or they are isolated nodes, which are distant from all other nodes (including, by definition, metro stations). For $T=0.2$, we find that nodes are in classes (1)--(3). As before, nodes in class (2) are either 2--3 edges from a metro station or are isolated. The nodes in class (3)---which are the class of nodes that are typically not reached by WTM contagions initialized with cluster seeding---are all relatively isolated, so one can construe them as peripheral nodes in the network \cite{Rombach2014}. Interestingly, our experimental results suggest that the inability to reach a node [i.e., nodes in classes (3) and (4)] is related to a global network property (i.e., whether it is ``isolated''), whereas the inability to seed a large contagion [i.e., nodes in classes (2) and (4)] depends on both local and global network properties.

In the bottom row of {Supplementary} Fig.~\ref{fig:london_008}, we show properties of the metro proximities $\{\psi_i\}$ for the London transit network. In panel (d), we show a histogram of the frequencies of nodes at a given metro proximity $\psi_i$, and we note that most nodes are 5--20 edges from a metro station. In panel (e), we give a scatter plot of the nodes' total degrees $\{d_i\}$ versus their metro proximities $\{\psi_i\}$. Note that the metro stations (for which $\psi_i=0$) have large degrees relative to the other nodes: their mean degree is $5$, whereas the mean degree of all nodes is approximately $2.59$. In panel (f), we show that isolated nodes, which by definition are distant from metro stations, also tend to be distant to other nodes in the London transit network. Specifically, for each node $i$, we plot the mean length of the shortest path from it to the remaining nodes $j\in\mathcal{V}\setminus\{i\}$ versus its metro proximity $\psi_i$. Nodes with large values of $\psi_i$ are also more distant (on average) to the other nodes. It is therefore appropriate to use the term ``isolated'' to describe these nodes.

Combining the results from {Supplementary} Figs.~\ref{fig:london_007} and \ref{fig:london_008}, we find when we ignore the activation times of infinity that WTM maps have larger values of $\rho$ when $T$ is in the approximate interval $(0.1,0.2)$ than when $T$ takes other values. The activation times of infinity result from the existence of a few nodes $i$ such that WTM contagions that we  initialize with cluster seeding centered at those nodes tend not to spread very far. These nodes tend to be in class (2), and they are often either 2--3 edges from metro stations or are isolated nodes. Finally, when $T$ is sufficiently large so that nodes belong to class (3) (e.g., as occurs for $T>0.2$), then the values of $\rho$ are comparatively very small. Recall that nodes in class (3), which almost never adopt contagions, are relatively isolated nodes in the network.

%%%%%%%%%%%%%%%%%%%%%%%%%%%%%%%%%%%%%%%%
~\\{\bf { \Large ``Egocentric'' Analysis of Geometry}}\\
%%%%%%%%%%%%%%%%%%%%%%%%%%%%%%%%%%%%%%%%

Thus far, we have studied geometry through the Pearson correlation coefficient $\rho$ given by Eq.~\eqref{eq:RHO}. As we discussed in {Sec.~III~D of the main text (and also see \ref{sec:geometry})}, $\rho$ describes the correlation between node-to-node distances $\{m(i,j)\}$ for the intrinsic locations $\{\bold w^{(i)}\}$ [see Eq.~\eqref{eq:distance1}] and node-to-node distances $\{m^{\rm{(WTM)}}(i,j)\}$ for the point clouds $\{\bold x^{(i)}\}$, $\{\bold y^{(i)}\}$, or $\{\bold z^{(i)}\}$ that result from a WTM map [see Eq.~\eqref{eq:distance_wtm}]. We calculate the correlation $\rho$ using the $N(N-1)/2$ unordered pairs of nodes $(i,j)\in\mathcal{V}\times\mathcal{V}$ (where $i\not=j$), and one can interpret it as comparing the geometry of these two point clouds at a ``network level.'' To gain further insight, we now compare the geometry of the two point clouds at a ``node level'' by computing ``egocentric'' correlation coefficients that consider only node-to-node distances that involve a particular node $i$. Specifically, we study a set of Pearson correlation coefficients $\{\hat\rho_i(T)\}$ for a given $i\in\mathcal{V}$.

\begin{figure*}[t!]
\centering
\includegraphics[width=.9\linewidth]{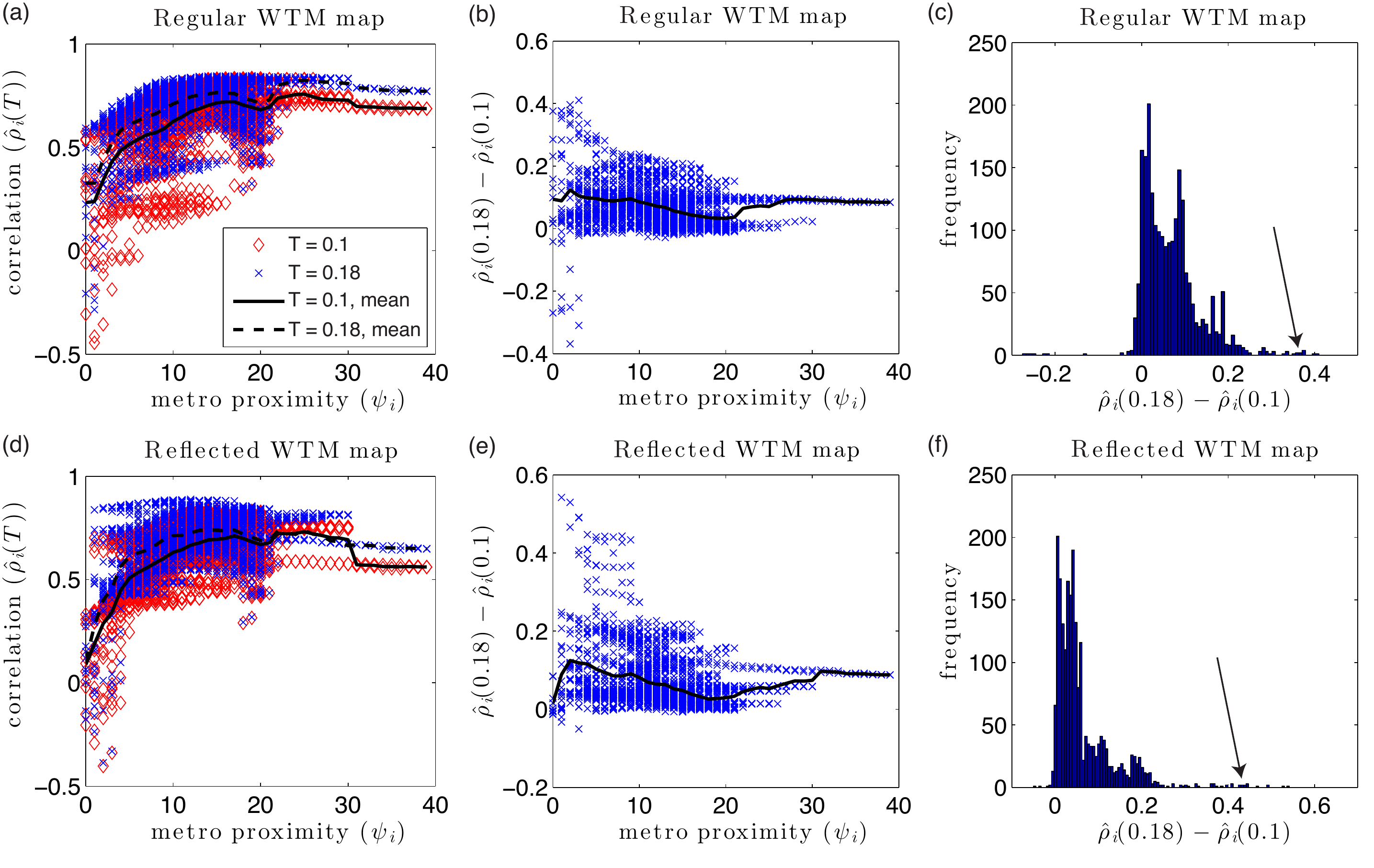}
\caption{
Egocentric correlation coefficients $\{\hat\rho_i(T)\}$ for WTM maps applied to the London transit network. We show results for (top panels) the map $\mathcal{V}\mapsto\{\bold x^{(i)}\}$ and (bottom panels) the map $\mathcal{V}\mapsto\{\bold y^{(i)}\}$. For both maps, we neglect activation times that are infinite (see \ref{sec:London}). 
(Left column) In panels (a) and (d), we plot the egocentric correlations $\{\hat\rho_i(T)\}$ versus the metro proximities $\{\psi_i\}$ for thresholds $T\in\{0.1,0.18\}$. We only show values for nodes that are not in class (2). The solid curve indicates the mean of $\hat\rho_i(0.1)$ for a given value of $\psi_i$, and the dashed curve indicates the mean of $\hat\rho_i(0.18)$ for a given value of $\psi_i$. Note that increasing $T$ typically leads to an increase of $\hat\rho_i(T)$ in both panels (a) and (d). 
(Center column) We plot $[\hat\rho_i(0.18)-\hat\rho_i(0.1)]$ versus $\psi_i$. The solid curves indicate the mean values for a given $\psi_i$. Observe that the values of $\hat\rho_i(T)$ are typically larger for $T=0.18$ than they are for $T=0.1$. 
(Right column) We plot the observed frequencies of $[\hat\rho_i(0.18)-\hat\rho_i(0.1)]$ for nodes $j \in \mathcal{V}$. Note that the frequencies are rather heterogeneous, and they appear to have a heavy tail (see the arrows): although $[\hat\rho_i(0.18)-\hat\rho_i(0.1)]$ tends to be small and positive for most nodes $j\in\mathcal{V}$, there are some nodes for which $[\hat\rho_i(0.18)-\hat\rho_i(0.1)]$ is rather large. See \ref{sec:London} for further discussion.
}
\label{fig:london_1}
\end{figure*}

We introduce the egocentric correlation coefficient $\hat\rho_i(T)$ for the regular WTM map $\mathcal{V}\mapsto\{\bold x^{(i)}\}$, and we note that one can apply it to any version of a WTM map. For each node $i$, we study the Pearson correlation coefficient $\hat \rho_i(T)$ that relates node-to-node distances $\{m(i,j)\}$ from node $i$ to all nodes $j\in\mathcal{V}$ with respect to the intrinsic locations $\{\bold w^{(i)}\}$ [see Eq.~\eqref{eq:distance1}] to the node-to-node distances $\{m^{\rm{(WTM)}}(i,j)\}$ from node $i$ to all nodes $\{j\}\in\mathcal{V}$ for a point cloud $\{\bold x^{(i)}\}$ that results from a WTM map [see Eq.~\eqref{eq:distance_wtm}]. Specifically, we compute
\begin{equation}
	\hat\rho_i(T) = \frac{\sum_{j=1}^N \big[m(i,j)-\overline{m(i,j)} \big]\big[m^{\rm{(WTM)}}(i,j)-\overline{m^{\rm{(WTM)}}(i,j)} \big]}{
\sqrt{\sum_{j=1}^N \big[m(i,j)-\overline{m(i,j)} \big]^2}
\sqrt{\sum_{j=1}^N \big[m^{\rm{(WTM)}}(i,j)-\overline{m^{\rm{(WTM)}}(i,j) }\big]^2}}\label{eq:RHO3} \,,
\end{equation}
where the bar above a variable indicates that we are taking its mean for all nodes $j\in\mathcal{V}$. Note the strong similarity between Eq.~\eqref{eq:RHO3} and Eq.~\eqref{eq:RHO}; the only difference is that the summations in Eq.~\eqref{eq:RHO3} are over $j$ rather than over both $j$ and $i$. 

In Supplementary Fig.~\ref{fig:london_1}, we study egocentric correlation coefficients $\{\hat\rho_i(T)\}$ for WTM maps on the London transit network for two values of the threshold $T$. In the top panels, we show results for the map $\mathcal{V}\mapsto\{\bold x^{(i)}\}$; in the bottom panels, we show results for the map $\mathcal{V}\mapsto\{\bold y^{(i)}\}$. For both maps, we handle the activation times of infinity by neglecting them. In the left column, we plot egocentric Pearson correlation coefficients $\{\hat\rho_i(T)\}$ versus the metro proximities $\{\psi_i\}$ for the threshold values $T\in\{0.1,0.18\}$. We show values only for nodes that are not in class (2). Note that the larger value of $T$ tends to have larger values of $\hat\rho_i(T)$ in both panels (a) and (d). We highlight this feature further in the center column by plotting $[\hat\rho_i(0.18)-\hat\rho_i(0.1)]$ versus $\psi_i$.  The solid and dashed curves, respectively, indicate the mean values for $T = 0.1$ and $T = 0.18$ for a given $\psi_i$. In the right column of {Supplementary} Fig.~\ref{fig:london_1}, we plot histograms of the frequencies of observed values $[\hat\rho_i(0.18)-\hat\rho_i(0.1)]$, and we note that they appear to have heavy tails: $[\hat\rho_i(0.18)-\hat\rho_i(0.1)]$ tends to be small and positive for most nodes $j\in\mathcal{V}$, but there are some nodes for which $[\hat\rho_i(0.18)-\hat\rho_i(0.1)]$ is rather large.

%%%%%%%%%%%%%%%%%%%%%%%%%%%%%%%%%%%%%%%%
~\\{\bf {\Large Summary of Experiments with the London Transit Network}}\\
%%%%%%%%%%%%%%%%%%%%%%%%%%%%%%%%%%%%%%%%

We studied WTM contagions on a London transit network in which nodes are intersections that are connected either by roads (which we interpreted as geometric edges) or by metro lines (which we interpreted as non-geometric edges). Similar to our study of WTM contagions on synthetic networks, we found that WFP and ANC arise for WTM contagions on this empirical network, and the type of epidemic propagation depends significantly on the contagion threshold $T$. We studied these WFP and ANC by analyzing the geometry of WTM maps, and we observed that the geometry of point clouds that result from WTM maps agree better with the geometry of the nodes' intrinsic locations on Earth's surface for values of $T$ in the approximate range $(0.1,0.2)$ than for other values of $T$. To obtain this result, we examined situations with activation times of infinity in two different ways: (1) setting those times to be $2N$, as in {the synthetic examples} in the main text; and (2) ignoring these values in our subsequent calculations.  We found the latter approach to be more useful for the London transit network. Our investigation led us to assign nodes into four classes based on their ability to initiate large contagions and consistently adopt contagions, and our calculations yielded an interesting connection between the proximity of nodes to metro stations and their behavior with respect to WTM contagions.

\clearpage

%%%%%%%%%%%%%%%%%%%%%%%%%%%%%%%%%%%%%%%%
%%%%%%%%%%%%%%%%%%%%%%%%%%%%%%%%%%%%%%%%
\section{Supplementary Note 2: Denoising Networks with WTM Maps}\label{sec:denoising}
%%%%%%%%%%%%%%%%%%%%%%%%%%%%%%%%%%%%%%%%
%%%%%%%%%%%%%%%%%%%%%%%%%%%%%%%%%%%%%%%%

The embedding of a network into a metric space has numerous applications, ranging from the control and optimization of dynamics to network ``denoising'' (i.e., the identification of spurious and missing edges). In this section, we highlight one application of WTM maps: the identification of noisy edges. 

Our methodology for denoising proceeds as follows. Given the WTM map for a network, we determine the length $m^{\textrm{(WTM)}}(i,j)$ [given by Eq.~\eqref{eq:distance_wtm} in \ref{sec:geometry}] in the embedding space of each edge $(i,j)\in\mathcal{E}$. Because we expect non-geometric edges to have larger lengths than geometric edges, examining the set of edge lengths $\{m^{\textrm{(WTM)}}(i,j)\}_{(i,j)\in\mathcal{E}}$ allows one to infer edge type. For example, by studying the distribution of edge lengths, one can choose a \emph{partitioning threshold} to partition the edges into classes (i.e., geometric and non-geometric) by comparing their lengths to the partitioning threshold.  There exist various heuristic approaches for selecting such a partitioning threshold, so we will consider all possible partitioning thresholds in our experiments. To do this, we construct \emph{receiver operating characteristic} (ROC) curves that examine the fraction of false positives and false negatives as the partitioning threshold is increased from the smallest edge length to the largest edge length. 

To gauge the performance of this approach for denoising networks, we compare our results to a popular approach based on subgraph statistics. For each edge $(i,j)\in\mathcal{E}$, we compute the 
Jaccard index $|\mathcal{N}_i \cap \mathcal{N}_j | / |\mathcal{N}_i \cup \mathcal{N}_j |$ to measure the overlap of the set $\mathcal{N}_i\triangleq \{k\in\mathcal{V}: A_{ik}\not=0\}$ of nodes that neighbor node $i$ with the set $\mathcal{N}_j\triangleq \{k\in\mathcal{V}: A_{jk}\not=0\}$ of nodes that neighbor node $j$ \cite{Goldberg2003}. Similar to our approach of comparing the edge lengths to some partitioning threshold, one can compare the edges' Jaccard indices to a partitioning threshold and then vary the partitioning threshold to yield a ROC curve. This allows for a direct comparison between the two approaches. 

Note that the approach of Ref.~\cite{Goldberg2003} is ``local''---i.e., each edge is classified based on the properties of the subgraph that contains nodes and edges that are adjacent to that edge---whereas denoising based on WTM maps is a ``global'' approach that uses an entire network for the denoising procedure.

%%%%%%%%%%%%%%%%%%%%%%%%%%%%%%%%%%%%%%%%
~\\{\bf {\Large Denoising the London Transit Network}}\\
%%%%%%%%%%%%%%%%%%%%%%%%%%%%%%%%%%%%%%%%

In our first experiment, we examine the utility of WTM maps for identifying the metropolitan lines in the London transit system that we study in Sec.~I~F of the main text. Because this noisy geometric network results from the merging of two network layers---a road network and the metropolitan system---our aim in this context is to disaggregate the two network layers based on the assumption that metropolitan lines connect nodes that are farther apart than those that are connected by roads. In this experiment, we purposely do not utilize the known node locations, as we are interested in the ability of WTM maps to identify the metro lines based on the network structure alone. 

In Supplementary Fig.~\ref{fig:london_denoise}(a), we plot ROC curves for symmetric WTM maps that we construct with various choices of the WTM threshold $T$. For these maps, we set the activation times of infinity to $2N$. Perfect inference of the noisy edges would correspond to a ROC curve in which the true-positive rate is always $1$ for any nonzero false-positive rate. We also note that the ROC curves for the WTM depend strongly on $T$. For $T=0.1$, the curve shows poor performance, similar to what one obtains using a Jaccard-index approach \cite{Goldberg2003}. For larger $T$ (i.e., $T\gtrapprox 0.1$), denoising based on WTM maps outperforms this other approach. Recall for the London transit network that when $T$ surpasses $0.1$, we observe an increase in WFP [which is indicated by the larger values of $\rho$ in {Supplementary} Fig.~\ref{fig:london_007}(c) when $T$ surpasses $0.1$]. The fact the ROC curves are still very high in {Supplementary} Fig.~\ref{fig:london_denoise}(a) when $T>0.2$ is somewhat unexpected, because we previously observed that there is disagreement between the geometry of WTM maps and the geometry of the actual London transit network for this range [see the drop in $\rho$ values that occurs in {Supplementary} Fig.~\ref{fig:london_007}(c) as $T$ surpasses $0.2$]. Interestingly, these results imply that the length of edges in the WTM map can still be very predictive for classifying edges as geometric or non-geometric even when the geometry of the WTM map disagrees with that of the actual network.

Here, we perform additional experiments to explore network properties that can help to shed light on these results.

%%%%%%%%%%%%%%%%%%%%%%%%%%%%%%%%%%%%%%%%
~\\{\bf {\Large Denoising Noisy Two-Dimensional Square Lattices}}\\
%%%%%%%%%%%%%%%%%%%%%%%%%%%%%%%%%%%%%%%%

We now do an experiment to highlight that local algorithms based on an assumption about the local network structure---e.g., a prevalence of 3-cycles (i.e., triangles) \cite{Goldberg2003}---can be very inaccurate if that assumption is invalid. In particular, modern road networks are known to exhibit a prevalence of subgraphs other than 3-cycles \cite{Strano2012} (e.g., city blocks can give rise to $4$-cycles), and we therefore expect the lack of 3-cycles to be a significant factor that influences the results in Supplementary Fig.~\ref{fig:london_denoise}(a).

In our experiment, we examine the inference of noisy edges for a synthetic noisy geometric network with $N=40^2$ nodes that are embedded as a 2D lattice with periodic boundary conditions. To construct the ``substrate'' geometric network, we place $d^{\textrm{(G)}}=4$ geometric edges for each node to connect nearest-neighbor nodes both horizontally and vertically. We then add $40^2$ non-geometric (i.e., ``noisy'' edges) uniformly at random to pairs of nodes that are not already connected by a geometric edge. Therefore, each node has $d^{\textrm{(NG)}}=2$ non-geometric edges on average. We note that this procedure---which adds non-geometric edges to a network that already has geometric edges---is identical to the procedure that we used for families {\bf(b)} and {\bf(d)} of the noisy ring networks in \ref{sec:models}.

\begin{figure*}[t]
\centering
\includegraphics[width=\linewidth]{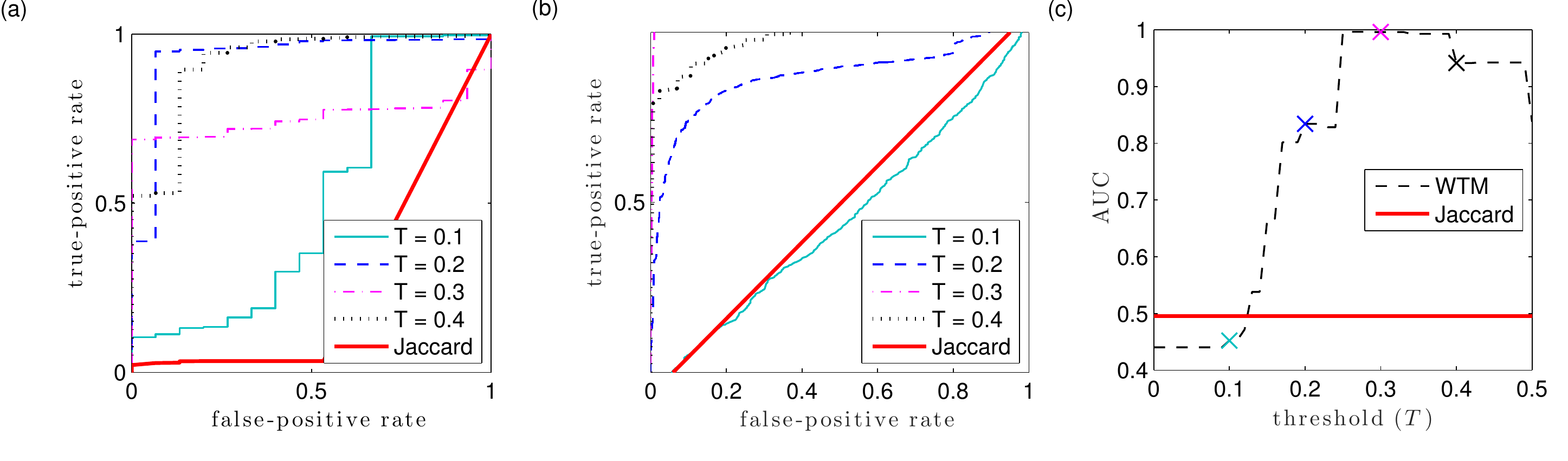}
\caption{
Inference of noisy edges in the London transit network that we studied in Sec.~I~F of the main manuscript. We show results for WTM maps using various values of the threshold parameter $T$.
(a) We show receiver operating characteristic (ROC) curves for the classification of edges in the London transit network as geometric or non-geometric based on WTM maps in addition to a ``local'' approach based on the Jaccard index \cite{Goldberg2003}.  
(b) ROC curves for inferring the noisy edges in a noisy 2D square lattice. 
(c) We plot the the areas under the ROC curves (AUC) for various values of $T$. We use crosses to indicate the values of $T$ that we used in panel (b). See \ref{sec:denoising} for further discussion.
}
\label{fig:london_denoise}
\end{figure*}

\begin{figure*}[b!]
\centering
\includegraphics[width=\linewidth]{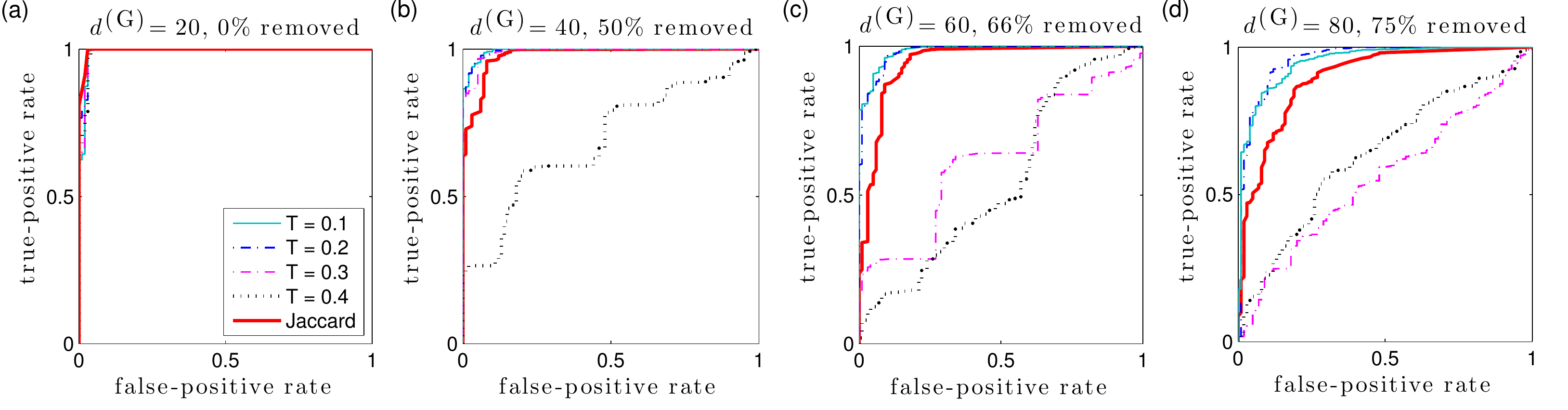}
\caption{
ROC curves for the classification of edges} as geometric or non-geometric based on WTM maps with various WTM thresholds $T$. The four panels (a)--(d) indicate results for four networks, which we generate so that their geometric edges have a tunable amount of stochasticity, which we implement by creating geometric edges and then removing some percentage of them uniformly at random. Note that an increased amount of stochasticity generally decreases the inference accuracy. {See \ref{sec:denoising} for further discussion.
}
\label{fig:removals}
\end{figure*}

In {Supplementary} Fig.~\ref{fig:london_denoise}(b), we plot ROC curves for the inference of noisy edges via symmetric WTM maps using  various values of the WTM threshold $T$. As before, we set the activation times of infinity to $2N$. Note that the best ROC curve corresponds to $T=0.3$, and that the ROC curves for WTM maps with $T\in\{0.2,0.3,0.,4\}$ are much better than that for the Jaccard-index approach.

To more precisely compare the different ROC curves for different $T$, in {Supplementary} Fig.~\ref{fig:london_denoise}(c), we plot the \emph{area under the ROC curve} (AUC) as a function of $T$ (dashed curve). We use crosses to indicate the values of $T$ that we used to generate {Supplementary} Fig.~\ref{fig:london_denoise}(b). To gauge the performance of inference using WTM maps, we note that the best attainable AUC is a value of 1 (which is almost reached for the WTM map with $T=0.3$). Using the horizontal red line, we show the AUC for the Jaccard-index approach. Its value is approximately $0.5$, indicating that it is comparable to random guessing in this scenario.

%%%%%%%%%%%%%%%%%%%%%%%%%%%%%%%%%%%%%%%%
~\\{\bf { \Large {Denoising} Noisy Ring Lattices with Removal of Geometric Edges}}\\%\label{sec:removal}
%%%%%%%%%%%%%%%%%%%%%%%%%%%%%%%%%%%%%%%%

In our final experiment, we examine the effect of of stochasticity on the inference of noisy edges. In particular, we explore the inference of noisy geometric networks in which we have removed some percentage of the geometric edges. We consider family {\bf(a)} (see \ref{sec:models}) of the noisy ring lattices, in which nodes are evenly placed on the unit circle in $\mathbb{R}^2$. We construct networks with $N=200$ nodes, where each node has $d^{\textrm{(NG)}}=1$ non-geometric edge and we consider various choices of geometric degree $d^{\textrm{(G)}}$. We then remove some percentage of the $Nd^{\textrm{(G)}}/2$ geometric edges---chosen uniformly at random---to include stochasticity. A nice feature of this experiment is that we can simultaneous increase $d^{\textrm{(G)}}$ and increase the edge-removal percentage so that the expected number of geometric edges (after removals) remains constant. 
In this procedure, note that although the number of geometric edges after removal is constant by construction, the mean length of the geometric edges tends to increase as we consider higher levels of stochasticity (i.e., by adding and then removing a larger number of edges). 

In Supplementary Fig.~\ref{fig:removals}, we plot ROC curves for the inference of noisy edges using symmetric WTM maps in which we set the activation times of infinity to $2N$. In panels (a)--(d), we show results for four networks, which we construct using progressively larger values of $d^{\textrm{(G)}}$ and in which remove an associated larger percentage of geometric edges so that, on average, every node has $d^{\textrm{(G)}}=20$ geometric edges after the removals. In each panel, we depict ROC curves for several values of the WTM threshold $T$ as well as for the Jaccard-index approach. Note that the ROC curves generally become lower as one moves from panel (a) to (d). In panel (a), for example, WTM maps for all $T$ values and the Jaccard-index approach lead to the accurate inference of the noisy edges. In panel (d), however, we find that the WTM map with $T=0.2$ leads to the best ROC curve. Our main finding is that incorporating stochasticity into the presence of geometric edges inhibits the successful inference of noisy edges. Depending on the value of $T$ and the network parameters, denoising based on WTM maps can perform either better or worse than a local approach based on the Jaccard index.

%%%%%%%%%%%%%%%%%%%%%%%%%%%%%%%%%%%%%%%%
~\\{\bf {\Large Summary of Experiments for Denoising Networks}}\\
%%%%%%%%%%%%%%%%%%%%%%%%%%%%%%%%%%%%%%%%

Our experiments highlight the use of WTM maps for the denoising of networks. We now briefly discuss the advantages and drawbacks of this novel technique in comparison to other approaches; an in-depth exploration would be very interesting, but it is well beyond the scope of the present paper. One class of previous approaches are ``local'' approaches that make an assumption about local network properties, such as a prevalence of 3-cycles (i.e., triangles), and then infer ``noisy'' edges to be the ones that do not follow this assumption. One can attempt to infer whether a particular edge is consistent with such an assumption by examining a Jaccard index or another subgraph statistic \cite{Goldberg2003,Singer2011,Singer2013}. Because these are ``local'' approaches, they have the advantage of being fast and straightforward to compute. In contrast, our approach based on WTM maps is reminiscent of ``global'' approaches that leverage a model for an entire network (i.e., as opposed to a model for the local subgraph structure) to find edges that do not adhere to the model \cite{Clauset2008,Liben2007,Guimera2009}. We note that these prior efforts often have focused on the problem of identifying missing (rather than spurious) edges, although these problems are closely related \cite{Guimera2009}.

In our experiments, we have illustrated examples of noisy geometric networks in which a global approach based on WTM maps can be advantageous to a local approach. We demonstrated that the global perspective of the WTM can be beneficial for denoising networks that fail to have a sufficient prevalence of 3-cycles (so that methods based on, e.g., the Jaccard index, do not perform well in those scenarios). We have demonstrated this situation both for the London transit network and for noisy 2D square lattices. Furthermore, even in scenarios in which 3-cycles are prevalent, we found that the WTM and Jaccard-index approaches show similar levels of performance [see {Supplementary} Fig.~\ref{fig:removals}(a)]. For noisy ring lattices that also include stochasticity in the geometric edges, we found (depending on the value of the WTM threshold $T$) that an approach based on WTM maps can lead to either higher or lower AUC values than an approach based on the Jaccard index.

\clearpage

%%%%%%%%%%%%%%%%%%%%%%%%%%%%%%%%%%%%%%%%
%%%%%%%%%%%%%%%%%%%%%%%%%%%%%%%%%%%%%%%%
\section{Supplementary Note 3: Generalizations of the Noisy Ring Lattice}\label{sec:models}
%%%%%%%%%%%%%%%%%%%%%%%%%%%%%%%%%%%%%%%%
%%%%%%%%%%%%%%%%%%%%%%%%%%%%%%%%%%%%%%%%

In the main text, we analyzed the WTM on noisy ring lattices. In this section, we review our construction of noisy ring lattices and introduce three additional families of noisy geometric networks that use an underlying ring manifold. In these families, we introduce heterogeneity into the nodes' geometric and non-geometric degrees, which we now denote, respectively, by $d_i^{\rm{(G)}}$ and $d_i^{\rm{(NG)}}$ for a given node $i$. We denote their means over the nodes by $\langle d_i^{\rm{(G)}}\rangle = N^{-1} \sum_i d_i^{\rm{(G)}}$ and $\langle d_i^{\rm{(NG)}}\rangle=N^{-1}\sum_i d_i^{\rm{(NG)}}$, respectively. We therefore adjust our definition of the ratio $\alpha$ to denote the ratio of the mean non-geometric degree to the mean geometric degree: 
\begin{equation}
	\alpha=\langle d_i^{\rm{(NG)}}\rangle/\langle d_i^{\rm{(G)}}\rangle\,. 
\end{equation}	
%For the network families that we define in Sec.~\ref{sec:models},
We note that it is equivalent to state that $\alpha$ denotes the ratio of the number of non-geometric edges to the number of geometric edges in a given network.

%%%%%%%%%%%%%%%%%%%%%%%%%%%%%%%%%%%%%%%%
~\\{\bf {\Large Families of Noisy Geometric Networks on a Ring Manifold}}\\
%%%%%%%%%%%%%%%%%%%%%%%%%%%%%%%%%%%%%%%%

We now define four families of noisy geometric networks on a ring manifold {given by the unit circle in $\mathbb{R}^2$}. We label these families as {\bf(a)}, {\bf(b)}, {\bf(c)}, and {\bf(d)}. In {Supplementary} Fig.~\ref{fig:network_models}, we illustrate an example network for each family and plot its corresponding adjacency matrix and degree distribution.

\begin{itemize}

{\small
\item{Family \bf(a)}. To generate the noisy ring lattice that we studied in the main text, we place $N$ nodes evenly on the unit circle in $\mathbb{R}^2$ so that each node $i$ has location $\bold w^{(i)}=[\cos(\theta_i),\sin(\theta_i)]^T$ with $\theta_i=2\pi i / N$. We then add geometric edges between neighboring node pairs $(i,j)\in\mathcal{V}\times\mathcal{V}$, so that each node $i$ has exactly $d_i^{\rm{(G)}}=\langle d_i^{\rm{(G)}}\rangle$ geometric edges. That is, we connect each node to its nearest $d_i^{\rm{(G)}}/2$ neighbors on each side, and we note that $d_i^{\rm{(G)}}$ is even because of symmetry. We then assign non-geometric edges randomly using (a slight modification of) the configuration model \cite{bollobasrandom} so that each node has exactly $d_i^{\rm{(NG)}}=\langle d_i^{\rm{(NG)}}\rangle$ non-geometric edges. As in the configuration model, we connect ends of edges (i.e., ``stubs'') to each other uniformly at random, but we disallow self-edges and multi-edges. Our implementation of the configuration model is a slight modification of the original version, because we want to guarantee that the set of geometric edges is disjoint from the set of non-geometric edges. Specifically, if we propose a candidate edge between two nodes that would lead to a disallowed situation (i.e., it would lead to a self-edge, multi-edge, or an edge that is already a geometric edge), then we discard the candidate edge, and we propose a new candidate edge as prescribed by the configuration model. The resulting network is a {$(\langle d_i^{\rm{(G)}}\rangle+\langle d_i^{\rm{(NG)}}\rangle)$-regular network} that contains $N\langle d_i^{\rm{(G)}}\rangle/2$ geometric edges and $N\langle d_i^{\rm{(NG)}}\rangle/2$ non-geometric edges. The geometric edges form a deterministic backbone (as in the Newman-Watts variant \cite{Newman2000,Porter2012} of the Watts-Strogatz model \cite{Watts1998}), whereas we obtain non-geometric edges through a {stochastic} process.
}

{\small
\item {Family \bf (b)}. Our first generalization of the noisy ring lattice in family {\bf (a)} is to allow heterogeneity in the number of non-geometric edges that are incident to a given node $i$ (i.e., its non-geometric degree $d_i^{\rm{(NG)}}$). The total number of non-geometric edges is still equal to the constant $N\langle d_i^{\rm{(NG)}}\rangle/2$, but we now distribute them uniformly at random among the $\frac{N\cdot (N-1-d^{\rm{(G)}})}{2}$ possible edge locations that are unoccupied by geometric edges. Hence, the subgraph that consists only of non-geometric edges limits to an Erd\H{o}s-R\'{e}nyi (ER) network when $N\gg d^{\rm{(G)}}$ \cite{bollobasrandom}. The distribution of non-geometric degrees is thus a binomial distribution that is centered at $\langle d_i^{\rm{(NG)}}\rangle$. 
}

\begin{figure}[thb!]
\centering
\includegraphics[width=.9\linewidth]{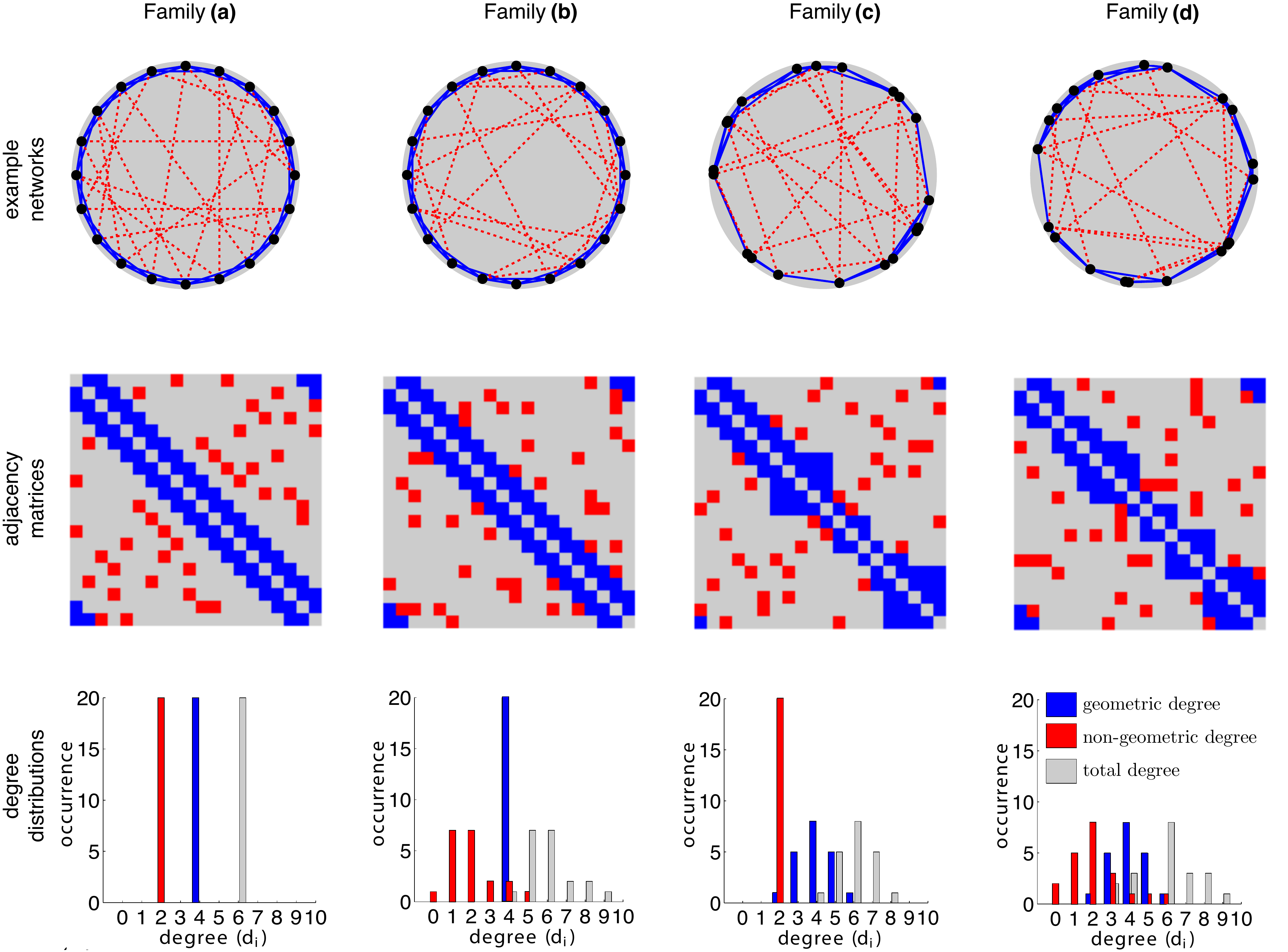}
\vspace{0cm}
\caption{
Example networks with $N=20$ nodes, a mean geometric degree of {$\langle d_i^{\rm(G)}\rangle=4$}, and a mean non-geometric degree of $\langle d_i^{\rm(NG)}\rangle=2$, for four families of noisy geometric networks on a ring manifold: family {\bf(a)}, the noisy ring lattice (which we also discuss in the main manuscript), for which nodes are evenly spaced and have constant geometric and non-geometric degrees; family {\bf(b)}, for which the nodes are evenly spaced, have constant geometric degrees, and have heterogeneous non-geometric degrees; family~{\bf(c)}, for which we sample the node locations from the {unit circle in $\mathbb{R}^2$} using a {stochastic} process (see the text), and the nodes have heterogeneous geometric degree and constant non-geometric degrees; and family {\bf(d)}, for which we {randomly} sample the node locations from the {unit circle in $\mathbb{R}^2$}, and the nodes have heterogeneous geometric and non-geometric degrees. (As we discuss in the text, we do the sampling uniformly at random.) The top row depicts example networks, where blue solid and red dashed lines indicate geometric and non-geometric edges, respectively. The center row depicts the corresponding adjacency matrices; blue pixels {indicate} geometric edges {that} align along the diagonal, whereas red pixels {indicate} non-geometric edges {that} arise randomly. The bottom row depicts the corresponding distributions for the geometric (red), non-geometric (blue), and total (grey) degrees. Note that the geometric degrees are identical for families {\bf{(a)}} and {\bf{(b)}}, with $d_i^{\rm{(G)}}=4$, whereas they are heterogeneous with mean $\langle d_i^{\rm{(G)}}\rangle=4$ for families {\bf{(c)}} and {\bf{(d)}}. For families {\bf{(a)}} and {\bf{(c)}}, the non-geometric degrees are identical ($d_i^{\rm{(NG)}}=2$), whereas they are heterogeneous with mean $\langle d_i^{\rm{(NG)}}\rangle=2$ for families {\bf{(b)}} and {\bf{(d)}}. {See \ref{sec:models} for further discussion.}
}
\label{fig:network_models}
\end{figure}

{\small
\item {Family \bf (c)}. Our second generalization of {\bf (a)} is to allow heterogeneity in the node locations on the unit circle in $\mathbb{R}^2$. Constraining geometric edges by distance, in turn, leads to heterogeneity in the number of geometric edges that are incident to a given node $i$ (and hence in its geometric degree $d_i^{\rm{(G)}}$). To make such a generalization in a tunable manner, we assign the node locations (or, equivalently, the angles $\{\theta_i\}$ in the case of the unit circle) to be evenly spaced as for family {\bf (a)}, and we then perturb these locations using a random variable $\delta \theta_i$, so that the location for each node $i$ is given by $[\cos(\theta_i+\delta\theta_i),\sin(\theta_i+\delta\theta_i)]^T$. We consider a Gaussian-distributed random variable $\delta \theta_i\sim\mathcal{N}\left(0,(s\frac{2\pi}{N})^2\right)$, where one can vary $s$ to adjust the amount of heterogeneity in node location along {a} ring manifold. The choice $s=0$ recovers the original node locations, and $s\to\infty$ corresponds to sampling locations on the unit circle uniformly at random. Unless we specify otherwise, we use $s=1/2$. To generate geometric edges, we choose a parameter $\epsilon>0$ and place edges between all pairs of nodes $i$ and $j$ such that $|\theta_i-\theta_j|<\epsilon$. To compare networks from family {\bf (c)} to networks from families {\bf (a)} and {\bf (b)}, for which the nodes have the identical geometric degree $d_i^{\rm{(G)}}=\langle d_i^{\rm{(G)}}\rangle$, we choose the parameter $\epsilon$ so that each network in family {\bf (c)} has exactly $N\langle d_i^{\rm{(G)}}\rangle/2$ edges. 
}

{\small
\item {Family \bf (d)}. Our final network family combines the generalizations from families {\bf (b)} and {\bf (c)} so that there is heterogeneity in the non-geometric degrees $\{d_i^{\rm{(NG)}}\}$ (where the mean is $\langle d_i^{\rm{(NG)}}\rangle$), the geometric degrees $\{d_i^{\rm{(G)}}\}$ (where the mean is $\langle d_i^{\rm{(G)}}\rangle$), and the node locations $\{\bold w^{(i)}\}$.  
}
\end{itemize}

%%%%%%%%%%%%%%%%%%%%%%%%%%%%%%%%%%%%%%%%
~\\{\bf {\Large Perturbed Bifurcation Results}} \\
%%%%%%%%%%%%%%%%%%%%%%%%%%%%%%%%%%%%%%%%

Equations (1) and (2) in the main text give sequences of critical thresholds that determine WFP and ANC for large networks of family {\bf(a)}. Recall that the degrees $d_i^{\rm{(G)}}$ and $d_i^{\rm{(NG)}}$ for each node $i$ are deterministic and constant for family {\bf(a)}.  However, {here} we introduce various types of stochasticity (and hence heterogeneity) for these degrees in network families {\bf (b)}--{\bf(d)}. Because of such heterogeneity, the critical thresholds that we derived previously for network family {\bf(a)} no longer accurately describe the WTM contagion dynamics. However, based on numerical experiments, we find that {Eqs.~(1) and (2) in the main text} still describe contagion dynamics at a given node $i$ if we use the correct geometric and non-geometric degrees. Specifically, the ability of node $i$ to adopt a contagion via WFP when it has no infected non-geometric neighbors is given approximately by Eq.~\eqref{eq:WFP_crits} with the substitutions $d^{\rm{(G)}}\mapsto d_i^{\rm{(G)}}$ and $d^{\rm{(NG)}}\mapsto d_i^{\rm{(NG)}}$. Similarly, the ability of node $i$ to adopt a contagion via ANC is given approximately by Eq.~\eqref{eq:ANC_crits} with the substitutions $d^{\rm{(G)}}\mapsto d_i^{\rm{(G)}}$ and $d^{\rm{(NG)}}\mapsto d_i^{\rm{(NG)}}$. Hence, for each node $i\in\mathcal{V}$, there are sequences of critical thresholds, $\{T_k^{\rm{(WFP)}}\}$ and $\{T_k^{\rm{(ANC)}}\}$, that are (potentially) specific to that node. Consequently, the nodes can exhibit qualitatively dissimilar contagion dynamics with respect to WFP and ANC. For example, for a given threshold $T$, some nodes can have geometric and non-geometric degrees that support WFP but no ANC, whereas other nodes can have degrees that support both WFP and ANC.  Nevertheless, one can construe the bifurcation analysis that we developed for family {\bf{(a)}} as an approximate bifurcation analysis for the other families. In this light, note that if the degree heterogeneities are sufficiently small compared to the mean degrees, then we still identify four different qualitative regimes of WTM contagion dynamics that are marked by the absence versus presence of WFP and ANC. However, the boundaries that separate these regimes are perturbations of what we found for family {\bf{(a)}}.

More precisely, for each node $i$, let $\delta_i^{\rm{(G)}}=d_i^{\rm{(G)}}-\langle d_i^{\rm{(G)}}\rangle$ denote the difference between its geometric degree and the mean geometric degree. Similarly, let $\delta_i^{\rm{(NG)}}=d_i^{\rm{(NG)}}-\langle d_i^{\rm{(NG)}}\rangle$ denote the difference between its non-geometric degree and the mean non-geometric degree. Restricting our attention to the critical thresholds given by {Eqs.~(1) and (2)} in the main text for $k=0$ (although one can write similar expressions for other values of $k$), we can express the critical thresholds in terms of $\delta_i^{\rm{(G)}}$ and $\delta_i^{\rm{(NG)}}$ as
\begin{align}
	T_0^{\rm{(ANC)}} (\delta_i^{\rm{(G)}},\delta_i^{\rm{(NG)}}) &\triangleq \frac{ 1 + \frac{\delta_i^{\rm{(NG)}} }{ \langle d_i^{\mathrm{(NG)}}\rangle }}
	{{\alpha^{-1}+1 + \frac{(\delta_i^{\rm{(G)}} + \delta_i^{\rm{(NG)}}) }{ \langle  d^{\rm{(NG)}} \rangle  }}}\label{eq:ANC_crits2}\\
		T_0^{\rm{(WFP)}}(\delta_i^{\rm{(G)}},\delta_i^{\rm{(NG)}}) &\triangleq \frac{1 + \frac{\delta_i^{\rm{(G) }}}{\langle d_i^{\mathrm{(G)}}\rangle}}
	{{2+2\alpha +\frac{2(\delta_i^{\rm{(G)}} + \delta_i^{\rm{(NG)}})}{\langle d_i^{\mathrm{(G)}}\rangle }}}\label{eq:WFP_crits2}.
\end{align}
Expressions \eqref{eq:ANC_crits2} and \eqref{eq:WFP_crits2} summarize the effect of degree heterogeneity on the critical thresholds that describe a WTM contagion at a given node $i$. When there is no degree heterogeneity (e.g., family {\bf(a)}, where $\delta_i^{\rm{(G)}}=\delta_i^{\rm{(NG)}}=0$), we recover our results from the main text: $T_0^{\rm{(ANC)}}(0,0)=1/(\alpha^{-1}+1)$ and $T_0^{\rm{(WFP)}}(0,0)=1/(2+2\alpha)$. {Meanwhile}, when $\delta_i^{\rm{(G)}}$ and $\delta_i^{\rm{(NG)}}$ are both small compared to their respective means $\langle d_i^{\rm{(G)}}\rangle $ and $\langle d_i^{\rm{(NG)}}\rangle$, the perturbed critical thresholds are approximately equal to those for the mean degree. In other words, $T_0^{\rm{(ANC)}}(\delta_i^{\rm{(G)}},\delta_i^{\rm{(NG)}})\approx T_0^{\rm{(ANC)}}(0,0)$ when $(\delta_i^{\rm{(G)}}+\delta_i^{\rm{(NG)}})/\langle d_i^{\rm{(NG)}}\rangle$ is small, and $T_0^{\rm{(WFP)}}(\delta_i^{\rm{(G)}},\delta_i^{\rm{(NG)}})\approx T_0^{\rm{(WFP)}}(0,0)$ when $(\delta_i^{\rm{(G)}}+\delta_i^{\rm{(NG)}})/\langle d_i^{\rm{(G)}}\rangle $ is small. We therefore interpret our bifurcation analysis for network family {\bf{(a)}} as an approximate bifurcation analysis for network families {\bf{(b)}}--{\bf{(d)}}. We expect our interpretation to be increasingly accurate as the mean degrees become larger relative to the heterogeneity in the degrees.

\begin{figure*}[ht]
\centering
\includegraphics[width=.8\textwidth]{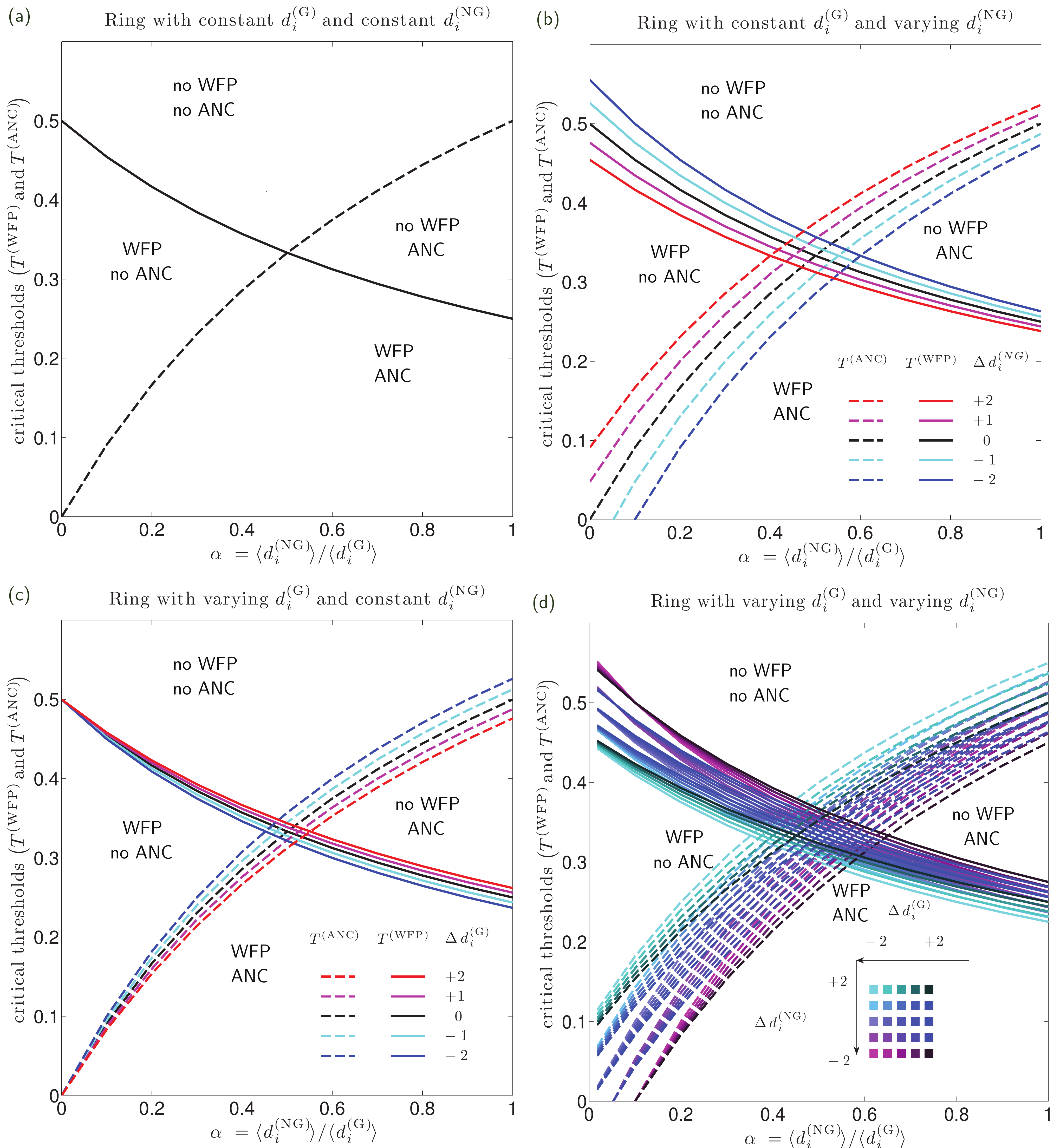}
\caption{
Node-specific critical thresholds for the $(T,\alpha)$ parameter plane, where $\alpha$ denotes the ratio of the mean non-geometric degree $\langle d_i^{\rm{(NG)}}\rangle$ to the mean geometric degree $\langle d_i^{\rm{(G)}}\rangle$. We plot Eqs.~\eqref{eq:ANC_crits2} and \eqref{eq:WFP_crits2} for nodes with $|\delta_i^{\rm{(G)}}|\le2$ and $|\delta_i^{\rm{(NG)}}|\le2$. We show these {curves} for the example degrees $d_i^{\rm{(G)}}=20$ and $d_i^{\rm{(NG)}} \in [0,20]$. Panels (a)--(d), respectively, depict the appropriate choices for $\delta_i^{\rm{(G)}}$ and $\delta_i^{\rm{(NG)}}$ for the heterogeneities in network families {\bf(a)}--{\bf(d)}. If the {perturbations of the} node degrees are both sufficiently small compared to the nodes' degrees, then we still obtain four qualitatively different contagion regimes for all nodes. We again characterize these different regimes based on the presence versus absence of WFP and ANC. However, because of the heterogeneity of the nodes' degrees, transitions between these regimes in the $(T,\alpha)$ parameter plane occur at different values for different nodes. That is, the boundaries between the WTM contagion regimes have ``thickened.'' We note for any fixed $|\delta^{\rm{(G)}}|,|\delta^{\rm{(NG)}}|>0$ that the perturbed curves approach those that correspond to $\delta^{\rm{(G)}}=\delta^{\rm{(NG)}}=0$ as the mean geometric and non-geometric degrees increase. {See \ref{sec:models} for further discussion.}
}
\label{fig:analytical_thresholds}
\end{figure*}

In {Supplementary} Fig.~\ref{fig:analytical_thresholds}, we plot curves that indicate the node-specific critical thresholds given by Eqs.~\eqref{eq:ANC_crits2} and \eqref{eq:WFP_crits2} for the $(T,\alpha)$ parameter plane. We show results for $d^{\rm{(G)}}=20$ and $d^{\rm{(NG)}}\in [0,20]$ for several choices of $\delta_i^{\rm{(G)}}$ and $\delta_i^{\rm{(NG)}}$ such that $|\delta_i^{\rm{(G)}}|\le2$ and $|\delta_i^{\rm{(NG)}}|\le2$. Panels (a)--(d), respectively, demonstrate the heterogeneity that arises in $\delta_i^{\rm{(G)}}$ and $\delta_i^{\rm{(NG)}}$ from network families {\bf(a)}--{\bf(d)}. Note for all panels that there exist parameter regimes in which the nodes support similar contagion phenomena (i.e., WFP and no ANC, WFP and ANC, no WFP and ANC, or no WFP and no ANC) even through their degrees are heterogeneous. Therefore, one can construe the set of curves that one obtains for multiple values of $\delta_i^{\rm{(G)}}$ and $\delta_i^{\rm{(NG)}}$ as a ``thickening'' of the boundary between regions of qualitatively different dynamics.  In other words, as we vary parameters, we see that the transitions between regions of different dynamics can occur {for slightly different parameter values} for different nodes in a network. Note, however, that this interpretation does not take into account the distribution of node degrees, as we have only shown the critical threshold curves in {Supplementary} Fig.~\ref{fig:analytical_thresholds} for degrees that are near the mean degrees (i.e., for $|\delta_i^{\rm{(G)}}|\le2$ and $|\delta_i^{\rm{(NG)}}|\le2$). As we discussed in {this Supplementary Note}, families {\bf{(b)}}--{\bf{(d)}} lead to networks with heterogeneous node degrees, and it is then possible that $|\delta_i^{\rm{(G)}}|>2$. (For example, the non-geometric degrees follow a binomial distribution for families {\bf{(b)}} and {\bf{(d)}}.) Therefore, one can construe our bifurcation analysis for family {\bf{(a)}} as an approximate bifurcation analysis for families {\bf{(b)}}--{\bf{(d)}} only when the heterogeneities in the two types of degrees are both sufficiently small when compared to the means $\langle d_i^{\rm{(G)}}\rangle$ and $\langle d_i^{\rm{(NG)}}\rangle$.

\clearpage
%%%%%%%%%%%%%%%%%%%%%%%%%%%%%%%%%%%%%%%%
%%%%%%%%%%%%%%%%%%%%%%%%%%%%%%%%%%%%%%%%
\section{Supplementary Note 4: A Set of Filtrations Defines a Metric}\label{sec:filtration}
%%%%%%%%%%%%%%%%%%%%%%%%%%%%%%%%%%%%%%%%
%%%%%%%%%%%%%%%%%%%%%%%%%%%%%%%%%%%%%%%%

In this section, we prove that the set of activation times for a WTM contagion with threshold $T$ on a network induces a metric on the node set $\mathcal{V}=\{1,2,\dots, N\}$. Let $\tilde x^{(i)}_j$ denote the activation time (which we assume to be finite for all node pairs $(i,j)\in\mathcal{V}\times\mathcal{V}$) for node $i$ for a contagion initialized with the seed node $j$. We will show that $\left(\mathcal{V},m^{\rm(WTM)}(i,j)\right)$ is a discrete metric space with metric $m^{\rm(WTM)}(i,j) = \tilde x^{(i)}_j + \tilde x^{(j)}_i$. However, rather than showing this result for the specific case of a WTM contagion, we will prove a more general result using the observation that the growing set of infected nodes during one realization of a WTM contagion defines a {``filtration''} of the node set $\mathcal{V}$. We will therefore prove that any ``complete'' and ``consistent'' set of filtrations (see the definitions below) on a finite set $\mathcal{V}$ induces a metric on $\mathcal{V}$. Subsequently, because the filtrations that result from realizations of a WTM contagion on a given network with contagion seeds $\{j\}$ for $j \in \{1,\dots,N\}$ satisfy the conditions of completeness and consistency, it follows that $m^{\rm(WTM)}(i,j) = \tilde x^{(i)}_j + \tilde x^{(j)}_i$ is a metric on $\mathcal{V}$ whenever $\tilde x^{(j)}_i<\infty$ for every $i,j\in\mathcal{V}$.\\

%%%%%%%%%%%%%%%%%%%%%%%%%%%%%%%%%%%%%%%%%%%%%%%%%%%
~\\{\bf \Large Complete and Consistent Filtrations}\\
%%%%%%%%%%%%%%%%%%%%%%%%%%%%%%%%%%%%%%%%%%%%%%%%%%%

Before proving that the set of activation times---and, more generally, any ``complete'' and ``consistent'' set of ``filtrations''---leads to a metric, we give a few definitions.\\

\noindent{\bf Definition:} \emph{Filtration}.\\
Consider a sequence of sets $\mathcal{N}_{t}$ for $t \in \{1,2,\dots\}$. The sequence of sets is called a {\it filtration} \cite{eh,edels2010,carlsson} if it has the property that $\mathcal{N}_t \subseteq \mathcal{N}_{t+1} $ for all $t$\footnote{One can also define a filtration sequence of filtrations using the superset relation $\mathcal{N}_t \supseteq \mathcal{N}_{t+1}$.  Additionally, one can also generalize the notion of a filtration to concepts like ``zigzag filtrations,'' which allow both subset and superset relations amidst the sequence of sets \cite{Carlsson2}.}.\\

\noindent{\bf Definition:} \emph{Completeness}.\\
Let $\mathcal{V}$ be a finite set with cardinality $|\mathcal{V}|=N$. We define a set of filtrations to be {\it complete} if there are $N$ filtrations of the following form: for every $j\in\mathcal{V}$, there exists a filtration such that
\begin{equation}
\{ j\}=\mathcal{N}_0(j) \subseteq  \mathcal{N}_1(j) \subseteq  \mathcal{N}_2(j)\subseteq  \dots \subseteq  \mathcal{N}_{t^*_j}(j) =\mathcal{V}.\label{eq:completeness}
 \end{equation}
 Note that the filtration $\{\mathcal{N}_t(j)\}$ consists of nested sets $\mathcal{N}_t(j)$, where the innermost set is the element $\{j\}$ and the outermost set (i.e., the $t_j^*$th set) is the complete set $\mathcal{V}$ of indices.\\

\noindent{\bf Definition:} \emph{Consistency}.\\
Let $\mathcal{V}$ be a finite set with cardinality $|\mathcal{V}|=N$, {and consider a set of filtrations in which the $j$th filtration corresponds to nested sets $\{\mathcal{N}_t(j)\}$ that are indexed by $t$.} We define the set of filtrations to be \emph{consistent} if, for any two filtrations $\{\mathcal{N}_t(i)\}$ and $\{\mathcal{N}_t(j)\}$ {from the set}, the following is true:
\begin{equation}
	\mathcal{N}_t(i) \subseteq \mathcal{N}_{\tau}(j) \implies \mathcal{N}_{t+1}(i) \subseteq \mathcal{N}_{\tau+1}(j)\,,\label{eq:consistent}
\end{equation}
where the indices $t$ and $\tau$ can be different from each other. 

\clearpage
%%%%%%%%%%%%%%%%%%%%%%%%%%%%%%%%%%%%%%%%%%%%%%%%%%%
~\\{\bf \Large Filtration-Induced Metrics}\\
%%%%%%%%%%%%%%%%%%%%%%%%%%%%%%%%%%%%%%%%%%%%%%%%%%%

\noindent{\bf Theorem:} \emph{A Metric Induced by Filtrations}.\\\\
Let $\mathcal{V}$ be a finite set with cardinality $|\mathcal{V}|=N$, and let $\mathcal{N}_{t}(j)$ denote sets that define a complete and consistent set of filtrations on $\mathcal{V}$. Additionally, let $t^{(i)}_j$ denote the smallest index $t$ such that $i\in \mathcal{N}_{t}(j)$. It then follows that $m(i,j) = t^{(j)}_i + t^{(i)}_j $ defines a metric on the set $\mathcal{V}$.\\

\noindent{\it Proof:}
\noindent First, we show that $m(i,j)\ge 0$ and {that} $m(i,j)=0 $ {implies} $ i=j$. Note that $t_j^{(i)} \geq 0$ for any $i,j\in\mathcal{V}$; this implies that $m(i,j)\ge0$. Similarly, $m(i,j)=0$ requires $t_j^{(i)}=t_i^{(j)}=0$, which in turn requires that $\mathcal{N}_{0}(j)=\{i\}$ (and $\mathcal{N}_{0}(i)=\{j\}$). However, we know by definition that $\{j\}=\mathcal{N}_{0}(j)$ (and $\{i\}=\mathcal{N}_{0}(i)$), so it must be the case that $i=j$. It is trivial to show that $m(i,j)=m(j,i)$. Finally, we complete the proof by showing that $m(i,j)$ satisfies the triangle inequality: $m(i,j)\le m(i,k)+m(k,j)$. This step is a bit more complicated, and it relies on the completeness and consistency of the set of filtrations. Using the definition of $m(i,j)$, it suffices to show that $t^{(i)}_j\le t^{(k)}_j + t^{(i)}_k$. Using the notation $a={t^{(i)}_j}$, $b={t^{(k)}_j}$, and $c={t^{(i)}_k}$, we will prove that $a\le b+c$. Because the result is trivial when $b\ge a$ due to the non-negativity of $c$, we can assume that $a> b$. By definition, it must be the case that $i\in\mathcal{N}_{a}(j)$, $k\in\mathcal{N}_{b}(j)$, and $i\in\mathcal{N}_{c}(k)$. Because $b < a$, it must also be the case that $\mathcal{N}_{b}(j) \subseteq \mathcal{N}_{a}(j)$.  We now consider $\{k\}=\mathcal{N}_0(k) \subseteq \mathcal{N}_{b}(j)$, which uses the completeness of the filtrations. Using the consistency property, it follows that $\mathcal{N}_{1}(k) \subseteq \mathcal{N}_{b+1}(j)$, $\mathcal{N}_{2}(k) \subseteq \mathcal{N}_{b+2}(j)$, and so on. Repeating this procedure demonstrates that $\mathcal{N}_{c}(k) \subseteq \mathcal{N}_{b+c}(j)$. Noting that $i\in\mathcal{N}_{c}(k)$, it follows that $i\in \mathcal{N}_{b+c}(j)$. It follows, in turn, that $t^{(i)}_j\le b+c$, which is equivalent to $a\le b+c$. \qed\\

\noindent{\bf{Corollary}:} \emph{A Metric Induced by WTM Contagions}.\\
Consider a network with node set $\mathcal{V}$ and edge set $\mathcal{E}$ that consists of a single connected component. Let $ \tilde x_j^{(i)}$ denote the activation time of node $i$ for a WTM contagion with seed $\{j\}$.  As before, we assume that $\tilde x^{(i)}_j$ is finite for all node pairs $(i,j)\in\mathcal{V}\times\mathcal{V}$. {It then follows that} $m(i,j)=\tilde x_i^{(j)}+ \tilde x_j^{(i)}$ is a metric on the node set $\mathcal{V}$.\\

\noindent{\it Proof:}
\noindent  It suffices to show that $N$ realizations of a WTM contagion with the set of contagion seeds $\mathcal{S}^{(j)}=\{j\}$ (for $j\in\mathcal{V}$) produces a complete and consistent set of filtrations on $\mathcal{V}$. It will be convenient to use the notation $t^{(i)}_j= \tilde x^{(i)}_j$. We first prove completeness. Let $\mathcal{N}_t(j)$ denote the set of nodes for realization $j$ that have adopted the contagion by time $t$. Note that $\mathcal{N}_0(j)=\mathcal{S}^{(j)}=\{j\}$ for each $j$. Additionally, $\mathcal{N}_t(j) \subseteq \mathcal{N}_{t+1}(j)$ for any $t$, as nodes cannot unadopt a contagion during a time step. Therefore, the sequence $\{\mathcal{N}_t(j)\}_{t=0}^{t^*_j}$ yields a filtration of the node set $\mathcal{V}$ that satisfies Eq.~\eqref{eq:completeness}. It follows that the set of filtrations of the form $\{\mathcal{N}_t(j)\}_{t=0}^{t^*_i}$ (for $j\in\mathcal{V}$) is a complete set of filtrations. 
We now prove consistency. Consider two realizations of a WTM contagion on a single network. Let $\mathcal{N}_{t_i}{(i)}\subset\mathcal{V}$ denote the set of nodes that are adopters at time $t_i$ for the $i$th realization, and let $\mathcal{N}_{t_j}{(j)}\subset\mathcal{V}$ denote the set of nodes that are adopters at time $t_j$ for the $j$th realization. To have consistency, it must be true that $\mathcal{N}_{t_i+1}{(i)}\subseteq \mathcal{N}_{t_j+1}{(j)}$ if $\mathcal{N}_{t_i}{(i)}\subseteq \mathcal{N}_{t_j}{(j)}$. Suppose that $t_i$ and $t_j$ are times such that $\mathcal{N}_{t_i}{(i)}\subseteq \mathcal{N}_{t_j}{(j)}$, and consider the spreading that occurs for a WTM contagion during one time step. By definition, the update rule for each node is identical across all realizations of a WTM contagion. (In other words, for a node $k$, the fraction of infected neighbors $f_k$ must surpass $T$ for adoption.) Additionally, for any node $k\in\mathcal{V}$, increasing the infection size can only increase $f_k$. Hence, if $f_k>T$ for some node $k$ when nodes $\mathcal{N}_{t_i}{(i)}$ are infected, then $f_k>T$ is also true if we instead consider a superset of $\mathcal{N}_{t_i}{(i)}$ to be infected. Thus, the set $\mathcal{N}_{t_i+1}{(i)}$ of adopters at time step $t = t_i+1$ must satisfy $\mathcal{N}_{t_i+1}{(i)}\subseteq \mathcal{N}_{t_j+1}{(j)}$. The $N$ realizations of a WTM contagion with seeds $\mathcal{S}^{(j)}=\{j\}$ (where $j\in\mathcal{V}$) thus produce to a complete and consistent set of filtrations on $\mathcal{V}$ for which $t^{(i)}_j=x^{(i)}_j$. It follows that the activation times define a metric on the node set $\mathcal{V}$. \qed

\clearpage

%%%%%%%%%%%%%%%%%%%%%%%%%%%%%%%%%%%%%%%%
%%%%%%%%%%%%%%%%%%%%%%%%%%%%%%%%%%%%%%%%
\section{Supplementary Note 5: Algorithm for Constructing WTM Maps}\label{sec:algorithm}
%%%%%%%%%%%%%%%%%%%%%%%%%%%%%%%%%%%%%%%%
%%%%%%%%%%%%%%%%%%%%%%%%%%%%%%%%%%%%%%%%

In this section, we describe our algorithm for constructing WTM maps and discuss its computational complexity. We also conduct numerical simulations to illustrate the scaling of computation time with respect to network size $N$ and mean node degree $d =N^{-1}\sum_{i,j}A_{ij}$. We thereby confirm that the typically observed computational cost of our algorithm scales quadratically with $N$ and linearly with $ d $. That is, for $N$ nodes and $M=Nd/2$ edges, the typical computational cost is $\mathcal{O}(NM)$.

%%%%%%%%%%%%%%%%%%%%%%%%%%%%%%%%%%%%%%%%
~\\{\bf \Large Algorithm and Computational Complexity}\\
%%%%%%%%%%%%%%%%%%%%%%%%%%%%%%%%%%%%%%%%

We begin by describing our algorithm for constructing a WTM map. (See the pseudocode in Algorithm~\ref{alg:WTM} for a summary.) For a WTM map of a network with $N$ nodes, we implement $N$ realizations of a WTM contagion. We simulate the $j$th realization with cluster seeding by initializing the nodes in the contagion seed $\mathcal{S}^{(j)}=\{j\}\cup \{k|A_{jk}\not=0\}$ (i.e., node $j$ and its neighbors) as infected and all other nodes as uninfected. The activation time for the seed nodes is $t = 0$. That is, $x_j^{(i)}=0$ for $i\in\mathcal{S}^{(j)}$. After initialization, we simulate a WTM contagion for time steps $t=1,2,\dots$ until the dynamics reaches equilibrium. In other words, we reach a time step in which no new node becomes infected; this is guaranteed to occur at some time $t<N$.
When considering the set of nodes that can become infected during a given time step $t$, it is sufficient to check only the subset of nodes $i\subset\mathcal{V}$ that are (1) not yet infected and (2) adjacent to a node that was infected during the previous time step (i.e., at time $t-1$). Therefore, as the contagion spreads, it is important to record which nodes adopt the contagion during each time step. Given such a list, upon reaching time step $t$, we examine the neighbors of all nodes that became infected at time $t-1$. Any uninfected node $i$ (among those neighbors) whose ratio $f_i$ of infected neighbors to total neighbors satisfies $f_i>T$ then becomes infected at time $t$ (which we record as its activation time).

We now comment on the computational complexity of {Supplementary} Algorithm~\ref{alg:WTM}. There are $N$ different contagions (because of cluster seeding centered at node $j \in \mathcal{V}$). For each one, we need to calculate the activation time of every node $i\in\mathcal{V}$. Therefore, the computational complexity of computing a WTM map is at least $\mathcal{O}(N^2)$. Because we examine the neighbors of recently infected nodes at each time step, our algorithm also scales linearly with the node degree $ d = d^{\textrm{(G)}}+d^{\textrm{(NG)}}$ (which is identical for every node $i\in\mathcal{V}$ in the experiments below), giving a total complexity of $\mathcal{O}(N^2d)$.  However, there is scope to speed up the construction of WTM maps. If one constructs the dissimilarity matrix that encodes shortest paths between nodes (e.g., as required by Isomap \cite{Tenenbaum2000}) using a ``naive'' method, then its computational complexity is also $\mathcal{O}(N^2)$. However, one can speed up the problem of computing shortest paths using Djikstra's algorithm \cite{Djikstra}, and we expect that similar improvements are possible for WTM maps. 

Before we numerically support the $\mathcal{O}(N^2d)$ computational complexity, we comment on the worst-case scenario, which has a complexity of $\mathcal{O}(N^3)$. This situation corresponds to a network in which every node is connected to every other node and exactly one node adopts the contagion at each time step for every contagion. Although such a scenario is technically feasible with general WTM contagion dynamics \cite{Watts2002}, this cannot arise in the WTM contagions that we study (and we also note that one can also analyze such a pathological situation using mean-field theory \cite{porter2014}) because we set $T_i=T$ for all $i\in\mathcal{V}$ in our experiments. Finally, although our implementation of {Supplementary} Algorithm~\ref{alg:WTM} is sufficiently fast for the purposes of the present paper, we note that one can speed it up further by parallelizing it because the different initial conditions are independent of each other.

\begin{algorithm}[t]
\begin{algorithmic}[1]
\For{each node $j \in \mathcal{V} = \{1, \dots, N\}$}
\State Initialize cluster seeding by infecting $j$ and its neighbors; record their activation times as $0$
\State Run WTM contagion dynamics:
\While{dynamics has not reached equilibrium}
%\State Check all neighbors of nodes that got activated at last timestep
\For{$i$ is a neighbor of a node that was infected during the last time step}
\If{$i$ is still uninfected}
\If{fraction of activated neighbors $f_i>T$}
\State infect node $i$ and record its activation time
%\Else
%\State node $i$ remains susceptible
\EndIf 
\EndIf   
\EndFor
\EndWhile
\EndFor
\end{algorithmic}
\caption{
Construction of a WTM map with threshold $T$ for a network with $N$ nodes
}
\label{alg:WTM}
\end{algorithm}

~\\

%%%%%%%%%%%%%%%%%%%%%%%%%%%%%%%%%%%%%%%%
~\\{\bf {\Large Numerical Investigation of Computational Cost}}\\
%%%%%%%%%%%%%%%%%%%%%%%%%%%%%%%%%%%%%%%%

We implement Supplementary Algorithm~\ref{alg:WTM} in both {\sc Matlab} and {\sc c++}. In our discussion, we focus on our \textsc{c++} implementation (which we have made publicly available, as we discussed in Sec.~III~A of the main text). We conduct numerical experiments to study the scaling behavior of the computational cost with respect to $N$ and $\langle d \rangle$.  All of the results that we report in the present section are mean values that we compute using 10 realizations for a particular choice of parameter values. We run these simulations on a computer with the following specifications: Debian GNU/Linux 7 operating system; 32 GB RAM memory; and 8 processor cores (Intel Core i7-4770 CPU @ 3.40GHz).

In Supplementary Fig.~\ref{fig:comp_time}(a), we show the run times $\delta t$ (in seconds) of our algorithm. These give computational costs for constructing WTM maps for noisy ring lattices with various sizes $N \in[32,31623]$, which we construct while keeping the node degrees fixed at $(d^{\textrm{(G)}},d^{\textrm{(NG)}})=(10,2)$. We show results for thresholds $T=\{0.05,0.2,0.35,0.45\}$. Note for these values of $T$ that the dependence of $\delta t$ on $N$ is much stronger than the dependence of $\delta t$ on $T$. The symbols in {Supplementary} Fig.~\ref{fig:comp_time}(a) give the observed computation times, and the solid lines give the inferred scaling behavior, which we assume takes the form $\delta t= 10^{\Gamma}  N^{\zeta}$ for some constants $\zeta$ and $\Gamma$. In {Supplementary} Table \ref{tab:com_scaling_fit}, we summarize the fitted values for exponent $\zeta$ and the prefactor $\Gamma$ for various values of $T$.  (As we discuss in the table caption, we use a least-squares fit.)
We find that ${\zeta}\approx 2$ for all $T$, supporting our claim of quadratic scaling behavior. We neglect the observed values of $\delta t$ for $N=32$ for our fitting  procedure because we are interested in the scaling as $N\to\infty$.

In {Supplementary} Fig.~\ref{fig:comp_time}(b), we investigate the dependence of the computational cost on node degree $d$. In this experiment, we fix $N=2000$ and $T=0.35$ [which yields the largest values of $\delta t$ in {Supplementary} Fig.~\ref{fig:comp_time}(a)], and we vary $d  \in \{12,24,\dots,96\}$. We plot the observed values of $\delta t$ versus $d$ for several choices of the ratio $\alpha = d^{\textrm{(NG)}}/d^{\textrm{(NG)}}$. For all values of $\alpha$, we observe positive scaling with $d$ that we expect to be linear. As expected, we find that $\delta t$ is much smaller when $\alpha\ge1/2$ than when $\alpha<1/2$. For large values of $\alpha$, WTM contagions tend to either not spread at all (e.g., when $T$ is large) or spread very quickly due to frequent ANC (e.g., when $T$ is small). Therefore, the number of time steps that are necessary to reach equilibrium is small. This, in turn, yields a small value of $\delta t$.

\begin{figure*}[t!]
\centering
\includegraphics[width=.9\linewidth]{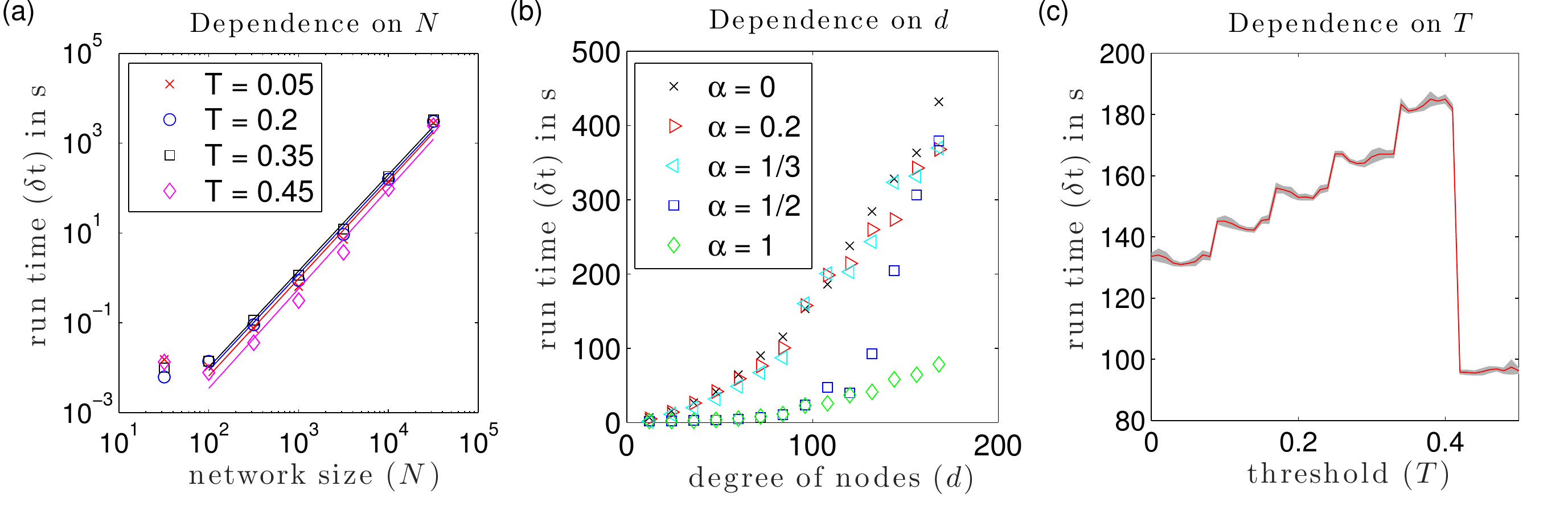}
\caption{
Computational costs from experiments with {Supplementary} Algorithm~\ref{alg:WTM}.  We use the run time $\delta t$ (in seconds) to measure the cost of constructing WTM maps with threshold $T$ for noisy ring lattices with $N$ nodes.
(a) We plot the observed values of $\delta t$ (symbols) versus $N$ for several choices of $T$. Note that $\delta t$ varies only slightly with respect to $T$, whereas the dependence on $N$ is much stronger. We show results for $(d^{\textrm{(G)}},d^{\textrm{(NG)}})=(10,2)$; that is, $\alpha = 0.2$ and $d=12$. The solid lines indicate the inferred scaling behaviors $\mathcal{O}(N^\zeta)$; as we illustrate in {Supplementary} Table \ref{tab:com_scaling_fit}, the scalings are approximately quadratic $\mathcal{O}(N^2)$. 
(b) We plot observed values of $\delta t$ for a noisy ring lattice with $N=2000$ nodes with various choices for the node degree  $ d = d^{\textrm{(G)}}+d^{\textrm{(NG)}}$. 
We show results for various $d$ for several choices of the ratio $\alpha = d^{\textrm{(NG)}}/d^{\textrm{(NG)}}$. For all values of $\alpha$, we observe that $\delta t$ scales approximately linearly with $d$. Networks with large values of $\alpha$ promote transmission via ANC, which saturates the network in fewer time steps than that for smaller $\alpha$, which subsequently leads to considerably smaller run times (e.g., see the results for $\alpha\ge 1/2$ versus $\alpha<1/2$). 
(c) The solid curve indicates $\delta t$ versus the WTM threshold $T$; the shaded region near the curve indicates the standard deviation over 10 realizations for a given threshold $T$. When a contagion saturates the network, so that all nodes eventually adopt the contagion (i.e., $T\le T_0^{\textrm{(WFP)}}\approx0.4167$), observe that $\delta t$ tends to increase with $T$. By contrast, when a contagion does not spread (i.e., $T\gtrapprox0.4167$), then $\delta t$ is very small compared to the values when the contagion does spread. We also note that the abrupt jumps in $\delta t$ are well-aligned with the critical thresholds given by Eqs.~\eqref{eq:ANC_crits} and \eqref{eq:WFP_crits}. The shaded region near the curve indicates the standard deviation (in units of time $\delta t$) over 10 realizations for a given threshold $T$. See \ref{sec:algorithm} for further discussion.
}
\label{fig:comp_time}
\end{figure*}

In Supplementary Fig.~\ref{fig:comp_time}(c), we further explore the dependence of $\delta t$ on WFP and ANC by plotting $\delta t$ versus threshold $T$ for a noisy ring lattice with $N=10000$ nodes and $(d^{\textrm{(G)}},d^{\textrm{(NG)}})=(2,10)$.  (In this case, $\alpha=0.2$.) First, note that there is a very large drop in $\delta t$ near $T=T_0^{\textrm{(WFP)}}\approx0.4167$ that corresponds to the bifurcation that separates the region in which the contagion spreads from the one in which it does not. For $T\gtrapprox0.4167$, the contagion spreads to just a few additional nodes (or no additional nodes), so it requires very few time steps for to reach an equilibrium state. For $T\le 0.4$, the contagion spreads faster as $T$ increases, which leads in turn to larger values of $\delta t$. Finally, note that there are several sharp jumps in $\delta t$; these correspond to the bifurcations in the contagion dynamics [see Eqs.~\eqref{eq:ANC_crits} and \eqref{eq:WFP_crits}].

%\clearpage

\begin{table}[h]
\begin{center}
Inferred run time scaling: $\delta t= 10^{\Gamma}  N^{\zeta}$\\
%\hline
\begin{tabular}{c||c|c}
\hline\hline
 %& \multicolumn{2}{c}{} \\ \hline
~~~ threshold $T~~~$ &~~~exponent $\zeta~~~$ & ~~~prefactor $\Gamma~~~$ \\ \hline
0.05 & 2.1675  & $-6.5207$ \\
0.2 & 2.1407  & $-6.3416$ \\
0.35 & 2.1440 &  $-6.2792$ \\
0.45 & 2.2195  & $-6.8987$
\end{tabular}
\caption{
{
For various choices of threshold $T$, we infer the scaling behavior that relates the computational cost (i.e., the run time $\delta t$) to the network size $N$. For our inference procedure, we assume a power-law relationship $\delta t= 10^{\Gamma} N^{\zeta}$, and we fit the constants $\Gamma$ and $\zeta$ using a least-squares fit.  In this fit, the horizontal coordinates are $\log(N)$, and the vertical coordinates are $\log(\delta t)$ (We neglect the results for $N=32$ in our fitting procedure.) Note that the exponents are approximately quadratic: $\zeta\approx2$. {See \ref{sec:algorithm} for further discussion.}
}
}
\label{tab:com_scaling_fit}
\end{center}
\end{table}

\clearpage

%%%%%%%%%%%%%%%%%%%%%%%%%%%%%%%%%%%%%%%%
%%%%%%%%%%%%%%%%%%%%%%%%%%%%%%%%%%%%%%%%
\section{Supplementary Note 6: Additional Theory {for Noisy Ring Lattices}}
%%%%%%%%%%%%%%%%%%%%%%%%%%%%%%%%%%%%%%%%
%%%%%%%%%%%%%%%%%%%%%%%%%%%%%%%%%%%%%%%%

In this section, we extend the bifurcation analysis that {we presented in Sec.~III~B} of the main text. In particular, we provide further details on our analysis for two contagion phenomena: {\it wavefront propagation} (WFP) along a network's underlying manifold and the {\it appearance of new clusters} (ANC) of a contagion due to transmission across non-geometric edges.

%%%%%%%%%%%%%%%%%%%%%%%%%%%%%%%%%%%%%%%%%%%%%%%%%%%
~\\{\bf \Large Appearance of New Contagion Clusters (ANC)}\\
%%%%%%%%%%%%%%%%%%%%%%%%%%%%%%%%%%%%%%%%%%%%%%%%%%%

{ANC describes a contagion transmission in which a node becomes infected exclusively due to exposure via non-geometric edges. That is, the node's neighbors from geometric edges must not already be infected. As we discussed in {Secs.~I~D and III~B of} the main text, we are able to describe this phenomenon with a sequence of critical thresholds:}
\begin{equation}
	T^{\rm{(ANC)}}_k \triangleq \frac{d^{\rm{(NG)}}-k}{d^{\rm{(G)}}+d^{\rm{(NG)}}}\label{eq:ANC_crits}\,,
	~~~~~~~~~~
	k=0,1,\dots,{d^{\textrm{(NG)}}}\,,
\end{equation}
where $d^{\mathrm{(G)}}$ and $d^{\rm{(NG)}}$, respectively, denote a node's geometric and non-geometric degree for the noisy ring lattice with $N$ nodes. (A node's ``geometric degree'' is its number of geometric stubs, that is, the number of its stubs that obey the original geometric space constraints, and a node's ``non-geometric degree'' is its number of non-geometric stubs.) 
For $T \in \left[T_{k+1}^{\rm{(ANC)}},T_k^{\rm{(ANC)}}\right)$, a node adopts a contagion if at least $(d^{\rm{(NG)}}-k)$ non-geometric neighbors are infected. For $T\ge T^{\rm{(ANC)}}_0$, the contagion cannot spread exclusively by exposure to the contagion via non-geometric edges. In this section, we show by considering spreading exclusively on the subgraph that includes all nodes but only non-geometric edges that the rate of ANC of a WTM contagion increases as the contagion threshold $T$ decreases.

We first consider the probability that a given node has exactly $k$ infected non-geometric neighbors, given that $q(t)$ of the $N$ nodes are infected at time step $t$. First, consider the case $k=1$, in which a node $i$ has exactly one infected non-geometric neighbor. Given that node $i$ has $d^{\rm{(NG)}}$ non-geometric edges (which we label as $e_1,\dots,e_{d^{\rm{(NG)}}}$), there are $d^{\rm{(NG)}}$ possible outcomes with $k=1$. For example, $e_1$ is incident to an infected node and the remaining edges are incident to uninfected nodes, $e_2$ is incident to an infected node and the remaining edges are incident to uninfected nodes, and so on. {Recalling that we place non-geometric edges uniformly at random for the noisy ring lattice,} the probability that edge $e_1$ is incident to an infected node is $\frac{q(t)}{N-1}$, as there are $q(t)$ such potential infected nodes and there are $N-1$ other nodes (because there are no self-edges). If edge $e_1$ is incident to an infected node, then the probability that edge $e_2$ is incident to an uninfected node is $\frac{N-1-q(t)}{N-2}$. If edges $e_1$ and $e_2$ are incident, respectively, to an infected node and an uninfected node, then the probability that edge $e_3$ is incident to an uninfected node is $\frac{N-1-q(t)-1}{N-3}$. We can continue arguing similarly for the other edges. Taking into account that there are $d^{\rm{(NG)}}$ possible outcomes in which the $d^{\rm{(NG)}}$ edges are incident to exactly one infected node, the probability that a node has exactly one infected non-geometric neighbor is
\begin{equation}
 	P(1) = d^{\rm{(NG)}}\times \frac{q(t)\left(\prod_{k'=0}^{d^{\rm{(NG)}}-2}N-1-q(t)-k'\right)}{\prod_{k'=0}^{d^{\rm{(NG)}}-1}\left(N-1-k'\right)}\,.
\end{equation}
More generally, the probability that a node has exactly $k$ infected non-geometric neighbors is
\begin{equation}
	 P(k) = \left(d^{\rm{(NG)}}\atop k\right) \frac{\left(\prod_{k'=0}^{k-1}q(t)-k'\right)\left(\prod_{k'=0}^{d^{\rm{(NG)}}-1-k}N-1-q(t)-k'\right)}{\prod_{k'=0}^{d^{\rm{(NG)}}-1}\left(N-1-k'\right)}\,.\label{eq:Pk1}
\end{equation}
For fixed $d^{\rm{(NG)}}\ll N$ and $q(t)=\mathcal{O}(N)$, Eq.~(\ref{eq:Pk1}) simplifies to
\begin{equation}
	 P(k) \approx \left(d^{\rm{(NG)}}\atop k\right) \left[\frac{q(t)}{N}\right]^k \left[1-\frac{q(t)}{N}\right]^{{d^{\rm{(NG)}}-k}} .
	 \label{eq:Pk2}
\end{equation}

We now estimate the expected {contagion size $g(t)$} of a WTM contagion that spreads exclusively via ANC. In other words, we neglect exposures to the contagion from geometric edges, as we are assuming that they do not contribute to spreading. We define
\begin{equation}
	k^{\rm{(ANC)}}(T) = \max\left\{k' ~|~ T^{\rm{(ANC)}}_{k'} > T\right\}\,.\label{eq:ANC_k_100}
\end{equation}
It follows that the minimum number of non-geometric neighbors that need to be infected for a node $i$ to adopt the WTM contagion is $(d^{\rm{(NG)}}-k^{\rm{(ANC)}})$. Using Eq.~\eqref{eq:Pk1} and Eq.~\eqref{eq:ANC_k_100}, we estimate that the expected contagion growth satisfies
\begin{equation}
	g(t+1)  = g(t) + [N-g(t)] \sum_{k'=0}^{k^{\rm{(ANC)}}} P(d^{\rm{(NG)}}-k')\,, \label{eq:g}
\end{equation}
where we calculate the expectation for $g(t)$ over the ensemble of noisy ring lattices. We again stress that Eq.~\eqref{eq:g} estimates the size of a WTM contagion for ANC independent of WFP and does not account for the joint effect of spreading via both geometric and non-geometric edges. It therefore gives a lower bound for the size of the contagion {[i.e., $q(t)$]} for the regime that exhibits ANC but no WFP.

%%%%%%%%%%%%%%%%%%%%%%%%%%%%%%%%%%%%%%%%%%%%%%%%%%%
~\\{\bf \Large Wavefront Propagation (WFP)}\\
%%%%%%%%%%%%%%%%%%%%%%%%%%%%%%%%%%%%%%%%%%%%%%%%%%%

{WFP describes the situation in which a contagion cluster expands because a node in its ``boundary,'' which we define as the set of nodes that are adjacent via a geometric edge to an infected node in the contagion cluster, becomes infected at time step $t$. 
In the main text, }we found that WFP has the following sequence of critical thresholds:
\begin{equation}
	T_k^{\rm{(WFP)}} \triangleq \frac{d^{\mathrm{(G)}}/2-k}{{d^{\mathrm{(G)}}+d^{\rm{(NG)}}}}\,,
	~~~~~~~~~~
	k=0,1,\dots,\frac{d^{\textrm{(G)}}}{2}\, .
	\label{eq:WFP_crits}
\end{equation}
{Assuming that the non-geometric edges of nodes in the contagion cluster's boundary are incident to nodes that are not infected, a} 
wavefront propagates with a speed of $k+1$ nodes per time step for $T \in \left[T_{k+1}^{\rm{(WFP)}},T_k^{\rm{(WFP)}}\right)$. For $T\ge T_0^{\rm{(WFP)}}$, there is no WFP. For a contagion that consists of a single cluster that is expanding via WFP in both directions along a noisy ring lattice, the size $q(t)$ of the contagion (i.e., the number of nodes that have adopted the contagion) for time $t\in \{0,1,2,\dots\}$ has a lower bound of
\begin{equation}
	h(t) = \underbrace{(1+d^{\rm{(G)}}+d^{\rm{(NG)}})}_{\rm{seed\  nodes}} + 2k^{\rm{(WFP)}}t\,,\label{eq:size_small_q}
\end{equation}
where 
\begin{equation}
	k^{\rm{(WFP)}} \triangleq 1+\max \{k' ~|~ T^{\rm{(WFP)}}_{k'} > T\}\label{eq:WFP_time}
\end{equation}
and the factor of 2 accounts for WFP in both directions along the ring.

Note that $h(t)$ is a lower bound for $q(t)$ because we have assumed that the non-geometric edges of nodes in the contagion cluster's boundary are incident to nodes that are not infected. This assumption is not always valid, so Eq.~\eqref{eq:size_small_q} is a lower bound because the invalidation of this assumption can only increase the rate of WFP. That is, nodes in a contagion cluster's boundary will adopt a contagion even if the number of geometric neighbors that are infected is smaller than what is required by Eq.~\eqref{eq:WFP_crits}. 

Above we showed that the expected probability that a non-geometric edge of a node is incident to an infected node is $q(t)/(N-1)\approx q(t)/N$. Similarly, for a node with a non-geometric degree of $d^{\textrm{(NG)}}$, the expected probability that none of its non-geometric edges are incident to an infected node is approximately $[1-q(t)/N]^{d^{\textrm{(NG)}}}$. This is therefore the probability that our WFP analysis given by Eq.~\eqref{eq:WFP_crits} is valid for a given node at a given time step $t$. For large networks (i.e., $N\gg1$) and early stages of a contagion (i.e., $q(t)\ll N$), the probability that our assumption is valid is approximately equal to 1. In this situation, Eqs.~\eqref{eq:WFP_crits}--\eqref{eq:WFP_time} accurately describe WFP (and the spread of the contagion). However, when $q(t)\approx N$, our assumption is almost certainly invalid, and we observe accelerated speeds of WFP. Interestingly, for large networks (i.e., $N\gg1$), we find that such acceleration occurs infrequently early in a WTM contagion and that it occurs rather frequently towards the end of a contagion [i.e., just before $q(t)\to N$, which is when a contagion saturates a network]. Accelerated WFP is improbable [because $q(t)$ is small, but $N$ is large] in the early stages of a contagion on a large network. When $t = 0$, for example, $q(0)=d^{\rm{(G)}} + d^{\rm{(NG)}} \ll N$. However, during the late stages of a contagion [i.e., $q(t)\approx N$], accelerated WFP is very likely at every time step. Therefore, for small $q(t)$, Eq.~\eqref{eq:size_small_q} is both a lower bound for $q(t)$ and an approximation for it. In general, the speed of WFP increases with time until it reaches an upper bound of $d^{\rm{(G)}}/2$ nodes per time step. This bound corresponds to the situation in which all nodes that are incident via geometric edges to one side of a contagion cluster become infected during each time step. Note that there is no acceleration of WFP when $k^{\rm{(WFP)}}=d^{\rm{(G)}}/2$, as the wavefront is already propagating at its fastest rate.

\clearpage
%%%%%%%%%%%%%%%%%%%%%%%%%%%%%%%%%%%%%%%%
%%%%%%%%%%%%%%%%%%%%%%%%%%%%%%%%%%%%%%%%
\section{Supplementary Note 7: Extended Discussion of Point-Cloud {Analyses}}\label{sec:geometry}
%%%%%%%%%%%%%%%%%%%%%%%%%%%%%%%%%%%%%%%%
%%%%%%%%%%%%%%%%%%%%%%%%%%%%%%%%%%%%%%%%
 
In this section, we provide further details on {our approach to analyzing the point clouds that result from} WTM maps. {In particular, we provide a detailed discussion of the} following three items:
\begin{enumerate}
\item The \emph{Pearson correlation coefficient $\rho$}, which we use to investigate a point cloud's geometry. %(Sec.~\ref{sec:geometry}). 
\item The \emph{embedding dimension $P$}, which we use to investigate a point cloud's dimensionality.% (Sec.~\ref{sec:dimension}). 
\item The \emph{difference $\Delta=\l_1-l_2$ in lifetimes of the two most persistent 1-cycles} [i.e., one-dimensional (1D) holes], which we use to investigate a point cloud's topology.% (Sec.~\ref{sec:topology}). 
\end{enumerate}
{We restrict our discussion of the above items for a point cloud that results from a regular WTM map $\mathcal{V}\mapsto\{\bold x^{(i)}\}$, but one can apply the same techniques to any point cloud, including one that results from a reflected WTM map $\mathcal{V}\mapsto\{\bold y^{(i)}\}$ or a symmetric WTM map $\mathcal{V}\mapsto\{\bold z^{(i)}\}$. (See {Sec.~I~C} of the main text for further discussion of these maps.)}

We find for certain WTM contagion parameters that the structure of the point cloud that results from a WTM maps can reveal manifold structure in the original network and that one can quantify such structure using the values of $\rho$, $P$, and $\Delta$. Importantly, one can thus use our approach to study not only manifold structure in networks but also the WTM contagion dynamics itself (e.g., uncovering the extent to which WFP dominates ANC or vice versa).

%%%%%%%%%%%%%%%%%%%%%%%%%%%%%%%%%%%%%%%%%%%%%%%%%%%
~\\{\bf \Large Analysis of Geometry}\\
%%%%%%%%%%%%%%%%%%%%%%%%%%%%%%%%%%%%%%%%%%%%%%%%%%%

Studying the geometry of {a point cloud such as $\{\bold x^{(i)}\}_{i\in\mathcal{V}}$ that results from} a WTM map can reveal the extent to which the geometry of a WTM contagion follows the underlying geometry of a network. We investigate the extent to which the distance between two {nodes in a point cloud that results from a} WTM map relates to the distance between those nodes in the original metric-space embedding of the noisy geometric network. Specifically, we restrict our attention to noisy geometric networks in which the nodes $\mathcal{V}$ have intrinsic locations $\{\bold w^{(i)}\}_{i\in\mathcal{V}}\in\mathcal{M}$ on a manifold $\mathcal{M}\subset\mathbb{R}^p$. {That is, they lie in a $p$-dimensional ambient space $\mathbb{R}^p$ that we equip} with the Euclidean norm $\|\bold w\|_2 = \sqrt{\sum_{k=1}^p w_k^2}$. We require the dimension $p$ to be equal to the point cloud's ``embedding dimension''. In other words, there is no subspace of dimension smaller than $p$ that one can define by a hyperplane that captures the manifold. Using the Euclidean metric, the distance between nodes $i$ and $j$ in the ambient space is 
\begin{equation}
	m(i,j) = \sqrt{\sum_{k=1}^p \left(w^{(i)}_k - w^{(j)}_k\right)^2}\,.\label{eq:distance1}
\end{equation}

We also use the Euclidean norm for the point cloud $\{\bold x^{(i)}\}\in\mathbb{R}^N$ that results from a WTM map. 
The distance between {node $i$ and node $j$ in such a point cloud is thus given by
\begin{equation}
	 m^{(\mathrm{WTM})}(i,j) = \sqrt{\sum_{k=1}^N \left(x^{(i)}_k - x^{(j)}_k \right)^2}\,.\label{eq:distance_wtm}
\end{equation}
Given two {sets of} distances $m$ and $m^{\rm{(WTM})}$, we compute the Pearson correlation coefficient
\begin{equation}
	\rho = \frac{\sum_{i=1}^N\sum_{j=i+1}^N \big[m(i,j)-\overline{m}(i,j) \big]\big[m^{(\mathrm{WTM})}(i,j)-\overline{m^{(\mathrm{WTM})}}(i,j) \big]}{
\sqrt{\sum_{i=1}^N\sum_{j=i+1}^N \big[m(i,j)-\overline{m}(i,j) \big]^2}
\sqrt{\sum_{i=1}^N\sum_{j=i+1}^N \big[m^{(\mathrm{WTM})}(i,j)-\overline{m^{(\mathrm{WTM})}}(i,j) \big]^2}}\label{eq:RHO}
\end{equation}
between {all non-identical, unordered pairs $(i,j)\in\mathcal{V}\times\mathcal{V}$.} Because $i \neq j$ for distinct nodes, there are $N(N-1)/2$ such pairs.

Note that calculating Eq.~(\ref{eq:RHO}) requires the activation time $x^{(i)}_j$ to be finite for all nodes $i$ and realizations $j$ of a WTM contagion. Unfortunately, this is not the case whenever there is a node that never becomes activated. Indeed, $x^{(i)}_j=\infty$ for all nodes other than the seed if the threshold $T$ is too large. (For instance, $T\ge\max\{T_0^{(\mathrm{WFP})},T_0^{(\mathrm{ANC})}\}$ is too large for the example of the noisy ring lattice.) For practical purposes, we set $x^{(i)}_j=2N$ in such cases, where we note that $x^{(i)}_j \leq N-1$ for any numerical simulation in which node $i$ eventually becomes infected. In \ref{sec:London}, we discuss other methods for handling situations with activation times of infinity.

%%%%%%%%%%%%%%%%%%%%%%%%%%%%%%%%%%%%%%%%%%%%%%%%%%%
~\\{\bf \Large Analysis of Dimensionality}\\
%\subsection{Analysis of Dimensionality}\label{sec:dimension}
%%%%%%%%%%%%%%%%%%%%%%%%%%%%%%%%%%%%%%%%%%%%%%%%%%%

We study the dimensionality of {a point cloud that results from a WTM map} by exploring {its} {\it embedding dimension}. For a manifold $\mathcal{M}\subset \mathbb{R}^p$, we define the embedding dimension $P$ as the minimum hyperplane dimension over all hyperplanes that span the manifold $\mathcal{M}$. Because a point cloud typically contains noise, which can potentially increase the dimensionality above that of an underlying manifold, we estimate embedding dimension using residual variance \cite{Cox2010,Tenenbaum2000}.

Given the {set of points} $\{\bold x^{(i)}\}_{i\in\mathcal{V}}\in\mathbb{R}^N$, {we consider each $i\in\mathcal{V}$ and} let $\{\bold{\hat x}^{(i)}(p)\}$ denote the linear projection onto $\mathbb{R}^p$ that we obtain from principal component analysis (PCA)\cite{Cox2010,Tenenbaum2000}. Let $\rho^{(p)}$ denote the Pearson correlation coefficient that relates node-to-node distances $m^{\mathrm{(WTM)}}$ from the original point cloud to node-to-node distances
\begin{equation}
	 m^{(p)}(i,j) = \sqrt{\sum_{k=1}^p \left[\hat x^{(i)}_k(p) - \hat x^{(j)}_k(p) \right]^2} \label{eq:distance_p}
\end{equation}
in the projected point cloud. {It follows that} $\rho^{(p)}$ is given by Eq.~\eqref{eq:RHO} with the {substitution} $m(i,j)\mapsto m^{(p)}(i,j)$. 

The residual variance of such a linear dimension reduction is $R_p = 1-(\rho^{(p)})^2$. We estimate the embedding dimension as the smallest dimension $P$ such that the residual variance is (strictly) less than $0.05$. That is, $P=\min\{p|R_p<0.05\}$. For our calculations of embedding dimension, we only consider dimensions up to $P=20$, as this simplifies the computational overhead of calculating $P$. Our motivation for this simplification (besides reducing computational cost) is that we are particularly interested in determining whether or not $P$ is close to the known embedding dimension (e.g., $P=2$ for the unit circle in $\mathbb{R}^2$, in which our noisy ring lattices are embedded).

\clearpage

%%%%%%%%%%%%%%%%%%%%%%%%%%%%%%%%%%%%%%%%%%%%%%%%%%%
~\\{\bf { \Large Analysis of Topology}}\\
%%%%%%%%%%%%%%%%%%%%%%%%%%%%%%%%%%%%%%%%%%%%%%%%%%%

In this section, we explain how to analyze the topology of a {point cloud that results from a WTM map}. We present our {analysis} for a general point cloud $\mathcal{U}=\{\bold u^{(i)}\}_{i=1}^n\in\mathbb{R}^J$ (i.e., there are $n$ points $\bold u^{(i)}$ in $J$ dimensions). {We note for a typical WTM map, for which we map all $N$ nodes based on $N$ contagions, that one obtains a point cloud $\{\bold u^{(i)}\}$ with $n=J=N$}.

A set $\mathcal{U}$ has a very simple topology. If $\bold u^{(i)} \neq \bold u^{(j)}$ for $i \neq j$, then $\mathcal{U}$ consists of $N$ distinct connected components that correspond to the points $\{\bold u^{(i)}\}$. There are no 1-cycles in $\mathcal{U}$. To infer the topology of a meaningful underlying manifold (if present) that gives rise to a point cloud, we consider its topology across different spatial scales. In particular, we are interested in the topology of the sets
\begin{equation}
 	\mathcal{U}^{(r)} = \bigcup_{ i \in \{1,\dots, n\}}\{\bold u \in \mathbb{R}^J:  || \bold u^{(i)} - \bold u ||_2 \leq r\}
\end{equation}	
for different values of $r \in [0,\infty)$. That is, we study the topology of sets that we construct as the union of radius-$r$ balls centered at points $\bold u^{(i)}\in \mathcal{U}$. Note that $\mathcal{U}^{(0)}=\mathcal{U}$. {We choose to use the Euclidean norm,} but it is also possible to use other norms.

\begin{figure*}[b!]
\centering
\includegraphics[width=.9\linewidth]{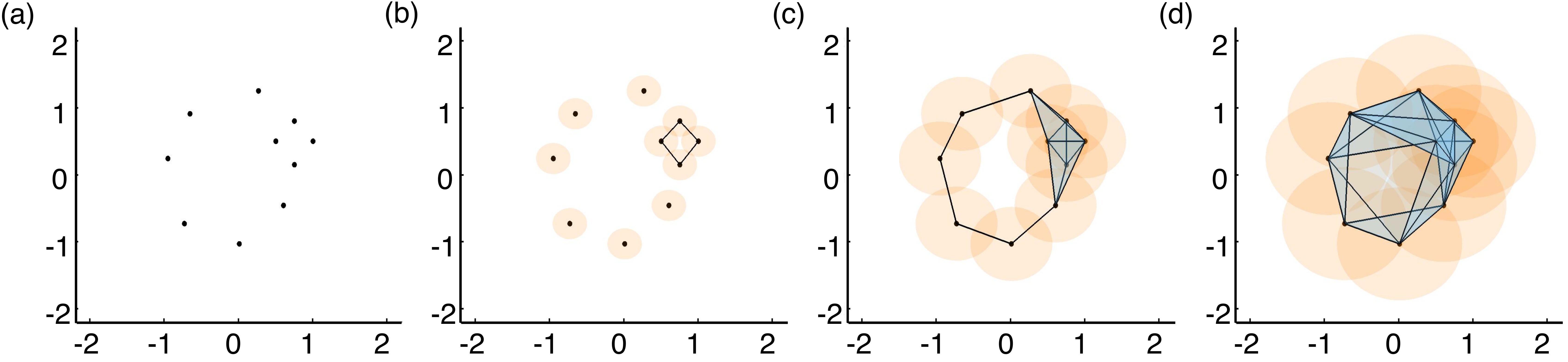}
\caption{
We study the topology of a point cloud $\mathcal{U}$ by examining the persistent homology that is induced by a Vietoris-Rips filtration. This entails examining simplicial complexes that are created by forming, for every set of points, a simplex (e.g., an edge, a triangle, a tetrahedron, etc.) whose diameter is at most $r$. Increasing $r$ from $0$ and considering how a simplicial complex evolves yields a filtration. In panel (a), we show a point cloud $\mathcal{U}=\{\bold u^{(i)}\}$ that consists of a noisy sample of the unit circle. In this example, there are $n=10$ points in $J=2$ dimensions. In panels (b)--(d), we show $\mathcal{U}^{(r)}$ for $r \in\{  0.22, 0.6, 0.85\}$. One can approximate the homology of $\mathcal{U}^{(r)}$ using a Vietoris-Rips complex that is given by the nodes, edges, and triangles that we show in the panels. The first 1-cycle in $\mathcal{U}^{(r)}$ occurs at $r = 0.22$. It is a result of the noisy sampling, and it is filled in almost immediately {as $r$ increases}.  In panel (c), we show the dominant 1-cycle (i.e., the 1-cycle that corresponds to the ring and persists across many spatial scales). It is born at $r = 0.5$ and persists until {$r \approx 0.81$}. Identifying a single persistent 1-cycle indicates that the point cloud lies on a ring manifold. See \ref{sec:geometry} for further discussion.
}
\label{fig:RipsCircle}
\end{figure*}

We start with an example. In {Supplementary} Fig.~\ref{fig:RipsCircle}, we show a noisy point cloud that we sample from a ring manifold. In particular, we sample the points uniformly from a unit circle in $\mathbb{R}^2$, and we add a small amount of noise to their locations in the embedding space $\mathbb{R}^2$. When $r = 0$, there are $10$ distinct connected {components, which correspond to the individual points}. As we increase $r$, four of the components merge to create a 1-cycle [see {Supplementary} Fig.~\ref{fig:RipsCircle}(b)]. As we continue to increase $r$, this 1-cycle fills in very soon after its birth. After it is filled in, another 1-cycle appears when $r=0.5$ [see {Supplementary} Fig.~\ref{fig:RipsCircle}(c)]. This 1-cycle persists for a larger range of $r$ values than the first 1-cycle, and it appears to correspond to {a} ring manifold that underlies the point cloud. This illustrates that one can study the topology of a point cloud by examining 1-cycles that persist across different spatial scales. To make this statement more quantitative, we employ tools from persistent homology \cite{eh,edels2010,carlsson}. 

For every set $\mathcal{U}^{(r)}$, one can assign homology groups $H_c(\mathcal{U}^{(r)})$, where $c\in \{0,1,2,\dots\}$. The rank $\beta_c$ of the group $H_c(\mathcal{U}^{(r)})$ counts the number of $c$-dimensional topological features that are present in $\mathcal{U}^{(r)}$. In particular, $\beta_0$ counts the number of connected components, $\beta_1$ counts the number of 1-cycles (which one can construe as a 1D hole or loop), and $\beta_2$ counts the number of cavities [i.e., two-dimensional (2D) holes]. The fact that $\mathcal{U}^{(r)} \subseteq \mathcal{U}^{{(r')}}$ for $r \leq r'$ is very important. As we discussed earlier in this section, a sequence of sets with this property is a filtration. Thus, for any sequence $\{r_i\}$ that satisfies $r_i\le r_{i+1}$ for $i\in \{1,2,\dots\}$, the sequence of sets $\{\mathcal{U}^{(r_i)}\}$ forms a filtration of $\mathbb{R}^2$. Examining changes of the topological features across the different elements of $\{\mathcal{U}^{(r_i)}\}$ reveals multiple-scale topological features of the point cloud $\{\bold u^{(i)}\}$.

\begin{figure}[b!]
\centering
{\includegraphics[width=.45\linewidth]{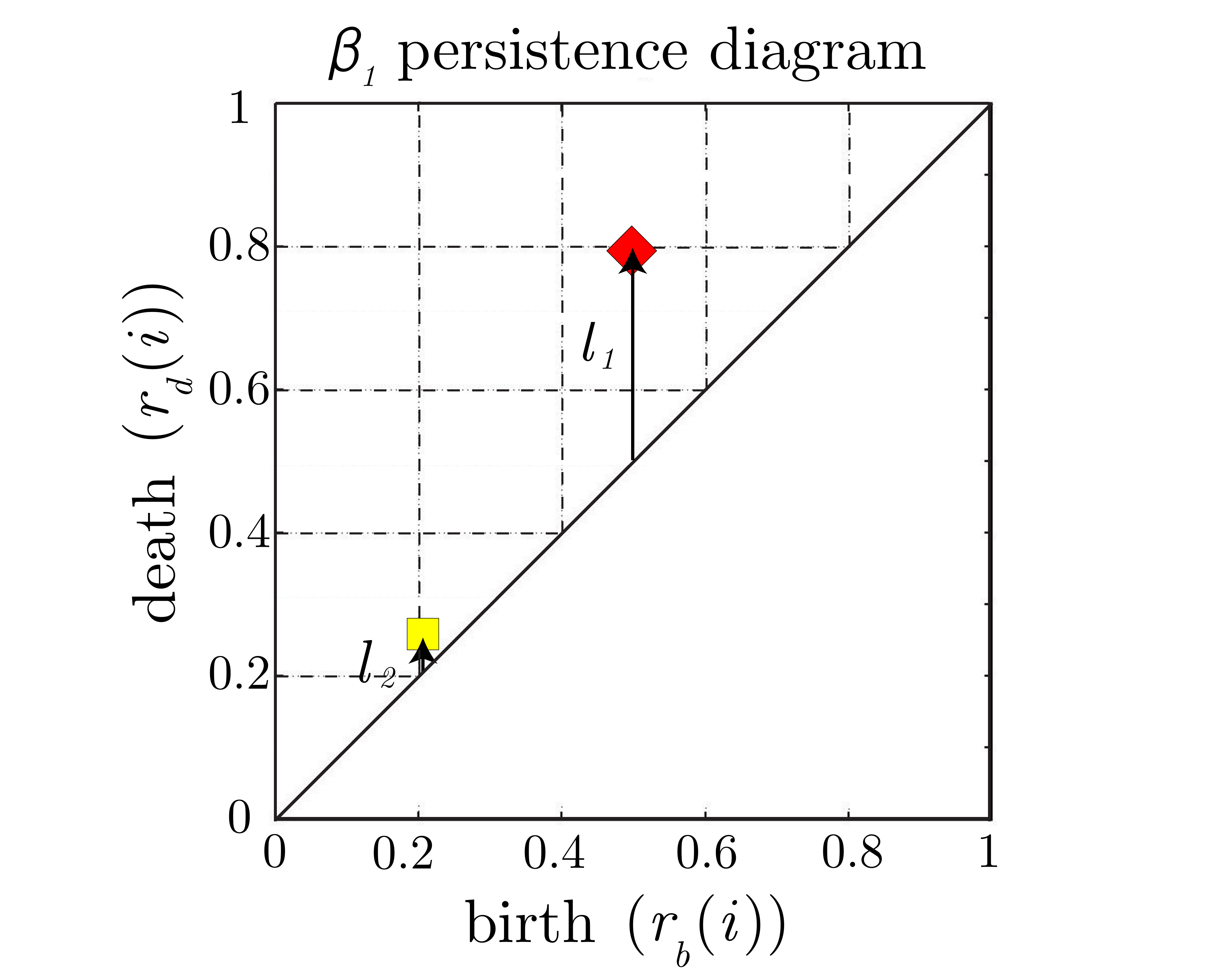}}
\caption{ 
A $\beta_1$ persistence diagram that summarizes the 1D features (i.e., 1-cycles) that are revealed by the filtration $\mathcal{U}^{(r)}$ in {Supplementary} Fig.~\ref{fig:RipsCircle}. It contains two points, which correspond to the two observed 1-cycles. One point (the red diamond) indicates a 1-cycle that persists over a long range of spatial scales. Its lifetime $l_1=r_d(1)-r_b(1)$ is thus large. The second point (the yellow square) indicates another 1-cycle. Its small lifetime $l_2=r_d(2)-r_b(2)$ indicates that it dies a short time after it is born, so it does not persist over many spatial scales. The large difference $\Delta=l_1-l_2$ in the top two lifetimes indicates that the point cloud contains a single dominant 1-cycle and offers strong evidence that the point cloud lies on a ring manifold. {See \ref{sec:geometry} for further discussion.}
%{\bf map: do we ever explicitly define the $b$ and $d$ subscripts in the SI text? please check this and add one sentence to define it if so}
}
\label{fig:PersistenceRipsCircle}
\end{figure}

In the present paper, we are interested in understanding the birth and death of 1-cycles of $\mathcal{U}^{(r)}$ as we vary $r$. The quantity $\beta_1$ encodes such information, which one can summarize by drawing a \emph{persistence diagram}. In {Supplementary} Fig.~\ref{fig:PersistenceRipsCircle}, we show the $\beta_1$ persistence diagram for the point cloud in {Supplementary} Fig.~\ref{fig:RipsCircle}. The diagram contains two points, which correspond to the two 1-cycles that we discussed previously. The horizontal (``birth'') axis of the point is the value of $r$ at which the 1-cycle corresponding to this point first appears in $\mathcal{U}^{(r)}$, and the vertical (``death'') axis indicates when the 1-cycle is filled in. Enumerating the points $i=1,2,\dots$ for every point $i$ with coordinates $(r_b(i), r_d(i))$ in the persistence diagram (where $r_b$ denotes when a feature is born and $r_d$ denotes when a feature dies), we define the ``lifetime'' $l_i = r_d(i) - r_b(i)$, and we denote the set of lifetimes of all points by $L=\{l_1,l_2,\dots \}$ (which we order such that $l_1\ge l_2\ge\dots$).  Topological features with longer lifetimes (i.e., ones that are more persistent) indicate more dominant features in a point cloud. In our example, there is one point with a very short lifetime that corresponds to a 1-cycle that arises for a single spatial scale due to the noisy sampling. The other point has a much larger lifetime, which indicates that its associated 1-cycle persists across many spatial scales. We thereby identify the ring structure of the sampled manifold. For the purpose of identifying whether or not a point cloud lies on a ring manifold, we summarize persistence diagrams by using the difference $\Delta=l_1-l_2$ between the most persistent lifetimes. Large values of $\Delta$ correspond to persistence diagrams that consist of a single point with a large lifetime, as we expect for a point cloud that lies on a ring manifold.

In practice, computing the persistent homology of a set $\mathcal{U}^{(r)}$ is complicated. However, the so-called ``Nerve Theorem'' \cite{borsuk} guarantees that the homology of $\mathcal{U}^{(r)}$ is the same as the homology of a corresponding \v{C}ech complex, which simplifies analysis but is computationally expensive to construct. Therefore, we study an approximation of the \v{C}ech complex that is known as the Vietoris-Rips complex. For a given point cloud $\mathcal{U} = \{\bold u^{(1)}, \bold u^{(2)}, \dots, \bold u^{(n)} \}  \in \mathbb{R}^J$ and $r \in \mathbb{R}$, the \emph{Vietoris-Rips complex} $\rm{VR}^{(r)}$ consists of the simplices $(\bold u^{(s_1)}, \bold u^{(s_2)}, \dots, \bold u^{(s_k)})$ such that $\| \bold u^{(s_i)} - \bold u^{(s_j)} \|_2 \leq r$ for all $s_i$ and $s_j$. In the present paper, we are interested only in identifying the 1-cycles in $\rm{VR}^{(r)}$, so it is sufficient for us to use only 0-simplices (i.e., points), 1-simplices (i.e., line segments), and 2-simplices (i.e., triangles).

To compute persistent homology, we use the software package {\sc Perseus} \cite{Mischaikow2013} (version 3.0 Beta), and we also check some of our results using the {\sc javaPlex} Persistent Homology Library \cite{javaplex}. To construct Vietoris-Rips filtrations $\rm{VR}^{(r)}$ for a point cloud that results from a WTM map (e.g., $\{\bold x^{(i)}\}$), we use Eq.~\eqref{eq:distance_wtm} to define distances between points. As an input to {\sc Perseus}, we use the dissimilarity matrix in which the entry in the $i$th row and $j$th column encodes the distance between nodes $i$ and $j$ given by Eq.~\eqref{eq:distance_wtm}.

%{\bf map: we put software in italics above, but normally there is a different font for that ; e.g. {\sc this one}; I think it would be best to use italics for definitions and small caps for software, software commands, etc.; if you change this, please do so consistently throughout both documents}

In {Supplementary} Fig.~\ref{fig:persistence_diagrams}, we study $\beta_1$ persistence diagrams for point clouds that result from the application of WTM maps to noisy ring lattices. We thereby reveal the absence versus presence of 1-cycles in the point cloud. We analyze the $\beta_1$ persistence diagrams for several values of the WTM threshold $T\in[0,0.5]$ and several choices for non-geometric degrees $d^{\rm{(NG)}}\in[0,20]$ for networks with $N=200$ nodes and a geometric degree of $d^{\rm{(G)}}=20$ (which implies that $\alpha=d^{\rm{(NG)}}/d^{\rm{(G)}}\in[0,1]$). 
A red diamond in {Supplementary} Fig.~\ref{fig:persistence_diagrams} represents the point that corresponds to the most persistent 1-cycle. We indicate the second-most persistent 1-cycle using a yellow square, and we mark the remaining points in the persistence diagram using white circles. If there is only one dominant 1-cycle, then the separation between the red diamond and the other points is large. To measure this separation, we calculate $\Delta =l_1 - l_2$, where $l_1$ and $l_2$ are the lifetimes of the dominant and the second most dominant 1-cycle, respectively. The background coloration reflects the value of $\Delta$. To construct a filtration using various values of $r$, we consider 100 evenly-spaced values of $r$ that range from $0$ up to the maximum distance distance $r_{\text{max}}\triangleq \max_{i,j\in\mathcal{V}} || \bold z^{(i)}-\bold z^{(j)}||_2$ between any two points. For our plots, we normalize all $r$ values by $r_{\text{max}}$, so $\Delta \in [0,1]$. It follows that $\Delta\approx 1$ indicates the presence of the ring topology, whereas small values of $\Delta$ indicates its absence.

\begin{figure*}[t!]
\centering
\includegraphics[width=.75\linewidth,angle=0]{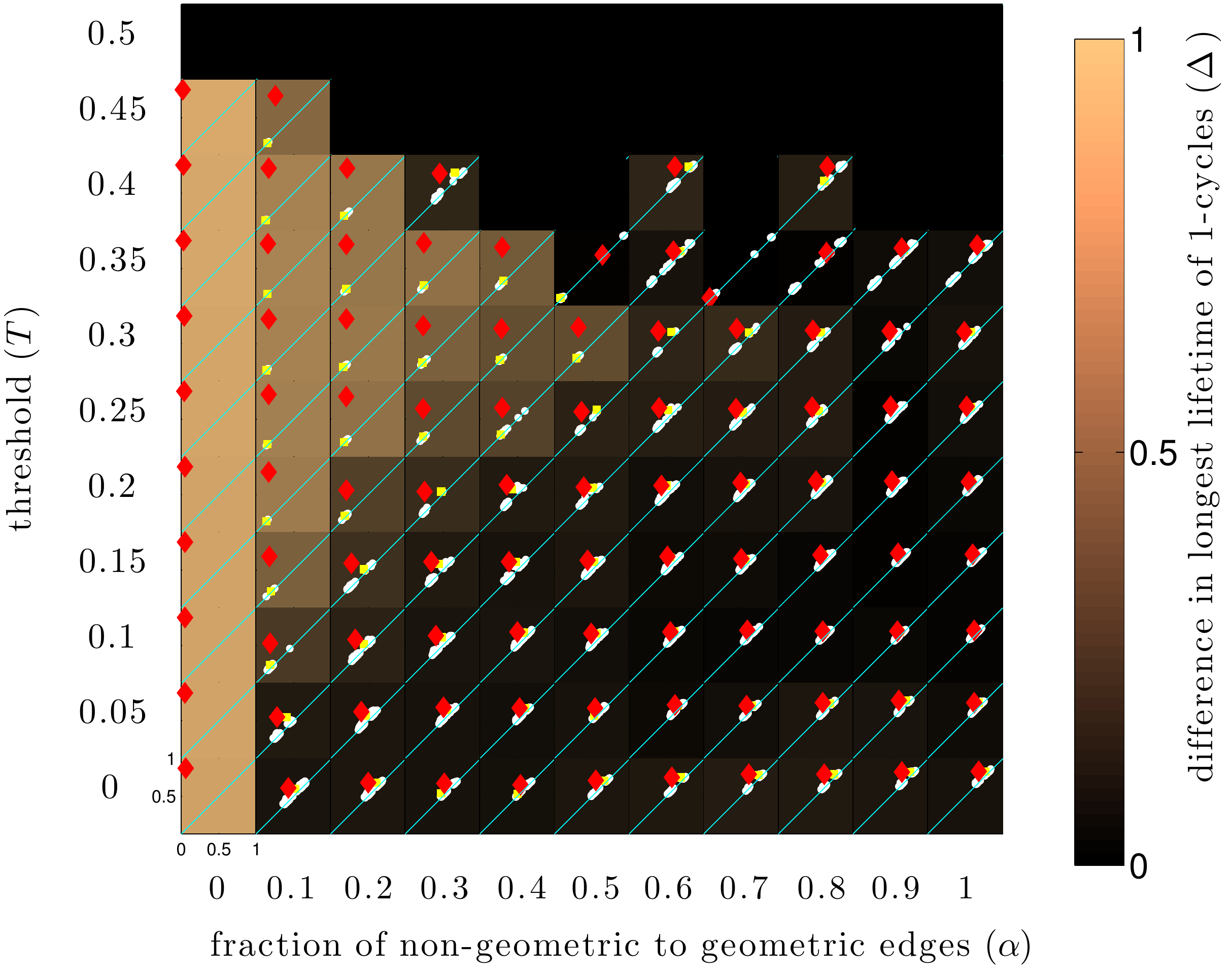}
\caption{ 
We show a grid of $\beta_1$ persistence diagrams for point clouds that result from the application of WTM maps with various values of the threshold $T$ to noisy ring lattices for various values of the ratio $\alpha=d^{\rm{(NG)}}/d^{\rm{(G)}}$ of non-geometric edges to geometric edges. We show results for $T=0,0.05,\dots,0.5$ for networks with $N=200$ nodes each with $d^{\rm{(G)}}=20$ geometric edges and $d^{\rm{(NG)}}\in\{0,2,\dots,20\}$ non-geometric edges. For a given point cloud, we apply a Vietoris-Rips filtration to yield the $\beta_1$ persistence diagram that summarizes the multiscale 1D features (i.e., 1-cycles or loops). In each persistence diagram, we use a red diamond to mark the most persistent 1-cycle, a yellow square to mark the second most persistent 1-cycle, and white circles to indicate the remaining 1-cycles that we find. Note that the lifetime $l_i$ of a given point $i$ corresponds to the height above the diagonal (the cyan lines). We shade the background color of each persistence diagram according to the difference $\Delta=l_1-l_2$ between the two largest lifetimes. Note that $\Delta \in [0,1]$ due to {normalization. (See Sec.~III~F of the main manuscript and \ref{sec:geometry}.)} The magnitude of $\Delta$ provides strong evidence regarding whether or not a given point cloud arises from a 1D ring topology.  In the main manuscript, we thus summarize our topological analysis with the parameter {$\Delta$. [For example, see Fig.~6(c) of the main manuscript.]} Note that we do not do any calculations (as indicated by the empty squares) for WTM maps in which any node has an activation time of infinity [i.e., when there is at least one pair $(i,j)$ such that $x_j^{(i)}=\infty$]. {See \ref{sec:geometry} for further discussion.}
}
\label{fig:persistence_diagrams}
\end{figure*}

\clearpage

%%%%%%%%%%%%%%%%%%%%%%%%%%%%%%%%%%%%%%%%
\section{Supplementary Note 8: Complex Contagions on a Ring Manifold \label{sec:numerics}}
%%%%%%%%%%%%%%%%%%%%%%%%%%%%%%%%%%%%%%%%

In this section, we give results for numerical experiments in which we study the geometry, dimensionality, and topology of point clouds that result from the application of symmetric WTM maps $\mathcal{V}\mapsto\{\bold z^{(i)}\}$ to noisy geometric networks generated by network families {\bf(a)}--{\bf(d)}, which we defined in \ref{sec:models}. We thereby reveal the extent to which a WTM contagion exhibits WFP that follows the underlying ring manifold (i.e., the extent to which spreading occurs across a network subgraph that contains exclusively geometric edges) versus ANC. In particular, WFP is more prevalent than ANC when one can identify the properties of the underlying manifold in the point cloud that results from a WTM map. 

To give some perspective for our numerical experiments, we compare our results for point clouds produced by WTM maps to results from two well-known methods of mapping network nodes as a point cloud: a Laplacian eigenmap \cite{Belkin2003} and Isomap \cite{Tenenbaum2000}. In particular, we consider a 2D Laplacian eigenmap in which we map each node $i$ to $[v_i^{(2)},v_i^{(3)}]^T\in\mathbb{R}^2$, where $\bold v^{(j)}$ is the eigenvector that corresponds to the $j$th eigenvalue $\lambda_j$ of the unnormalized Laplacian matrix $L$ (i.e., $L\bold v^{(j)}=\lambda_j\bold v^{(j)}$) and we have ordered the eigenvalues so that $0=\lambda_1<\lambda_2\le\lambda_3\le\dots\le\lambda_N$. The unnormalized Laplacian matrix has the form $L=\text{diag}(d_1,d_2,\dots,d_N)-A$, where $d_i=d_i^{\rm{(G)}}+d_i^{\rm{(NG)}}$ is the total degree of node $i$ and $A$ is the adjacency matrix. As we discussed in {Sec.~I~C} of the main text, Isomap entails mapping network nodes based on the shortest paths between nodes. It corresponds to a WTM map with $T=0$ if we initialize the contagions with node seeding rather than cluster seeding. As we will see, when assessing the extent to which point clouds that result from WTM maps resemble the underlying ring manifold, we typically find a range of threshold values for which the geometry, dimensionality, and topology of the manifold is more apparent in WTM maps than for Laplacian eigenmap and Isomap methods.  For other threshold values, the manifold is less apparent for WTM maps than for the other methods. 

Note that Laplacian eigenmaps and Isomap were introduced originally for the purpose of nonlinear dimension reduction of point-cloud data} rather than for network analysis. They were developed to map a high-dimensional point cloud to a network and then to map that network to a low-dimensional point cloud. Therefore, applying a Laplacian eigenmap or Isomap directly to a network---especially one that is unweighted---is different from what they were designed to do. In particular, for networks that arise from high-dimensional data---e.g., ones with nodes that are connected to each other by applying a $k$-nearest-neighbor algorithm---one often weights network edges based on distances in the original, high-dimensional point cloud. Incorporating such additional information can, of course, improve the results of dimension reduction (e.g., when attempting to ``learn'' manifold attributes such as topology, geometry, and dimensionality). Finally, when considering dimension reduction such as manifold learning in networks (i.e., rather than point clouds), one should determine the approach to dimension reduction (e.g., whether the algorithm is based on diffusion, shortest paths, or contagion dynamics) based on the application at hand. (For example, one might be more interested in conservative processes in some situations and in non-conservative processes in others.)

%%%%%%%%%%%%%%%%%%%%%%%%%%%%%%%%%%%%%%%%
~\\{\bf \Large Numerical Results for Geometry}\\
%%%%%%%%%%%%%%%%%%%%%%%%%%%%%%%%%%%%%%%%

\begin{figure*}[ht!]
\centering
\includegraphics[width=.4\linewidth]{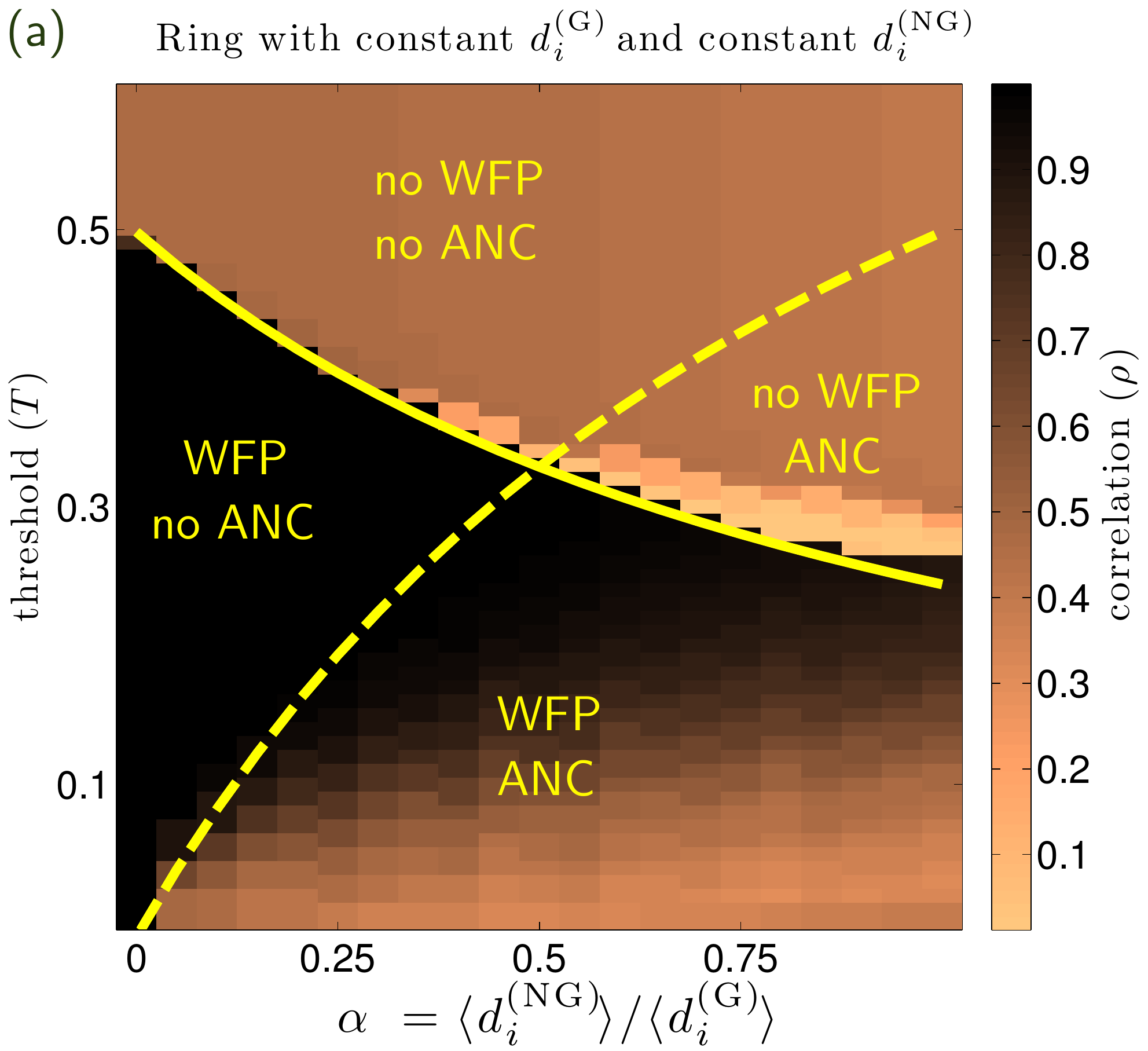}
\includegraphics[width=.4\linewidth]{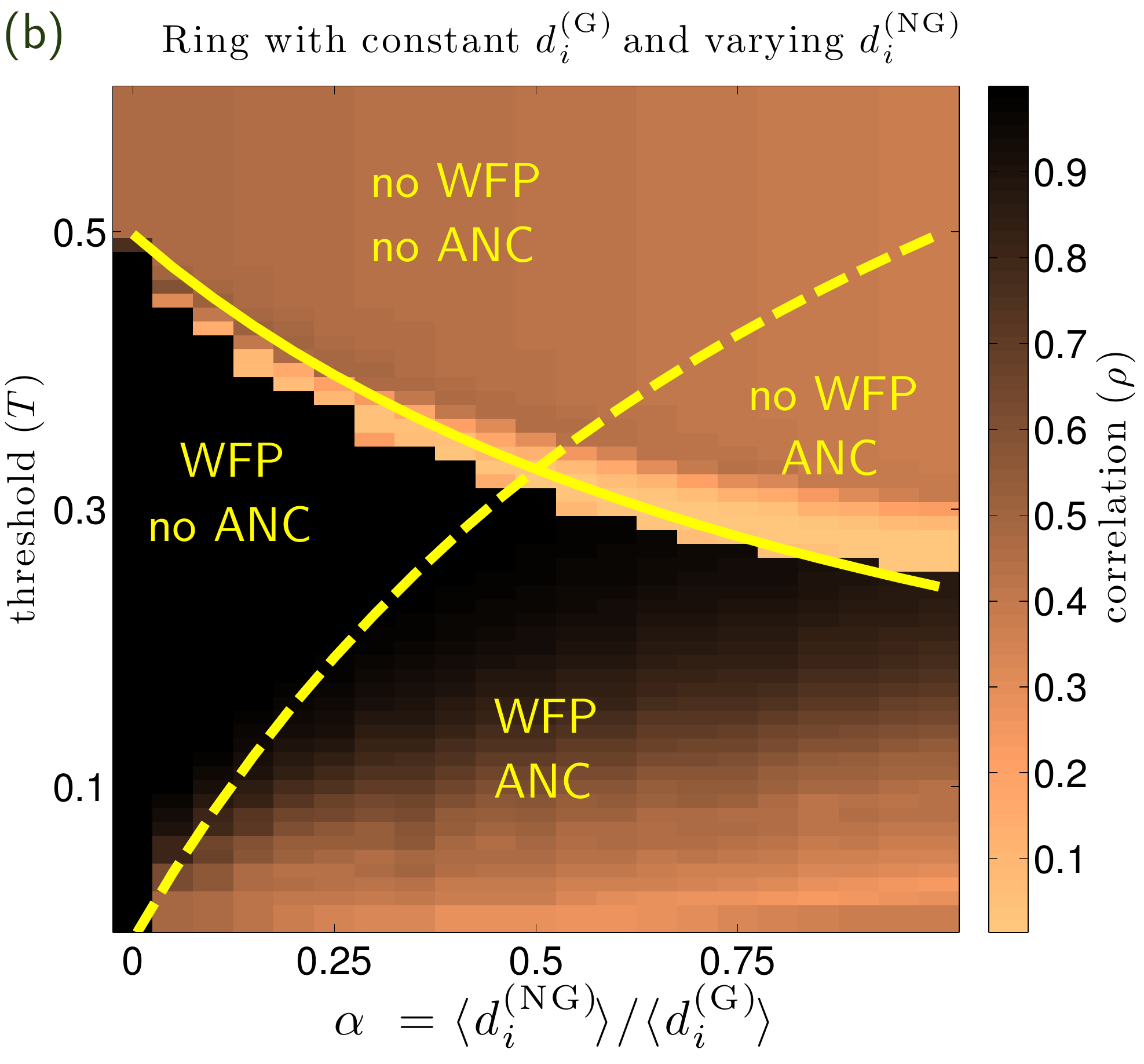}
\includegraphics[width=.4\linewidth]{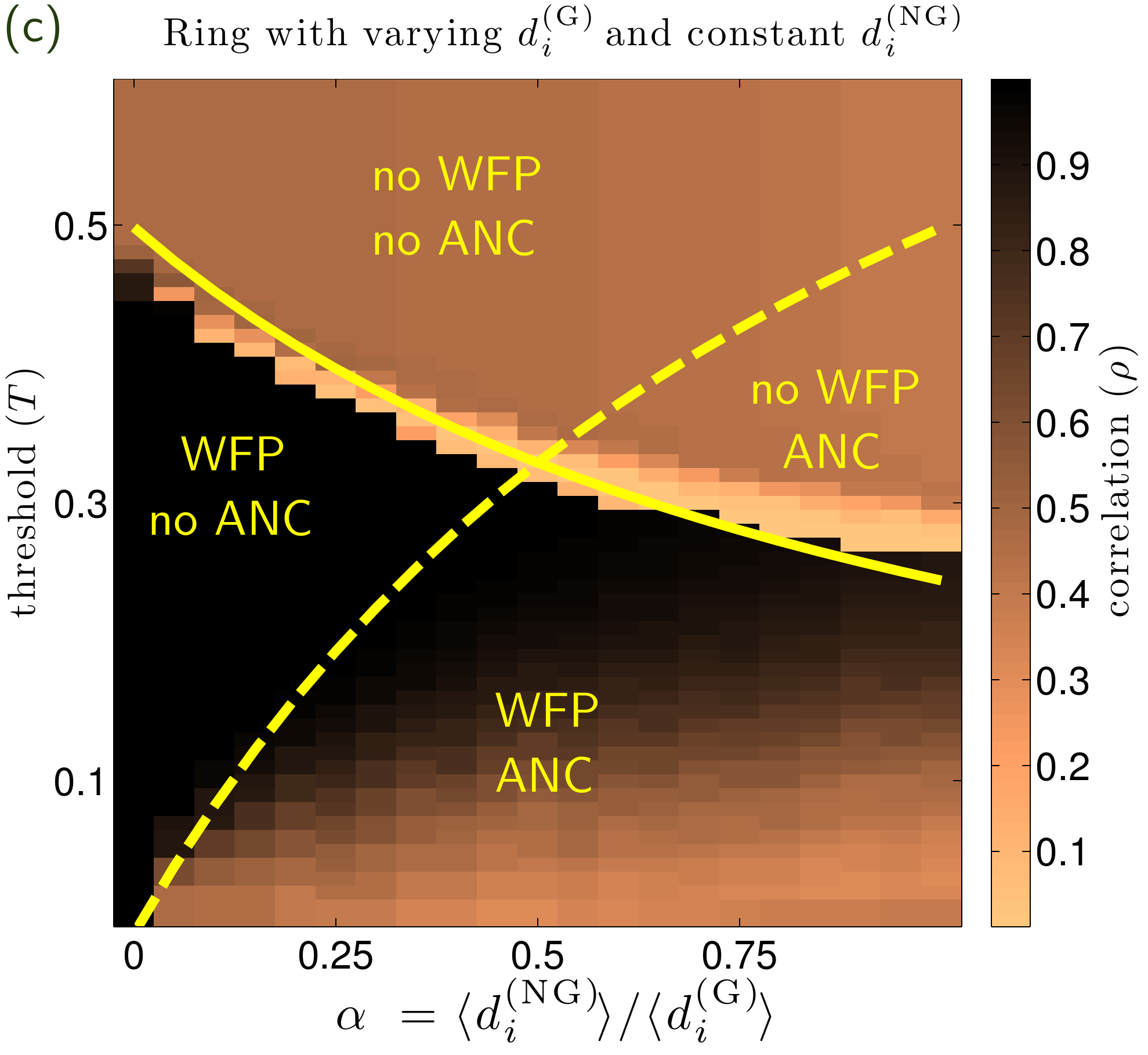}
\includegraphics[width=.4\linewidth]{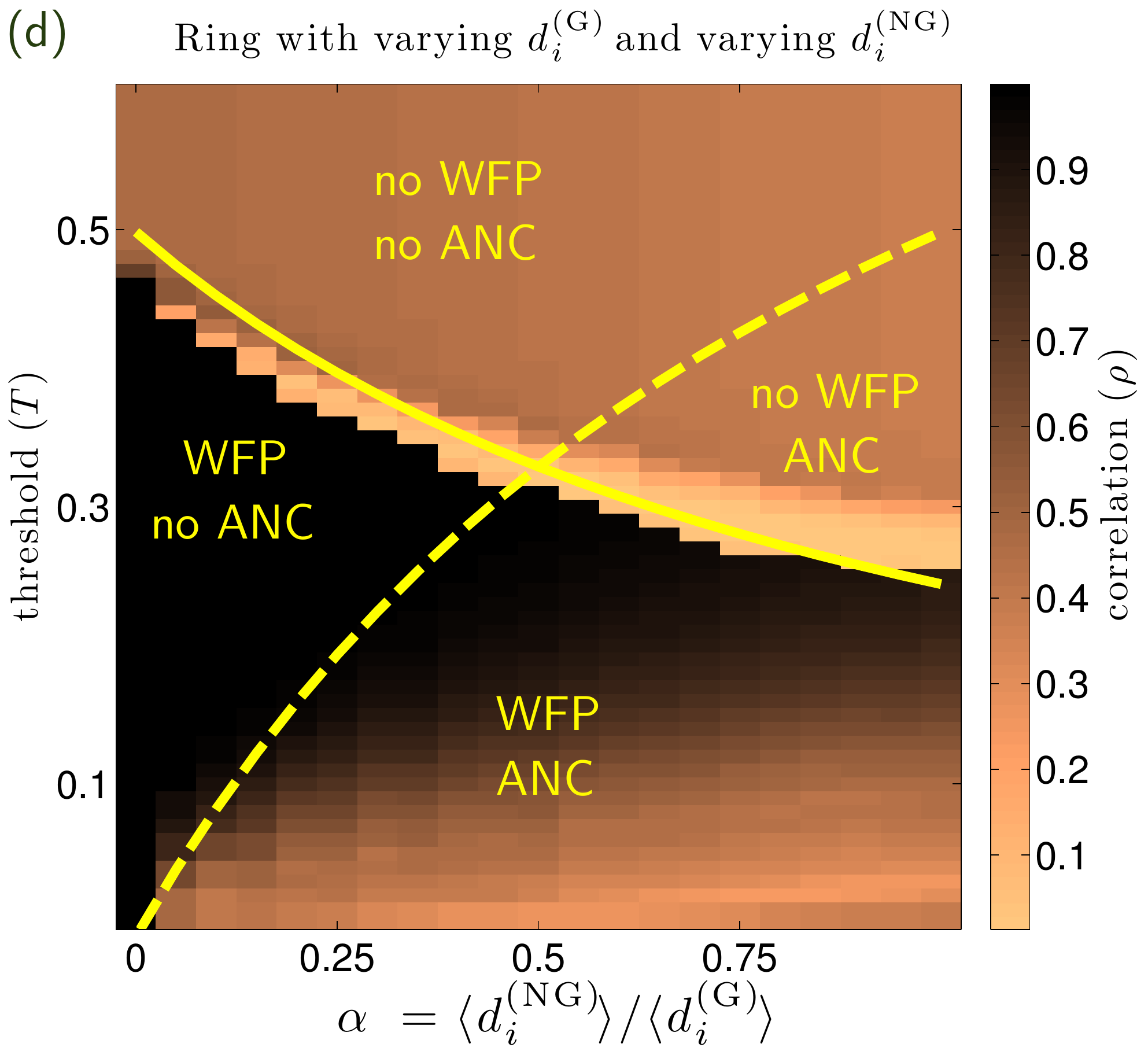}
\caption{ 
We study the geometry of symmetric WTM maps by calculating a Pearson correlation coefficient $\rho$ to compare node-to-node distances for the WTM map $\{\bold z^{(i)}\}\in\mathbb{R}^N$ to those for the node locations $\{\bold w^{(i)}\}\in\mathbb{R}^2$ on the ring {manifold. (See Sec.~III~D of the main manuscript.)} We show these values of $\rho$ in the $(T,\alpha)$ parameter plane, where $\alpha$ is the ratio of the number $N\langle d_i^{\rm{(NG)}} \rangle/2$ of non-geometric edges to the number $N\langle d_i^{\rm{(G)}} \rangle /2$ of geometric edges. Panels (a)--(d), respectively, illustrate results for network families {\bf(a)}--{\bf(d)}. For each panel, we construct a noisy ring network with $N=1000$ nodes and mean geometric degree $\langle d_i^{G}\rangle=40$, and we vary the mean non-geometric degree $\langle d_i^{\rm{(NG)}} \rangle \in [0,40]$ to study the parameter range $\alpha\in[0,1]$. The solid and dashed curves, respectively, give the theoretical approximations from Eqs.~\eqref{eq:ANC_crits2} and \eqref{eq:WFP_crits2} with $\delta_i^{\rm{(G)}}=\delta_i^{\rm{(NG)}}=0$. {See \ref{sec:numerics} for further discussion.}
}
\label{fig:correlation_contour}
\end{figure*}

\begin{figure*}[ht!]
\centering
\includegraphics[width=.4\linewidth]{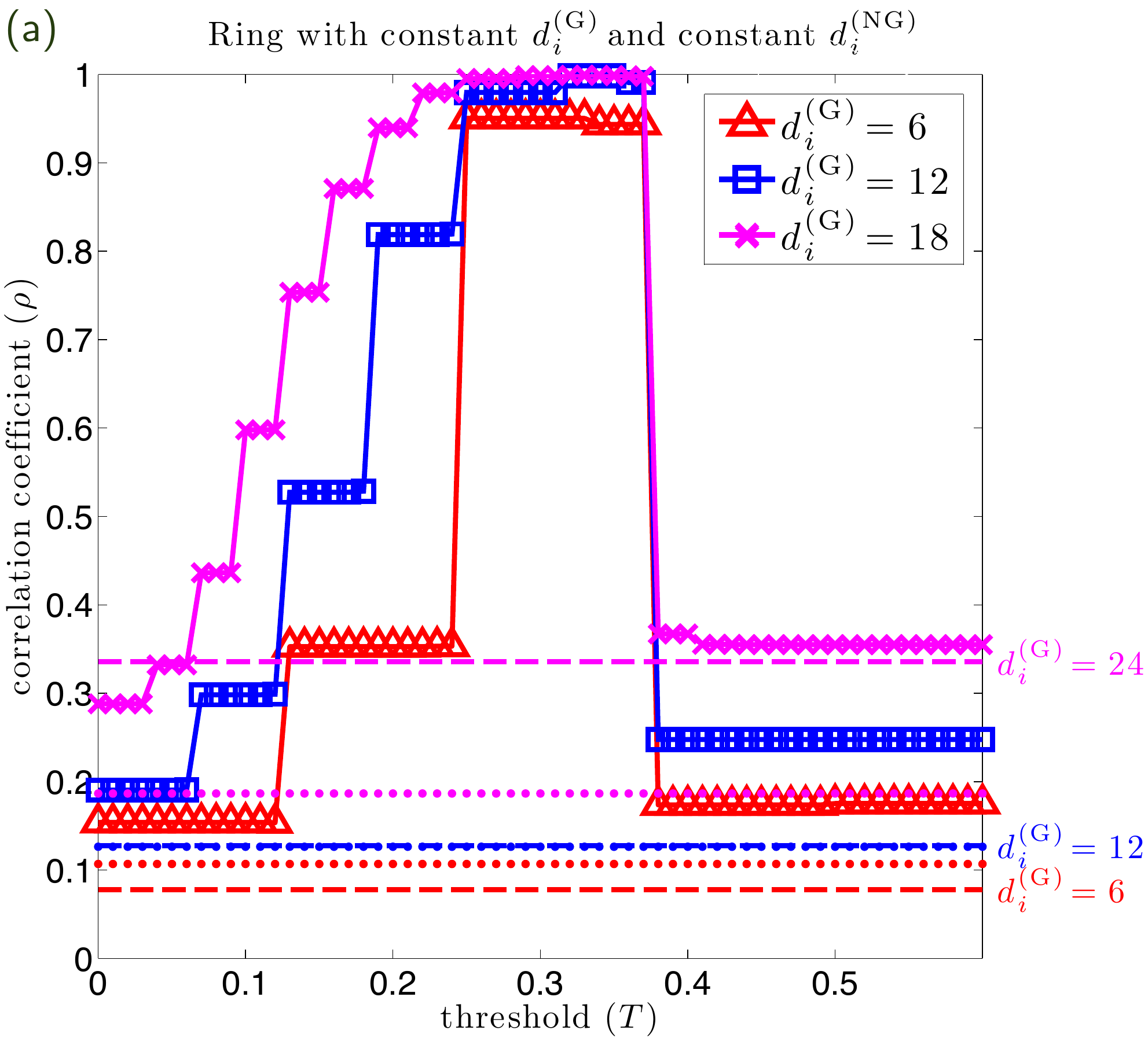}
\includegraphics[width=.4\linewidth]{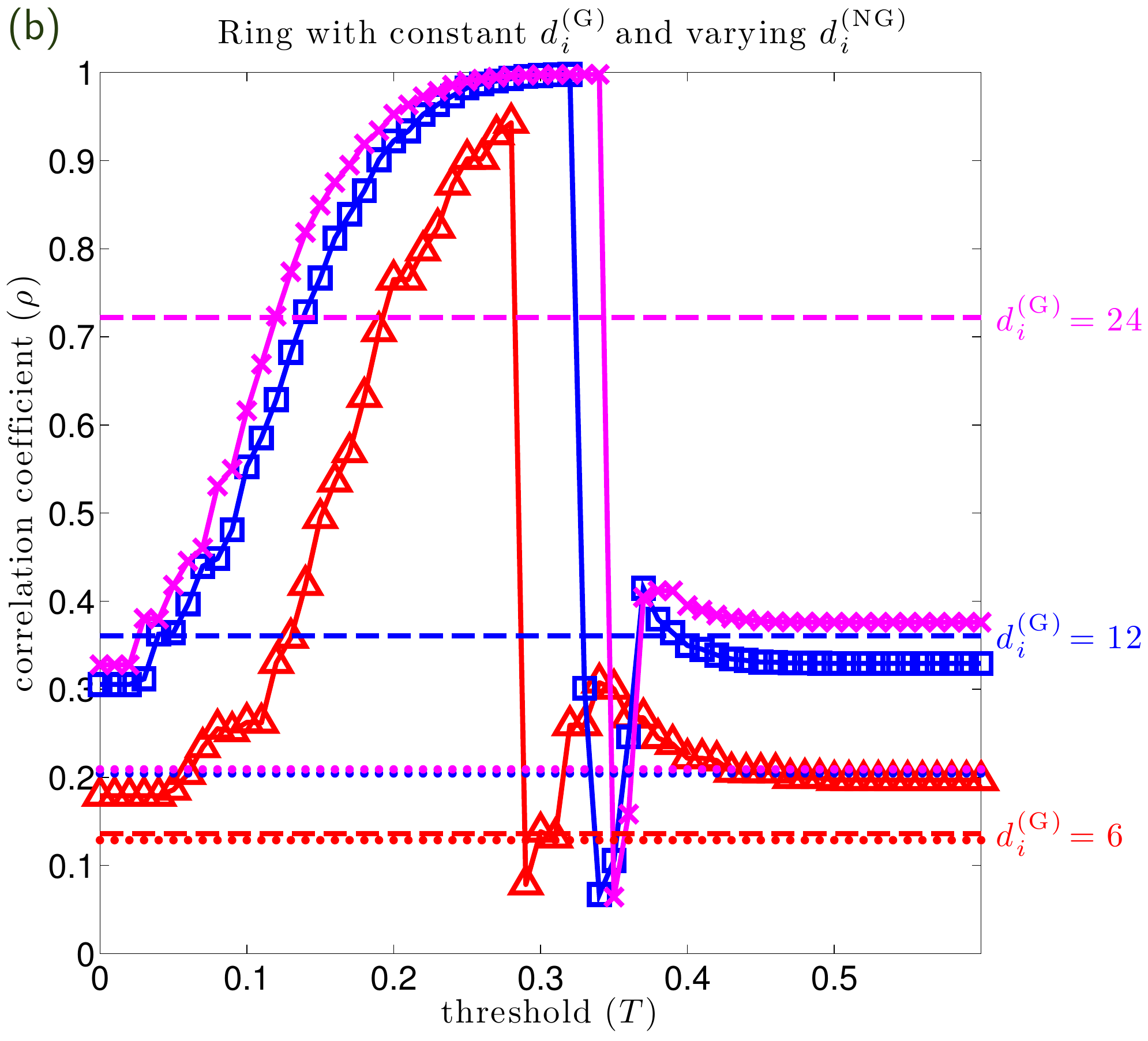}
\includegraphics[width=.4\linewidth]{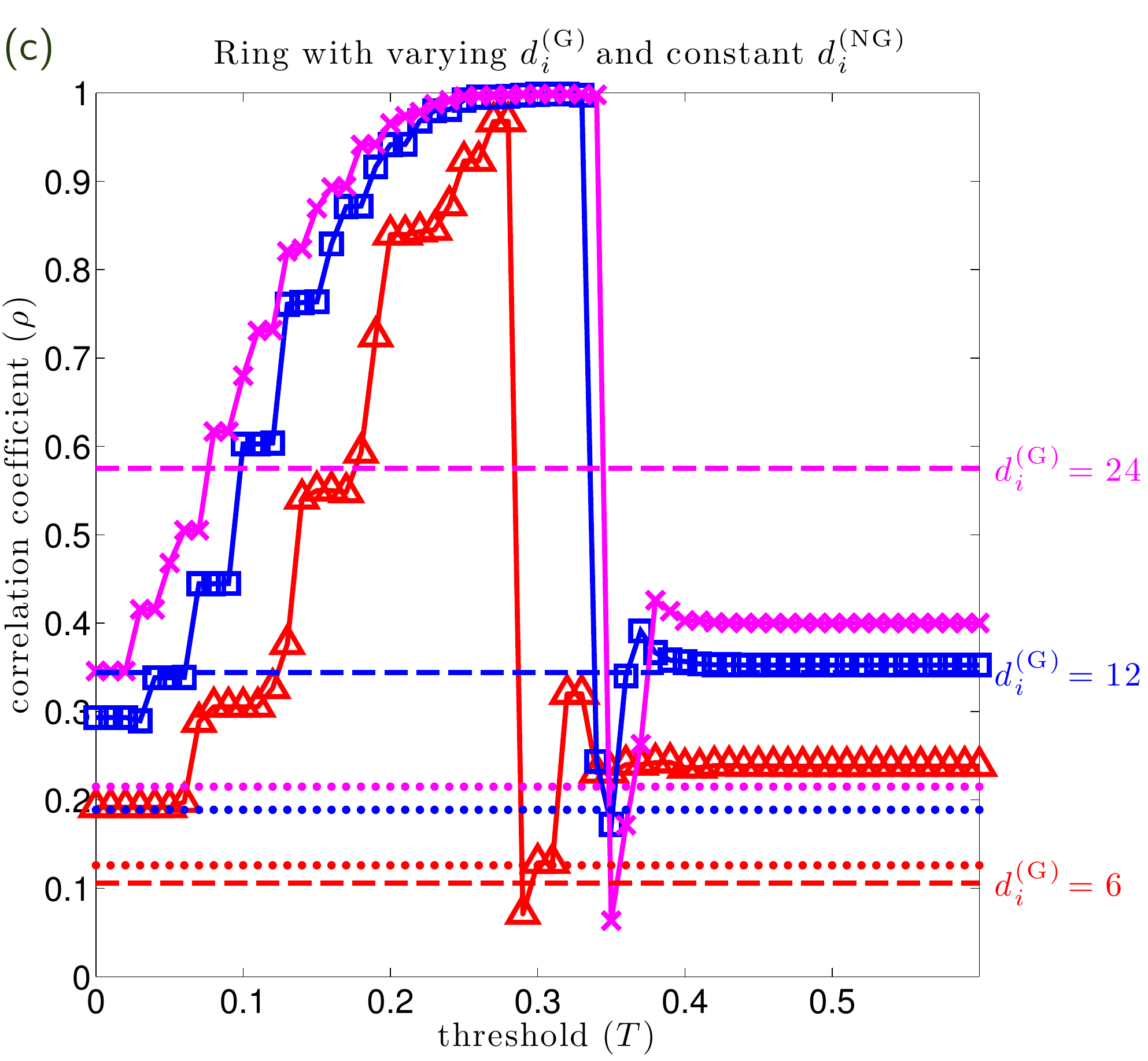}
\includegraphics[width=.4\linewidth]{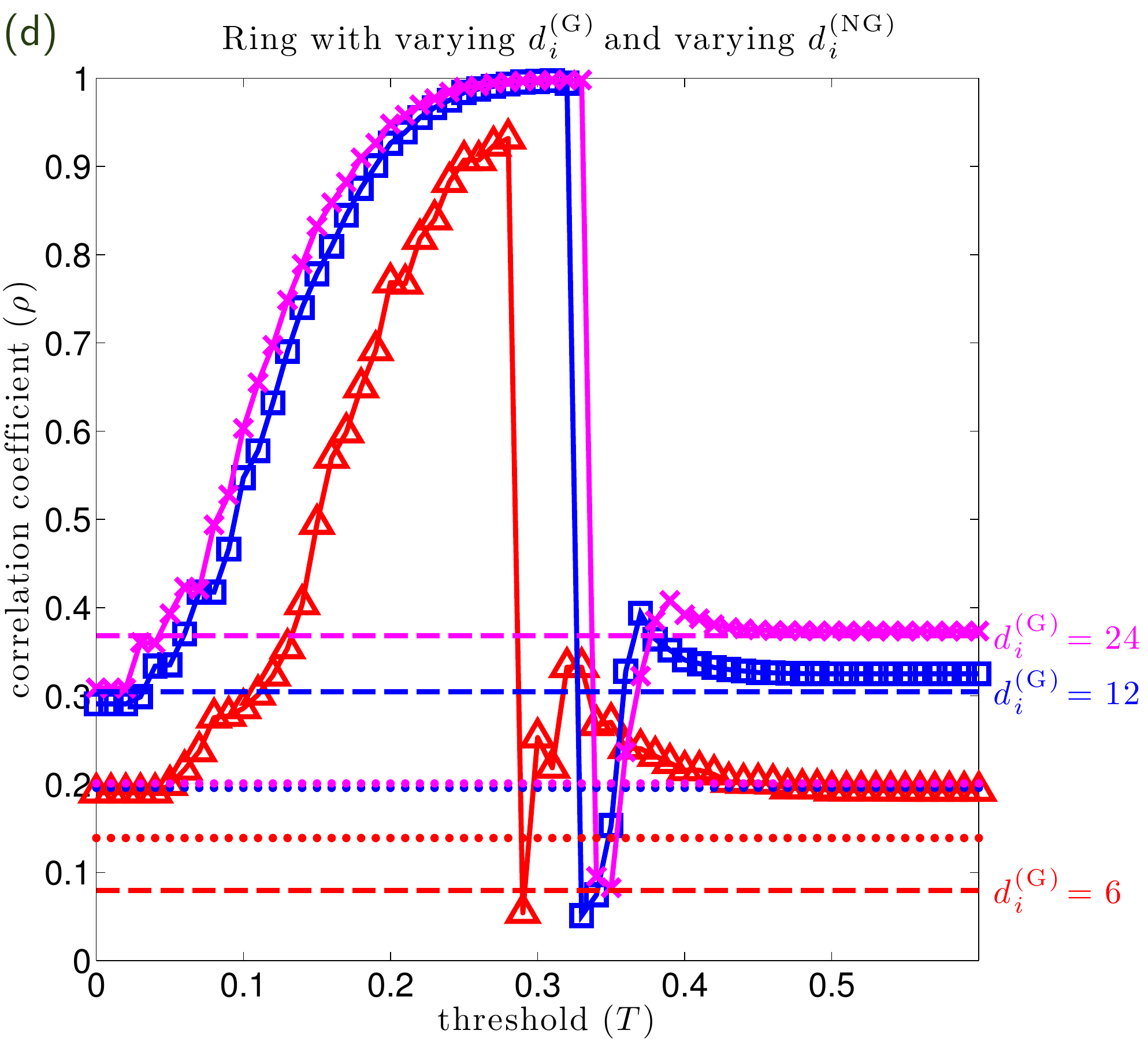}
\caption{
We study the geometry of symmetric WTM maps by calculating a Pearson correlation coefficient $\rho$ as a function of WTM threshold $T$ to compare node-to-node distances for the WTM map $\{\bold z^{(i)}\}\in\mathbb{R}^N$ to those for the node locations $\{\bold w^{(i)}\}\in\mathbb{R}^2$ on {a} ring manifold. Panels (a)--(d), respectively, illustrate results for network families {\bf(a)}--{\bf(d)}, and they amount to vertical cross sections of the corresponding contour plots in {Supplementary} Fig.~\ref{fig:correlation_contour} (i.e., for a constant value of $\alpha$). In each panel, we study WTM maps on a noisy ring network with $N=1000$ nodes with $\alpha =1/3$ for several choices of mean node degrees:
$(\langle d_i^{\rm{(G)}}\rangle,\langle d_i^{\rm{(NG)}}\rangle)=(6,2)$ (red triangles), $(\langle d_i^{\rm{(G)}}\rangle,\langle d_i^{\rm{(NG)}}\rangle)=(12,4)$ (blue squares), and $(\langle d_i^{\rm{(G)}}\rangle,\langle d_i^{\rm{(NG)}}\rangle)=(24,8)$ (magenta $\times$ symbols). We also plot $\rho$ for a 2D Laplacian eigenmap \cite{Belkin2003} (dashed lines) and for the Isomap algorithm \cite{Tenenbaum2000} (dotted lines). In all panels and for all mapping algorithms, increasing mean node degree tends to increase $\rho$, so the ability of the maps to translate the underlying ring manifold's geometry to a point cloud improves with increasing mean node degree for these experiments. Additionally, note that the curves for the largest mean degree (magenta $\times$ symbols) remain more consistent across the panels. {See \ref{sec:numerics} for further discussion.}
}
\label{fig:correlation_alpha_fix}
\end{figure*}

\begin{figure*}[ht!]
\centering
\includegraphics[width=.4\linewidth]{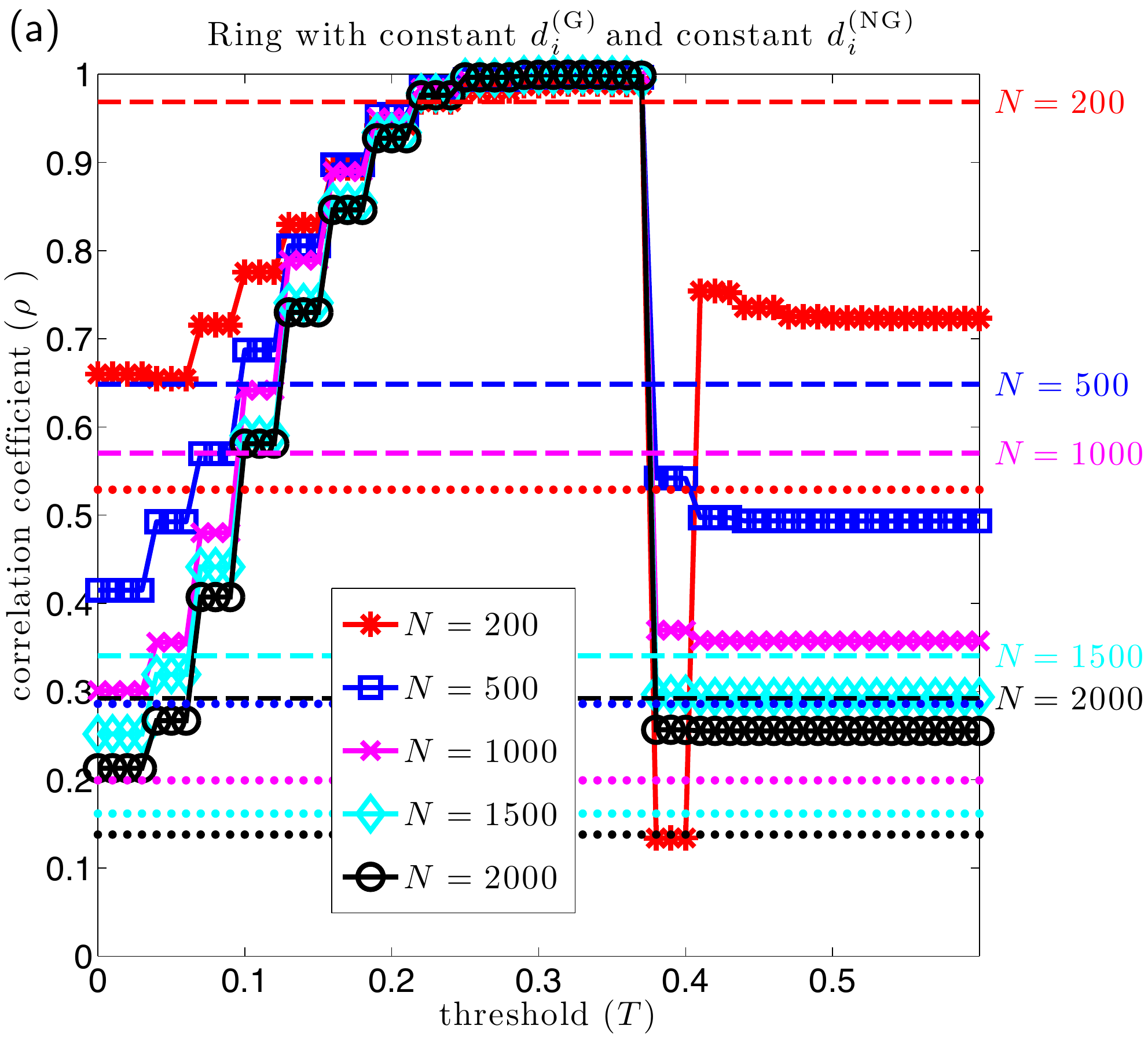}
\includegraphics[width=.4\linewidth]{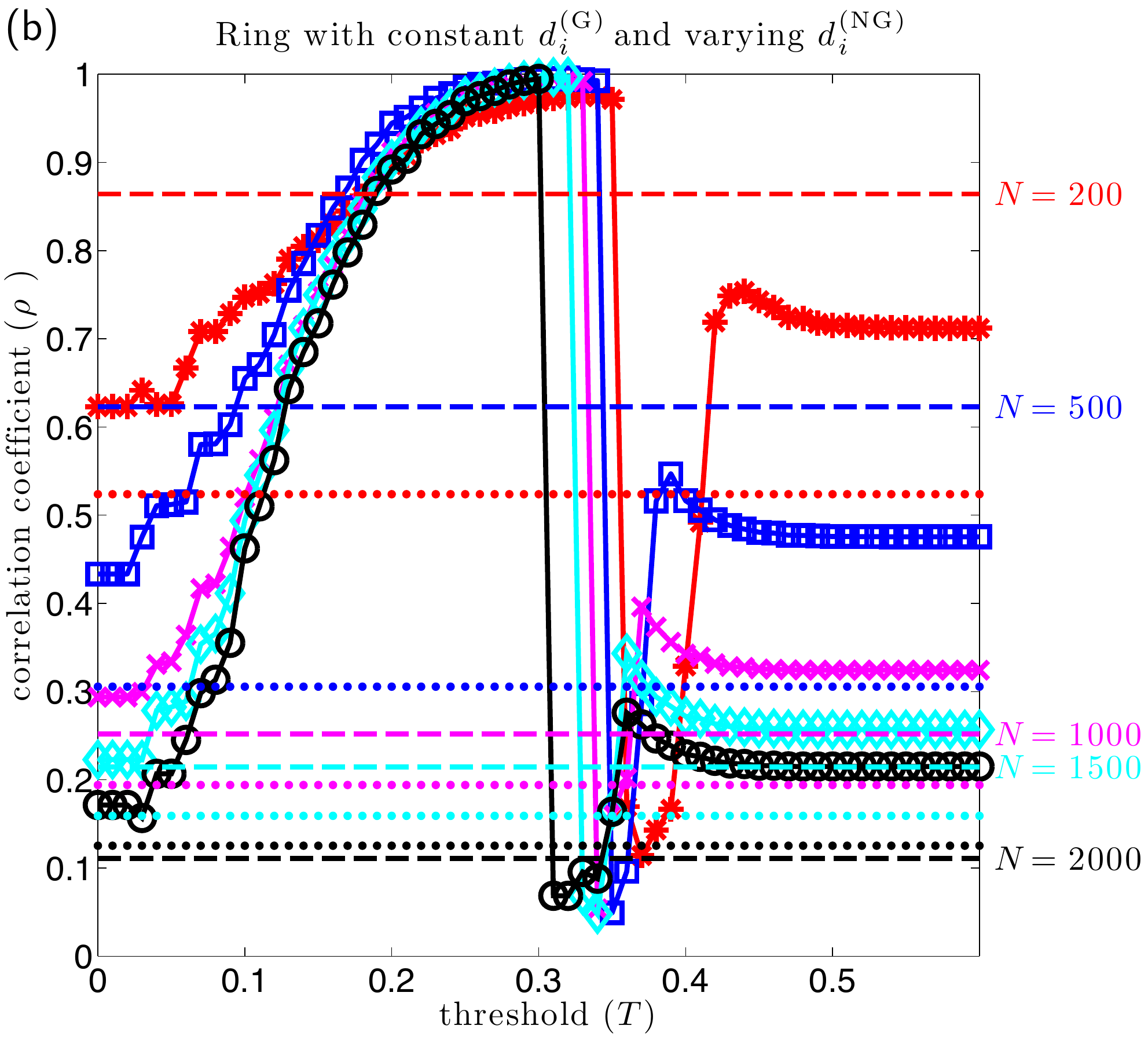}
\includegraphics[width=.4\linewidth]{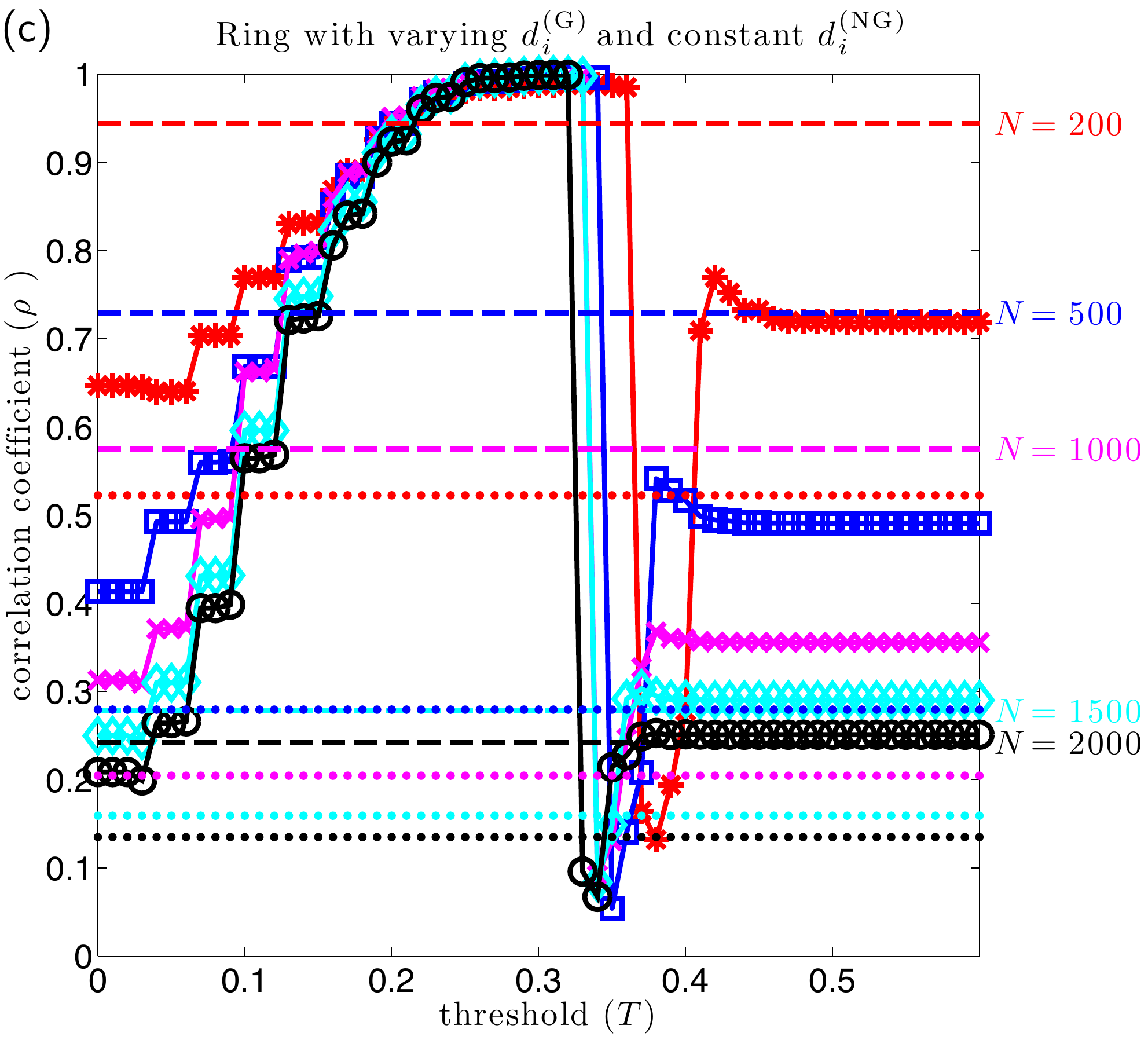}
\includegraphics[width=.4\linewidth]{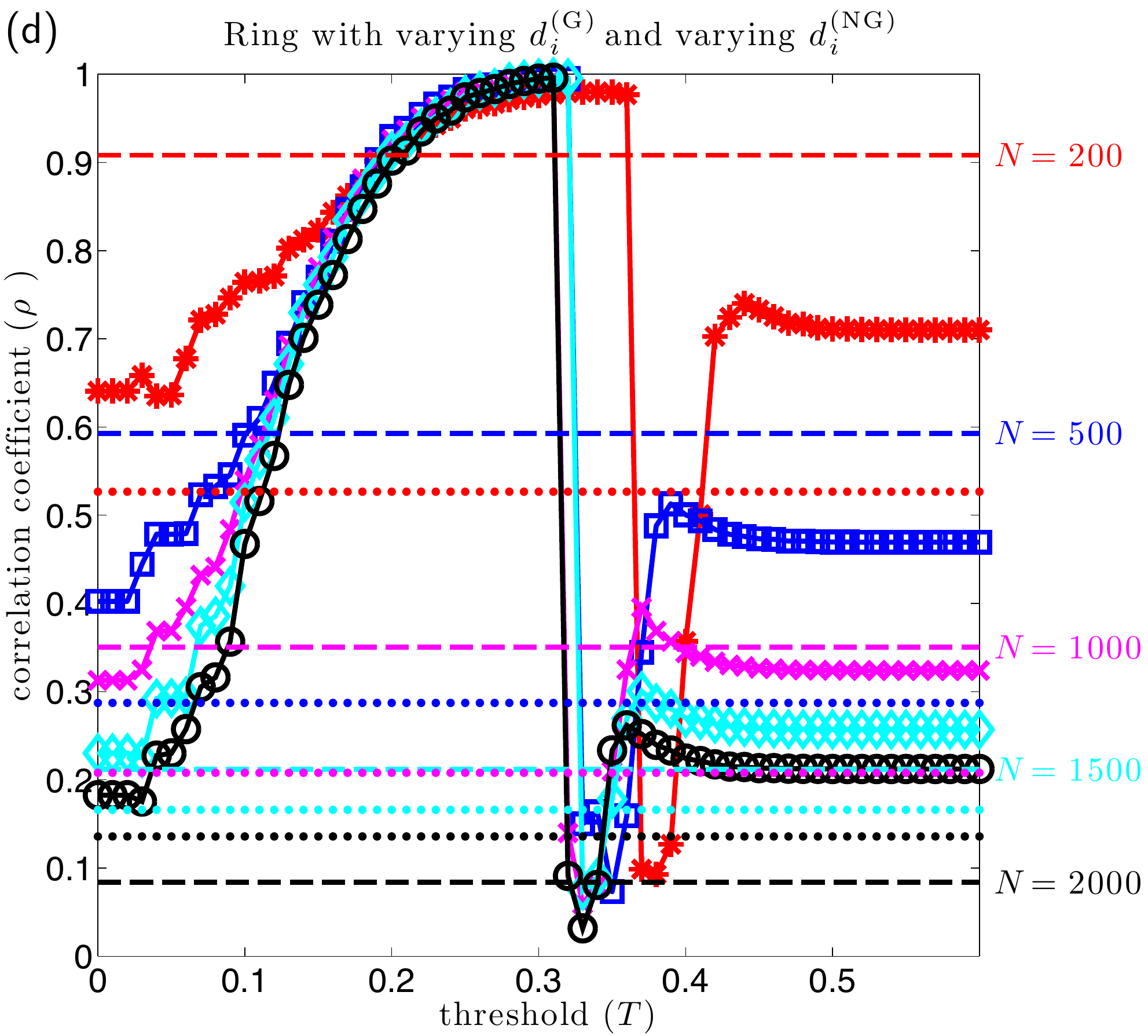}
\caption{
We study the geometry of symmetric WTM maps by calculating a Pearson correlation coefficient $\rho$ as a function of WTM threshold $T$ to compare node-to-node distances for the WTM map $\{\bold z^{(i)}\}\in\mathbb{R}^N$ to those for the node locations $\{\bold w^{(i)}\}\in\mathbb{R}^2$ on {a} ring manifold. Panels (a)--(d), respectively, illustrate results for network families {\bf(a)}--{\bf(d)}, and they amount to vertical cross sections of the corresponding contour plots in {Supplementary} Fig.~\ref{fig:correlation_contour} (i.e., for a constant value of $\alpha$), although we show results for several choices of network size $N$. In each panel, we study WTM maps on a noisy ring network with $\langle d_i^{{\rm(G)}}\rangle = 24$ and $\langle d_i^{\rm{(NG)}}\rangle=8$ (i.e., $\alpha = 1/3$), and we show curves of $\rho$ versus $T$ for networks of sizes $N=\{ 200, 500, 1000, 1500, 2000\}$. We also plot $\rho$ for a 2D Laplacian eigenmap \cite{Belkin2003} (dashed lines) and for the Isomap algorithm \cite{Tenenbaum2000} (dotted lines). In all panels and for all mapping algorithms, increasing the network size $N$ tends to decrease $\rho$, except for WTM maps that are characterized by WFP and little (or no) ANC. {See \ref{sec:numerics} for further discussion.}
}
\label{fig:correlation_N_vary}
\end{figure*}

In this section, we compare the geometry of symmetric WTM maps for networks in the families {\bf(a)}--{\bf(d)}, {which we defined in \ref{sec:models}}, via calculating a Pearson correlation coefficient $\rho$ to compare WTM distances to distances in an underlying manifold. We also investigate the effects on WTM maps of varying the mean geometric and non-geometric degrees and the network size $N$ (when we hold other parameters constant). We show our results in {Supplementary} Figs.~\ref{fig:correlation_contour}--\ref{fig:correlation_N_vary}. Panels (a)--(d), respectively, give our results for network families {\bf (a)}--{\bf (d)}. Unless we indicate otherwise, we show results for one network from each family in these and subsequent figures.

In Supplementary Fig.~\ref{fig:correlation_contour}, we plot $\rho$ for point clouds that result from symmetric WTM maps for the $(T,\alpha)$ parameter plane. The solid and dashed curves yield approximate bifurcation curves, which we obtain from Eqs.~\eqref{eq:ANC_crits2} and \eqref{eq:WFP_crits2} with $\delta_i^{\rm{(G)}}=\delta_i^{\rm{(NG)}}=0$. Note that panel (a) depicts similar information as {Supplementary} Fig.~6(a) of the main text, although the results that we now show are for a larger network with larger degrees. For all panels, the curve given by $T_0^{\mathrm{(WFP)}}$ agrees very well with a relatively abrupt transition that one can observe by examining the geometry of the WTM maps via the coefficient $\rho$. In contrast, the curve for $T_0^{\mathrm{(ANC)}}$ appears to not be as closely related to $\rho$. To illustrate this, fix $\alpha=0.25$ and consider increasing values of $T$ in any panel. As $T$ surpasses $T_0^{\rm{(WFP)}}$, there is a large drop in the value of $\rho$. By contrast, $\rho$ changes only slightly when we cross $T\approx T_0^{\mathrm{(ANC)}}$. 
Comparing the four panels to each another, we find that the agreement between $T_0^{\mathrm{(WFP)}}$ and the observed shifts in $\rho$ decreases as the node degrees become more heterogeneous. In particular, the transition that is imposed by $T_0^{\mathrm{(WFP)}}$ appears to shift to smaller values of $T$, so the heterogeneities that we introduce in network families {\bf(b)}--{\bf(d)} mostly affect the regime in which $T\approx T_0^{\mathrm{(WFP)}}$. However, the qualitative behavior of WTM contagions in the $(T,\alpha)$ parameter plane is similar for all four families of networks.

In Supplementary Fig.~\ref{fig:correlation_alpha_fix}, we study the effect on $\rho$ for symmetric WTM maps when we increase the mean node degrees $\langle d_i^{\rm{(G)}}\rangle$ and $\langle d_i^{\rm{(NG)}}\rangle$. Fixing $\alpha=1/3$, we plot $\rho$ as a function of the threshold $T$. This amounts to examining vertical cross sections of the four panels in {Supplementary} Fig.~\ref{fig:correlation_contour}. We study the effect of varying mean node degree by showing results for $(\langle d_i^{\rm{(G)}}\rangle,\langle d_i^{\rm{(NG)}}\rangle)=(6,2)$ (red triangles), $(\langle d_i^{\rm{(G)}}\rangle, \langle d_i^{\rm{(NG)\rangle}})=(12,4)$ (blue squares), and $(\langle d_i^{\rm{(G)}}\rangle,\langle d_i^{\rm{(NG)}}\rangle)=(24,8)$ (magenta $\times$ symbols). We also show results for a 2D Laplacian eigenmap \cite{Belkin2003} (horizontal dashed lines) and Isomap \cite{Tenenbaum2000} (horizontal dotted lines) applied to our (unweighted) networks.
Comparing the $\rho$ values for the WTM maps with various mean degrees, we note that increasing the mean degrees smoothens the dependence of $\rho$ versus $T$. Specifically, for smaller values of $T$ (e.g., for $T<0.3$), the discontinuous jumps in $\rho$ become smaller as the mean degrees increase. Interestingly, increasing heterogeneity in the node degrees also smoothens the curves of $\rho$ versus $T$. For example, the curves are smoother for network families {\bf (b)}--{\bf (d)} than they are for family {\bf{(a)}}. 
Additionally, note in all panels that we observe an abrupt drop in $\rho$ for $T\approx T_0^{\rm{(WFP)}}=1/(2+2\alpha)=3/8$. However, in more heterogeneous situations, this abrupt drop can shift to smaller values of $T$. This is most apparent when comparing the four panels for $(\langle d_i^{\rm{(G)}}\rangle,\langle d_i^{\rm{(NG)}}\rangle)=(6,2)$ (red triangles): the drop occurs at $T=3/8=0.375$ in panel (a), whereas it occurs at approximately at $T\approx0.29$ in panels (b)--(d). To contrast this large shift, when comparing the panels for $(\langle d_i^{\rm{(G)}}\rangle,\langle d_i^{\rm{(NG)}}\rangle)=(24,8)$ (magenta $\times$ symbols), we observe that the change is smaller [i.e.,the abrupt drop in $\rho$ is at $T=0.375$ in panel (a), whereas it occurs at $T\approx0.35$ in panel (d)]. Finally, note in all panels that $\rho$ 
%in general %{\bf map: it seemed to not exactly be true for a couple specific values, so "in general" which mathematically can mean 'all the time' depending on how it's written/parsed doesn't quite seem appropriate}
increases as we increase the mean degrees, and we observe similar increases in $\rho$ for the Laplacian-eigenmap and Isomap algorithms. Thus, in this series of experiments, increasing mean node degree improves the ability of the maps to translate the geometry of the underlying manifold to the resulting WTM point cloud.  %For larger mean degrees relative to a fixed network size (i.e., for a network with a fixed number of nodes), we expect more complicated behavior to occur.

In {Supplementary} Fig.~\ref{fig:correlation_N_vary}, we study the geometry of symmetric WTM maps by plotting $\rho$ versus $T$ for networks of various sizes $N$. We fix $\langle d_i^{\rm{(G)}}\rangle=24$ and $\langle d_i^{\rm{(NG)}}\rangle=8$ (that is, $\alpha=1/3$) and plot $\rho$ versus the threshold $T$ for $T\in[0,0.6]$. In each panel, we show results for several choices of network size $N\in[200,2000]$ to illustrate how $\rho$ depends on $N$. As $N$ increases, we observe that $\rho$ systematically decreases for WTM maps that correspond to contagions in which WFP is not the dominant phenomenon. However, for WTM maps in which WFP dominates (e.g., when $T\in[0.2,0.25]$), we find that $\rho$ remains above $0.85$. This provides strong evidence that, for this parameter regime, WTM maps translate the geometry of the underlying ring manifold to the resulting point cloud for a wide range of network sizes (with the other parameters held constant). One does \emph{not} obtain such independence with network size when using a Laplacian eigenmap or Isomap. In those cases, we find that $\rho$ systematically decreases as $N$ increases (with the other parameters held constant).

\clearpage

%%%%%%%%%%%%%%%%%%%%%%%%%%%%%%%%%%%%%%%%
~\\{\bf \Large Numerical Results for Dimensionality}\\
%\subsection{Numerical Results for Dimensionality}\label{sec:result_dim}
%%%%%%%%%%%%%%%%%%%%%%%%%%%%%%%%%%%%%%%%

In this section, we examine the dimensionality of point clouds that result from symmetric WTM maps that we apply to networks on {a} ring manifold. As {we} discussed in {Sec.~III~E and } \ref{sec:geometry}, we {study their} ``embedding dimension'' $P$, {which we define for a point cloud} to be the smallest dimension $p$ such that the residual variance $R_p$ for the projection onto $\mathbb{R}^p$ is small. In practice, we use PCA for such projections, and we specify ``small'' as being (strictly) less than $0.05$. (In other words, we lose less than 5\% of the variance after the projection.) 
{Importantly, if the point cloud is a noisy sample of points on a manifold, then $P$ is an approximation for the embedding dimension of the manifold.% (see \ref{sec:geometry}).}

We show our results for embedding dimension of point clouds resulting from WTM maps in {Supplementary} Figs.~\ref{fig:dimension_contour}--\ref{fig:dimension_alpha_fix}. Panels (a)--(d), respectively, give our results for network families {\bf (a)}--{\bf (d)}. 
In {Supplementary} Fig.~\ref{fig:dimension_contour}, we plot $P$ in the $(T,\alpha)$ parameter plane for networks with $N = 200$ nodes, mean geometric degree $\langle d_i^{\rm{(G)}}\rangle=20$, and various mean non-geometric degrees $\langle d_i^{\rm{(NG)}}\rangle$. We also plot the approximate bifurcation curves given by Eqs.~\eqref{eq:ANC_crits2} and \eqref{eq:WFP_crits2} with $\delta_i^{\rm{(G)}}=\delta_i^{\rm{(NG)}}=0$. Note that panel (a) is similar to the plot in {Supplementary} Fig.~6(b) of the main text. We observe in all panels that WTM maps for the contagion regime that we expect to exhibit WFP but no ANC yield point clouds $\{\bold z^{(i)}\}$ with a small embedding dimension of $P\approx2$. This result is {expected}, because {a} ring manifold is exactly the unit circle in $\mathbb{R}^2$. {That is, it is a one-dimensional manifold that requires at least two dimensions to be embedded in a Euclidean space.}
Note that this low dimensionality persists into the regime that we expect to exhibit both WFP and ANC, although the embedding dimension $P$ increases as one moves away from the regime exhibiting WFP and no ANC.

%{\bf map: is it "increases" or should it be "tends to increase" ? (i.e., are there exceptions?)}; DRT: no exceptions so increases

In {Supplementary} Fig.~\ref{fig:dimension_alpha_fix}, we continue our investigation of the dimensionality of point clouds that result from the application of symmetric WTM maps to networks on {a} ring manifold by showing their embedding dimension $P$ as a function of threshold $T$. One can construe the curves of $P$ versus $T$ as a vertical cross section of the contour plots in {Supplementary} Fig.~\ref{fig:dimension_contour}; we show results for several choices of mean node degrees: $(\langle d_i^{\rm{(G)}}\rangle,\langle d_i^{\rm{(NG)}}\rangle)=(6,2)$ (red triangles), $(\langle d_i^{\rm{(G)}}\rangle,\langle d_i^{\rm{(NG)}}\rangle)=(12,4)$ (blue squares), and $(\langle d_i^{\rm{(G)}}\rangle,\langle d_i^{\rm{(NG)}}\rangle)=(18,6)$ (magenta $\times$ symbols). We also show values (horizontal dotted lines) of $P$ versus $T$ for the point clouds that we obtain by applying Isomap \cite{Tenenbaum2000} to the networks.  We obtain horizontal lines because Isomap does not include any dependence on $T$. We do not investigate the dimensionality of the 2D Laplacian eigenmaps, as we fix their dimension to 2 in our study.

Note that the curves in panels of {Supplementary} Fig.~\ref{fig:dimension_alpha_fix} are rather similar to each other. In particular, for all panels, we find the smallest embedding dimension $P$ for the regime in which we expect a WTM contagion to exhibit WFP without ANC [i.e., for $T\in(1/4,3/8)$]. Additionally, we consistently identify the correct embedding dimension (i.e., $P=2$) for this regime as long as mean degrees are sufficiently large (e.g., see the magenta $\times$ symbols).  For smaller mean degrees, we still observe that $P$ is small for a similar range of the threshold $T$. However, the curves of $P$ versus $T$
%in general
 tend to suggest that smaller mean degrees lead to larger embedding dimensions in our numerical experiments. 
%{\bf map: it was written "in general", and I have the same comment on that as I did above; check my wording change}
For Isomap (in which we map nodes based on shortest paths), we observe in our experiments that the embedding dimension $P$ is always at least 10. Additionally, the embedding dimension $P$ for Isomap appears to decrease systematically as the mean degrees increase. Thus, using shortest paths to map nodes for network families {\bf (a)}--{\bf (d)} leads to point clouds with a dimensionality that is higher than $P=2$; however, it might be possible to recover the correct embedding dimension of {a} ring manifold when the mean degrees are sufficiently large (keeping all other parameters fixed). Finally, note that $P\le20$  in all panels. Recall that this is the maximum value of $P$ that we can observe because it is the largest projection that we consider.

\begin{figure*}[ht!]
\centering
\includegraphics[width=.4\linewidth]{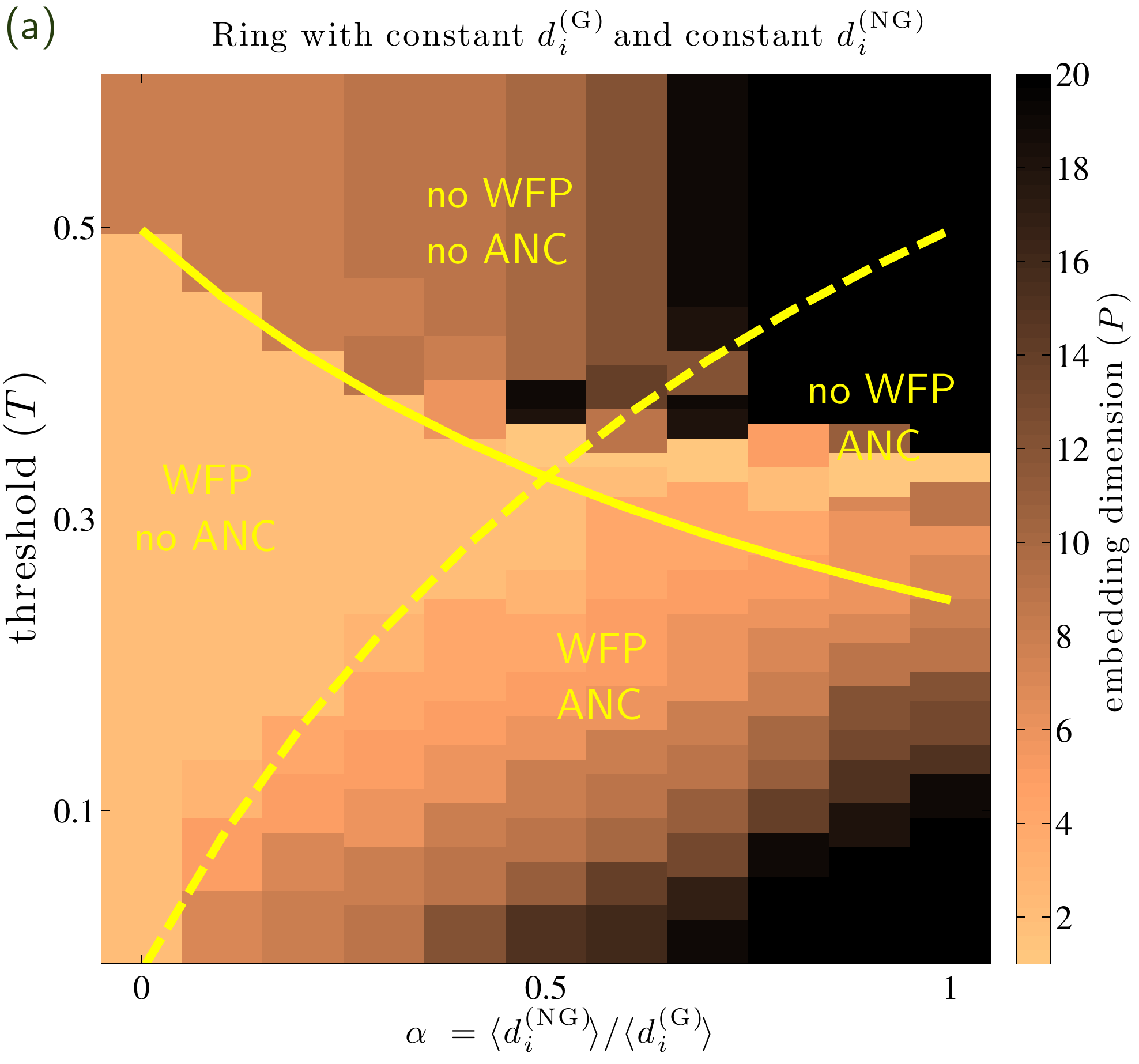}
\includegraphics[width=.4\linewidth]{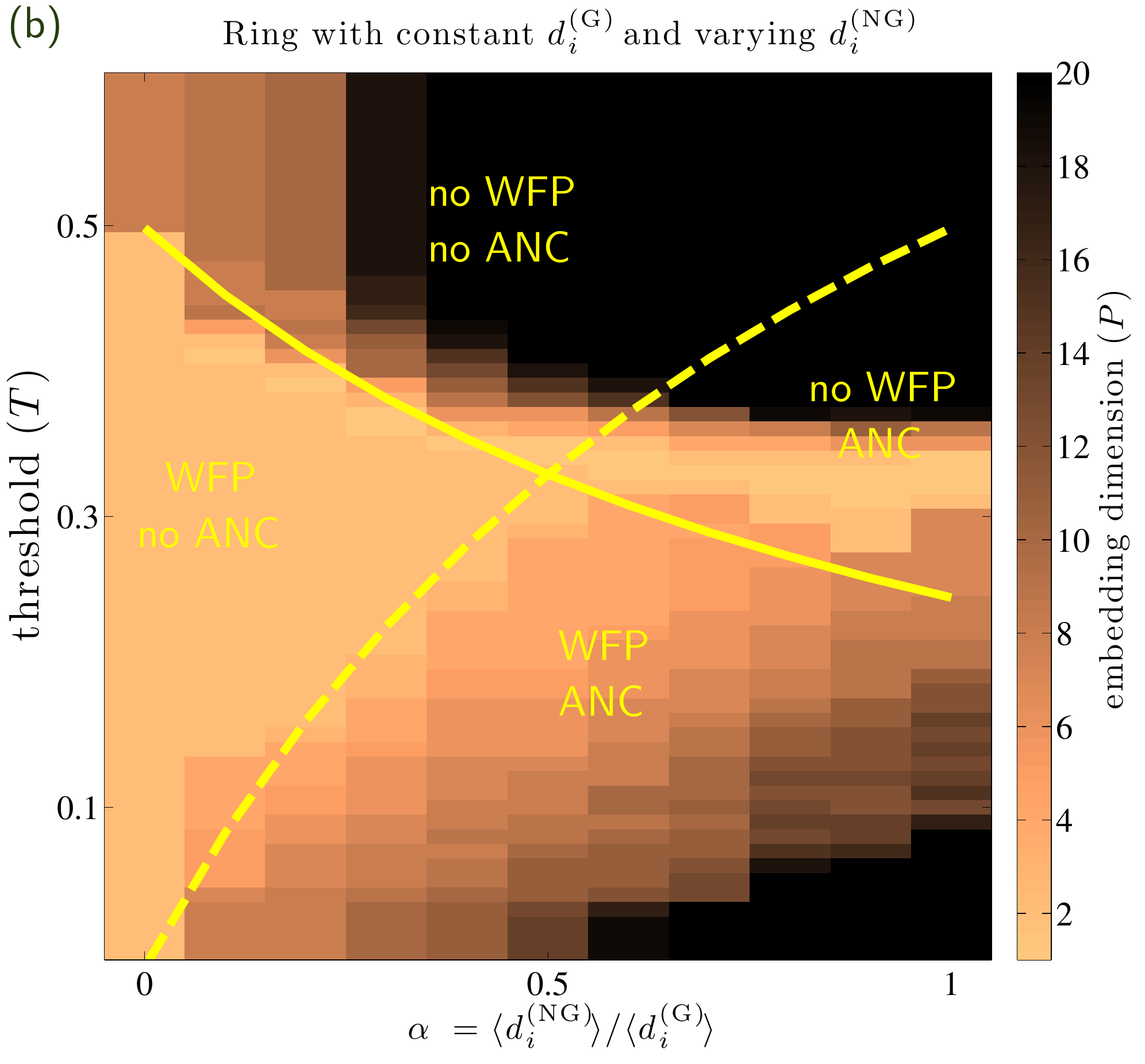}
\includegraphics[width=.4\linewidth]{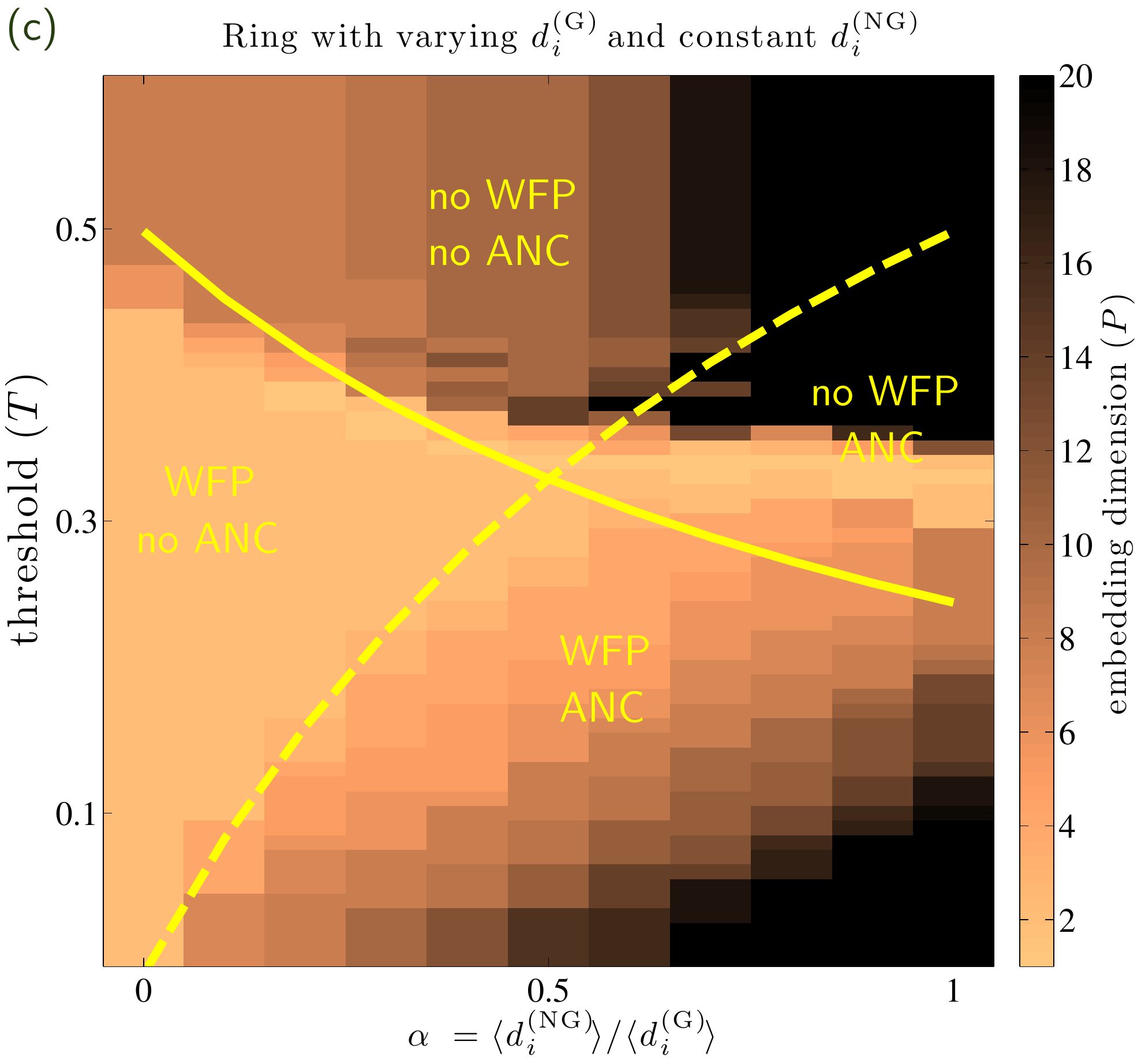}
\includegraphics[width=.4\linewidth]{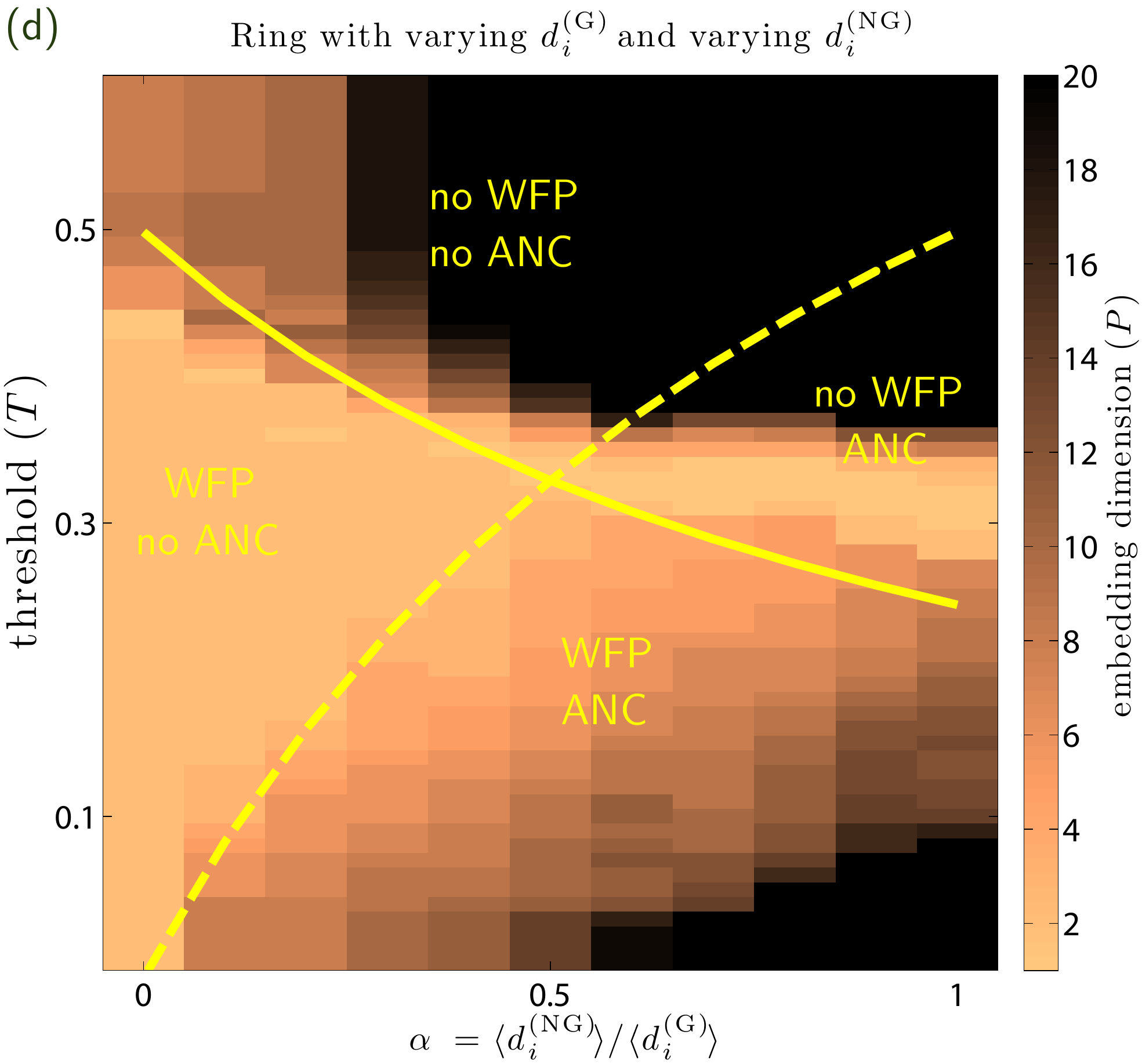}
\vspace{0cm}
\caption{ 
We examine the dimensionality of point clouds that result from symmetric WTM maps that we apply to networks on {a} ring manifold by studying their embedding dimension $P=\min\{p|R_p<0.05\}$, where $R_p$ denotes the residual variance for the projection onto {$\mathbb{R}^p$. (See Sec.~III~E of the main manuscript.)} We plot $P$ in the $(T,\alpha)$ parameter plane for networks with $N = 200$ nodes, mean geometric degree $\langle d_i^{\rm{(G)}}\rangle=20$, and various values of the non-geometric degree $\langle d_i^{\rm{(NG)}}\rangle$. As before, panels (a)--(d), respectively, illustrate the results for network families {\bf(a)}--{\bf(d)}. We use solid and dashed curves, respectively, to indicate the theoretical critical threshold values given by Eqs.~\eqref{eq:ANC_crits2} and \eqref{eq:WFP_crits2} with $\delta_i^{\rm{(G)}}=\delta_i^{\rm{(NG)}}=0$. In all panels, we see that WTM maps for the contagion regime that we predict to be characterized by WFP but no ANC yield point clouds with an embedding dimension of $P\approx2$, which agrees with the fact that {a} ring manifold is embedded in $\mathbb{R}^2$. This low-dimensional structure persists into the regime that we predict has both WFP and ANC, although the embedding dimension $P$ increases as one moves away from the regime that exhibits WFP and no ANC. {See \ref{sec:numerics} for further discussion.}
%{\bf map: per Yannis's comment, "no ANC" is an asymptotic thing, right, because of the stochasticity of those edges? I don't think this needs discussion in the caption, but I feel like the text and SI need us to spell this out explicitly (I don't think we're explicit about this) to convey things clearly}
%DRT: Important clarification: no ANC is not an asymptotic thing. no acceleration of WFP is an asymptotic thing
}
\label{fig:dimension_contour}
\end{figure*}

\begin{figure*}[ht!]
\centering
\includegraphics[width=.35\linewidth]{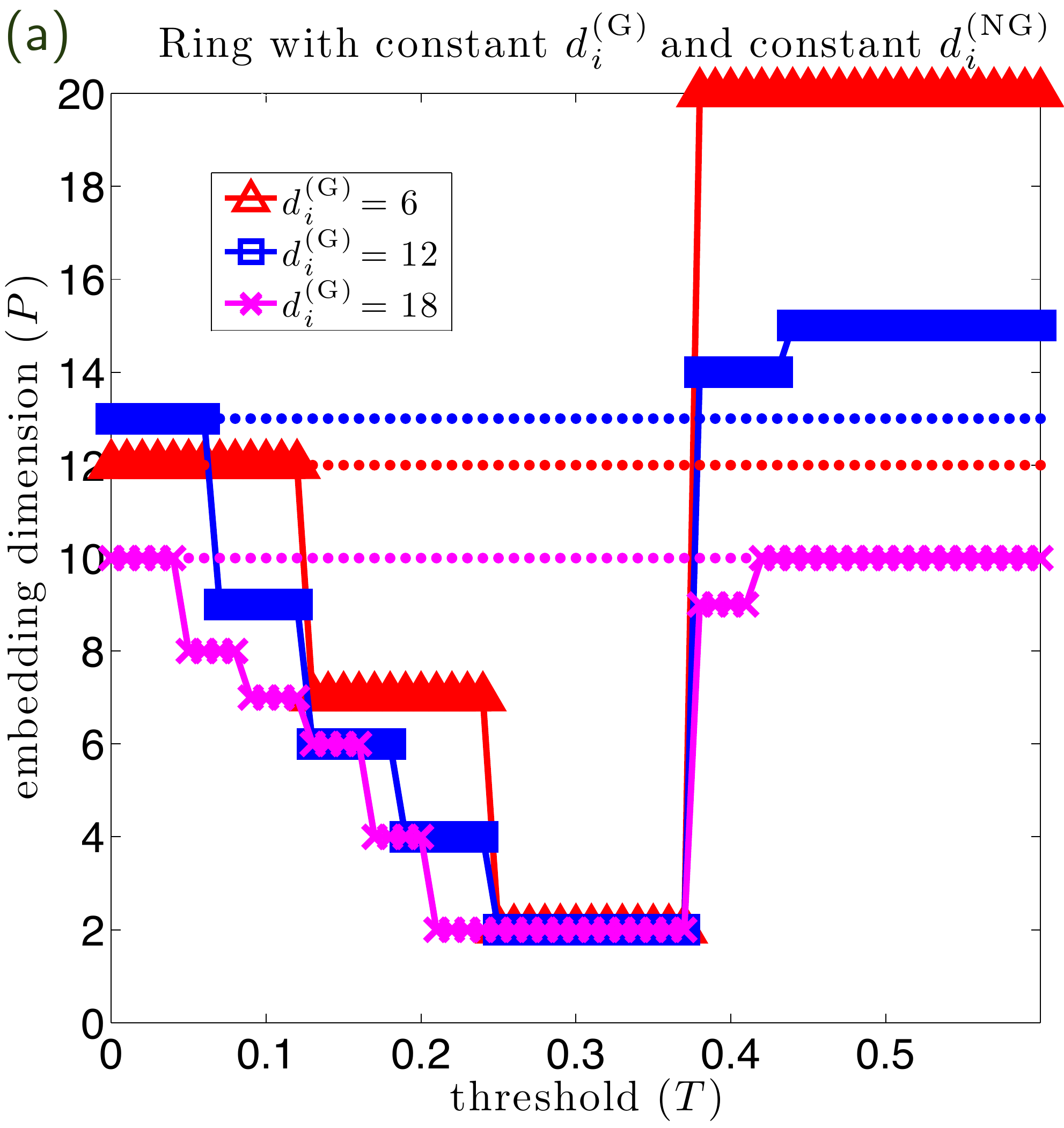}
\includegraphics[width=.35\linewidth]{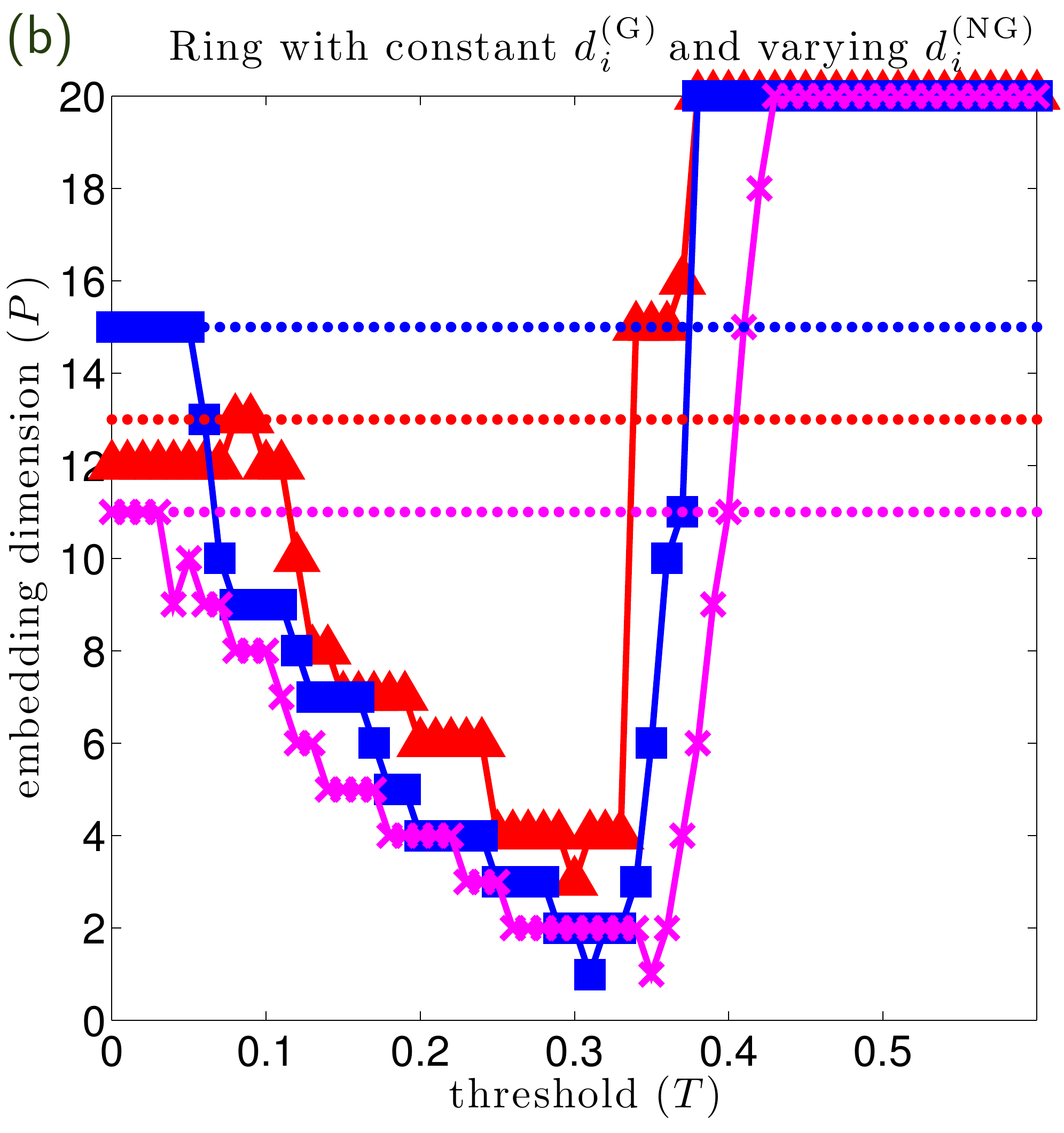}
\includegraphics[width=.35\linewidth]{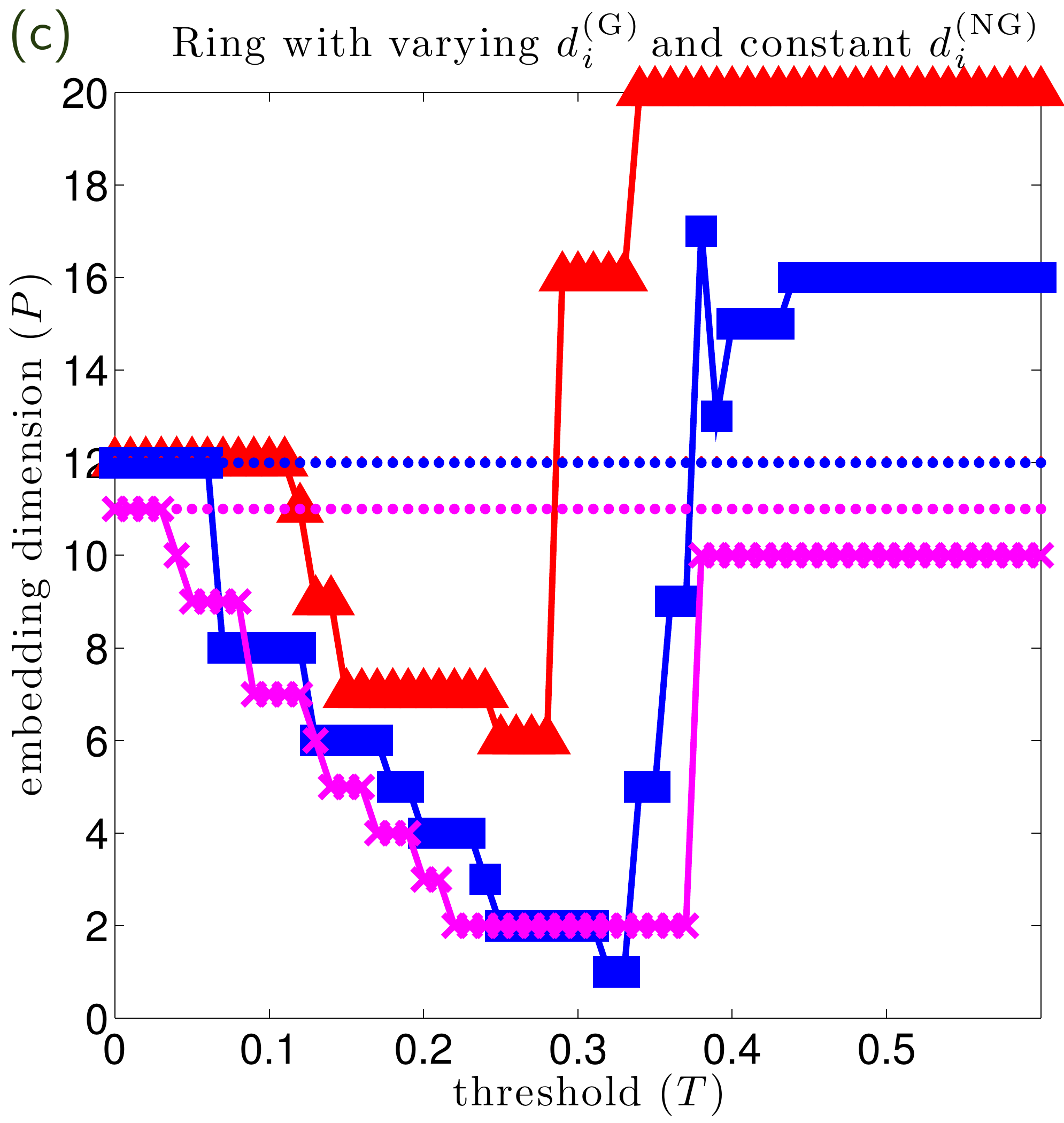}
\includegraphics[width=.35\linewidth]{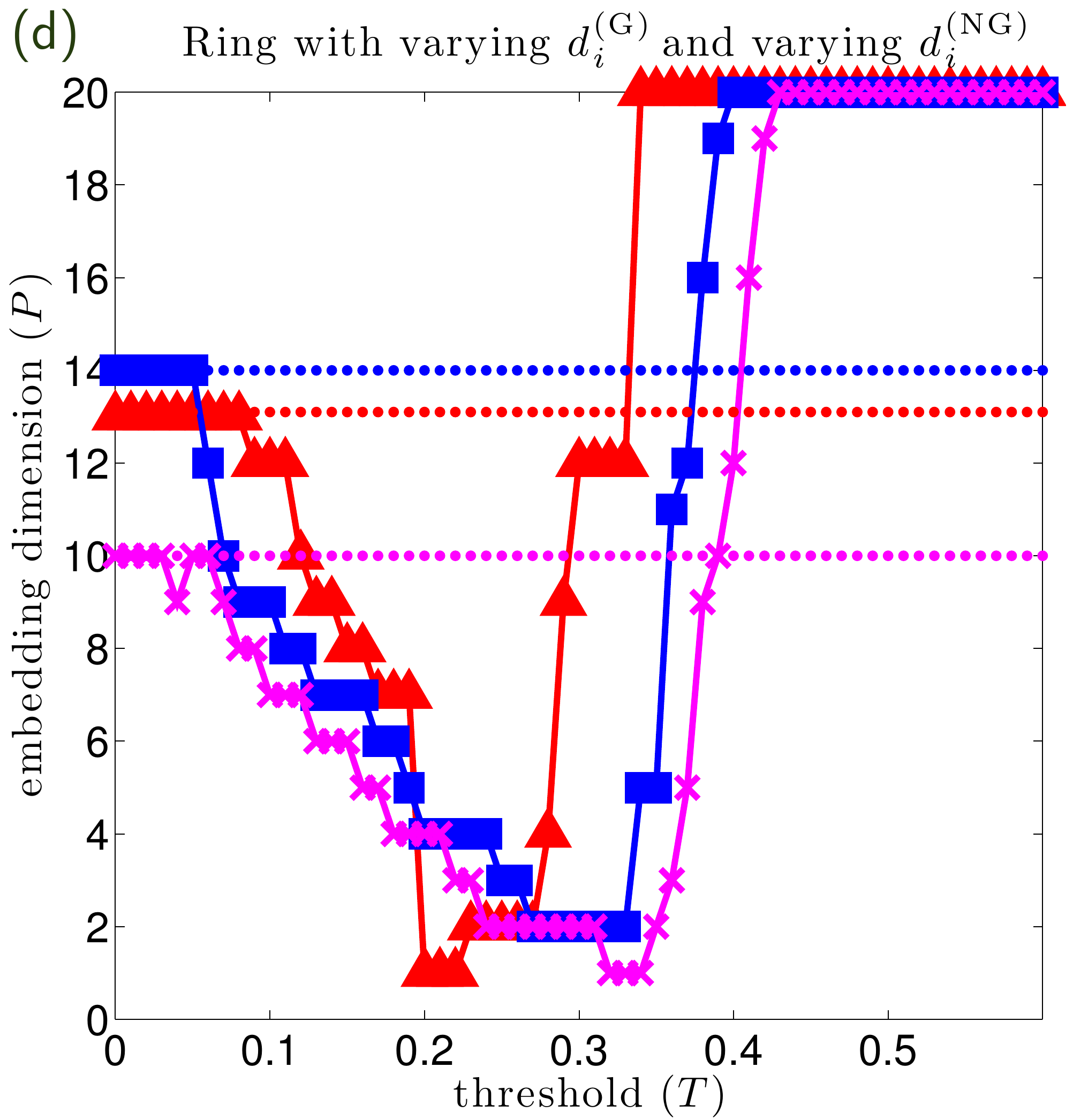}
\caption{  
We study the dimensionality of point clouds that result from symmetric WTM maps by showing their embedding dimension $P$ as a function of $T$ for networks with $N=200$ and $\alpha=1/3$.  One can construe these curves of $P$ versus $T$ as vertical cross sections for the contour plots in {Supplementary} Fig.~\ref{fig:dimension_contour}; we show results for several choices of mean node degrees: $(\langle d_i^{\rm{(G)}}\rangle,\langle d_i^{\rm{(NG)}}\rangle)=(6,2)$ (red triangles), $(\langle d_i^{\rm{(G)}}\rangle,\langle d_i^{\rm{(NG)}}\rangle)=(12,4)$ (blue squares), and $(\langle d_i^{\rm{(G)}}\rangle,\langle d_i^{\rm{(NG)}}\rangle)=(18,6)$ (magenta $\times$ symbols). As before, panels (a)--(d) correspond to network families {\bf(a)}--{\bf(d)}. In all panels, we identify the correct dimension (i.e., $P=2$) for the regime that we expect WTM contagion to exhibit WFP without ANC [i.e., $T\in(1/4,3/8)$] if the mean node degrees are sufficiently large (see magenta $\times$ symbols). The curves for the other mean degrees also consistently depict small values of $P$ for a similar range of threshold $T$. We also plot $P$ versus $T$ for the point clouds that we obtain by mapping the nodes based on shortest paths, as in Isomap \cite{Tenenbaum2000} (horizontal dotted lines). For these experiments, $P \geq 10$ from Isomap in all panels, although it appears to decrease systematically with increasing mean node degrees. {See \ref{sec:numerics} for further discussion.}
}
\label{fig:dimension_alpha_fix}
\end{figure*}

\clearpage

%%%%%%%%%%%%%%%%%%%%%%%%%%%%%%%%%%%%%%%%
%\subsection{Numerical Results for Topology}\label{sec:result_topo}
~\\{\bf \Large Numerical Results for Topology}\\
%%%%%%%%%%%%%%%%%%%%%%%%%%%%%%%%%%%%%%%%

In this section, we study the topology of point clouds that result from symmetric WTM maps applied to noisy geometric networks on {a} ring manifold. As {we} discussed in {Sec.~III~F and} \ref{sec:geometry}, we examine the difference $\Delta=l_1-l_2$ between the largest lifetimes for 1D features (i.e., 1-cycles). We determine the persistence of these 1-cycles across spacial scales using a Vietoris-Rips filtration of the point cloud \cite{carlsson,edels2010,eh}. We normalize the difference in lifetimes so that $\Delta\in[0,1]$. We show our results in {Supplementary} Figs.~\ref{fig:topology_contour}--\ref{fig:topology_alpha_fix}. Panels (a)--(d), respectively, give our results for network families {\bf (a)}--{\bf (d)}.

In {Supplementary} Fig.~\ref{fig:topology_contour}, we plot $\Delta$ in the $(T,\alpha)$ parameter plane. We show results for networks with $N=200$ nodes, mean geometric degree of $\langle d_i^{\rm{(G)}}\rangle=20$ and various mean non-geometric degrees $\langle d_i^{\rm{(NG)}}\rangle\in[0,20]$. In each panel, the solid and dashed curves indicate, respectively, our approximate bifurcation curves from Eqs.~\eqref{eq:ANC_crits2} and \eqref{eq:WFP_crits2} with $\delta_i^{\rm{(G)}}=\delta_i^{\rm{(NG)}}=0$. Note that panel (a) is similar to {Supplementary} Fig.~6(c) from the main text. Variations in $\Delta$ appear to correspond closely with the theoretical curves. For example, fixing $\alpha<0.5$ and increasing $T$, we observe an increase in $\Delta$ as $T$ surpasses $T_0^{\rm{(ANC)}}$ and a decrease in $\Delta$ as $T$ surpasses $T_0^{\rm{(WFP)}}$. 
%{\bf map: the word "exceed" is not quite being used correctly; do you mean that it increases (in the first case) precisely at the point that it passes that quantity? I have changed it to "surpasses"; please check that that is correct; what I am trying to distinguish is if you are talking about the entire regime in which one is larger than the other or if you are talking about the specific point where one surpasses the other; your prior phrasing did not make it clear which one you meant, so I changed the wording to the one I thought you meant based on other context}; DRT: nice rewording
For all four network families, $\Delta$ is largest for WTM maps that correspond to the contagion regime that we predict to be characterized by WFP without ANC. By comparing the panels, we see that the identifiability of the underlying ring topology (as indicated by $\Delta\approx1$) decreases as we increase the heterogeneity in the nodes' degrees. For example, panel (a) includes parameter values $(T,\alpha)$ for which $\Delta> 0.9$, but {$\Delta<0.6$} in panel (d) for the same {portion} of the parameter plane $(T,\alpha)$.

In {Supplementary} Fig.~\ref{fig:topology_alpha_fix}, we continue our investigation of the topology of the point clouds that result from symmetric WTM maps by fixing $\alpha=1/3$ and plotting $\Delta$ as a function of $T$. One can construe these curves of $\Delta$ versus $T$ as examining vertical cross sections from the panels in {Supplementary} Fig.~\ref{fig:topology_contour} with several choices of mean degrees: $(\langle d_i^{\rm{(G)}}\rangle,\langle d_i^{\rm{(NG)}}\rangle)=(6,2)$ (red triangles), $(\langle d_i^{\rm{(G)}}\rangle,\langle d_i^{\rm{(NG)}}\rangle)=(12,4)$ (blue squares), and $(\langle d_i^{\rm{(G)}}\rangle,\langle d_i^{\rm{(NG)}}\rangle)=(24,8)$ (magenta $\times$ symbols).  As before, the networks have $N=200$ nodes. 
Observe in {Supplementary} Fig.~\ref{fig:topology_alpha_fix} that $\Delta$ tends to decrease as the heterogeneity of the network increases. For example, the values of $\Delta$ in panels (b) and (c) tend to be smaller than those in panel (a), and the $\Delta$ values in panel (d) tend to be even smaller than those in panels (b) and (c). This decrease makes it harder to successfully identify the ring topology in the point clouds. This is most evident for the curves that correspond to $(\langle d_i^{\rm{(G)}}\rangle,\langle d_i^{\rm{(NG)}}\rangle)=(6,2)$ (red triangles). Although we observe large values of $\Delta$ in panels (a) and (b) for the point clouds for the contagion regime that we predict to exhibit WFP without ANC [i.e., $T\in(1/4,3/8)$], we find that the values of $\Delta$ for this regime are much smaller in panels (c) and (d). In this experiment, $\Delta$ does not depend on mean node degrees in a simple manner. In network family {\bf (a)}, for example, when comparing the $\Delta$ versus $T$ curve for $(\langle d_i^{\rm{(G)}}\rangle,\langle d_i^{\rm{(NG)}}\rangle)=(12,4)$ to that for $(\langle d_i^{\rm{(G)}}\rangle,\langle d_i^{\rm{(NG)}}\rangle)=(6,2)$, we observe larger $\Delta$ values when increasing the mean degrees. However, restricting out attention to the range $T\in(1/4,3/8)$, the curve of $\Delta$ versus $T$ for $(\langle d_i^{\rm{(G)}}\rangle,\langle d_i^{\rm{(NG)}}\rangle)=(24,8)$ yields $\Delta$ values that are smaller than those for $(\langle d_i^{\rm{(G)}}\rangle,\langle d_i^{\rm{(NG)}}\rangle)=(12,4)$. This nontrivial behavior might be due to the relatively small differences in magnitude between the mean node degrees and the network size ($N=200$) that we use for this experiment. In this experiment, we also compute $\Delta$ for Isomap, and we find that $\Delta \approx 0$ in all cases.  We thus omit these results from {Supplementary} Fig.~\ref{fig:topology_alpha_fix}.

\begin{figure*}[ht!]
\centering
\includegraphics[width=.4\linewidth]{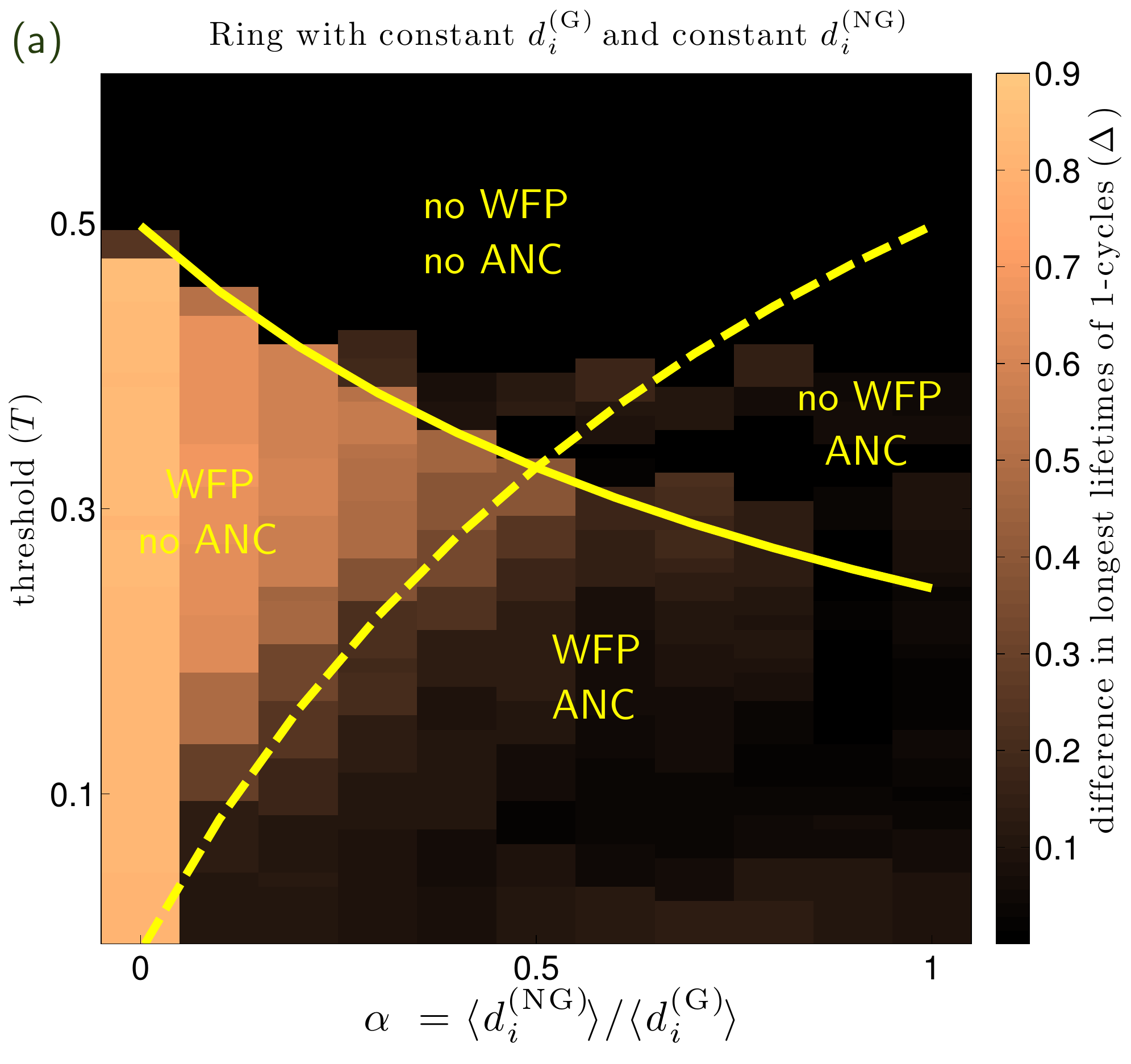}
\includegraphics[width=.4\linewidth]{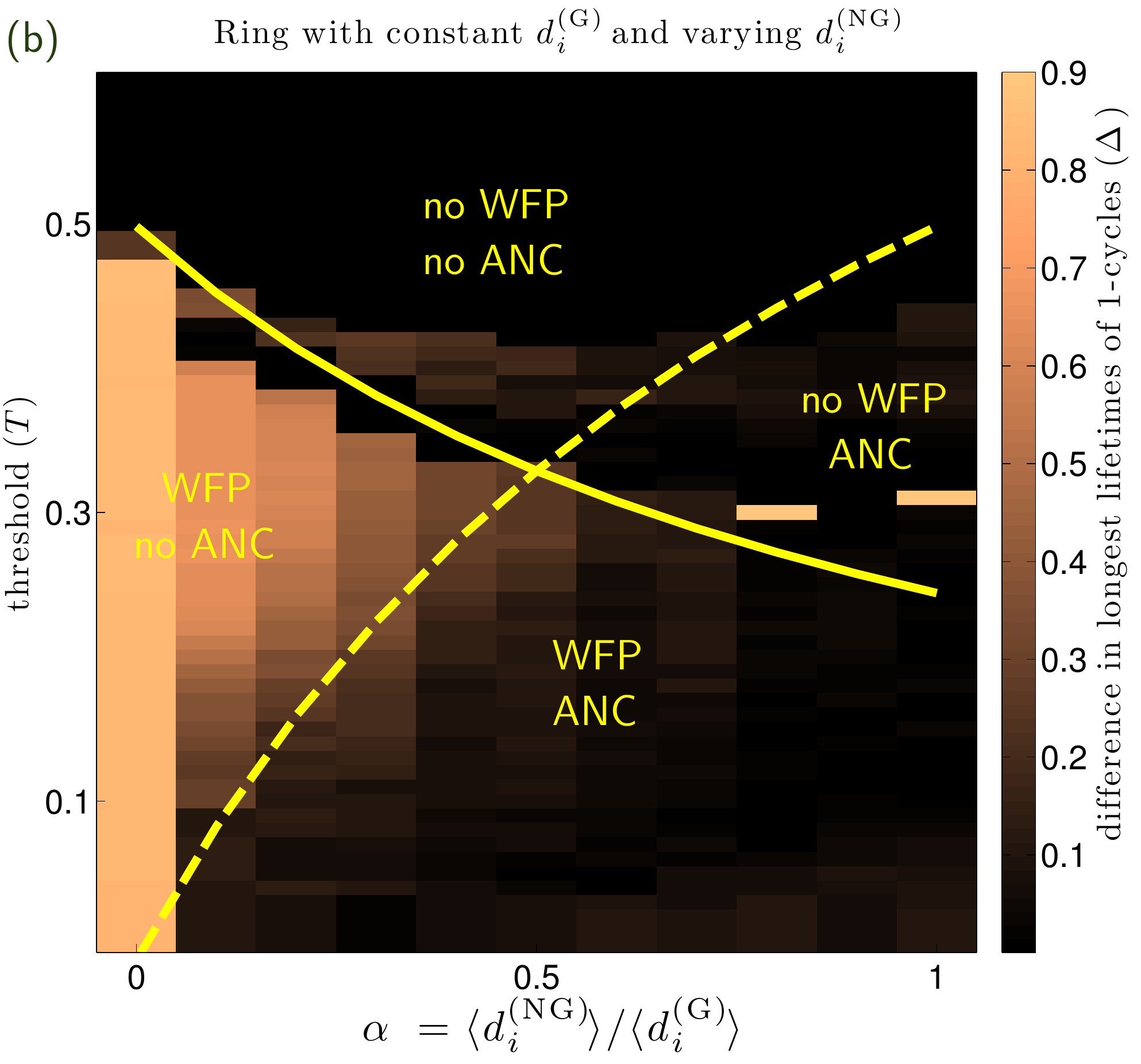}
\includegraphics[width=.4\linewidth]{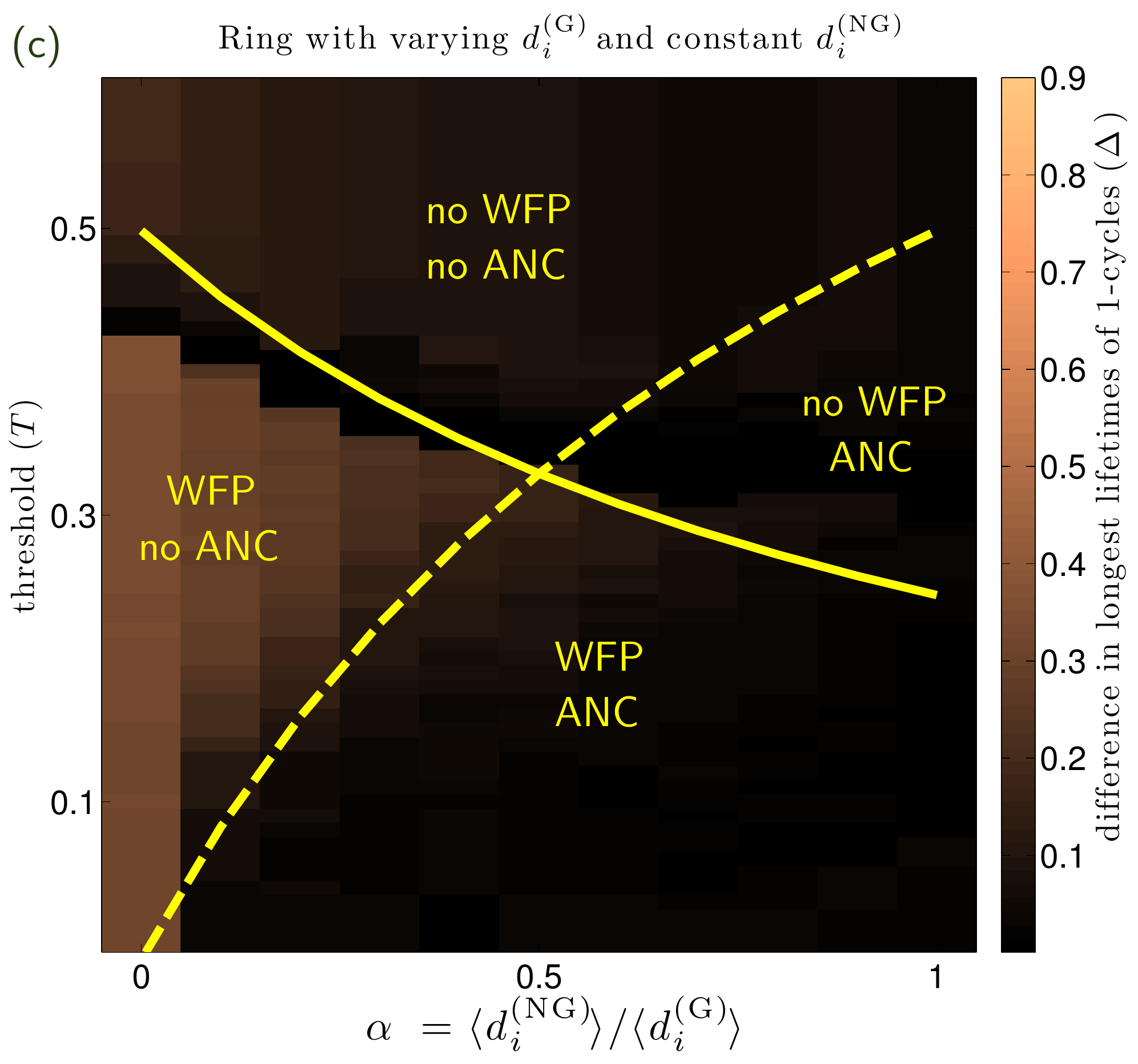}
\includegraphics[width=.4\linewidth]{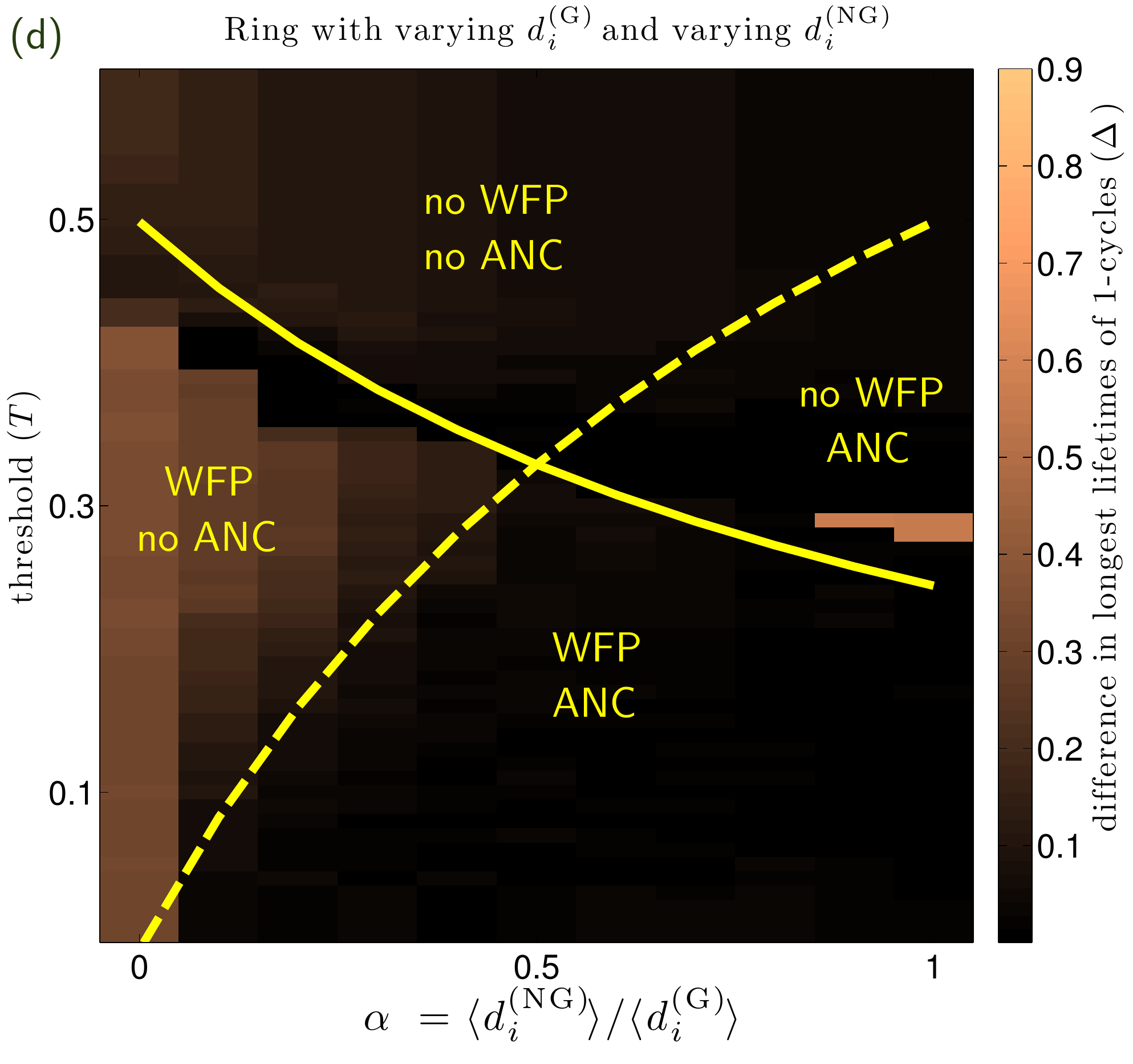}
\caption{  
We study the topology of symmetric WTM maps by calculating the difference $\Delta=l_1-l_2$ between the largest lifetimes for 1D features {(i.e., for 1-cycles); see Sec.~III~F of the main manuscript.} We normalize the lifetime difference so that $\Delta\in[0,1]$. Panels (a)--(d), respectively, show results for network families {\bf(a)}--{\bf(d)}. Each network has $N=200$ nodes, a mean geometric degree of $\langle d_i^{\rm{(G)}}\rangle=20$, and various mean non-geometric degrees $\langle d_i^{\rm{(NG)}}\rangle\in[0,20]$. The solid and dashed curves, respectively, give the approximate bifurcation curves from Eqs.~\eqref{eq:ANC_crits2} and \eqref{eq:WFP_crits2} with $\delta_i^{\rm{(G)}}=\delta_i^{\rm{(NG)}}=0$. Panel (a) is similar to Fig.~6(c) of the main manuscript. In all panels, we observe evidence that the point clouds that result from WTM maps that correspond to contagions with WFP but no ANC lie on a ring manifold. However, this evidence becomes weaker (as indicated by smaller values of $\Delta$) as the networks become more heterogeneous [e.g., compare panel (d) to panel (a)]. {See \ref{sec:numerics} for further discussion.}
}
\label{fig:topology_contour}
\end{figure*}

\begin{figure*}[ht!]
\centering
\includegraphics[width=.4\linewidth]{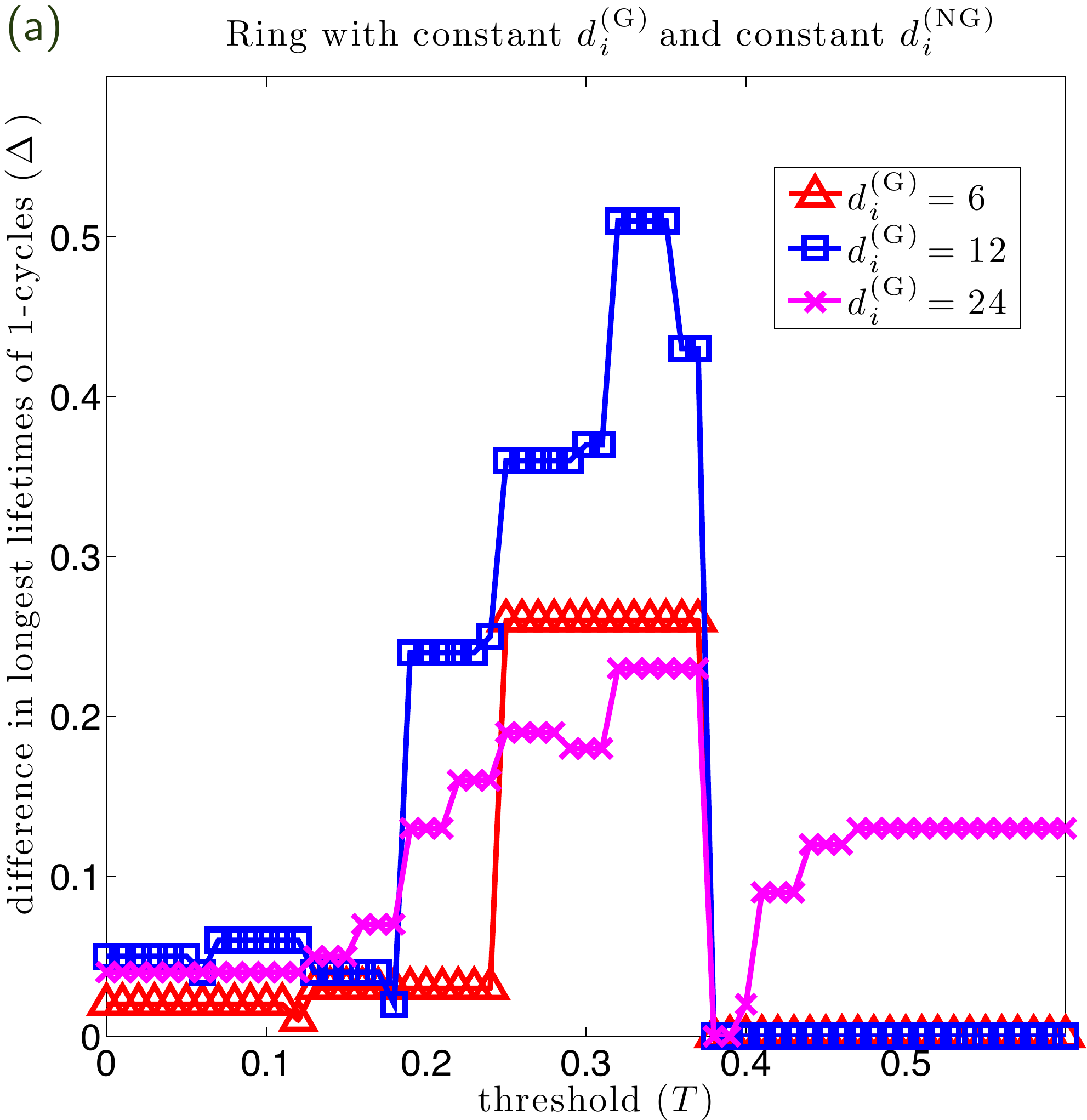}
\includegraphics[width=.4\linewidth]{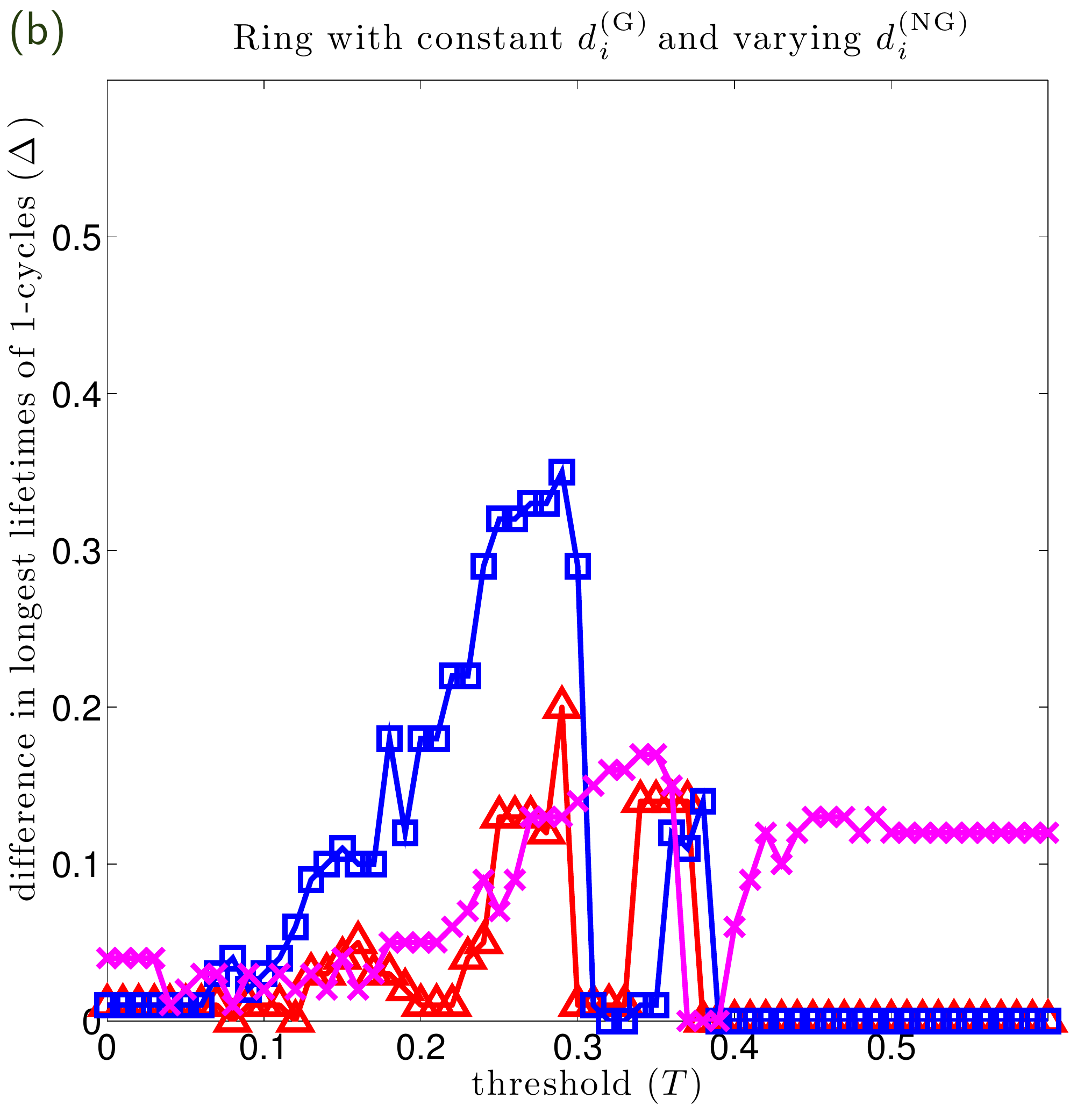}
\includegraphics[width=.4\linewidth]{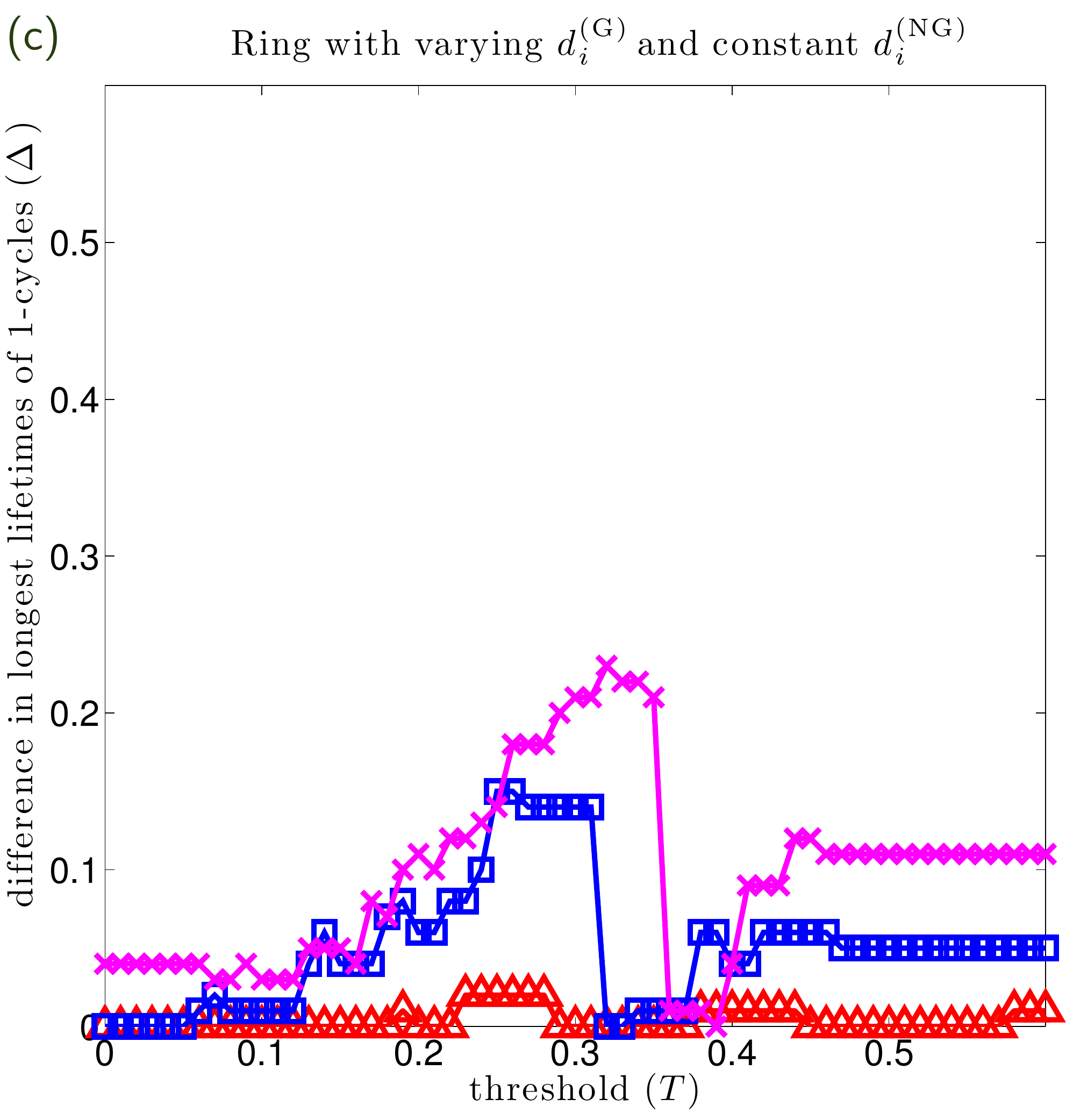}
\includegraphics[width=.4\linewidth]{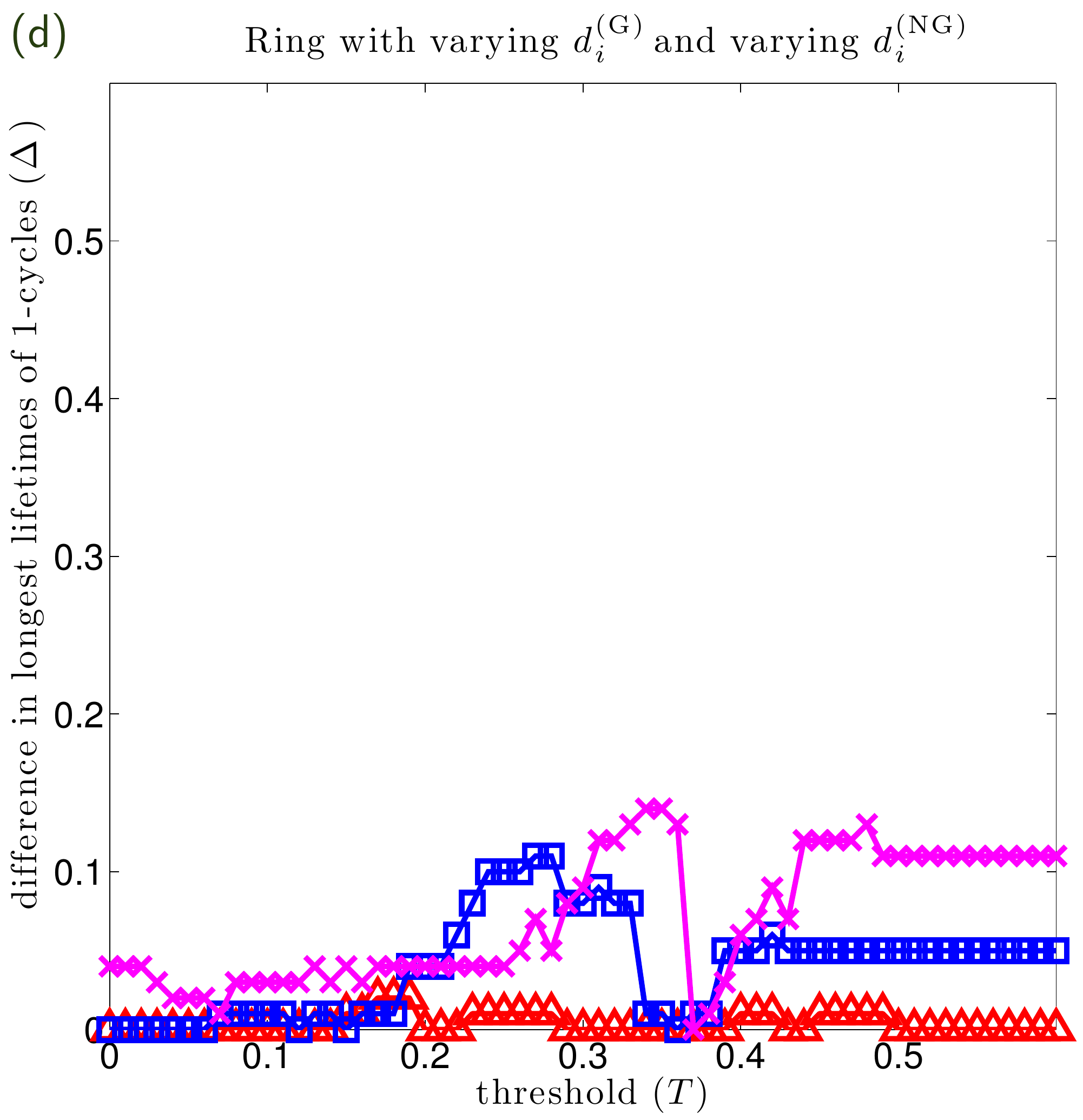}
\caption{ 
We study the topology of symmetric WTM maps by plotting $\Delta$ as a function of $T$ for $N=200$ and $\alpha=1/3$. As before, panels (a)--(d) correspond, respectively, to network families {\bf(a)}--{\bf(d)}. One can construe the curves of $\Delta$ versus $T$ as vertical cross sections of the contour plots in {Supplementary} Fig.~\ref{fig:topology_contour}; we consider several different choices of mean node degrees: $(\langle d_i^{\rm{(G)}}\rangle,\langle d_i^{\rm{(NG)}}\rangle)=(6,2)$ (red triangles), $(\langle d_i^{\rm{(G)}}\rangle,\langle d_i^{\rm{(NG)}}\rangle)=(12,4)$ (blue squares), and $(\langle d_i^{\rm{(G)}}\rangle,\langle d_i^{\rm{(NG)}}\rangle)=(24,8)$ (magenta $\times$ symbols). Note that 
%in general, 
introducing heterogeneity tends to decrease the ability to identify the ring topology in the point cloud with $\Delta$. 
%{\bf map: I commented out 'in general', because of the mathematical meaning of 'always'; please check the current phraing; should it be 'decreases' or 'tends to decrease'? I had a similar question (see my notes) in some other panels}
For example, note that the values of $\Delta$ in panels (b) and (c) are smaller than those in panel (a), and the values of $\Delta$ in panel (d) are smaller than those in panels (b) and (c). Additionally, in panel (c) and panel (d), we see that when the mean degrees are too small (e.g., see red triangles), then $\Delta\approx0$ for all thresholds $T$. Thus, we do not find evidence of the ring topology for these point clouds. {See \ref{sec:numerics} for further discussion.}
}
\label{fig:topology_alpha_fix}
\end{figure*}

\clearpage 

%%%%%%%%%%%%%%%%%%%%%%%%%%%%%%%%%%%
%\subsection{Sampling the Ring Manifold}
~\\{\bf \Large {Non-Uniform} Sampling of {a} Ring Manifold}\\
%%%%%%%%%%%%%%%%%%%%%%%%%%%%%%%%%%%

%{\bf {singular vs plural in the title above}}%{we only sample the unit circle, singular}

In our numerical experiments thus far, we have investigated symmetric WTM maps for four families of noisy geometric networks on a ring manifold. {(See \ref{sec:models} for their descriptions.)} Network families {\bf(c)} and {\bf(d)} allow heterogeneity in the node locations along {a} ring manifold through the placement of nodes via unevenly-spaced angles $\{\theta_i\}$ along the unit circle. Recall that each node $i$ has an associated angle $\theta_i=\frac{2\pi i }{N}+\delta\theta_i$, where we draw $\delta\theta_i\sim\mathcal{N}(0,(s\frac{2\pi}{N})^2)$ from a Gaussian distribution with a standard deviation of $s\frac{2\pi}{N}$. Note that $\frac{2\pi}{N}$ is the spacing between the $N$ nodes if they are spaced uniformly on the ring. Consequently, by varying the parameter $s$, one can tune the level of heterogeneity in node location and thus the heterogeneity of the geometric degrees $\{d_i^{\rm{(G)}}\}$. Recall that $s\to\infty$ corresponds to sampling locations on the unit circle uniformly at random. In our previous experiments, we let $s=1/2$ for network families {\bf(c)} and {\bf(d)}. In this section, we investigate the effect of varying $s$. Because $s>0$ introduces heterogeneity in the geometric degrees, we consider both the case in which the nodes' non-geometric degrees are identical and the case in which they are heterogeneous. That is, the networks that we now consider are generalizations of network families {{\bf(c)} and {\bf(d)}, but we now also vary the level of heterogeneity in the geographic spacing of nodes on the ring.}

In {Supplementary} Fig.~\ref{fig:noise_vary}, we show results for the (left column) geometry, (center column) dimensionality, and (right column) topology of symmetric WTM maps, where we fix $\alpha=1/3$ and $N=200$ and we vary the threshold $T$. We consider networks with $N=200$ nodes, mean geometric degree of $\langle d_i^{\rm{(G)}}\rangle=24$, mean non-geometric degree of $\langle d_i^{\rm{(NG)}}\rangle=8$, and $s\in\{0,1/2,1,3/2,2,5/2,\infty\}$. The top row corresponds to generating noisy geometric edges so that every node has exactly $d_i^{\rm{(NG)}}=\langle d_i^{\rm{(NG)}}\rangle$ non-geometric edges, and the bottom row corresponds to generating noisy geometric edges uniformly at random so that the non-geometric degree $d_i^{\rm{(NG)}}$ of a node $i$ is a {binomially}-distributed random variable. See the descriptions of the network families in \ref{sec:models}. Using horizontal dashed lines, we show results for the mapping of nodes for Isomap (i.e., based on shortest paths). We omit these results from panels (c) and (f), because we obtain $\Delta \approx 0$ in these cases. The dashed lines in panels (a) and (d) give values of $\rho$ for a 2D Laplacian eigenmap. (It is 2D by construction, so we do not investigate its embedding dimension $P$.)

Increasing network heterogeneity by increasing $s$ has a significant effect on the structure of the point clouds that result from symmetric WTM maps. For example, we see in panels (a) and (d) that increasing $s$ shifts the abrupt drop-off in the Pearson correlation coefficient $\rho$, which originally occurs near its expected value of $T_0^{\rm{(WFP)}}=3/8$, to progressively smaller values of $T$. In fact, we see in all panels that increasing $s$ causes the curves of $\rho$ versus $T$ to shift to the left. Additionally, in panels (a) and (d), we see for sufficiently large $s$ that there is a regime in which $\rho$ is small for all threshold values $T$. In panels (b) and (e), we still obtain regimes in which the WTM maps are low-dimensional (i.e., $P\approx2$). However, as $s$ increases, the range of $T$ values for which $P$ indicates low dimensionality becomes smaller and shifts to the left. In panels (c) and (f), one can also observe that the ability to identify the ring topology becomes more difficult with increasing $s$. In panel (c), we obtain large 1-cycle lifetimes $\Delta$ when $T\in(1/4,3/8)$ for $s=0$; this provides strong evidence that the point cloud lies on a ring manifold. For small $s$ (e.g., $0\le s\le 3/2$), we also obtain large values of $\Delta$, but the range of thresholds $T$ that produce large $\Delta$ are smaller and have shifted to the left. However, when $s$ is large (e.g., $s=\infty$), $\Delta$ remains small for all threshold values $T$ in panels (c). There is even less evidence of the ring topology in panel (f), as $\Delta$ remains small for all values of $s$ and $T$.

\begin{figure*}[ht!]
\centering
\includegraphics[width=.329\linewidth,height=.344\linewidth]{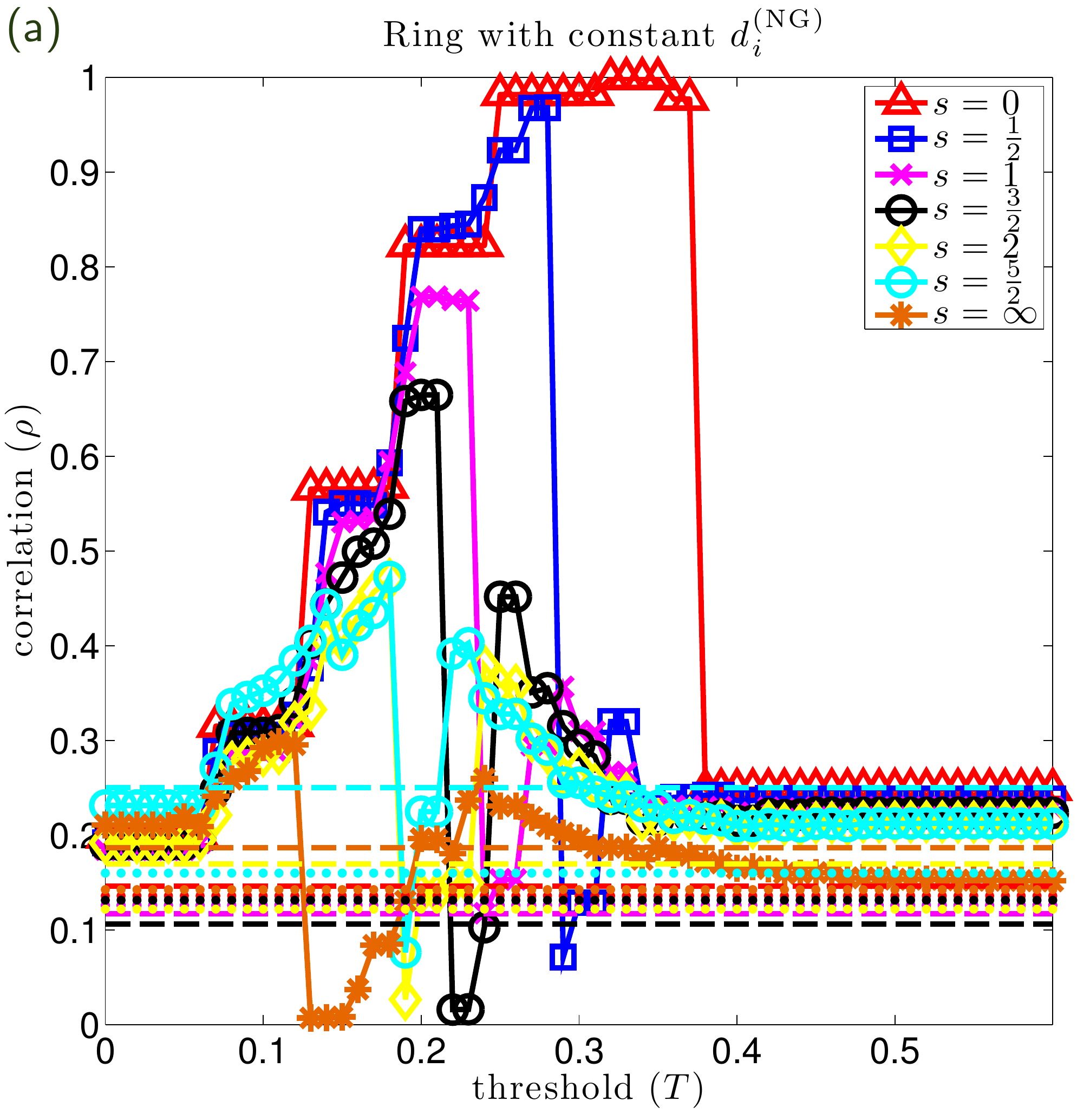}
\includegraphics[width=.329\linewidth]{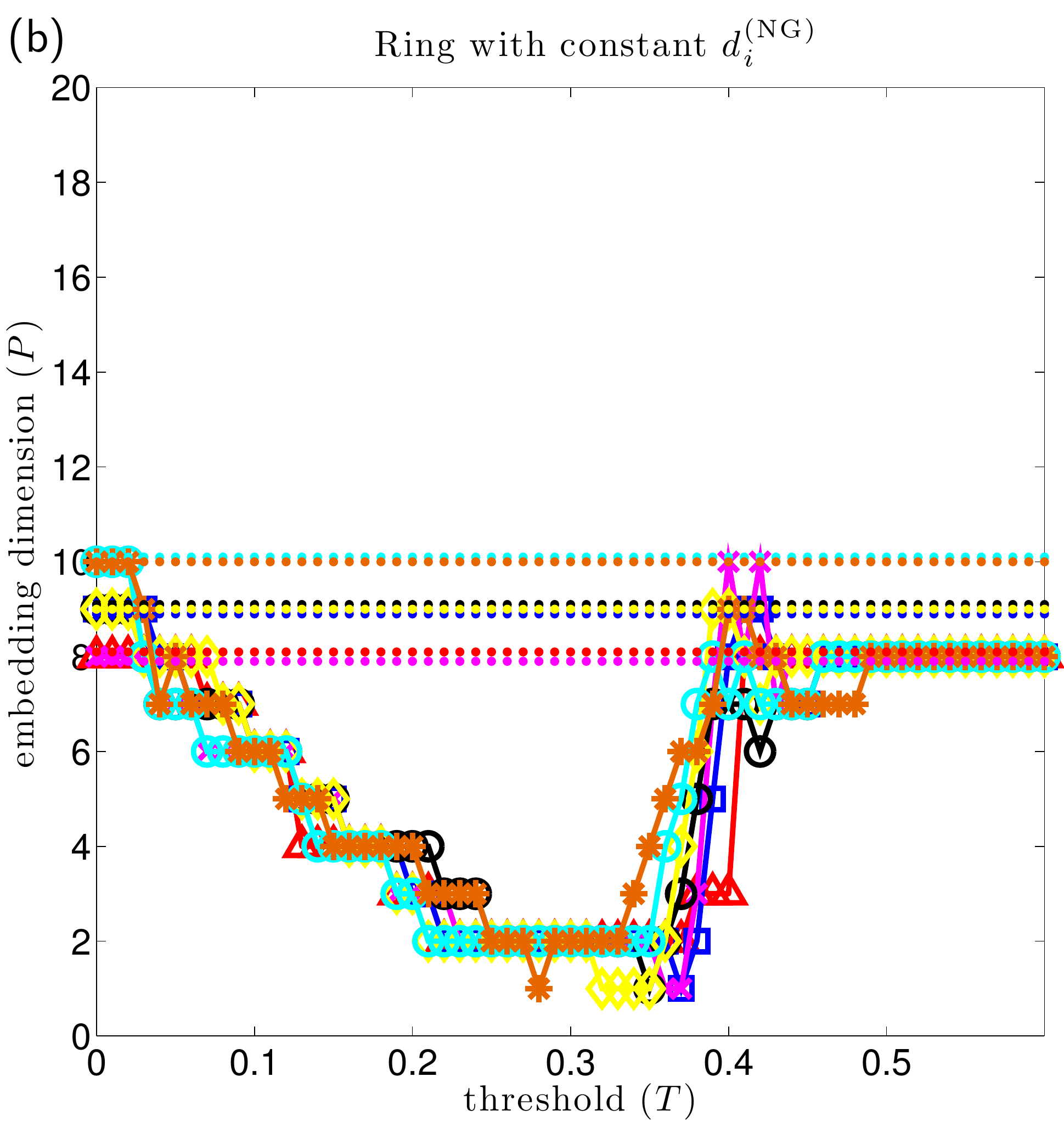}
\includegraphics[width=.331\linewidth]{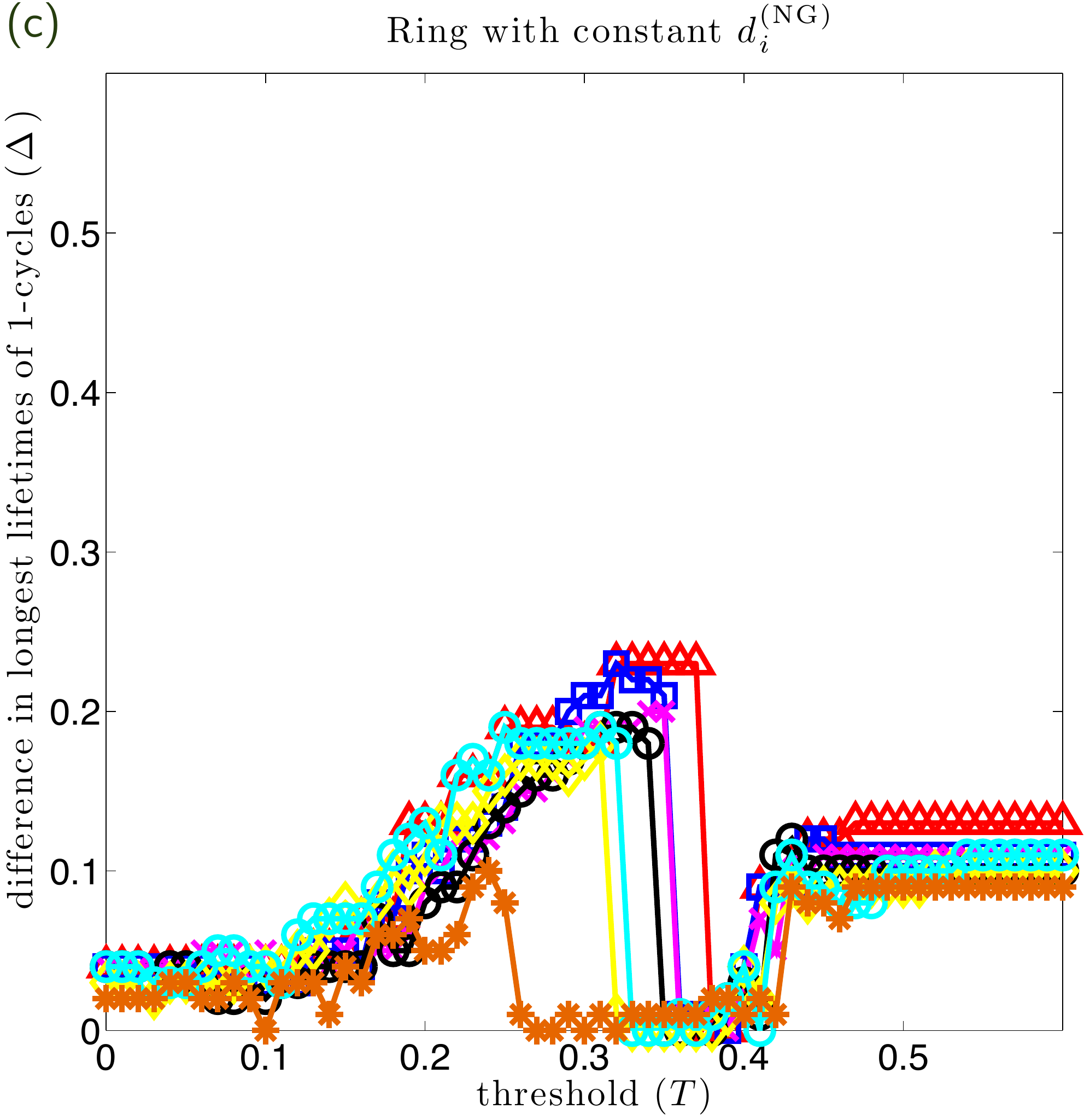}\\
\includegraphics[width=.329\linewidth]{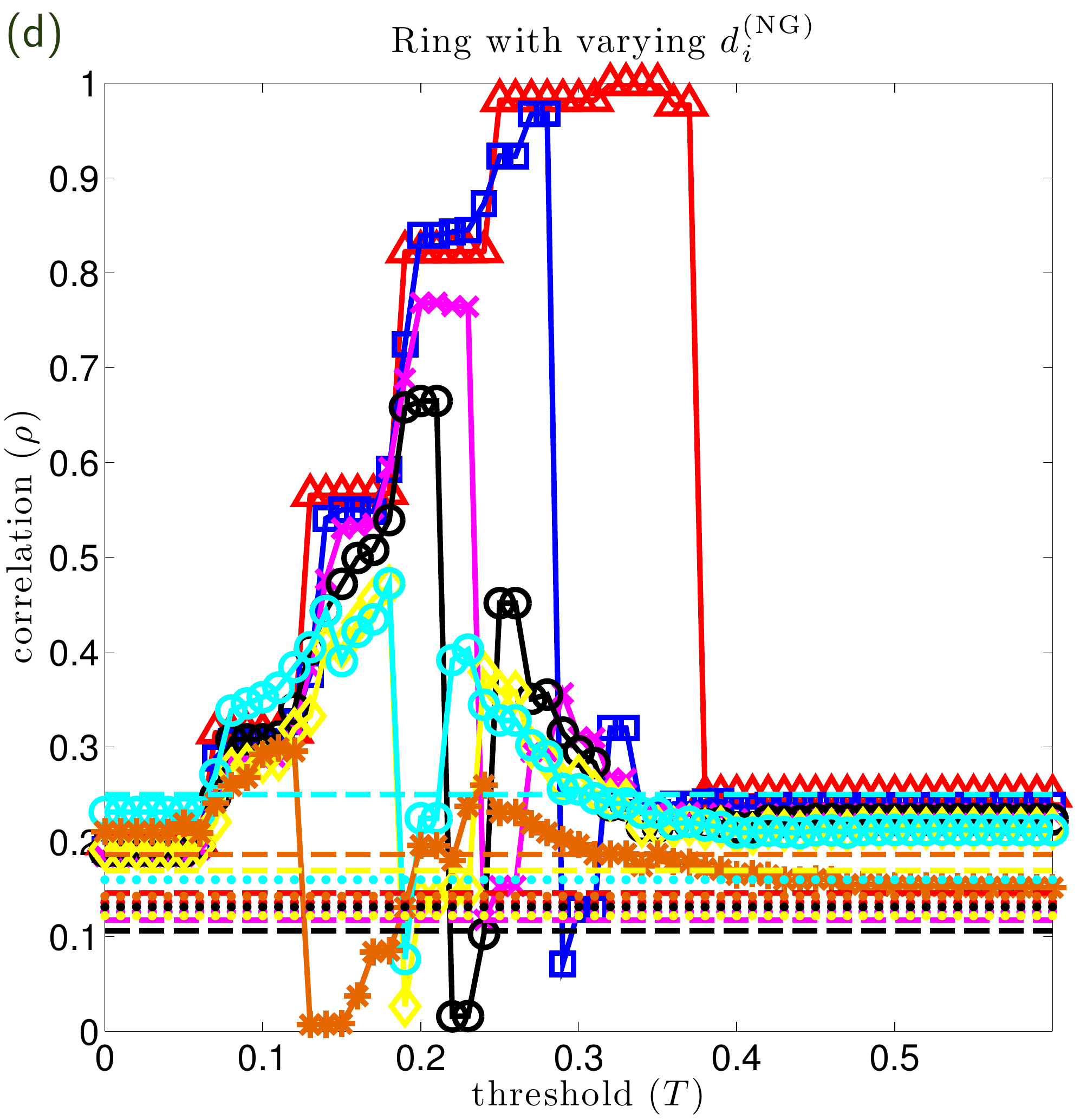}
\includegraphics[width=.329\linewidth]{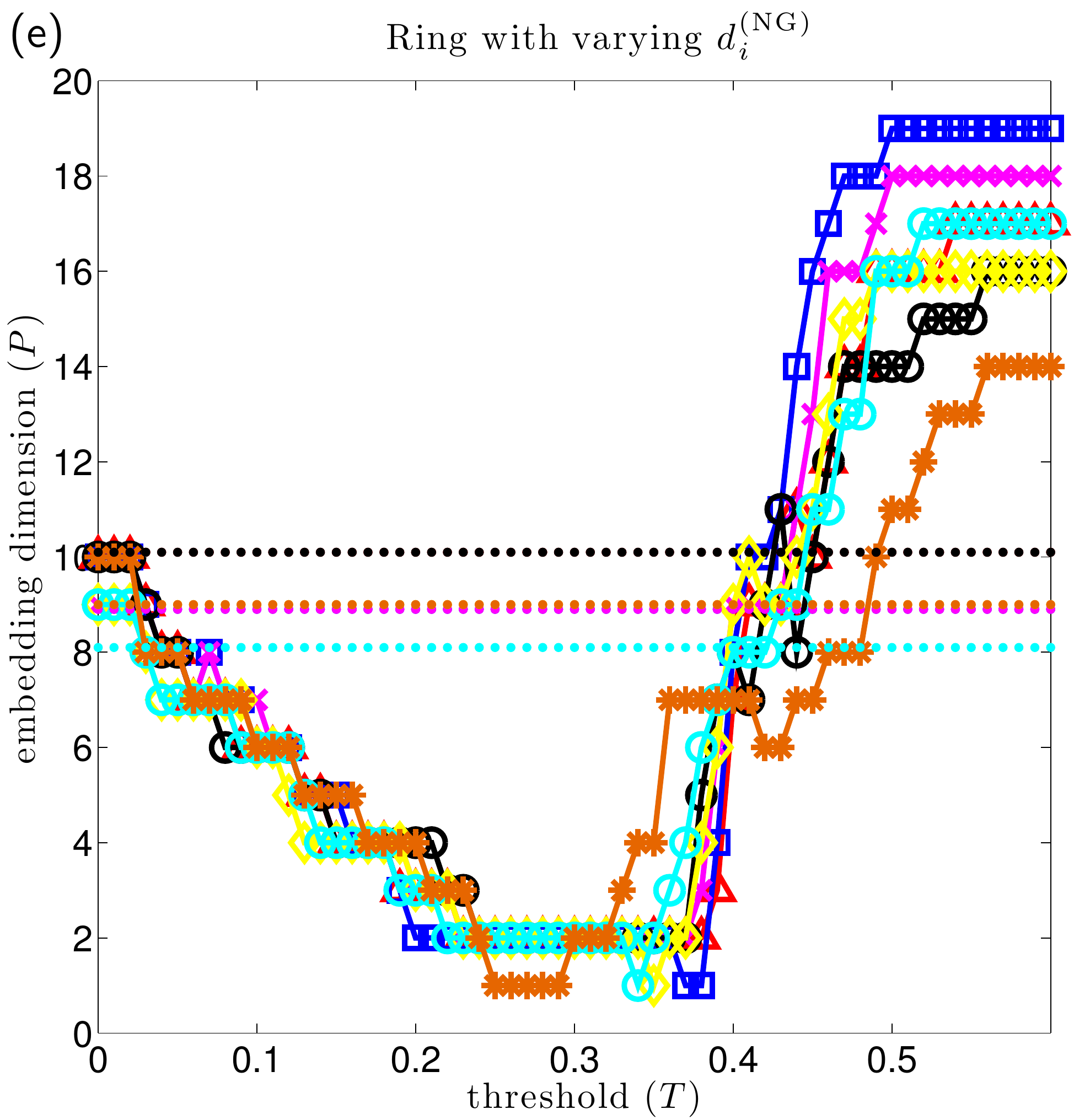}
\includegraphics[width=.331\linewidth]{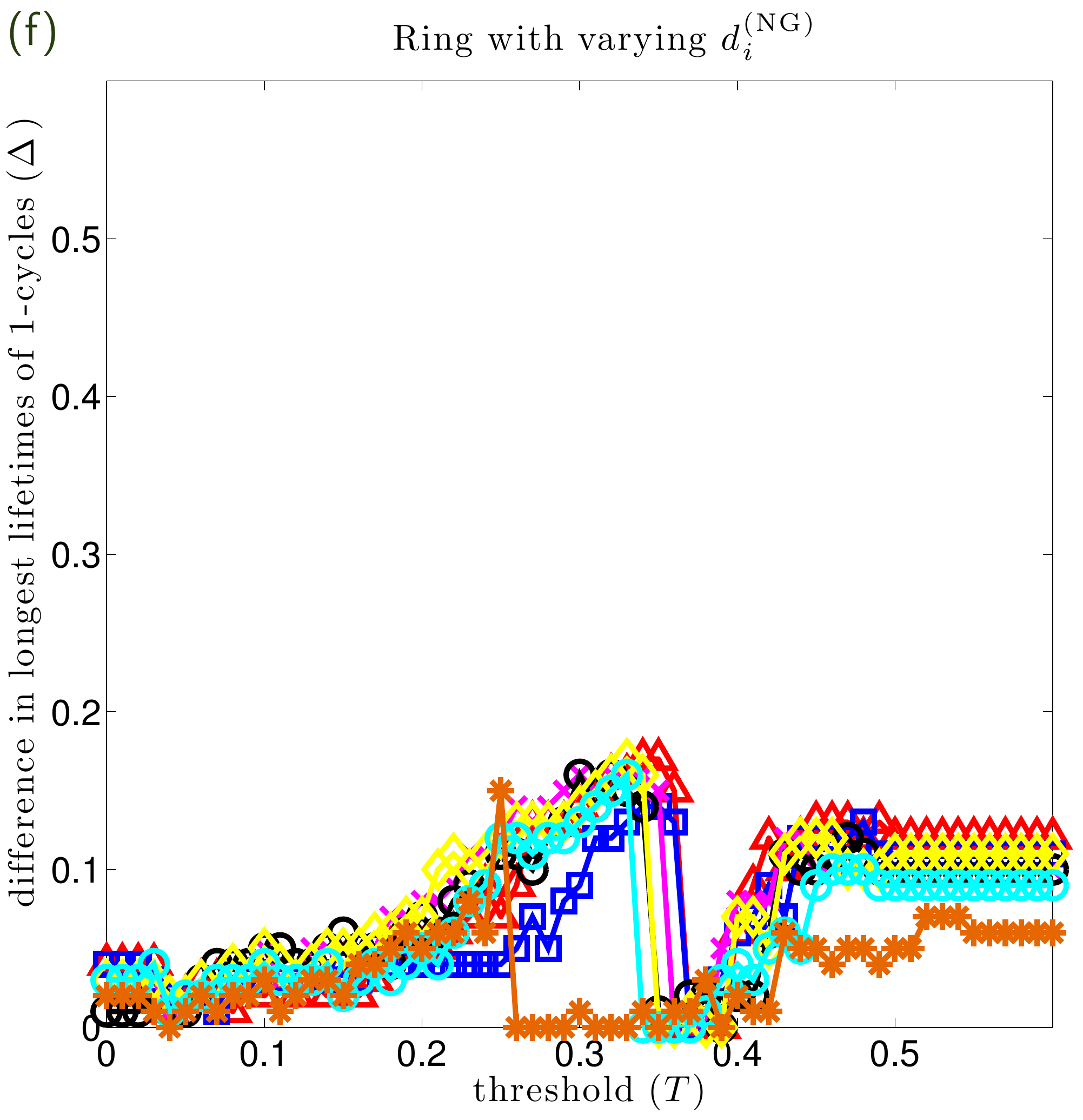}
\caption{
We study the (left column) geometry, (center column) dimensionality, and (right column) topology for symmetric WTM maps applied to noisy geometric networks with node locations $\{\bold w^{(i)}\}$ that we sample randomly from a ring manifold, where the parameter $s$ controls the amount of heterogeneity that we introduce into the spacing of the nodes along the ring. (Recall that we used $s=1/2$ to define network families {\bf{(c)}} and {\bf{(d)}}.) We show results for various values of the threshold $T$ for networks with $N=200$ nodes, mean geometric degree $\langle d_i^{\rm{(G)}}\rangle=24$, and mean non-geometric degree $\langle d_i^{\rm{(NG)}}\rangle=8$. We consider both non-geometric degrees that are (top row) constant across the nodes, as in network family {\bf{(c)}} and (bottom row) heterogeneous, as in family {\bf{(d)}}. 
%{\bf map: you write "similar", but how is this not "the same" with respect to the stated feature; we shouldn't write "similar" for a feature that is the same; that is confusing}
%In both cases, the nodes have a mean of $\langle d_i^{\rm{(NG)}}\rangle=8$ non-geometric edges. 
Note that increasing heterogeneity in node spacing on the manifold, which in turn increases the heterogeneity in the geometric degrees $\{d_i^{\rm{(G)}}\}$ (although their mean is constant), tends to lead to a decrease in the ability of the symmetric WTM maps to recover the properties of the underlying manifold in the resulting point cloud. One sees this mostly clearly when examining the geometry and topology, as there are significant drops in $\rho$ and $\Delta$ as $s$ increases. The dotted lines in panels (a), (b),(d), and (e) indicate the values that we observe for the mapping of nodes based on shortest paths, as in the Isomap algorithm. The dashed lines in panels (a) and (d) indicate values for a 2D Laplacian eigenmap. 
%See our discussion of these methods at the beginning of \ref{sec:numerics}. 
{See \ref{sec:numerics} for further discussion.}
}
\label{fig:noise_vary}
\end{figure*}

\clearpage

%%%%%%%%%%%%%%%%%%%%%%%%%%%%%%%%%%%%%%%%
~\\{\bf \Large Summary of Experiments with { a Ring Manifold}}\\
%\subsection{Summary of Experiments with Synthetic Networks}
%%%%%%%%%%%%%%%%%%%%%%%%%%%%%%%%%%%%%%%%

We have conducted an extensive investigation of the geometry, dimensionality, and topology of symmetric WTM maps for several families of noisy geometric networks on a ring manifold. We now briefly summarize our results. 

We demonstrated that the structure (e.g., geometry, dimensionality, and topology) of WTM maps depends strongly on the contagion threshold $T$ and the network parameters (e.g., the number of nodes $N$, the geometric degrees $\{d_i^{\rm{(G)}}\}$, and the non-geometric degrees $\{d_i^{\rm{(G)}}\}$). Consequently, the extent to which a WTM contagion exhibits wavefront propagation (WFP) versus the appearance of new clusters (ANC) of contagions also depends on these parameters. Bifurcation analysis did a good job of predicting which parameter regimes have similar point-cloud structures. This is particularly evident in the $(T,\alpha)$ parameter plane in {Supplementary} Figs.~\ref{fig:correlation_contour}, \ref{fig:dimension_contour}, and \ref{fig:topology_contour}, where we observed that the geometry, dimensionality, and topology of WTM maps align well with our theoretical predictions for the occurrence of bifurcations in the dynamics of a WTM contagion. We found such agreement even for networks with heterogeneity in geometric degrees, non-geometric degrees, and/or node locations along {a} ring manifold. As we discussed {in \ref{sec:models}}, we interpret our bifurcation analysis for the noisy ring lattice as an approximate bifurcation analysis for the networks with heterogeneous structures. As expected, we also observed that the accuracy of our approximation increases as the mean node degrees increase. However, its accuracy is sensitive to a variety of factors---including the threshold $T$, the network size $N$, and the particular type of structural heterogeneity in the network. In many of our numerical experiments, we compared the structure of point clouds that result from WTM maps to those that result from a 2D Laplacian eigenmap \cite{Belkin2003} and Isomap \cite{Tenenbaum2000}, which map the network nodes based on diffusion dynamics and shortest paths, {respectively}.  Our approach provides a nice complement to these methods.

%{\bf {are we able to be more specific with "above"? in the pointer above}}

%%%
%%%

\clearpage

{

%~\\{\bf \Large 
\section{{Supplementary} Discussion}

%We conclude by discussing 
In this {Supplementary Discussion, we further consider} the implications of our study for three research areas that have diverse motivations and goals but which share a common interest in understanding spreading processes on networks.

%\smallskip
%{The first application area is high-dimensional data analysis of contagions and other spreading processes.} 
~\\{\bf \Large High-Dimensional Data Analysis of Contagions and Other Dynamics}\\

Research on network epidemiology \cite{Brockmann2013,Colizza2006,Hufnagel2004,He2014} underscores the importance of the perspective that we have taken in the present paper. For example, Brockmann and Helbing \cite{Brockmann2013} recently defined node-to-node distances based on a stochastic model for contagions that takes into account human mobility patterns in the worldwide airline network, and they reported that such a notion of distance did a good job of predicting global contagions. In their study, Brockmann and Helbing reported that node-to-node distances are insensitive to the contagion parameters in their model. By contrast, we find that the geometry, dimensionality, and topology of contagions depends sensitively on the contagion parameters (e.g., the threshold $T$) of the WTM. This appears to arise from the thresholding process, so we expect it to be relevant for complex contagions in general because of the importance of social reinforcement \cite{Centola2010,Centola2007,Centola2007b,Ghasemiesfeh2013}. 

Our perspective can be applied to study other spreading processes \cite{porter2014}, where it has the potential to offer insights into phenomena such as information seeding \cite{Kempe2003} and targeted immunization \cite{Pastor2002,Rhodes1997}.  Moreover, a large variety of other processes---including some of the most heavily investigated dynamical processes (e.g., $k$-core percolation and other types of percolation) \cite{porter2014,Gleeson2013}, more intricate complex-contagion models \cite{Melnik2013}, and even some local methods for community detection \cite{jeub2014}---also satisfy filtration conditions that are based on node states and the dynamics of such states. One can thus construct contagion maps for these processes and study them using the approach that we have illustrated. Computational homology offers a promising (and novel) approach for studying all of those situations.

%\smallskip
%{\noindent{\bf Dimension reduction in networks.} 
 %{ The second application area is dimension reduction of networks.} 
 ~\\{\bf \Large Dimension Reduction of Networks}\\

%{\bf {future vs past pointers in this section: I fixed several; double-check the whole section}}

In the present paper, we used the fact that WTM contagions satisfy a filtration condition. This makes it possible to study networks from the perspective of computational topology \cite{Petri2012,Petri2013,Kahle2013,Bobrowski2014}. {One can thus} construct a metric space based on when nodes adopt a contagion for different choices of initial conditions. (See  \ref{sec:filtration}.)  WTM contagions thereby allow the simultaneous study of network topology, geometry, and dimensionality. Such manifold learning has numerous applications, including inference of missing and spurious edges \cite{Singer2011,Singer2013,Clauset2008,Guimera2009,Liben2007}, efficient routing of information \cite{Boguna2010,Serrano2008}, and identification of attributes that are responsible for edge formation \cite{Hoff2002}. 
%It would be very interesting to study such applications using contagion maps.}
%, although we stress that for many scenarios, the particular method for dimension reduction (i.e., using contagions, diffusion, shortest-paths, or something else) should be dictated by the application of interest.}
To provide a step in this direction, in \ref{sec:denoising}, we {compared} the denoising of networks via WTM maps---a ``global'' approach for identifying spurious edges---to a popular ``local'' approach based on the statistics of subgraphs \cite{Goldberg2003}.
  
An important future direction is to improve the computational efficiency of constructing contagions maps. As we {discussed} in \ref{sec:algorithm}, the typical computational complexity for our construction of a WTM map with all possible initial conditions with clustering seeding is currently $\mathcal{O}(NM)$, where $M$ is the number of edges in a network. Approximation schemes based on ideas {such as} network sampling \cite{jeub2014} and random projections \cite{Bingham2001} offer promising approaches for improving computation speed.

~\\\\
 ~\\{\bf \Large Dimension Reduction of Point-Cloud Data}\\
 
Although we {focused} on manifold structure in networks, our approach extends naturally to point-cloud data {(e.g., images, videos, and time series)}---the traditional setting for manifold learning---if one first infers a proximity network using, {for example}, a $k$-nearest neighbor distance thresholding \cite{Belkin2003,Coifman2005,Gerber2007,Sorzano2014}. In this endeavor, a central pursuit has been the development of techniques that are robust to noise \cite{Gerber2007,Sorzano2014,Lafon2006}. It is well-known that diffusion distances are more robust than shortest-path distances to noisy edges, so maps that are based on diffusion \cite{Belkin2003,Coifman2005} can be preferable to the Isomap algorithm \cite{Tenenbaum2000} for noisy data \cite{Lafon2006}. However, noisy edges can still be problematic for diffusion distances, so some techniques attempt to denoise a network prior to mapping it \cite{Singer2013}. The robustness to noisy edges for WTM maps with contagions dominated by WFP makes them appealing, and it would be interesting to explore applications with noisy data. An important distinction of WTM maps from prior work is that our research is based on nonlinear and nonconservative dynamics (in particular, on complex contagions) rather than on {linear and} conservative dynamics such as diffusion (e.g., random walks)} \cite{Singer2011,Singer2013,Belkin2003,Coifman2005,Gerber2007,Sorzano2014}. These different classes of dynamics can behave very differently, and it is known that they give very different answers for questions like which nodes are most important \cite{ghosh2012} and what network structures constitute bottlenecks to such dynamics \cite{lerman2012pre} (which is closely related to which network structures yield dense communities of nodes \cite{jeub2014}). Comparing WTM maps to Laplacian eigenmaps \cite{Belkin2003} and Isomaps \cite{Tenenbaum2000} (see \ref{sec:numerics}) illustrates that these different dynamics lead to differences in the results of dimension reduction. It is thus important {to explore} dynamics other than diffusion for the analysis of point-cloud data.

%%%%%%%%%%%%%%%%%%%%%%%%%%%%%%%%%%%%%%%%
%%%%%%%%%%%%%%%%%%%%%%%%%%%%%%%%%%%%%%%%

}